\documentclass{Introductory/AimsEssay}
\begin{document}
\pagestyle{empty}
\maketitle
\pagenumbering{roman}
\graphicspath{{images/}}
\chapter*{\textbf{Declaration}}
\addcontentsline{toc}{chapter}{Declaration}

I, the undersigned, hereby declare that the work contained in this PhD thesis is my original work and that any work is done by others or by myself previously has been acknowledged and referenced accordingly.This thesis is submitted to the School of Physics, Faculty of Sciences, University of the Witwatersrand, Johannesburg, South Africa, in fulfillment of the requirements for the degree of Doctor of Philosophy in Physics. It has not been submitted before for any degree or examination in any other university.

\vspace{1.5cm}
 \includegraphics[height=1.0cm]{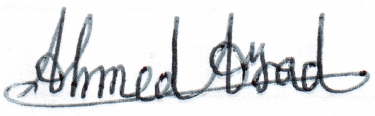}
\hrule
Ahmed Ayad Mohamed Ali \makebox[2in][r] 12 February, 2021
\newpage
\newpage
\chapter*{\textbf{Acknowledgments}}
\addcontentsline{toc}{chapter}{Acknowledgments}

This PhD thesis has been carried out at the University of the Witwatersrand, since January 2018. The research in this thesis was supported by the DST/NRF SKA post-graduate bursary initiative. Making this thesis come alive was my biggest dream in life. Even when I was experiencing hardship and depression due to sudden detours in my life, I still managed to persevere and complete my dream. It was not easy, but somehow, I made it through.  After Almighty God, many people deserve thanks for their support and help. It is, therefore, my greatest pleasure to express my gratitude to them all in this acknowledgment.

First and foremost, I would like to thank the Almighty God for giving me the strength, patience, and knowledge that enabled me to efficiently and effectively tackle this project. In the process of putting this thesis together, I realized how true this gift of writing is for me. You gave me the power to believe in my passion and pursue my dreams. I could never have done this without the faith I have in you.

Besides, I want to express my sincere gratitude to my supervisor Dr. G. Beck, for the patient guidance, encouragement, and advice he has provided throughout my time as his student. I have been extremely lucky to have a supervisor who cared so much about my work, responded to my questions and queries so promptly, and provided insightful and interesting questions and comments about the work.

My special thanks go to Prof. S. Colafrancesco, who, although no longer with us, continues to inspire by his example and dedication to the students he served over the course of his career. I would also like to thank Prof. A. Chen, Prof. R. de Mello Koch, Prof. K. Goldstein, and Prof. \'A. V\'eliz-Osorio for their excellent and patient technical assistance, for believing in my potential, and for the nice moments, we spent together.

Finally, I must express my very profound gratitude to my family, friends, and colleagues, for providing me with unfailing support and continuous encouragement throughout my study duration and through the process of researching and writing this dissertation. This accomplishment would not have been possible without them.

Thank you all for always being there for me. I have finally made it.

\newpage
\chapter*{\textbf{Dedication}} 
\addcontentsline{toc}{chapter}{Dedication}

\vspace*{\fill}
\begin{center}
\begin{minipage}{.8\textwidth}{\begin{center}
{\large \bf This thesis is dedicated to my Mother.} \\
{\bf \small The only one in the entire universe who never stopped believing in me.}
\end{center}} 
\end{minipage}
\end{center}
\vfill
\chapter*{\textbf{Abstract}} 
\addcontentsline{toc}{chapter}{Abstract}

Cosmology and particle physics are closer today than ever before, with several searches underway at the interface between cosmology, particle physics, and field theory. The mystery of dark matter (DM) is one of the greatest common unsolved problems between these fields. It is established now based on many astrophysical and cosmological observations that only a small fraction of the total matter content of the universe is made of baryonic matter, while the vast majority is constituted by dark matter. However, the nature of such a component is still unknown. One theoretically well-motivated approach to understanding the nature of dark matter would be through looking for light pseudo-scalar candidates for dark matter such as axions and axion-like particles (ALPs). Axions are hypothetical elementary particles resulting from the Peccei-Quinn (PQ) solution to the strong CP (charge-parity) problem in quantum chromodynamics (QCD). Furthermore, many theoretically well-motivated extensions to the standard model of particle physics (SMPP) predicted the existence of more pseudo-scalar particles similar to the QCD axion and called ALPs. Axions and ALPs are characterized by their coupling with two photons. While the coupling parameter for axions is related to the axion mass, there is no direct relation between the coupling parameter and the mass of ALPs. Nevertheless, it is expected that ALPs share the same phenomenology of axions. In the past years, axions and ALPs regained popularity and slowly became one of the most appealing candidates that possibly contribute to the dark matter density of the universe.

In this thesis, we start by illustrating the current status of axions and ALPs as dark matter candidates. One exciting aspect of axions and ALPs is that they can interact with photons very weakly. Therefore, we focus on studying the phenomenology of axions and ALPs interactions with photons to constrain some of their properties.

In this context, we consider a homogeneous cosmic ALP background (CAB) analogous to the cosmic microwave background (CMB) and motivated by many string theory models of the early universe. The coupling between the CAB ALPs traveling in cosmic magnetic fields and photons allows ALPs to oscillate into photons and vice versa. Using the M87 jet environment, we test the CAB model that is put forward to explain the soft X-ray excess in the Coma cluster due to CAB ALPs conversion into photons. Then we demonstrate the potential of the active galactic nuclei (AGNs) jet environment to probe low-mass ALP models and to potentially exclude the model proposed to explain the Coma cluster soft X-ray excess.

Further, we adopt a scenario in which ALPs may form a Bose-Einstein condensate (BEC) and, through their gravitational attraction and self-interactions, they can thermalize to spatially localized clumps. The coupling between ALPs and photons allows the spontaneous decay of ALPs into pairs of photons. For ALP condensates with very high occupation numbers, the stimulated decay of ALPs into photons is also possible, and thus the photon occupation number can receive Bose enhancement and grows exponentially. We study the evolution of the ALPs field due to their stimulated decays in the presence of an electromagnetic background, which exhibits an exponential increase in the photon occupation number by taking into account the role of the cosmic plasma in modifying the photon growth profile. In particular, we focus on quantifying the effect of the cosmic plasma on the stimulated decay of ALPs as this may have consequences on the detectability of the radio emissions produced from this process by the forthcoming radio telescopes such as the Square Kilometer Array (SKA) telescopes with the intention of detecting the cold dark matter (CDM) ALPs.

Finally, finding evidence for the presence of axions or axion-like particles would point to new physics beyond the standard model (BSM). This should have implications in developing our understanding of the nature of dark matter and the physics of the early universe evolution.

{\bf Keywords$:$} dark matter, axions, axion-like particles, strong CP problem, Peccei-Quinn solution, ALP-photon coupling, cosmic ALP background, Coma cluster soft X-ray excess, Bose-Einstein condensate, stimulated decay of ALPs, Square Kilometer Array, physics beyond the standard model
\renewcommand{\contentsname}{\textbf{Contents}}
\renewcommand{\listfigurename}{\textbf{List of Figures}}
\renewcommand{\listtablename}{\textbf{List of Tables}}
\renewcommand{\bibname}{\textbf{References}}
\tableofcontents
\listoffigures
\addcontentsline{toc}{chapter}{\listfigurename}
\listoftables
\addcontentsline{toc}{chapter}{\listtablename}
\hypersetup{linkcolor=blue}
\chapter*{\textbf{List of Publications}}
\addcontentsline{toc}{chapter}{List of Publications}

Some parts of this thesis have been submitted in the form of the following research papers to international journals for publication. In particular, the content of chapter \ref{ch5} is based on the released publications \cite{ayad2020probing, ayad2019phenomenology}. In addition, the results presented in chapter \ref{ch6} are based on the released publication \cite{ayad2020potential} and the forthcoming publication \cite{ayad2020quantifying}. These references are listed below for convenience.

\begin{itemize}
\item[1.] A. Ayad and G. Beck. \textit{Probing a cosmic axion-like particle background within the jets of active galactic nuclei}. Journal of Cosmology and Astroparticle Physics, 2020(04)$:\text{055}\text{--}\text{055}$, apr 2020. This work has been published in Journal of Cosmology and Astroparticle Physics (JCAP). ArXiv e-Print$:$1911.10078 [astro-ph.HE]. 

\item[2.] A. Ayad and G. Beck. \textit{Phenomenology of axion-like particles coupling with photons in the jets of active galactic nuclei}. This work has been accepted for puplication in the South African Institute of Physics (SAIP)-2019 Conference Proceedings. ArXiv e-Print$:$1911.10075 [astro-ph.HE].

\item[3-] A. Ayad and G. Beck. \textit{Potential of SKA to detect CDM ALPs with radio astronomy}. This work has been published in the International Conference on Neutrinos and Dark Matter (NDM)-2020 with the Andromeda Conference Proceedings.  ArXiv e-Print$:$2007.14262 [hep-ph]. 

\item[4-] A. Ayad and G. Beck. \textit{Quantifying the effect of cosmic plasma on the stimulated decay of axion-like particles}. This work has been submitted for possible publication in Journal of Cosmology and Astroparticle Physics (JCAP). ArXiv e-Print$:$2010.05773 [astro-ph.HE]. 

\end{itemize}

\newlist{abbrv}{itemize}{1}
\setlist[abbrv,1]{label=,labelwidth=1in,align=parleft,itemsep=0.1\baselineskip,leftmargin=!}
 
\chapter*{\textbf{List of Abbreviations}}
\addcontentsline{toc}{chapter}{List of Abbreviations}
\chaptermark{List of Abbreviations}
 
\begin{abbrv}
 
\item[ALPs]			Axion-like particles
\item[CP]		     	Charge-parity
\item[PT]            Parity-time
\item[PQ]			Peccei-Quinn
\item[DM]			Dark matter
\item[DE]			Dark energy
\item[CDM]			Cold dark matter
\item[HDM]			Hot dark matter
\item[WDM]			Warm dark matter
\item[FDM]	         Fuzzy dark matter
\item[AGNs]			Active galactic nuclei
\item[SMBH]			Supermassive black hole
\item[SMPP]			Standard model of particle physics
\item[SMC]			Standard model of cosmology
\item[SM]			Standard models of physics
\item[BSM]        Beyond the Standard Models
\item[QM]        Quantum mechanics
\item[GR]			General relativity
\item[GSW]	    Glashow, Salam, and Weinberg
\item[FRW]	     Friedmann-Robertson-Walker
\item[FLRW] Friedmann-Lema{\^\i}tre-Robertson-Walker
\item[QCD]			Quantum chromodynamics
\item[QED]  Quantum electrodynamics
\item[QFT]    Quantum field theory
\item[EW]     Electroweak
\item[CMB]  Cosmic microwave background
\item[CAB]  Cosmic axion or ALP background
\item[WIMP]  Weakly Interacting Massive Particles
\item[BEC]  Bose-Einstein condensate
\item[SKA]  Square kilometer array
\item[HERA]  Hydrogen epoch of reionisation array
\item[EDGES]  Global epoch of reionization signature
\item[BB]    Big bang
\item[BBN]    Big bang nucleosynthesis
\item[SUSY]  Supersymmetry
\item[MSSM] Minimal Supersymmetric Standard Model
\item[LSP]  Lightest superpartner
\item[LHC]  Large Hadron Collider
\item[CERN] European Organization for Nuclear Research
\item[NG]  Nambu-Goldstone
\item[NGB]  Nambu-Goldstone boson
\item[PNGB]  Pseudo-Nambu-Goldstone boson
\item[EDM]  Electric dipole moments
\item[PQWW]  Peccei-Quinn-Weinberg-Wilczek
\item[VEV] Vacuum expectation value
\item[KEK]  High Energy Accelerator Research Organization
\item[KSVZ]  Kim–Shifman–Vainshtein–Zakharov
\item[DFSZ]  Dine–Fischler–Srednicki–Zhitnitsky
\item[PBHs]   Primordial black holes
\item[QGP]           Quark-gluon plasma
\item[KK]           Kaluza-Klein
\item[HB]           Horizontal branch
\item[RGB]    Red-giant branch
\item[ICM]   Intracluster medium
\item[SI]  International system of units
\item[kpc]  Kiloparsec
\item[Mpc] Megaparsec

\end{abbrv}

\chapter*{\textbf{List of Symbols}}
\addcontentsline{toc}{chapter}{List of Symbols}
\chaptermark{List of Symbols}

\begin{abbrv}
 
\item[$s$]	                      Action
\item[$G$]			               Universal gravitational constant
\item[$H$]			               Hubble parameter
\item[$h$]		                   Dimensionless Hubble parameter
\item[$R$]			               Scale factor of the universe
\item[$c$]			              Speed of light in vacuum
\item[$T_{\mu \nu}$]      Energy momentum tensor
\item[$g_{\mu \nu}$]	    Metric tensor
\item[$G_{\mu \nu}$]		Einstein tensor
\item[$\mathfrak{R}$]			Ricci scalar
\item[$\mathfrak{R}_{\mu \nu}$]			Ricci tensor
 \item[$\mathrm{\Gamma}^{\alpha}_{\mu \nu}$]	     Christoffel symbol
\item[$R_c(t)$]  Radius of curvature
\item[$R_c$] Present radius of curvature
\item[$\mathrm{\Lambda}$]	                   Cosmological constant
\item[$\phi$]	    Scalar field
\item[$a$]	    Axion or ALP field
\item[$\rho$] Energy density of the universe
\item[$\rho_a$] Energy density of axions and ALPs
\item[$n_a$] Number density of axions and ALPs
\item[$m_a$] Mass of axions or ALPs
\item[$g_{a \gamma}$] Coupling parameter of axions or ALPs with photons
\item[$\tau_a$] Lifetime of axions or ALPs
\item[$m_e$] Mass of electrons
\item[$n_e$] Number density of electrons
\item[$P$]	 Pressure of the universe
\item[$\nu$]	Expanding velocity of the universe
\item[$\mathrm{\Omega}_{\mathrm{\Lambda}}$]	 Vacuum energy density parameter
\item[$\mathrm{\Omega_{m}}$]	 Matter energy density parameter
\item[$F_{\mu \nu}$]      Electromagnetic field tensor
\item[$\tilde{F}^{\mu \nu}$]      Dual electromagnetic field tensor
\item[$f_a$]	                     Energy scale of the PQ symmetry breaking
\item[$g_{a\gamma}$]	                      ALP-photon coupling parameter
\item[$\alpha$]	                     Fine structure constant
\item[$e$]	   Electric charge
\item[$E$]	  Electric field
\item[$B$]	  Magnetic field
\item[$T$]	                    Temperature
\item[$k_B$]	                Boltzmann constant
\item[$\mathrm{M}_{\odot}$] Solar mass

\end{abbrv}

\newpage
\pagenumbering{arabic}
\pagestyle{myheadings}
\chapter{\textbf{Introduction and Motivation}} \label{ch1}

The question of the fundamental origin of matter with explaining how nature works is one that has interested philosophers and scientists since the dawn of history. The answer to this question has been a subject of explanation in almost all civilizations and cultures until science has been able to give a version of the facts. The magnificent progress made in this regard particularly in the last few decades is undoubtedly one of the most important achievements of the human species throughout the ages to realize the nature of our reality. In theoretical physics, our present best understanding of the behavior of the universe is based upon the extraordinary successes of the standard model of particle physics (SMPP) \cite{glashow1961partial, weinberg1967model, salam1968weak} which describes the physics of the very small objects in terms of quantum mechanics (QM) \cite{planck1978gesetz}, together with the standard model of cosmology (SMC) \cite{gamow1946expanding, alpher1948evolution} which describes the physics of the very large objects in terms of the theory of general relativity (GR) \cite{einstein1915allgemeinen, einstein1916grundlage}. According to this investigation, we believe that the structure of the universe is explained in terms of a set of elementary particles interacting with each other through four fundamental forces. Gravity, electromagnetism, weak, and strong interactions are considered the four fundamental forces in nature, all of which are described based on symmetry principles \cite{maldacena2015symmetry}.

The SMC, and also called the $\mathrm{\Lambda} \text{CDM}$ model,  is essentially based on Einstein's general theory of relativity for the gravitational force, which improved upon Newton's theory of gravity \cite{newton2013philosophiae, bone1996sir}. Indeed it is a purely classical theory, as long as it does not incorporate any idea of quantum mechanics into the formulation. Today, general relativity is widely accepted as our best description of the physics of the gravitational field, and it has many successes in describing the structure of the universe on the macroscopic scale \cite{rovelli2004quantum, oriti2009approaches}. The SMPP, on the other hand,  is broadly accepted as the fundamental description of particle physics. It successfully handles the interactions of the elementary particles due to the other three fundamental forces, the electromagnetic, weak, and strong forces within the framework of quantum mechanics at the microscopic scale \cite{rovelli2004quantum, oriti2009approaches}.

Despite the many successes and the strong empirical support of both the two standard models (SM), at first sight, they appear to be incompatible since each of them is formulated based on principles that are explicitly contradicted by the other model. This contradiction leaves many foundational issues that are still poorly understood, and numerous basic questions remain active areas of current research. Although there is a lot of evidence to make one quite confident that many steps have been taken on the way to understand the behavior of matter and the structure of the universe, there are also many reasons to believe that something very basic is missing and fundamental understanding of the nature of matter and the current picture of the world is still incomplete. In the last few years, the connection between cosmology and particle physics has been developing very rapidly. The potential now exists to revolutionize our knowledge by looking for the possibility of new physics being discovered and this definitely requires new theories that will have to be developed or that existing theories will have to be amended to account for it.

\section{Standard model of cosmology}

The observed expansion of the universe is a natural result of any homogeneous and isotropic cosmological model based on general relativity \cite{slipher1915spectrographic, lundmark1924determination, hubble1931velocity}. In 1915, Einstein developed the general theory of relativity to improve Newton's theory of gravity in describing the gravitational interactions between matter. A comprehensive introduction to general relativity can be found in various textbooks, such as \cite{weinberg1972gravitation, hawking1973large, einstein2003meaning, carroll2004spacetime, wald2007general}. General relativity is a purely classical theory and does not incorporate any idea of quantum mechanics into the formulation. The critical aspect of the general relativity is the dynamical nature of spacetime, and it essentially depends upon the following two fundamental postulates$:$ 
\begin{itemize}
\item The principle of relativity, which states that all systems of reference are equivalent with respect to the formulation of the fundamental laws of physics.
\item The principle of equivalence, which states that in the vicinity of any point, a gravitational field is equivalent to an accelerated frame of reference in gravity-free space.
\end{itemize}
The consequences of these principles lead to the fundamental insight of the general relativity that is$:$ gravity can not only be regarded as a conventional force but rather as a manifestation of spacetime geometry. In other words, general relativity assumes that the gravitational force is a result of the curvature of spacetime. This should change our ideas about mass that came from Newtonian gravity, which implies that the mass is the source of gravity. In general relativity, the mass turns out to be a part of a more general quantity called the energy-momentum tensor ($T_{\mu \nu}$) \cite{schutz2009first}, which includes both energy and momentum densities and encodes how matter is distributed in spacetime.  It seems to be natural that the energy-momentum tensor is involved in the field equations for the general gravity, as we will see later. Thus, the energy and matter content alters the geometry of the spacetime, and the geometry of spacetime affects the motion of matter. Essentially as John Wheeler once said, ``matter tells spacetime how to curve, and spacetime tells matter how to move’’ \cite{wheeler2000geons}.

Mathematically, general relativity is defined by two central equations. The first is a set of  ten equations that give the relationship between spacetime and matter, known as the Einstein field equations \cite{einstein1915field}
\begin{equation} \label{eq.1.1}
\mathit{G}_{\mu \nu}= \mathfrak{R}_{\mu \nu} - \frac{1}{2} \mathfrak{R} \mathit{g}_{\mu \nu}= \frac{8 \pi \mathit{G}}{c^4} \mathit{T}_{\mu \nu} \:,
\end{equation}
where $\mathit{G}_{\mu \nu}$ is the Einstein tensor, $\mathfrak{R} _{\mu \nu}$  is the Ricci tensor \cite{parker1994mathtensor} that encodes information about the curvature of space-time given by the metric $\mathit{g}_{\mu \nu}$, $\mathfrak{R}$  is the Ricci scalar,  $\mathit{T}_{\mu \nu}$  is the energy-momentum tensor, $\mathit{G}$ is the universal gravitational constant, and $c$ is the speed of light. For an extensive review of tensors see \cite{sotiriou2010f}.

The second is the equation of the geodesic path, which governs how the trajectories of objects evolve in curved spacetime and it gives the equation of motion for freely falling particles in a specified coordinate system. In practice, this equation represents four second-order differential equations that determine $x^{\alpha}(\tau)$, given initial position and 4-velocity, where $\tau$ is proper time measured along the path of particle$:$
\begin{equation}
\dfrac{d^2 x^{\alpha}}{d \tau^2} + \mathrm{\Gamma}^{\alpha}_{\mu \nu} \left[ \dfrac{dx^{\mu}}{d \tau} \dfrac{dx^{\nu}}{d \tau} \right] = 0 \:,
\end{equation}

where $\mathrm{\Gamma}^{\alpha}_{\mu \nu}$ is known as a Christoffel symbol. One of the most spectacular successes of the general relativity is its role that leads to the birth of the standard model of big bang cosmology or simply the standard model of cosmology (SMC). For thorough reviews of this topic, see for example \cite{guth1982fluctuations, linde1982new, albrecht1982cosmology, bardeen1983spontaneous}. The formulation of the SMC was based on general relativity, and it has been very successful in explaining the observable properties of the cosmos \cite{oriti2009approaches}. In principle, general relativity provided a comprehensive and coherent description of space, time, gravity, and matter at the large scales \cite{davis1985evolution}. It was also capable of describing the cosmology of any given distribution of matter using the Einstein field equation.  Friedmann simplified the Einstein field equations by assuming the universe is spatially homogeneous and isotropic on large scales, and that is quite consistent with the recent observations \cite{friedmann1922125}. Together, homogeneity and isotropy lead to the cosmological principle, stating that on sufficiently large scales, the universe is homogeneous and isotropic, and essentially this means that all spatial positions in the universe are equivalent. The cosmological principle then restricts the metric to the Friedmann-Robertson-Walker (FRW) form \cite{friedman1922krummung, friedmann1924moglichkeit, robertson1935kinematics, robertson1936kinematicsII, robertson1936kinematicsIII, walker1937milne}
\begin{equation}
ds^2 = - dt^2 + R(t)^2 \left[ \frac{dr^2}{1-kr^2} + r^2 (d\theta^2 + \sin^2 \theta d\phi^2) \right]  \:,
\end{equation} 
in which $(r, \theta, \phi)$ are the comoving coordinates, and the dimensionless parameter $R(t) \equiv R_c(t) / R_c$ is called the scale factor of the universe that characterizes the size of the universe and hence its evolution. Here $R_c(t)$ is the radius of curvature of the 3-dimensional space at time $t$ and , by convention, $R_c$ is the radius of curvature at the present time $t_0$. If this metric equation is rewritten in terms of the conformal time $\tau$ instead of using the proper cosmic time $t$ as measured by a comoving observer, then it reduces to
\begin{equation}
ds^2 =  R(t)^2 \left[ - d \tau^2 + \frac{dr^2}{1-kr^2} + r^2 (d\theta^2 + \sin^2 \theta d\phi^2) \right]  \:.
\end{equation}
Here the dimensionless parameter $k \equiv \kappa / R_c$ determines the curvature of the space, where the number $\kappa$ is called the curvature constant, which takes only the discrete values$;$ $+1, 0, -1$, and distinguishes between the following different spatial geometries,
\begin{itemize}
\item \text{\boldmath $\kappa= +1$}, corresponds to a finite universe with spatially closed geometries, positively curved like a sphere.
\item \text{\boldmath $\kappa = 0$}, corresponds to an infinite universe with spatially flat geometries, uncurved like a plane.
\item \text{\boldmath $\kappa = -1$}, corresponds to an infinite universe with spatially open geometries, negatively curved like a hyperboloid.
\end{itemize}

By inserting the FRW metric into the Einstein field equations, we obtain a closed system of Friedmann equations, which describe the evolution of the scale factor$:$
\begin{align}
\left( \frac{\dot{R}}{R} \right)^2 &= \frac{8 \pi G}{3} \rho - \frac{k c^2}{R^2} \:, \\
\frac{\ddot{R}}{R} &= - \frac{4 \pi G}{3} \left( \rho + \frac{3 P}{c^2} \right) \:,
\end{align}
where $\rho$ is the energy density, and $P$ is the pressure of the universe. Most solutions to the Friedmann equations predict an expanding or a contracting universe, depending on some set of initial numbers, such as the total amount of matter in the universe. The solutions of the expanding universe lead to Hubble's law \cite{hubble1929relation}
 \begin{equation}
 v = H d \:,
\end{equation} 
where $v$ is expanding velocity, $d$ is the proper distance that is the distance to an object as measured in a surface of constant time, and $H$ is the rate of expansion of the universe, which known as the Hubble function
\begin{equation}
H \equiv \frac{\dot{R}}{R} \:.
\end{equation}
This allows establishing the age of the observable universe to be $\sim 14$ billion years. This equation is really important since it relates the empirical parameter $H$ discovered by Hubble to the expansion parameter of the Friedmann equation. The recent value for the today universe expansion rate is measured by the current value of the Hubble function and known as the Hubble constant $H_0$. Hubble initially overestimated the numerical value of $H_0$ and thought it is  $500 \ \text{km} \ \text{s}^{-1} \ \text{Mpc}^{-1}$.  The best current estimation of it, combining the results of different research groups, gives a value around $70 \pm 7 \ \text{km} \ \text{s}^{-1} \ \text{Mpc}^{-1}$ \cite{ryden2017introduction}. However, the initial value for the Hubble constant was much bigger than the recent one, but in any case, it was always big enough to make the Friedmann solution to the Einstein field equations inconsistent with the belief at this time that the universe is static and positively curved. The current value of the Hubble constant allows establishing the age of the observable universe to be about $14.0 \pm 1.4$ billion years. The discovery of the cosmic expansion then implies that the age of the universe is not infinite. Note that since the energy density of the universe $\rho$ must be a positive number, then the Friedmann to the Einstein field equations leads to the surprising result that the pressure of matter $P$ is negative. Einstein corrected for this by introducing the so-called cosmological constant $\mathrm{\Lambda}$ \cite{peebles2003cosmological} into the original field equations and forced it to escape the confusion with Friedmann solution for a static and positively curved universe \cite{riess1998observational, perlmutter1999measurements}. The modified field equations with the cosmological constant had a form
\begin{equation}
G_{\mu \nu} + \mathrm{\Lambda} g_{\mu \nu}= \mathfrak{R}_{\mu \nu} - \frac{1}{2} \mathfrak{R} g_{\mu \nu} + \mathrm{\Lambda} g_{\mu \nu}= \frac{8 \pi G}{c^4} T_{\mu \nu} \:.
\end{equation}
With the additional term of the cosmological constant to  the Einstein field equations, the Friedmann equations become
\begin{align}
\left( \frac{\dot{R}}{R} \right)^2 &= \frac{8 \pi G}{3} \rho - \frac{k c^2}{R^2(t)} + \frac{\mathrm{\Lambda} c^2}{3} \:, \\
\frac{\ddot{R}}{R} &= - \frac{4 \pi G}{3} \left( \rho + \frac{3 P}{c^2} \right) + \frac{\mathrm{\Lambda} c^2}{3}  \:.
\end{align}
In addition, conservation of the energy-momentum tensor yields a third equation which turns out to be dependent on the two Friedmann equations$:$
\begin{equation}
\dot{\rho} + 3 \frac{\dot{R}}{R} \left(\rho + \frac{P}{c^2}  \right) =0 \:.
\end{equation}
This is essentially the equation of state that relates the density $\rho$ to the pressure $P$. In the most common cases, the equation of state for the cosmological fluid is chosen to be of a mixture of non-interacting ideal fluids that is  given by the form
\begin{equation}
P_i = \omega_i \rho_i c^2 \:,
\end{equation}
where $P_i$ is the partial pressure, and $\rho_i$ is the partial energy density for each part of the cosmic fluid. While the factor $\omega_i$  is a constant whose value represents properties of different kinds of fluid,  for example, $\omega = 0$ gives a pressureless fluid, while $ \omega = 1/3$ corresponds to radiation, and it might take negative values to represent the negative pressure which could be caused by some other form of a perfect fluid. Indeed, the additional term with the cosmological constant $\mathrm{\Lambda}$ in the Einstein field equation should be the source of such a form of matter, which is known as dark energy.

As a final note on the Friedmann equations, dividing the first equation by factor $H^2$ gives  the sum of the energy densities of the various matter components of the universe
\begin{align}
1 &= \mathrm{\Omega}_i + \mathrm{\Omega}_{\mathrm{\Lambda}} + \mathrm{\Omega}_k \:, \\
\mathrm{\Omega}_i &= \frac{8 \pi G}{3 H^2} \rho_i \:, \quad  \mathrm{\Omega}_k = \frac{- k c^2}{a^2 H^2} \:, \quad \mathrm{\Omega}_\mathrm{\Lambda} = \frac{\mathrm{\Lambda} c^2}{3 H^2}  \:,
\end{align}
where $\mathrm{\Omega}$ is the energy density divided by the critical density $\rho_c = \frac{3 H^2}{8 \pi G}$, the energy density at which the universe is flat such as $\mathrm{\Omega}=1$. 

The general relativity and the relative SMC provide the most successful description of space, time, and gravity. This is supported by a number of experimental confirmations that have been found to agree with the theoretical predictions. The declaration in 2016 about the first direct detection of the gravitational-waves signal generated due to the merger of black holes has provided extraordinary evidence in support of the general relativity and the SMC \cite{abbott2016supplement}. In addition to the existence of gravitational waves, general relativity has been confirmed by other tests include the observations of the gravitational deflection of light, the anomalous advance in the perihelion of Mercury, and gravitational redshift \cite{treschman2019gravitational}.

\section{Standard model of particle physics} \label{Sec.1.2}

The SMPP is a very successful theory that considered our best current description of the known elementary particles of nature and their interactions. A wealth of canonical literature is available, see for example \cite{peskin2018introduction, sterman1993introduction, ellis2003qcd, burgess2006standard}. The model is capable of providing a quantitative description of three of the four fundamental interactions in nature, \ie electromagnetic, weak, and strong interaction. The electromagnetic and weak interactions are unified in the electroweak theory interaction that is described by the model of Glashow, Salam, and Weinberg (GSW), while the quantum chromodynamics (QCD) is the theory that describes the strong interaction. The fourth fundamental interaction is gravity, which is best described by the general relativity, as we already discussed in the previous section. The incorporation of the gravitational force into the SMPP framework is still an unresolved challenge \cite{salam1980gauge}.

The SMPP classifies the elementary particles into two main categories, \ie fermions and bosons. In this context, the ordinary matter\footnote{The terms ordinary matter and baryonic matter will be used to define the kind of matter described by the standard model of particle physics.} in the universe is basically made up of fermions that are held together by the fundamental forces through the exchange of bosons that mediate the forces between these fermions. The fermions are classified as particles with half-integer spins that obey the Pauli exclusion principle. They can be further subdivided into two basic classifications of elementary particles, \ie quarks and leptons, depending on which kind of interaction they are subjected to. Both classes consist of six particles, grouped into three doublets, called families or generations. The three quark doublets are$:$ up ($u$) and down ($d$), charm ($c$) and strange ($s$), top ($t$) and bottom ($b$). While the three lepton doublets consist of electron ($e$), muon ($\mu$), tau ($\tau$), and an associated neutrino to each of these leptons. Also, there are three additional doublets for each class, composed of leptons and quarks antiparticles. The antiparticles have the same mass as their partners but all quantum numbers opposite. On the other hand, bosons are classified as particles with integer spins and do not obey the Pauli exclusion principle. The photon ($\gamma$) is the gauge boson that mediates the electromagnetic interactions, and there are three such bosons, \ie $W^+$, $W^-$, $Z$ responsible for the weak interactions, while the strong interactions are mediated by 8 gluons ($G$) \cite{shifman1979qcd}. In addition, there is the Higgs boson \cite{atlas2012observation, bezrukov2012higgs} that gives the mass to all the standard model particles with which the field interacts through the Higgs mechanism \cite{higgs1966spontaneous}. A list of all the standard model particles and some of their properties are presented in table \ref{table1}.\noindent
\begin{table}[t!] 
\centering{
\scalebox{0.755}{
\begin{subtable}{1.5 \textwidth}
\begin{tabular}{|c||c|c|c|c||c|c|c|c|} 
\toprule \hline
\multicolumn{9}{|c|}{Fermions} \\ 
\hline 
\multicolumn{5}{|c||}{Quarks} &  \multicolumn{4}{c|}{Leptons} \\
\hline
\multirow{1}{*} { Generation}  & Name  & Symbol & Charge [$e$] & Mass [GeV] &  Name  & Symbol & Charge [$e$] & Mass [GeV]  \\  
\cline{1-9} 
\multirow{2}{*} {1st}  &  up  & u  &  $+2/3$ & $2.2^{+0.6}_{-0.4} \times 10^{-3}$& electron & e & $-1$ & $0.511 \times 10^{-3}$ \\ 
  &  down  & d  &  $-1/3$ & $4.7^{+0.5}_{-0.4} \times 10^{-3}$ & $e$-neutrino &  $\nu_e$ & $0$ & $< 2 \times 10^{-9}$  \\ 
\cline{1-9} 
\multirow{2}{*} {2nd}  & charm  & c  &  $+2/3$ & $1.27 \pm 0.03$& muon & $\mu$ & $-1$ & $0.106$  \\ 
  &  strange  & s  &  $-1/3$ & $96^{+8}_{-4} \times 10^{-3}$& $\mu$-neutrino& $\nu_{\mu}$ & $0$ & $< 0.19 \times 10^{-3}$   \\ 
 \cline{1-9} 
\multirow{2}{*} {3rd}  &  top  & t  &  $+2/3$ & $174.2 \pm 1.4$& tau & $\tau$ & $-1$ & $1.777$  \\ 
  &  bottom  & b  &  $-1/3$ & $4.18 \pm 0.04$ & $\tau$-neutrino & $\nu_{\tau}$ & $0$ & $< 0.0182$ \\ 
 \hline
\end{tabular}
\end{subtable}}} \\
\centering{\scalebox{0.8586}{
\begin{subtable}{1.5 \textwidth}
\begin{tabular}{|c|c|c|c|c|c|c|c|}
\hline
 \multicolumn{8}{|c|}{Bosons} \\
\hline 
Name  & Symbol & Charge [$e$]& Spin & Mass [GeV] & Interactions & Range & Interaction with \\
\hline 
Photon & $\gamma$ & $0$ &  $+1$ & $0$ & Electromagnetism& $\infty$& Charge \\ 
\cline{1-8} 
W-boson &$W^{\pm}$ & $\pm 1$ &  $+1$ & 80.4 & \multirow{2}{*} {Weak}  & \multirow{2}{*} {$10^{-18}$} & weak isospin \\ 
Z-boson&$Z^0$ & $0$ & $+1$ & 91.2 & &&  + hypercharge \\
\cline{1-8} 
Gluons &$G_{i=1, \dots, 8}$ &  $0$ &  $+1$ & $0$ & Strong & $10^{-15}$& color \\ 
\cline{1-8} 
Higgs &$H$ & $0$ & $0$ & 125 &&& \\ 
\hline \bottomrule
\end{tabular}
\end{subtable}}}
\caption{Overview over the standard model particles and some of their properties.}
\label{table1}
\end{table}

The six quarks in the standard model are classified by their so-called flavors.  The quarks three generations are ordered by increasing mass from the first to the third generation. The up-type quark in each generation has an electric charge equal to $+2/3$, while each down-type one carries an electric charge of $-1/3$.  Quarks are the only fundamental particles that experience all the four fundamental interactions. They participate in strong interactions because of their color charges, which came in three kinds, \ie red (r), green (g), blue (b) charges. They have never been observed yet in nature as free states, but only confined inbound states called hadrons \cite{gell2010quarks}.

Leptons come as well in six different flavors, and their three generations similarly are ordered by increasing mass. The electron type leptons carry $-1$ electric charge, while the associated neutrinos are electrically neutral. Unlike quarks, leptons do not carry color charge and do not participate in strong interactions. In turn, every generation has two chiral manifestations, the left-handed and right-handed one. Only left-handed particles can participate in weak interactions via $W^{\pm}$ bosons \cite{commins1983weak}. Both left-handed particles in a given quark or lepton generation are assigned a so-called weak-isospin quantum number, identifying them as partners of each other with respect to the weak interaction.

One important remark should be mention here, that only fermions from the first generation built up the observable stable matter in the universe. In contrast, the particles of the other two generations and compounds of them always decay into lighter particles from lower generations.  In fact, the role of these two last generations in describing the visible universe is not clearly understood yet.

Mathematically, the SMPP can be defined as a renormalizable quantum field theory (QFT) based on the following local gauge symmetry
\begin{equation}
G_{\text{SM}} = SU(3)_C \otimes SU(2)_L  \otimes  U(1)_Y \:.
\end{equation}
Each of these gauge groups is corresponding to model a different interaction of the three fundamental forces, which are incorporated into the SM framework$:$ the electromagnetic, the weak, and the strong forces. The electromagnetic interaction of the standard model particles is described by the theory of quantum electrodynamics (QED), which is a gauge theory based on a $U(1)_{\text{em}}$ symmetry group. Then the electroweak theory has been formulated to unify the electromagnetic and weak interactions between quarks and leptons to a single framework that is able to describe both the two interactions based on the electroweak gauge group $SU(2)_L \otimes U(1)_Y$. The $SU(2)_L$ group refers to the weak isospin charge $I$, while $U(1)_Y$ to the weak hypercharge $Y$ \cite{nishijima1955charge, gell1956interpretation}. Then the description of the electroweak interactions requires three massive gauge bosons $W^{\pm}$ and $Z$ in addition to the photon $\gamma$. In contrast, the theory of quantum chromodynamics describes the strong interaction based on the gauge group of local $SU(3)_C$ transformations of the quark-color fields. The QCD describes the strong interaction between quarks that arises from the exchange of the eight massless gluons that couple to the color charge of the fermions $G_{i=1, \dots, 8}$.

For further convenience, the standard model Lagrangian can be formed as the sum of four parts, which are the gauge interactions, fermions interactions, Higgs interaction, and  Yukawa interactions. Each of these four terms refers to one of the interaction mechanisms between the particles of the standard model. Therefore, the most general Lagrangian density of the standard model can be written as
\begin{align}
\L_{\text{SM}} & = \L_{\text{Gauge}} + \L_{\text{Dirac}} +\L_{\text{Higgs}} + \L_{\text{Yukawa}} \:.
\end{align}

The first term represents the kinetic terms for the gauge bosons and describes the interactions between them
\begin{equation}
\L_{\text{Gauge}} = - \frac{1}{4}  G^a_{\mu \nu} G_a^{\mu \nu} - \frac{1}{4}  W^i_{\mu \nu} W_i^{\mu \nu} - \frac{1}{4}  B_{\mu \nu} B^{\mu \nu}  \:,
\end{equation}
where $G^a_{\mu}(a=1, \dots,8)$ are the gluons of the strong interactions, and $W^i_{\mu}(i=1,2,3)$ and $B_{\mu}$ are the gauge bosons of the electroweak interactions. The covariant field strength tensors are defined as follows
\begin{align}
G^a_{\mu \nu} &= \partial_\mu G^a_\nu - \partial_\nu G^a_\mu - g_s f^{abc} G^b_\mu G^c_\nu \:,  \\
W^i_{\mu \nu} &= \partial_\mu W^i_\nu - \partial_\nu W^i_\mu - g \epsilon^{ijk} W^j_\mu W^k_\nu \:, \\
B_{\mu \nu} &= \partial_\mu B_\nu - \partial_\nu B_\mu  \:,
\end{align} 
where $g_s$ is the strong interaction coupling strength, and the structure constant $f^{abc}$  is defined as $[ \tau^a,\tau^b ] = i f^{abc} \tau^c$, $a, b, c$ run from $1$ to $8$ for  $\tau^a$ which are the generator for the  $SU(3)_C$ group. In contrast, the weak interaction coupling strength is $g$, and the structure constant is defined as $\epsilon^{abc}$ is defined as $[ \lambda^a,\lambda^b ] = i f^{abc} \lambda^c$, where $\lambda^a$ are the group $SU(2)_L$ generators with  $a, b, c$ run from $1$ to $3$.

The second term contains kinetic terms for the fermions, which describe the fermions interactions with the gauge bosons as well as their interactions with each other
\begin{equation}
\L_{\text{Dirac}} = \sum_f i \bar{\psi}_i \slashed{D} \psi_i + h.c. \:.
\end{equation}
The summation runs over all of the fermions, where $\psi_i$, and $\bar{\psi}_i$ describe the fermion field and the conjugate field. The covariant derivative $\slashed{D}$ featuring all the gauge bosons without self-interactions. The beauty of this term is that it contains the description of the electromagnetic, weak, and strong interactions, while the covariant derivative distinguishes between them. The $h.c.$ term represents the ``hermitian conjugate’’ of the Dirac interaction.

The third part of the Lagrangian is the Higgs part
\begin{equation}
\L_{\text{Higgs}} =  (D_\mu \phi)^\dagger (D_\mu \phi) - V(\phi^\dagger \phi) \:,
\end{equation}
where $\phi$ represents the Higgs field and the Higgs potential is given by
\begin{equation}
V(\phi^\dagger \phi) = \mu^2 \phi^\dagger \phi + \frac{\lambda}{2} (\phi^\dagger \phi)^2 \:.
\end{equation}
This part contains only the Higgs bosons and the electroweak gauge bosons. The first term describes how the gauge bosons couple to the Higgs field and forms its masses, while the second term represents the potential of the Higgs field.

The last part describes the Yukawa interactions \cite{yukawa1935interaction}, and consist of the most general possible couplings of the Higgs fields to the fermion fields$:$
\begin{equation}
\L_{\text{Yukawa}} = - y_{ij} \bar{\psi}_{Li} \psi_{Ri} \phi + h.c. \:,
\end{equation}
where $y_{ij} $ is the dimensionless Yukawa coupling. This part describes how the fermions couple to the Higgs field $\phi$ and thought to be responsible for giving fermions their masses when electroweak symmetry breaking occurs \cite{englert1964broken, higgs1964broken, higgs1964brokenn}. The character of neutrino masses is not yet known, and the $h.c.$ term is the hermitian conjugate of the Yukawa interaction that gives mass for antimatter particles.

Up to this point, we have constructed the standard model Lagrangian based on the mechanism of interactions. Schematically, it may also be useful to divide our approach to study the model Lagrangian into two sectors$:$
\begin{equation}
\L_{\text{SM}} = \L_{\text{EW}} + \L_{\text{QCD}} \:.
\end{equation}

The first part is the electroweak sector of the standard model, which is the subset of terms consisting of the $SU(2)_L$ and $U(1)_Y$ gauge fields as well as all the fermions with non-zero charges under these groups. Then the EW Lagrangian is of the general form
\begin{equation}
\L_{\text{EW}} = - \frac{1}{4}  W^i_{\mu \nu} W_i^{\mu \nu} - \frac{1}{4}  B_{\mu \nu} B^{\mu \nu} + \sum_f \bar{l}_f (i \slashed{D}_{\text{EW}} - m_f) l_f \:,
\end{equation}
The behavior of the electroweak Lagrangian is defined by the EW covariant derivative$:$
\begin{equation}
\slashed{D}_{\text{EW}} = \gamma^\mu (\partial_\mu + i g \lambda^a W^a_\mu + i  g' \frac{Y}{2} B_\mu) \:,
\end{equation}
where $W^a_\mu $ and $B_\mu$  the three gauge fields of the $SU(2)$ group and the $U(1)_Y$ group gauge field, respectively. While $g'$ is another coupling constant and $Y$ is called the hypercharge operator that is the generator of the $U(1)_Y$ group.

The strong interaction in the SMPP is described by the quantum chromodynamics sector, that composed of the $SU(3)$ gauge fields and non-singlet fields under this gauge group. Consequently, the most general gauge-invariant Lagrangian of QCD reads
\begin{equation}
\L_{\text{QCD}} =  - \frac{1}{4}  G^a_{\mu \nu} G_a^{\mu \nu}  + \sum_f \bar{q}_f (i \slashed{D}_s - m_f) q_f + h.c. + \L_{\theta}   \:.
\end{equation}
The summation runs over all the quark fields  $q_f$,  where $f$ is the quark flavor, and $m_f$ is the corresponding mass. The QCD sector is characterized by the  QCD covariant derivative $\slashed{D}_s$ that contains the coupling between the quarks and the gauge fields and defined as
\begin{equation}
\slashed{D}_s = \gamma^\mu (\partial_\mu + i g_s \tau^a G^a_\mu) \:,
\end{equation}
with $G^a_\mu$ represent the $8$ gluon fields of the strong interaction. As we discussed before, the term $h.c.$ is the hermitian conjugate to these sectors, and it is required to fix the gauge kinetic term. Lastly, the QCD gauge invariance allows for one additional term, which we have labeled the $\theta$-term \cite{baker2006improved}. This term has the following form
\begin{equation}
\L_\theta = \bar{\theta} \frac{g_s^2}{64 \pi^2} \epsilon^{\mu \nu \alpha \beta} G_a^{\mu \nu} G_a^{\alpha \beta} \:.
\end{equation}
Here $\epsilon^{\mu \nu \alpha \beta}$ is the totally antisymmetric tensor in four dimensions. The $\theta$-term appears as a consequence solution to the spontaneous breaking of the axial $U(1)_A$ symmetry in the QCD Lagrangian. Adding this term to the Lagrangian leads to another fundamental problem called the problem of strong CP (charge-parity) violation \cite{hooft1994symmetry, hooft1994computation}. The axion \cite{weinberg1978new, wilczek1978problem} is a very promising solution to such a problem and, at the same time, a possible dark matter candidate. We will discuss this problem with more details in chapter \ref{ch3} to understand why adding the $\theta$-term to the QCD Lagrangian is necessary and how axions can give rise to solutions to both the strong CP problem and the dark matter problem as well.

The SMPP is currently well accepted as the best description of nature at microscopic scales. Within the theoretical framework of the SMPP, a wide range of phenomena can be described to an impressive degree of accuracy, and its predictions as well have been verified experimentally to extraordinary accuracy. Perhaps the most significant success of the SMPP is the theoretical prediction of the Higgs boson, over $50$ years before its experimental detection in $2012$  by the ATLAS and CMS collaborations at the LHC \cite{atlas2012observation, chatrchyan2012observation}. Other successes of the model include the prediction of the $W$ and $Z$ bosons, the gluon, and the top and charm quark, also before they have even been observed.

\section{Problems with the standard model} 

As a matter of fact, most of our information about the structure of our universe came based on the SMPP together with the SMC, and for simplicity, let us call both of the two models the standard model (SM). Despite all these successes and more, the standard model does not provide a complete picture of nature, and it does not answer all questions. There is a number of theoretical and experimental reasons that lead to the belief that the standard model is not complete yet. Examples in this respect, a set of the major unsolved problems which can not be addressed by the standard model, are listed downward in this section.

\begin{itemize}
\item {\bf Gravity puzzle.} The SMPP is extremely successful in describing the electromagnetic, weak, and strong interactions, while gravity$;$ the last fundamental interaction, is only treated classically by the general relativity and not yet incorporated in the SMPP. However, there is a hypothesis that gravitational interactions are presumably mediated by a massless spin-2 particle called graviton, but this particle has not yet been observed due to the relative weakness of gravitation in comparison with the other fundamental forces. The possibility of a theory for massive gravitons is commonly referred to as Massive Gravity, but for the moment it can not be promoted into a renormalizable quantum theory \cite{schmidt2013classically}. For these reasons, both gravity and graviton are not included in the SMPP. Furthermore, there is is still a possibility that the general relativity does not hold precisely on very large scales or in very strong gravitational forces. In any case, the general relativity breaks down at the big bang, where quantum gravity effects became dominant. According to a naive interpretation of general relativity that ignores quantum mechanics, the initial state of the universe, at the beginning of the big bang, was a singularity. Seeking for reasonable explanations to these issues might be good enough reason to look for new physics beyond the general relativity \cite{rovelli2004quantum, oriti2009approaches}.

\item {\bf The gauge hierarchy problem.} The SMPP can not explain the large differences in the coupling constants of forces at low energy scales. In particular, there is no clarification of the mystery that why gravity is so much weaker than the other forces. In the same context, the mass of the Higgs boson has been measured as 125 GeV \cite{atlas2012observation, bezrukov2012higgs}, while the theoretical value of this mass that is calculated from the standard model is enormously larger than this experimental result. In a nutshell, the gauge hierarchy problem is about the question of why the physical Higgs boson mass is so small. It is possible to restore these values to the proper one through fine-tuning, but this is considered to be unnatural. It leads many to believe that there must be a better solution, but the problem is not yet settled \cite{gildener1976gauge, hatanaka1998gauge}.

\item {\bf Origin of the mass.} The Higgs Mechanism is introduced in the SMPP as the mechanism that generates the particle masses through a Yukawa-type interaction. The Higgs boson is the first scalar fundamental particle observed in nature. It gives masses to the fermions, $W$, and $Z$ bosons. However, the standard model does not tell us why this happens. It is also still not clear whether this particle is fundamental or composite, or if there are other Higgs bosons \cite{wilczek2012origins}.

\item {\bf Neutrino mass.} The SMPP predicts neutrinos to be massless\cite{pontecorvo1968neutrino}.  However, the experimental observation of neutrino oscillations implies that neutrinos are massive particles \cite{fukuda1998measurements}.  An extension of the standard model containing a massive right-handed, sterile neutrino can solve this problem. In such a model, the standard model neutrinos acquire mass, and the so-called seesaw mechanism \cite{yanagida1979horizontal, gell1979supergravity, minkowski1977mu} explains the smallness of their masses \cite{mohapatra1980neutrino}.

\item {\bf Flavor problem.} The flavor problem \cite{aranda2000u} is about the questions of why the SMPP contains precisely three copies of the fermions, and why are the masses of these fermions so hierarchical and are not in the same order. For example, it is not clear why the mass of the electron is about $0.511 \ \text{MeV}$, while the top quark has a mass of around $173 \ \text{GeV}$.

\item {\bf Matter-antimatter asymmetry.} According to the SMC, it is generally assumed that equal amounts of matter and antimatter should have been created after the big bang. However, the visible universe today appears to consist almost entirely of matter rather than antimatter. One of the current challenges in physics is to figure out what happened to cause this asymmetry between matter and antimatter \cite{ryden2017introduction}. A possible explanation could come from the study of CP-violation, which addresses a very fundamental question, are the laws of physics the same for matter and antimatter, or are matter and antimatter intrinsically different. The answer to this question may hold the key to solving the mystery of the matter-dominated universe.

\item {\bf The strong CP problem.} At the end of the previous section, we just mentioned that the strong CP problem results from the $\theta$-term in the QCD Lagrangian. This term contains the vacuum angle $\bar{\theta}$ with no apparent preferred value, while the current experimental limit sets a strong limit on its value$;$ that it must be $\bar{\theta} \lesssim 10^{-10}$. The problem of why the value of $\bar{\theta}$ is so small is known as the strong CP problem. It seems unlikely that the angle would be so close to zero by pure chance$;$ there should be a deeper explanation for this behavioral \cite{hooft1994symmetry, hooft1994computation}.

\item {\bf Inflation.} In cosmology, the better possibility to explain the puzzles that are the horizon problem, the flatness problem, and the origin of perturbations is to extend the SMC with inflation theory \cite{ryden2017introduction}. The theory assumes that the first fraction of a second of the big bang, the universe went through a stage of extremely rapid expansion called inflation. The theory usually requires adding a new heavy particle to the SMPP, provoking an acceleration of the universe expansion at its very early stages. This new particle is called the inflaton and is supposed to fill the whole universe, driving the dynamics of its expansion, before producing after a while the standard model particles that our world is made of today. Indeed, finding the correct theoretical description of inflation requires a new physics BSM, and it would not be easy to understand otherwise.

\item {\bf Dark matter and dark energy.} Because of the unexpected discovery that the acceleration of our universe expansion is not slowing down but instead speeding up, it became clear that our universe contains about $4.9\%$ of ordinary matter, $26.8\%$ of Dark Matter (DM), and $68.3\%$ of Dark Energy (DE) \cite{aghanim2018planck}. However, the nature of dark matter and dark energy, as well as the cause of the accelerated expansion of our universe, are still unknown \cite{bertone2005particle}. This problem represents one of the major unresolved issues in contemporary physics.
\end{itemize}

The standard model is unable to solve such questions, and these problems remain an open area for recent research, motivating us to continue looking for new physics beyond the standard model. At the time of writing this thesis, no evidence has been found for physics beyond the standard model. Nonetheless, the search for physics beyond the standard model is still an important guideline to new ideas proposed to answer these questions. In the following section, we discuss some of the approaches that are explored in this direction.

\section{Models beyond the standard model}

Despite the described criticism above, the incredible accuracy of the SMPP and the SMC leads to the suggestion that the standard model is simply incomplete rather than incorrect. Perhaps these models are only different effective parts or phases of a bigger picture that includes new physics hidden deep in the subatomic world or in the dark recesses of the universe. This is why a first step to build a new model that could address some of the problems is first to verify that it agrees with the predictions of the standard model. This is why many new models aim to expand the standard model rather than to provide an entirely new approach. Such models are typically called beyond standard model (BSM). There are a plethora of models that address the standard model problems in many different ways.

Accordingly, there is a lot of effort paid now in theoretical physics to introduce some new approaches beyond the standard model to solve some of its shortcomings. Hence we need to find an extension that tackles some or maybe all of these issues mentioned above to generalize the standard model. Some of them are enumerated below.

\begin{itemize}
\item  {\bf Supersymmetric theories.}  One of the most popular extensions to the standard model is supersymmetry, which is a fundamental symmetry between fermions and bosons introducing a set of new superpartners with opposite spin statistics for each standard model particle \cite{wess1989supergauge, wess1992supersymmetry}.  While the bosons have a positive contribution to the total vacuum energy, the fermionic contribution is negative. In non-supersymmetric theories, it is unreasonable that the fermions would exactly cancel the contributions of the bosons to give this small number. In supersymmetry, it is posited that every particle and its superpartner are degenerate in mass. Therefore, in addition to the usual term from quantum corrections to the Higgs mass from standard model particles, there would be a similar contribution to the Higgs mass with the opposite sign and the same magnitude from the superpartners. These two terms exactly cancel in the limit of exact supersymmetry, and there is thus no hierarchy problem.  Breaking of this symmetry at the electroweak scale could theoretically explain the small number. So far, though, no such symmetry has been found in nature.

\item {\bf Extra-dimensional theories.} Another exciting possible way to extend the standard model is by adding extra spatial dimensions to the common four-dimensional spacetime \cite{kaluza1921unitatsproblem, klein1991quantum}. These theories often reside at high energies and will, therefore, be manifest as effective theories at the low energy scale.  From the common four-dimensional point of view, particles that propagate through the extra dimensions will effectively be perceived as towers of heavy particles. The extra dimensions can be warped and provide an alternative solution to the hierarchy problem. In these models, the weak hierarchy is induced simply because the Planck scale is red-shifted to the weak scale by the warp factor. In general, these models explain the weakness of gravity by diluting gravity in a large bulk volume, or by localizing the graviton away from the standard model.

\item {\bf Grand unified theories.} This proposal to extend the standard model is exquisite because it attempts to unify the three gauge coupling of the standard model in one single one, and correspondingly the strong and electroweak forces unify into a single gauge theory \cite{ross1984grand}. This unification must necessarily take place at some high energy scale of order $10^{16} \ \text{GeV}$, and called the grand unified scale, at which all the three couplings become approximately equal. The central feature of these theories is that above the grand unified scale, the three gauge symmetries of the standard model tend to unify in one single gauge symmetry with a simple gauge group, and just one coupling constant. Below the grand unified scale, the symmetry is spontaneously broken to the standard model symmetries. Popular choices for the unifying group are the special unitary group in five dimensions $SU(5)$ and the special orthogonal group in ten dimensions $SO(10)$. Such theories leave unanswered most of the open questions above, except for the fact that it reduces the number of independent parameters due to the fact that there is only one gauge coupling at large energies. Unfortunately, with our present precision understanding of the gauge couplings and spectrum of the standard model, the running of the three gauge couplings does not unify at a single coupling at higher energies, but they cross each other at different energies and further model building is required\cite{allanach2016beyond}. In practice, further model building is required in order to make this work.

\item {\bf Theories of everything.} The idea of unifying the various forces of nature is not limited to the unification of the strong interaction with the electroweak one but extends to include gravity as well. Finding a theory of everything that thoroughly explains and links together all known physical phenomena is presently considered one of the most elusive goals of theoretical physics \cite{ellis1986superstring}. In practical terms, the immediate goal in this regard is to develop a theory that reconciles quantum mechanics and gravity and to build a quantum theory of gravity. Additional features, such as overcoming conceptual flaws in either theory or accurate prediction of particle masses, would be desired. The challenges in putting together such a theory are not just conceptual$;$ they include the experimental aspects of the very high energies needed to probe exotic realms. In the attempts to unify different interactions, several ideas and principles were proposed in this direction. String theory \cite{polchinski1998string} is one such reinvention, and many theoretical physicists think that such theories are the next theoretical step toward a true theory of everything.  Also, theories such as loop quantum gravity and others are thought by some to be promising candidates to the mathematical unification of quantum mechanics and gravity, requiring less drastic changes to existing theories. However, recent workplaces stringent limits on the putative effects of quantum gravity on the speed of light, and disfavors some current models of quantum gravity \cite{abdo2009limit}.

\end{itemize} 

\section{Motivation for an axion dark matter search}

At the present time, the dark matter mystery is one of the greatest common unsolved problems between the SMPP and the SMC. About $85\%$ of the universe's gravitating matter is nonluminous, and its nature and distribution are, for the most part, unknown \cite{komatsu2011seven}. Elucidating these issues is one of the most important problems in fundamental physics. Beginning with the nature of dark matter, one possibility is that it is made of new fundamental particles \cite{bertone2005particle}. We study dark matter scenarios in BSM physics and look for distinctive dark matter signature in direct and indirect dark matter searches, using astrophysical and cosmological probes.

A complete understanding of the nature of dark matter requires utilizing several branches of physics and astronomy. The creation of dark matter during the hot and rapid expansion of the early universe is understood through statistical mechanics and thermodynamics. Particle physics is necessary to propose candidates for dark matter content and explore its possible interactions with ordinary baryonic matter. General relativity, astrophysics, and cosmology dictate how dark matter acts on large-scales and how the universe may be viewed as a laboratory to study dark matter. Many other areas of physics come into play as well, making the study of dark matter a diverse and interdisciplinary field. Furthermore, the profusion of the ground and satellite-based measurements in recent years has rapidly advanced the field, making it dynamic and timely$;$ we are truly entering the era of precision cosmology \cite{garrett2011dark}.

In an attempt to explain the particle nature of the dark matter component, we study models of very light dark matter candidates called axions and axion-like particles. Axions are hypothetical elementary particles introduced by Peccei and Quinn to solve the CP problem of the strong interactions in quantum chromodynamics. Furthermore, there are more other particles that have similar properties to that of axions are postulated in many extensions of the standard model of particle physics and known as axion-like particles. One exciting aspect of axion and ALPs is that they might interact very weakly with the standard model particles. This property makes axions and ALPs plausible candidates that might contribute to the dark matter density of the universe. The capability of axions and ALPs to contribute to the discovery of the dark matter composition, in addition, to solve other problems in the standard model strongly encourages us to pursue researching for such interesting candidates of dark matter.
 
\section{Overview and outline of the thesis}

The standard model of particle physics, together with the standard model of cosmology, provides the best understanding of the origin of matter and the most acceptable explanation to the behavior of the universe. However, the shortcomings of the two standard models to solve some problems within their framework are promoting the search for new physics beyond the standard models. Dark matter is one of the highest motivating scenarios to go beyond the standard models. In this thesis, we focus on understanding the nature of dark matter through the study of the phenomenological aspects of axions and axion-like particles dark matter candidates, in cosmology and astrophysics.

In chapter \ref{ch2}, we review the current status of the research of the dark matter. We briefly explain the first hints that dark matter exists, elaborate on the strong evidence physicists and astronomers have accumulated in the past years, discuss possible dark matter candidates, and describe various detection methods used to probe the dark matter's mysterious properties. 

The theoretical backgrounds about the QCD axion, including the strong CP problem, the Peccei-Quinn solution, and the phenomenological models of the axion, are described. Then, the main properties of the invisible axions are briefly discussed in chapter \ref{ch3}. 

The possible role that axions and ALPs can play to explain the mystery of dark matter is the topic of chapter \ref{ch4}. To discuss whether they correctly explain the present abundance of dark matter, we investigate their production mechanism in the early universe. After, we discuss the recent astrophysical, cosmological, and laboratory bounds on the axion coupling with the ordinary matter.

In chapter \ref{ch5},  we consider a homogeneous cosmic ALP background (CAB) analogous to the cosmic microwave background (CMB) and motivated by many string theory models of the early universe. The coupling between the CAB ALPs traveling in cosmic magnetic fields and photons allows ALPs to oscillate into photons and vice versa. Using the M87 jet environment, we test the CAB model that is put forward to explain the soft X-ray excess in the Coma cluster due to CAB ALPs conversion into photons. Then we demonstrate the potential of the active galactic nuclei (AGNs) jet environment to probe low-mass ALP models and to potentially exclude the model proposed to explain the Coma cluster soft X-ray excess.
 
We turn our attention in chapter \ref{ch6} to consider a scenario in which ALPs may form Bose-Einstein condensate (BEC), and through their gravitational attraction and self-interactions, they can thermalize to spatially localized clumps. The coupling between ALPs and photons allows the spontaneous decay of ALPs into pairs of photons. For ALP condensates with very high occupation numbers, the stimulated decay of ALPs into photons is possible, and thus the photon occupation number can receive Bose enhancement and grows exponentially. We study the evolution of the ALPs field due to their stimulated decays in the presence of an electromagnetic background, which exhibits an exponential increase in the photon occupation number with taking into account the role of the cosmic plasma in modifying the photon growth profile. We focus on investigating the plasma effects in modifying the early universe stability of ALPs, as this may have consequences for attempts to detect their decay by the forthcoming radio telescopes such as the Square Kilometer Array (SKA) telescopes in an attempt to detect the cold dark matter (CDM) ALPs. 

Finally, chapter \ref{ch7} is devoted to summarize our arguments and to draw our conclusions. We point out that the research on axions and ALPs will be one of the main frontiers in the near future since the discovery of these particles can solve some of the common unresolved problems between particle physics and cosmology and take us a step forward towards understanding nature. For completeness, some useful notations and conversion relations are broadly outlined in the appendix.
\chapter{\textbf{General Context and Overview of Dark Matter}} \label{ch2}

During the last century, the development of astrophysical and cosmological observations have provided rich information that significantly improved our understanding of the universe. One of the most astounding revelations is that ordinary baryonic matter is not the dominant form of material in the universe, and instead, some strange invisible matter fills our universe \cite{garrett2011dark}. This new form of matter is known as the dark matter, and it is roughly five times more abundant than ordinary baryonic matter \cite{aghanim2018planck}. Although this strange dark matter has not been detected yet in the laboratory, there is a great deal of observational evidence that points to the necessity of its existence. The purpose of this chapter is to present a brief overview of the evidence for the existence of dark matter and study its properties, possible candidates, and detection methods.

\section{What is dark matter\boldmath$?$} \label{Sec.21}

In principle, direct information on cosmology can be obtained by measuring the spectrum of mass as it evolves with cosmic time, which could enable direct reconstruction of the present mass density of the universe \cite{cline2001sources}. But each type of observational test encounters degeneracies that can not be resolved without considering an additional sort of matter which behaves differently and can not be observed by all observational techniques. Maybe this new type of matter does not interact strongly enough with anything we can readily detect or see, and therefore it is basically invisible to us and is referred to as ``dark matter’’. The name dark matter is really just a label for the hole in our fundamental understanding of nature, that something is missing and we do not know what it is. Precisely, it is called dark because it does not emit, reflect, or absorb light, and since it has no identifiable form, so it is called by the most generic word, matter, as it behaves like matter. The only reason we believe dark matter makes up bout $85\%$ percent of the mass of the known universe is because of its observable gravitational effects \cite{lisanti2017lectures}. It is hardly the first time that scientists have invoked unseen entities to account for phenomena that seemed inexplicable at the time. Eventually, some such ghostly entities turned out to be real, and the rest were disproved \cite{livio2004dark}.

\section{ Evidence of existence of dark matter} \label{Sec.22}

In this section, we review the observational and astrophysical evidence for the presence of dark matter at a wide variety of scales from the scale of the smallest galaxies to clusters of galaxies and cosmological scales.

\subsection{The discovery of Neptune}

Let us here consider a historical precedent that might be related to dark matter$:$ the discovery of planet Neptune. Early in the 19th century, astronomers noticed that the newly discovered planet Uranus was not following the predicted orbit suggested by Newton's theory of gravitation. Some speculated that an unseen world was tugging on it. In 1846, Le Verrier assumed that this unseen world is an undiscovered planet, and accordingly, he calculated its expected location in the sky based upon Newton's laws \cite{le1846recherches}. Then this hypothetical planet was observed later in 1846 by Galle and is known now as planet Neptune \cite{galle1846account}.

We refer to this historical incident because it is similar to the current situation of the dark matter. Before the discovery of Neptune, it was just a theoretical-hypothetical to represent the unseen mass or ``dark matter’’. Of course, we know now that Neptune is not part of dark matter, but the idea of discovering something missed or unseen should be the same.

\subsection{Discovery of missing mass ``dark matter’’}

The evidence for dark matter comes for the first time in 1933, when the Swiss astronomer Fritz Zwicky studied the movement of galaxies in the nearby Coma cluster  (99 Mpc away from the Milky Way) \cite{zwicky1933rotverschiebung, babcock1939rotation}. The magnitudes of the velocities of galaxies with respect to each other were found to be much faster than the expected values that would be consistent with gravitational confinement based on the potential well arising from the visible matter alone. The argument was based on estimating the Coma cluster dynamical mass using the virial theorem \cite{jeans1921dynamical, hillstatistical}. The theory provides a general equation relating the average total kinetic energy $ \langle K \rangle$ with the  average total potential energy $\langle V \rangle$ of a abound system in equilibrium
\begin{equation} \label{eq.2.1}
2 \langle K \rangle + \langle V\rangle = 0 \:.
\end{equation}
For the moment, if the Coma cluster consists of $N$ galaxies, then its average total kinetic energy can be written as
\begin{equation} \label{eq.2.2}
\langle K \rangle = \frac{1}{2} \sum_i^N m_i v_i^2 = \frac{1}{2} v^2 \sum_i m_i = \frac{1}{2} M_{\text{c}} v^2 \:,
\end{equation}
where $v$ is the average velocity of the whole galaxies, and $M_{\text{c}}$ is the total mass of the Coma cluster. Then suppose that the Coma cluster is spherical, its average gravitational potential energy is approximately
\begin{equation} \label{eq.2.3}
\langle V \rangle = - \frac{1}{2} \sum_i^N  \sum_{j>i} \frac{G m_i m_j}{R_{ij}} = - \frac{3}{5} \frac{ GM^2_{\text{c}}}{R_{\text{c}}} \:,
\end{equation}
where the sum is taken over all possible pairs of galaxies. Here $R_{ij}$ is the effective distance between two galaxies and $R_{\text{c}}$ is the effective radius of the Coma cluster. Thus, we have obtained an expression for the two terms in equation \eqref{eq.2.1}, that leads to read
\begin{equation}
v^2 = \frac{3}{5} \frac{ G M_{\text{c}}}{R_{\text{c}}} \:.
\end{equation}
The last formula provides a method to estimate the total mass of the cluster in terms of its average velocity. The unexpected observations based on the total luminosity mass of the Coma cluster shows that the average velocity of the cluster galaxies was so high that
\begin{equation}
v^2 \gg \frac{3}{5} \frac{ G M_{\text{c}}}{R_{\text{c}}} \:.
\end{equation}
This implies that the Coma cluster may not obey the virial theorem, and the cluster is not even a gravitationally bound system. In such a case,  the system is dominated by kinetic energy and individual galaxies should escape from the cluster and hence the cluster must decay. But this situation is not consistent with observations. Another possible scenario is that the system is in virial equilibrium, but it contains a more gravitational potential, and accordingly, there is much more additional nonluminous matter. That said, the proof of the existence of such a nonluminous matter may only come with its direct detection \cite{sanders2010dark}. Therefore, Zwicky concluded that there must be a large amount of invisible matter within the cluster that he termed ``dark matter’’. Nevertheless, this being viewed as too unconventional, so the idea was not taken seriously and was ignored at that time.

The possibility that this missing matter would be non-baryonic was unthinkable at this time$;$ this evidence was pointed out as the first hint for the existence of dark matter. Zwicky's estimation of the discrepancy between the mean density of the Coma cluster obtained by his observation and the mean density derived from luminous matter was about a factor $400$ or $100$. Observations of X-ray emitting hot gas in the galaxy clusters reveal that most of baryonic content is in the form of such hot gas and its mass easily exceeds by a factor $3\textup{--}4$ the mass of stars in the individual galaxies \cite{jones1984structure}. Going back to Zwicky, who did not count on this hot gas, this means that a fraction of the missing matter was actually found. Modern observations confirm this discrepancy but were reduced to a factor of $5\textup{--}6$. We know that luminous stars make up only a tiny $1\%$ of the total cluster mass. The additional matter is $14\%$ in the form of a baryonic hot intracluster medium, and the remaining $ 85\%$ in the form of dark matter \cite{aghanim2018planck}.

\subsection{Rotation curves of spiral galaxies}

\begin{figure}[t!]
\centering
\includegraphics[scale=0.5]{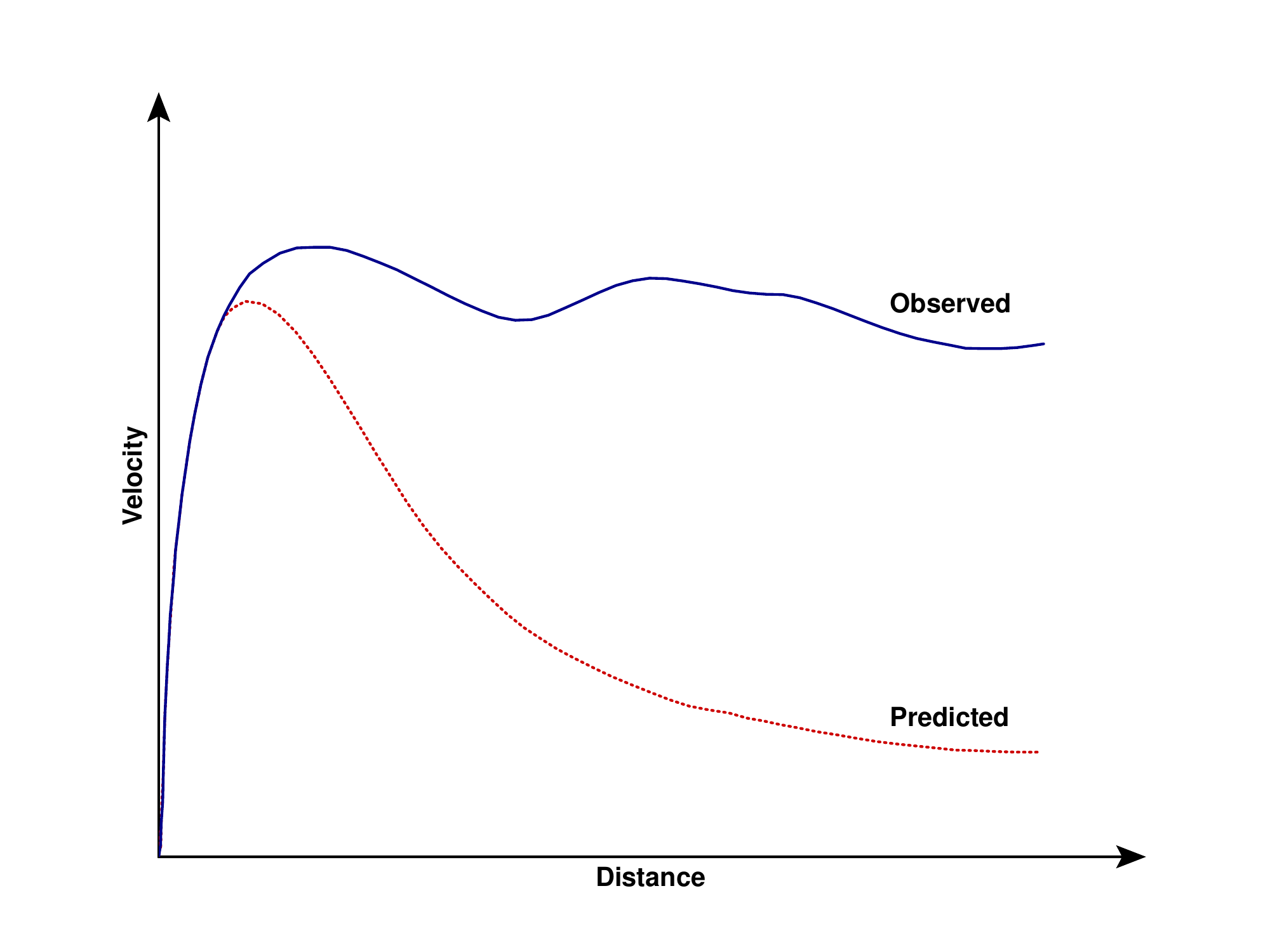} 
\caption{Rotation curve of a typical spiral galaxy, predicted and observed.}
\label{Fig.2.1}
\end{figure}

One of the most substantial evidence of the existence of dark matter has been found in the 1970s, because of the significant contributions of the astronomers Rubin, Ford, and Thonnard in studying the rotational motion of stars in spiral galaxies \cite{rubin1970rotation, rubin1978extended, rubin1980rotational}. Once again, explaining the results required the existence of vast amounts of invisible matter. In their work, they were measuring the rotational velocities of spiral galaxies via redshifts and used the data to calculate the corresponding mass distribution $M(r)$. Then they were comparing the results of the observations with the expected values that have been calculated by applying  Newton's law to a spherically symmetric distribution of mass. The comparison should give
\begin{equation}
F = \frac{G m M(r)}{r^2} = \frac{m v^2(r)}{r} \:,
\end{equation}
where $v(r)$ is the rotational velocity, $r$ represents the distance from the galactic center, and $M(r)$ is the mass enclosed within the distance $r$. The rotation curve is defined as the plot of the rotational velocity as a function of $r$, and it can be described by the following formula
\begin{equation} \label{eq.2.7}
v(r) = \sqrt{\frac{G M(r)}{r}} \:.
\end{equation}
Suppose the fact that spiral galaxies are composed of a central dense bulk and a thin disk in the outer region is considered. Therefore the great amount of the mass is located in the central bulk. If we assume a constant density of the bulk, so for $ r \ll R_c$, the mass increases as the volume $(\propto r^3)$,  while at large distances $r \gg R_c$ the mass should be independent of $r$. Inserting this information into equation \eqref{eq.2.7} gives us an expected rotation curve
\begin{equation}
v(r) \propto 
\begin{cases} 
\:  r      & \quad r \ll R_c \:, \\
\: r^{-1/2} & \quad r \gg R_c \:.
\end{cases}
\end{equation}
Thus, it is expected that the velocity should start increasing linearly with $r$ until reaching a maximum, and then it should fall off. But the general result does not agree with this expectation. Instead of falling off at large $r$, the observed rotation curves remain flat with increasing $r$, at least to values of $r$ comparable with the disk radius, see a typical example in figure \ref{Fig.2.1}. To explain this unexpected flat behavior, one could assume a modified theory of gravity or the existence of a more considerable amount of invisible mass, which only interacts gravitationally, extending further beyond the limits of the visible galaxy. Thus the dark matter explanation imposes itself as one of the most powerful solutions to the rotation curves problem in the galactic scale.

\subsection{Gravitational lensing} 

\begin{figure}[t!]
\centering
\includegraphics[scale=0.4]{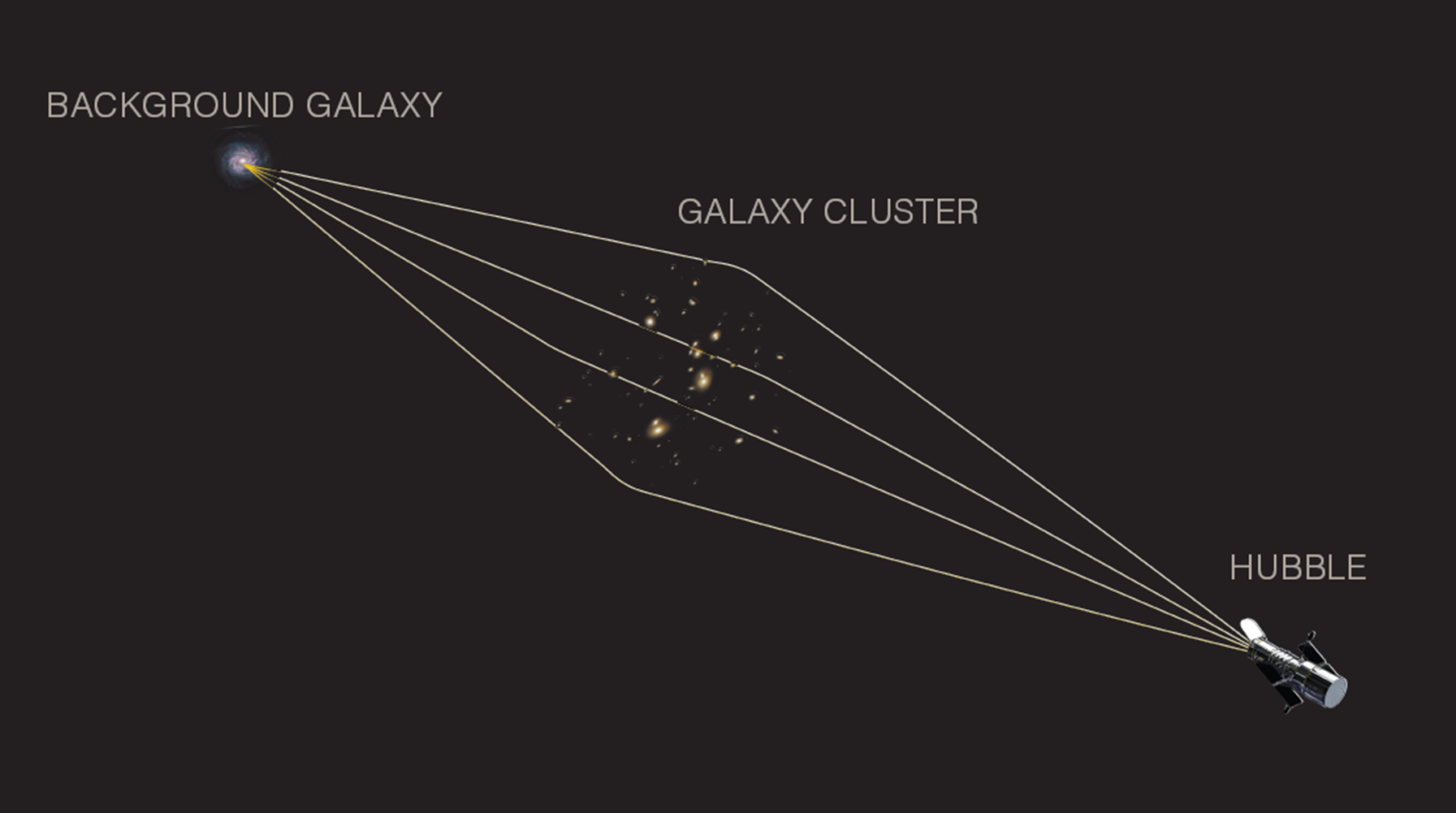}
\caption[An illustration of gravitational lensing. Light from a distant galaxy is bent by a foreground galaxy, and when this light is observed on Earth, we see virtual images of the distant galaxy at different sky positions. The light also gets magniﬁed, and this is visible in the top image, which is brighter than the lower image. The shape of the top images is also altered compared to the original shape of the galaxy. Image credit$:$ NASA/JPL-Caltech, 2010.]{An illustration of gravitational lensing. Light from a distant galaxy is bent by a foreground galaxy, and when this light is observed on Earth, we see virtual images of the distant galaxy at different sky positions. The light also gets magniﬁed, and this is visible in the top image, which is brighter than the lower image. The shape of the top images is also altered compared to the original shape of the galaxy. Image credit$:$ NASA/JPL-Caltech, 2010.}
 \label{Fig.2.2}
\end{figure}  

Another very solid evidence of the presence of dark matter comes from studying galaxy clusters. The method is used to make direct measurements of the total mass of the cluster based on the gravitational lensing effect \cite{einstein1911influence, einstein1936lens}. In particular, this technique exploits the main principle of general relativity that massive objectives cause curvature of spacetime. Therefore if clusters were replete with more dark matter, then this additional mass ought to produce a deeper divot in the fabric of spacetime, thereby a greater altering in the paths of light rays in the universe. Accordingly, the cluster would act as a giant lens, distorting the images of galaxies behind it. This phenomenon is known as strong gravitational lensing, see figure \ref{Fig.2.2}. The angular separation between the different images is
\begin{equation}
\theta = \sqrt{4GM \frac{d_{\text{LS}}}{d_L d_S}} \:,
\end{equation}
where $M$ is the mass of the object acting as the lens, $d_{\text{LS}}$ is the distance from the lens to the source and $d_L$, $d_S$ are the distances from the observer to the lens and the source, respectively. Hence, the size and the distances involved in these images captured by telescopes are directly linked to the mass of the lensing foreground cluster. When a large gravitational mass is located between a background source and the observer and leads to the bending of light, this effect is very apparent and is called strong lensing. But if the source that causes the bending of light is located exactly behind a massive circular object in the foreground, a complete ``Einstein ring’’ appears, in more complicated cases, like a background source that is slightly offset or a lens with a complex shape, one can still observe arcs or multiple images of the same source. The mass distribution of the lens can then be inferred by the measurement of the ``Einstein radius’’ or, more in general, by the positions and shapes of the source objects. Once again, the total measured mass of the lens is not in agreement with the evaluation of the luminosity mass. This leads once more to think that a large fraction of the mass of the clusters is composed of dark matter. This technique has also been used to create the first 3-dimensional maps of the dark matter distribution in the cosmic space and provides evidence for the large-scale structure of matter in the universe and constraints on cosmological parameters.

\subsection{Bullet cluster}

\begin{figure}[t!]
\centering
\includegraphics[scale=0.52]{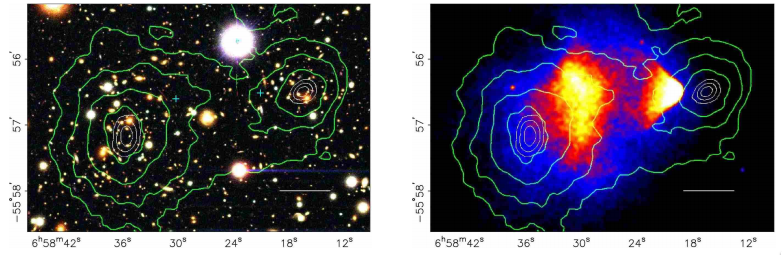} 
\caption[Matter distribution in X-rays (colors) and using the gravitational lensing (contours) of the Bullet cluster.]{Matter distribution in X-rays (colors) and using the gravitational lensing (contours) of the Bullet cluster. Figure taken from reference \cite{clowe2006direct}.}
\label{Fig.2.3}
\end{figure} 
 
Dark matter existence could also be inferred from the comparison between the luminosity mass of a cluster and the mass determined by the X-ray emission of its electron component$;$ for more detail about this method, see references \cite{sadat1997clusters, piffaretti2008total}. This allows inferring the temperature of the gas, which in turn gives information about its mass through the equation of hydrostatic equilibrium. For a system with spherical symmetry, the equation of hydrostatic equilibrium implies
\begin{equation}
\dfrac{d P}{d r} = - \frac{G M(r) \rho(r)}{r^2} \:,
\end{equation}
where $P$, $M(r)$, and $\rho$ respectively are the pressure, mass, and density of the gas at radius $r$. In order to rewrite this formula in a more suitable form in terms of the temperature $T$, we can use the equation of state for ideal gas, $PV = N k_B T$, where $V$ is the volume of the gas, $k_B$ is the Boltzmann constant, and $N$ is the total number of electrons, and ionized nuclei in the gas, which can be expressed as $N = M/m_p \mu$, where $M$ is the total mass of the gas, $m_p$ is the proton mass and $\mu \simeq 0.6$ is the average molecular weight. Since $M/V = \rho$, the equation of hydrostatic equilibrium for an ideal gas reads now
\begin{equation}
\dfrac{d \log \rho}{d \log r} + \dfrac{d \log T}{d \log r} = - \frac{G \mu m_p M(r)}{k_B T(r) r} \:.
\end{equation}
The temperature of clusters is roughly constant outside their cores, and the density profile of the observed gas at large radii roughly follows a power-law with an index between $-2$ and $-1.5$. Therefore, for the baryonic mass of a typical cluster,  the temperature should obey the relation
\begin{equation}
k_B T \approx (1.3 - 1.8) \ \text{keV} \left( \frac{M(r)}{10^{14} \mathrm{M}_{\odot}} \right) \left( \frac{1 \ \text{Mpc}}{r} \right) \:.
\end{equation}
The disparity between the temperature obtained using this calculation, and the corresponding observed temperature, $k_B T \approx 10 \ \text{keV}$, when $M(r)$ is identified with the baryonic mass, suggests the existence of a substantial amount of dark matter in clusters.

Based on this method, one of the most direct empirical evidence for the existence of dark matter can be extracted from studying the Bullet cluster, which is a cluster formed out of a collision of two smaller clusters \cite{clowe2006direct}. When the two galaxy clusters pass each other, the ordinary matter components collide and slow down, while on the other hand, the dark matter components pass each other without interaction and slowing down. It seems that the collision between the two galaxy clusters has led to the separation of dark matter and ordinary matter components of each cluster. This separation was detected by comparing X-ray images of the luminous matter taken with the Chandra X-ray Observatory with measurements of the cluster's total mass from gravitational lensing observations, see figure \ref{Fig.2.3}. This method not only gives evidence of the existence of dark matter but also allows finding the locations of the dark matter and ordinary matter in the cluster and reveals some differences between their behaviors. Although the two smaller clumps of ordinary matter are moving away from the center of the collision with low speeds, the two large clumps of dark matter are moving in front of them with higher speeds. This fact points out the collisionless behavior of the dark matter components, and this implies that the self-interactions of the dark matter components must be very weak. It seems interesting that this evidence can be counted as direct evidence for dark matter, as it is independent of the details of Newtonian gravitational laws.

\subsection{Evidence on cosmological scales}

The analysis of the cosmic microwave background \cite{fixsen1996cosmic, penzias1965measurement, dicke1965cosmic} is another useful tool not only for providing proof of dark matter but also for determining the total amount of dark matter in the universe. The CMB is the thermal radiation relic leftover from the early universe stages at redshift $z \sim 1100$ around $380, 000$ years after the big bang. The CMB consisted of photons emitted during the recombination era when the free electrons and protons\footnote{There is an approximation here that all baryons in the universe at this time are in the form of protons.} combine into neutral hydrogen atoms, thus allowing photons to decouple from matter and stream freely across the universe. In this way, the range of photons increases immediately from very short length scales to very long length scales and is quickly followed by their decoupling. The spectrum of the CMB is well described by a blackbody function at a temperature different from matter due to this decoupling. The blackbody radiation at the recombination temperature evolves into blackbody radiation in the present universe at a lower temperature. Because the temperature is proportional to the mean photon energy, which has redshifted with the expansion of the universe, the photons retain information about the state of the universe at the recombination timescale, and thus carries remnant information about the general properties of matter in the early universe.

\begin{figure}[t!]
\centering
\includegraphics[scale=.44]{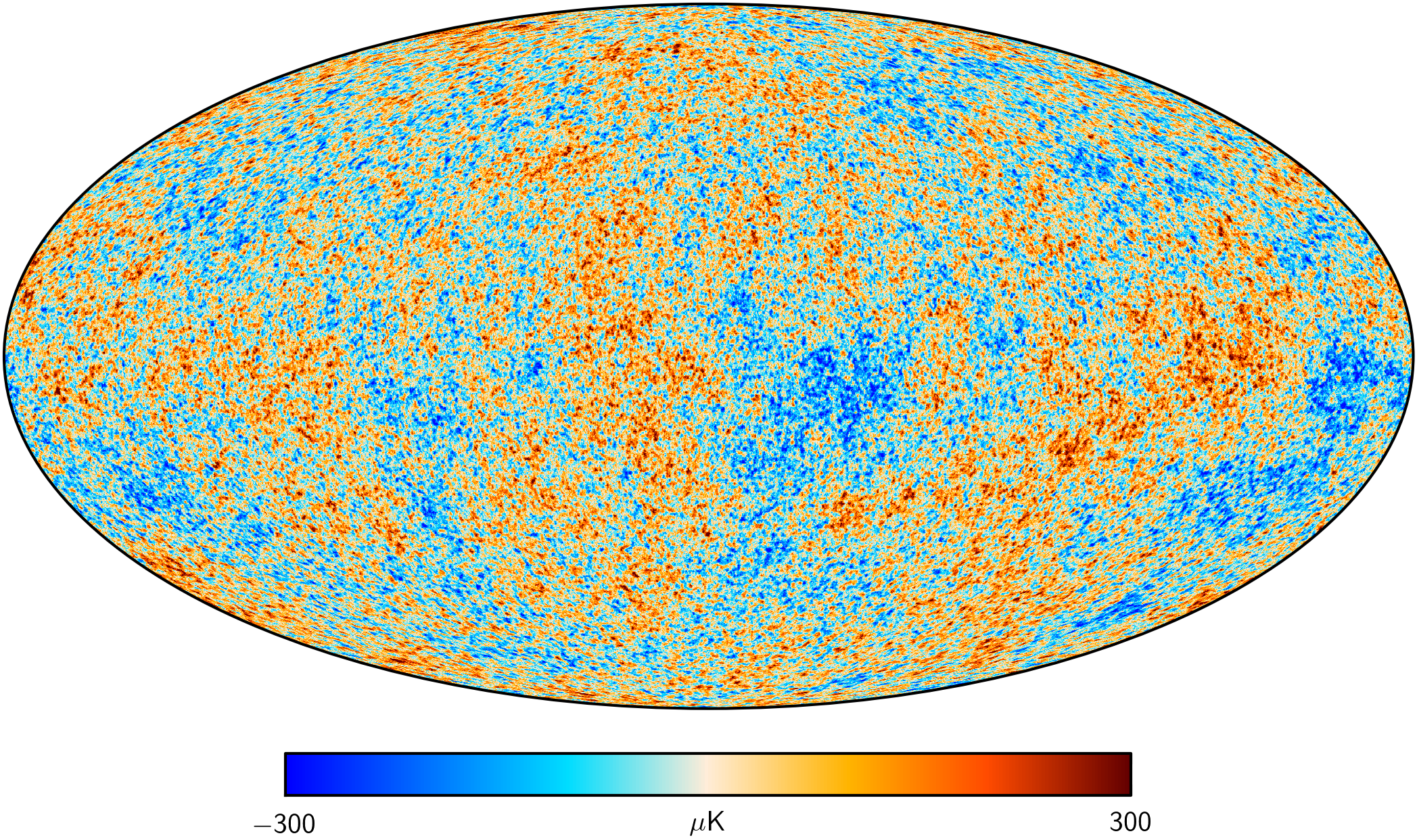} 
\includegraphics[scale=.44]{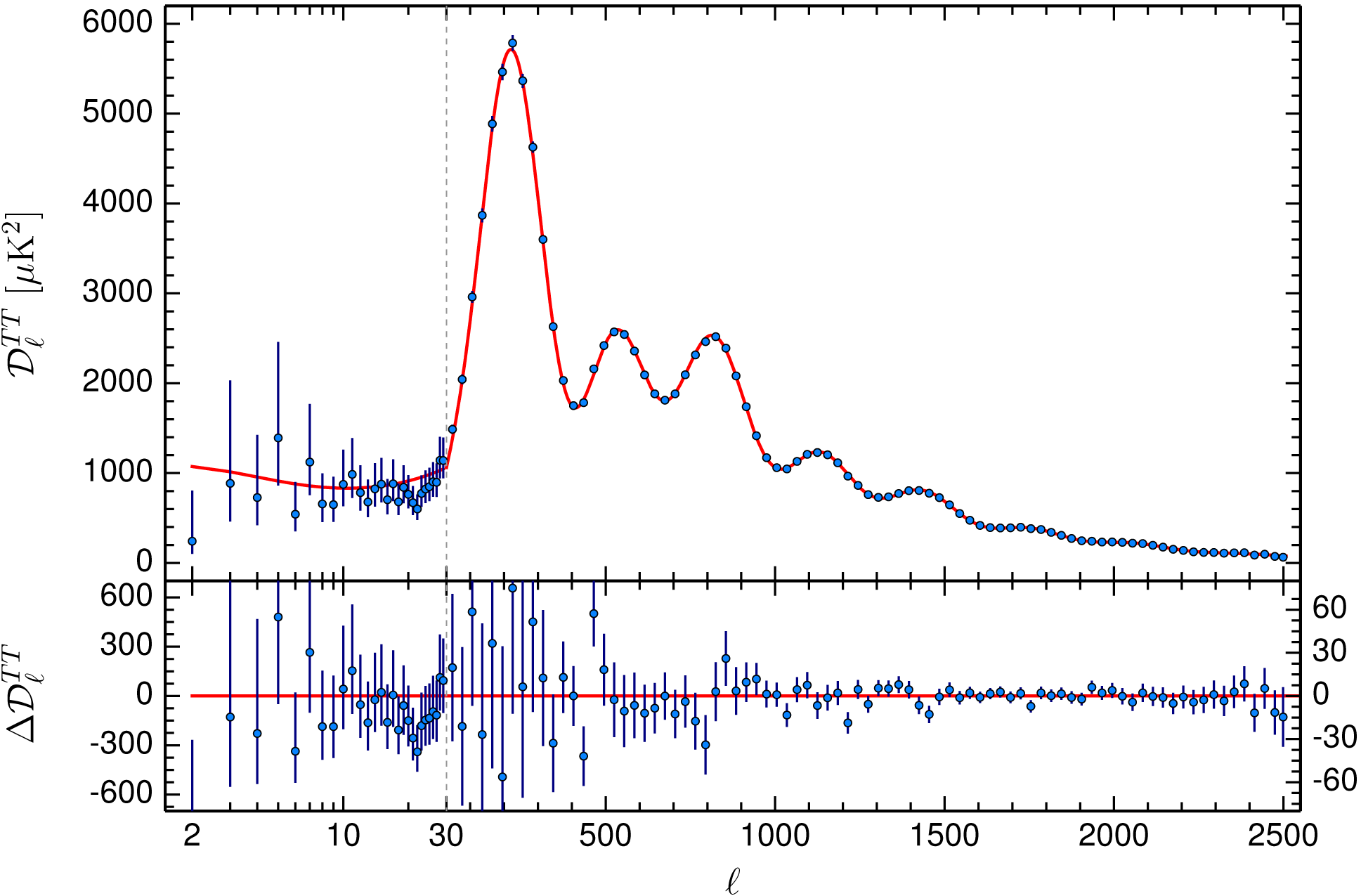} 
\caption[Left panel$:$ The full-sky map of the temperature anisotropies of the CMB as observed by Planck satellite in 2015. Right panel$:$ Planck 2015 temperature power spectrum data in blue and the best-fit base $\mathrm{\Lambda} \text{CDM}$ theoretical spectrum in red.]{Left panel$:$ The full-sky map of the temperature anisotropies of the CMB as observed by Planck satellite in 2015. Image credit$:$ ESA and the Planck Collaboration. Right panel$:$ Planck 2015 temperature power spectrum data in blue and the best-fit base $\mathrm{\Lambda} \text{CDM}$ theoretical spectrum in red. Image credit$:$ ESA and the Planck Collaboration.}
\label{Fig.2.4}
\end{figure}

At the present date, one can observe this CMB as a radiation with a perfect blackbody spectrum at temperature $T_0=2.725 \ \text{K}$ or energy $k_B T =2.35 \times10^{-4} \ \text{eV}$. Precision measurement of the anisotropies in the angular distribution of temperatures of the CMB sky map the presence of overdensities and under densities in the primordial plasma before recombination, see the left panel of figure \ref{Fig.2.4}. For this reason, one can read information on the baryon and matter distribution of the universe in the spectrum of CMB anisotropies. The observed temperature of the CMB as a function of the angular position in the sky $\theta$ only differs by a small amount from the mean, and therefore it represents anisotropies as a temperature difference
\begin{equation}
\frac{\mathrm{\Delta} T}{T} (\theta/ \phi) = \frac{T(\theta, \phi) - \bar{T}}{\bar{T}} \:.
\end{equation}
Represent this temperature difference as a function of position using an equivalent Fourier series in spherical coordinates, which are spherical harmonics, reads
\begin{equation}
\frac{\delta T}{T} (\theta, \phi) = \sum_{\ell=1}^{ \infty} \sum_{m=- \ell}^{\ell} a_{\ell m} Y_{\ell m} (\theta, \phi) \:,
\end{equation}
where $Y_{\ell m}(\theta, \phi)$ are spherical harmonics and $a_{\ell m}$ are the multipole moments. Taking into account the fact that on large scales the sky is extremely uniform, the anisotropies are extremely small$:$ $\delta T/T \sim 10^{-5}$, then the variance $C_\ell$ of a given moment can be defined as
\begin{equation}
C_\ell \equiv \langle \vert a_{\ell m} \vert^2 \rangle \equiv \frac{1}{2 \ell+1} \sum_{m=-\ell}^{\ell} \vert a_{\ell m} \vert^2 \:.
\end{equation}
On small sections of the sky, the universe is relatively flat, and therefore the spherical harmonic analysis becomes ordinary Fourier analysis in two dimensions. In this limit, $\ell$ becomes the Fourier wavenumber. Since the angular wavelength is $\theta= 2\pi/\ell$, large multipole moments correspond to small angular scales. From observations, it appears to be a reasonable approximation that the temperature fluctuations are Gaussian, see the right panel of figure \ref{Fig.2.4}. Consequently, the information from the CMB can be accurately represented as a function of the multipole moment.

Now it is essential to understand the causes and meaning of the underlying anisotropies, which give rise to the so-called acoustic peaks in the power spectrum in the right panel of figure \ref{Fig.2.4}. They are primarily the result of a competition between baryons and photons from the time of the baryon-photon plasma. The pressure of the relativistic photons works to erase temperature anisotropies, while the heavy non-relativistic baryons tend to form dense halos of matter, thus creating sizable local anisotropies. The competition between these two effects creates acoustic waves in the baryon-photon plasma and is responsible for the observed acoustic oscillations.

\begin{figure}[t!]
\centering
\includegraphics[scale=.50]{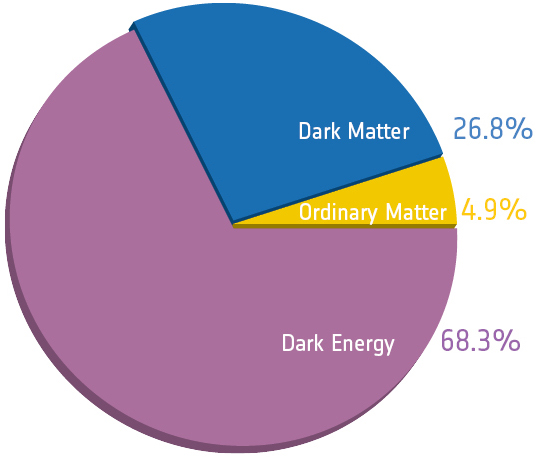} 
\caption[The contents of the universe, according to recent results from the Planck Satellite.]{The contents of the universe, according to recent results from the Planck Satellite. Image credit$:$ ESA and the Planck Collaboration.}
\label{Fig.2.5}
\end{figure}

Each peak of this distribution can be related to a cosmological parameter, thus providing a means of constraining cosmology through measurements of the CMB. The most recent such measurements come from the Plank Collaboration \cite{aghanim2018planck}.  According to this observation, the energy content of the universe is comprised of $68.3\%$ dark energy, $26.8\%$ dark matter, and $4.9\%$ baryonic matter, see figure \ref{Fig.2.5}. This provides compelling evidence for the existence of dark matter in large abundances throughout the universe.
 
\section{The need for non-baryonic dark matter}

For the moment, the evidence presented in the previous section already sufficient to conclude that most of the matter in the universe is in the form of dark matter. The nature of this dark matter, whether it is baryonic or non-baryonic, is not yet known. Although we are still open to the possibility that at least a portion of the dark matter content is baryonic, there are strong cosmological pieces of evidence that make us biased to the other hypothesis of non-baryonic dark matter. In this section, based on \cite{bertone2005particle, coc2004updated, gondolo1991cosmic, kolb1981early, sarkar2002high} we clarify this issue.

\subsection{Big bang nucleosynthesis}

The modeling of the early universe by the standard big bang model gives a scenario that led to the present cosmic abundances of elements outside the stars. According to the theory, our universe is thought to have begun as an infinitesimally small, infinitely hot, infinitely dense singularity point around $\sim 14$ billion years ago. Some theories suggest that immediately after the big bang, the universe was too hot for any matter to exist and filled with an unstable form of energy whose nature is not yet known. Directly after, the universe starts expanding and cooling down to the point where quarks could condense from the energy until a sort of thermal equilibrium existed between matter and energy. Then things were cool enough for quarks, leptons, and gauge bosons to associate with one another and form protons and neutrons, and more familiar particles like electrons and photons begin to appear. The temperature drops during the universe expansion until the neutrons and protons have frozen out. Then about one minute after the big bang, when the universe has cooled enough, the big bang nucleosynthesis (BBN) begins to create the light elements. A few hundred thousand years later, the universe had cooled enough for electrons to be bound to nuclei, forming atoms. This is known as the recombination era at which the universe first became transparent to radiation. Before that, photons were scattered by the free electrons, making the universe opaque. It is the radiation emitted during this recombination that makes up the cosmic microwave background radiation that we can still detect today. As things continued, the universe had cooled sufficiently that stars, galaxies, clusters, and superclusters have been formed during a long period of dark ages.
 
\begin{figure}[t!]
\centering
\includegraphics[scale=.50]{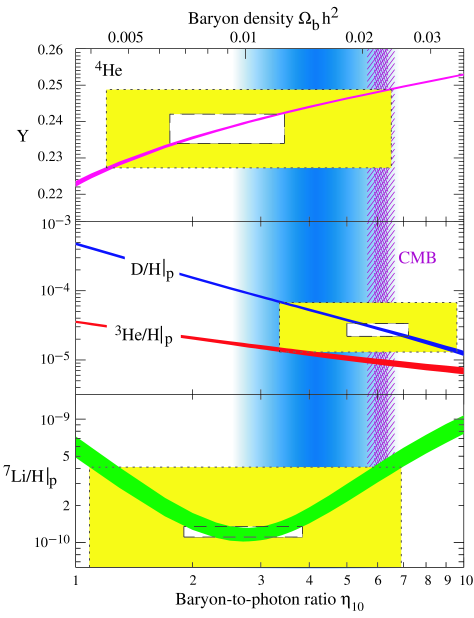} 
\caption[BBN predictions for the abundances of light-elements as a function of the baryon over photon ratio compared to observations.]{BBN predictions for the abundances of light-elements as a function of the baryon over photon ratio compared to observations. Figure is taken from reference \cite{bertone2005particle}.}
\label{Fig.2.6}
\end{figure}

What interests us here is that the production of the light nuclei ${}^2\text{H}, {}^3\text{He}, {}^4\text{He}$, and ${}^7\text{Li}$ occurred exclusively during the first few minutes of the big bang, while the heavier elements are thought to have their origins in the interiors of stars which formed much later in the history of the universe. Consequently, no more light elements can be formed after the BBN. Also, these elements can not be easily destroyed in stellar interiors. Therefore the baryonic contribution to the total mass contained in the universe can be determined based on the measurements of the present abundances of these light elements in cosmic rays. The resulting elemental abundances depend only on the nuclear reaction rate and the baryon-to-photon ration $\eta$ at the time. The nuclear reaction rate can be calculated using theoretical as well as laboratory analysis. While the parameter $\eta$ is the present ratio $\eta$ of the baryon number density $n_b$ to the photon number density $n_\gamma$. This quantity is directly related to the value of $\mathrm{\Omega}_b h^2$ deduced by Planck as
\begin{equation}
\eta \equiv \frac{n_b}{n_\gamma} \approx 2.738 \times 10^{-8} \: \mathrm{\Omega}_b h^2 = 6.11 \pm 0.04 \times 10^{-10} \:,
\end{equation}
where $\mathrm{\Omega}_b$ represents the baryon density of the universe, and $h \approx H_0/100 \ \text{km} \ \text{s}^{-1} \ \text{Mpc}^{-1} = 0.6727  \pm 0.0066$ is the dimensionless Hubble parameter. Matching the observed abundances of the light elements in today's universe to the prediction by the BBN requires satisfying the condition$:$
\begin{equation}
\mathrm{\Omega}_b h^2 = 0.02225 \pm 0.00016 \:.
\end{equation}
The BBN predictions for the abundances of the light elements as a function of the baryon over photon ratio compared to observations is shown in figure \ref{Fig.2.6}. This measurement exemplifies the best-fit BBN prediction and gives a good agreement with the values inferred from the CMB power spectrum that the baryon density is of $\mathrm{\Omega}_b \approx 0.04$. Combining this with the total matter density inferred from the CMB observations, BBN not only confirms the existence of dark matter but is also one more evidence for its non-baryonic nature. The independent measurements of the total mass density of the universe from large-scale structures provide that $\mathrm{\Omega}_m \approx 0.29$. This is directly implied that the remaining $\mathrm{\Omega}_{\textit{leftover}} \approx 0.25$ must be dark matter. Furthermore, we have another invaluable piece of information about its nature$:$ we see immediately that dark matter must be predominantly non-baryonic. Being non-baryonic allows the possibility that the dark matter is not capable of interacting with photons electromagnetically, which sits very well with the fact that it does indeed appear dark. This is also consistent with the dark matter being dissipationless$;$ were it able to absorb and re-emit photons, it would obscure stars within distant halos, as well as radiate away angular momentum and collapse with baryons to form stellar and galactic disks.

\subsection{Cosmic structure formation}

The final cosmological argument for the need for dark matter arises from the growth of cosmic structure from the nearly uniform distribution of baryonic matter imprinted on the CMB to the wide variety of structures observed today. A perfectly homogeneous expanding universe stays that way forever$;$ there will be no structures. However, this is not the case for our universe, as evident from the universe density map and galaxy redshift surveys. It appears that the present-day distribution of matter is very non-homogeneous, at least on scales up to about $100 \ \text{Mpc}$.

The standard theory for the formation of structure assumes that the universe was initially almost perfectly homogeneous, with a tiny modulation of its density field. The action of gravity then works to enhance the density contrast as time goes on, leading to the formation of galaxies or clusters when the self-gravity of an overdense region becomes large enough to decouple itself from the overall Hubble expansion. In this sense, one can use the value of the density fluctuations at the last scattering surface to estimate the recent value of it.

At a very early time, CMB indicates that the universe had a thermal distribution at the recombination era with temperature fluctuations $\delta T /T \sim  10^{-5}$. Because radiation and baryons were coupled before recombination, the distribution of the baryon density should also be with density fluctuations $\delta \rho /\rho \sim  10^{-5}$. Now let us look at how fluctuations in density evolve with time. For simplicity, we consider a flat and matter-dominated universe with $\mathrm{\Omega}_m = 1$ as a good approximation for the early times. The Friedmann equation can be written as
\begin{equation}
H^2 - \frac{8}{3} \pi G \rho = - \frac{kc^2}{R^2(t)} \:.
\end{equation}
If now we look at a small area of this universe which has a little more matter than the rest. Because of the excess matter, it evolves slightly differently as
\begin{equation} 
H^2 - \frac{8}{3} \pi G \rho' = - \frac{kc^2}{R^2(t)} \:.
\end{equation}
Subtracting one equation from the other, one easily gets the following expression for the fractional overdensity of the region$:$
\begin{equation}
\frac{\delta \rho}{\rho} \equiv \frac{\rho -  \rho'}{\rho} = - \frac{3 k c^2}{8 \pi G R^2 \rho} \:.
\end{equation}
This implies that the fractional overdensity of the region depends on the evolving quantities $R$ and $\rho$, as follows$:$
\begin{equation}
\frac{\delta \rho}{\rho} \sim \frac{1}{R^2 \rho} \sim \frac{1}{R^2 R^{-3}} \sim R  \:.
\end{equation}
Therefore as the universe gets bigger, the overdensity grows along with it. But since size and redshift are related $R= (1+z)^{-1}$, we can calculate the ratio between the today value of the expansion parameter and its value after the last scattering surface. The CMB is a picture of the universe at a redshift of $z \sim 1100$, while the present should have no redshift $z=0$. Thus,
\begin{equation}
\frac{R_{\text{today}}}{R_{\text{LSS}}} \sim 10^3 \:.
\end{equation} 
If there is only baryon matter and since $\delta \rho_b / \rho_b$ can not growth until the last scattering surface. Therefore
\begin{equation}
\left( \frac{\delta \rho}{\rho} \right)_{\text{today}} = 10^3 \times \left( \frac{\delta \rho}{\rho} \right)_{\text{LSS}} \approx 10^{-2} \:.
\end{equation}
This can not explain today's observations for the density fluctuation in the Milky way, for example, which is about $10^5$.  This implies that the density of a galaxy is much greater than the density of the universe as a whole, and taking into account only the baryonic content of matter in the universe can not explain the very large value of density fluctuations. The existence of such non-linear structures today implies that the growth of fluctuations must have been driven by the non-baryonic dark matter, which was not relativistic at recombination. After recombination, baryons decouple and fall into these overdensities before recombination, creating gravitational wells that the baryons can later collapse into after recombination.

\section{Basic properties of dark matter} \label{Sec.23}

Gradually it also becomes clear that dark matter is required not only to keep galaxies and clusters stable but also to structure the entire universe. It provides the scaffolding for the formation of stars, galaxies, and clusters. We strongly believe that the dynamic of astrophysical and cosmological systems, ranging from the size of the galaxy to the whole universe, can not be explained without assuming the existence of dark matter. At the same time, the identity of dark matter has far-reaching implications also in particle physics. It may lead to improve our understanding of the possible new physics beyond the standard model of particle physics. Therefore we discuss in this section some of the main properties that must characterize the identity of any possible dark matter candidate. The ideas in this section were discussed in \cite{sahni20045, lazarides2007particle, matarrese2011dark}.

\begin{itemize}

\item {\bf Relic abundance.} Astrophysical measurements indicate that the relic dark matter density accounts for about $85\%$ of the total matter density in the universe. It is possible that the dark matter particles were produced during the very early stages of the universe age after the big bang by either standard thermal production via scattering interactions in the thermal bath or nonthermal mechanisms. Therefore a good dark matter candidate must be able to be produced through such mechanisms under the early universe conditions with the correct abundance to account for this relic density.

\item {\bf Electrically neutral.} Dark matter is not observed to shine, and can not be detected by telescopes. For this reason, dark matter is considered to be optically dark. Therefore, a successful dark matter candidate must be electrically neutral or at least have very weak electromagnetic interactions. This required either the dark matter particle has vanishing, or at least very small, electric charge and electric and magnetic dipole moments, or the particles must be very heavy. The major consequence of this behavior is that the dark matter does not couple too strongly to photons and consequently can not cool by radiating photons. Thus, it will not collapse to the center of galaxies as the baryons do by radiating their energy away electromagnetically, and this means that the dark matter is very nearly dissipationless.

\item {\bf Interaction strength.} For the moment, all the astrophysical and cosmological observations indications for the presence on the dark matter is because of its gravitational effects. Therefore, dark matter particles should not interact or at least have very weak interactions with only photons and electrically charged particles, but must not couple too strongly to other electrically neutral particles as well. The reason for this is that all interactions between dark matter and baryons except gravity need to be very weak during the recombination epoch$;$ otherwise, there would be changes in the CMB acoustic peaks and would allow the dark matter to radiatively cool and affect the structure formation.

\item {\bf Collisionless.} Furthermore, as we just mentioned above, astrophysical and cosmological observations indicate that dark matter interacts through gravity only, and if the dark matter particles possess any other interactions, they must be very weak. In addition, the dark matter particles to be consistent with observations, they must be nearly collisionless and has no self-interaction, or at least several astrophysical constraints must be placed on their self-interactions.

\item {\bf Temperature.} Dark matter was probably non-relativistic (``cold’’) during the formation of large scale structures, as relativistic particles would cause the universe to be less clumpy than it is today. As a result, one can conclude that relativistic particles can not constitute the majority of the dark matter, instead, it should be dominated by non-relativistic species. Moreover, cold dark matter is capable of explaining the observed properties of galaxies quite well. However, this does not rule out totally the possibility that a fraction of dark matter consists of nearly massless particles moving at relativistic velocities (``hot’’). One more possibility is the modification of (``warm’’) dark matter particles. These particles were just relativistic at their decoupling temperature and cooled down to be non-relativistic at the time of matter-radiation equality. Another possibility can be a mixed dark matter composed of several distinct particle species with different temperatures.

\item {\bf Stability.} Dark matter must be either very stable on cosmological timescales or at least long-lived with lifetimes comparable to or longer than the present age of the universe. This is to make sure that they can survive from the early universe stage until now. Another possible scenario to make it stable is that its destruction and formation must be in equilibrium.

\item {\bf Non-baryonic nature.} Another important conclusion derived from observations is that the total contribution of the mutter content is about $29\%$ of the total energy density of the whole universe. In contrast, the results of the CMB, together with the predictions from BBN, suggest that only about $4\%$ of the total energy budget of the universe is made out of the ordinary baryonic matter. This directly implies that at least the majority of dark matter must be non-baryonic and constitutes about $2\%$ of the total energy-dense of the universe.

\item {\bf Fluid.} Dark matter must behave as a perfect fluid on large scales, which means that the granularity of the dark matter is sufficiently fine not to have been directly detected yet through various effects.

\end{itemize}

\section{Dark matter candidates} \label{Sec.26}

Despite the overwhelming success of dark matter in explaining cosmic phenomena, we do not know what particle or particles it is made of. There are a great number of dark matter candidate particles, some of which are considered more promising than others. Unfortunately, observations do not place stronger bounds on possible candidates. In this section, we outline some of the most popular candidates for the dark matter particles. While this list is by no means exhaustive, we will attempt to cover the range of possibilities that have been considered at least qualitatively. We will proceed very roughly from the smallest mass candidates at $10^{-22} \ \text{eV}$ to the largest at $10^{72} \ \text{eV}$ \cite{baltz2004dark}.

\subsection{WIMPs}

Weakly interacting massive particles (WIMPs) are suggested to be one of the very plausible candidates for the CDM component in the universe \cite{jungman1996supersymmetric, bergstrom2000non}. They represent a catch-all term for a bunch of particles that interact with a strength of the order of the weak-force interaction. Therefore they are supposed to interact with the ordinary matter only through their gravitational and the weak forces. In addition, it is suggested that they of large masses near the weak scale, that is between $10 \ \text{GeV}$ and a few $\text{TeV}$. This class of candidates is particularly interesting because of the WIMP miracle. The miracle is that the total mass of WIMPs has been independently predicted by a few different particle physics theories, and it is approximately the same mass as that required to explain all the extra gravitational force.  It is also probably the largest class of dark matter particles, firstly, as it consists of hundreds of suggested particles. Secondly, because its rest masses exceed those of the baryons, therefore could account for lots of dark matter if, as most theories predict, they are common in the universe.

\subsection{Axions and axion-like particles}

Axions arise in the Peccei-Quinn solution for the problem of CP violation in the theory of strong interactions \cite{peccei1977cp}. On the first hand, they represent an essential extension of the standard model by offering solutions to its internal problems. On the other hand, they are considered as a promising and well-motivated dark matter candidate. If axions do exist, they meet all the requirements of a successful cold dark matter candidate. They interact weakly with the baryons, leptons, and photons of the standard model. Also, they are non-relativistic during the time when the structure begins to form. Moreover, they are capable of providing some or even all of the CDM density. In addition, they are relatively light and electrically neutral. Although the axion mass is arbitrary over the range $10^{-6} \ \text{eV}$ to a few $\text{eV}$, the symmetry breaking occurs at a high energy scale and hence early in the universe. Axion theories predict that the universe is filled with a very cold Bose-Einstein condensate of primordial axions, which never comes into thermal equilibrium. The axions in this condensate are always non-relativistic, and therefore are CDM candidate and not HDM candidate as one would suppose according to their mass.

Furthermore, there are abundant theories beyond the standard model such as supersymmetric theories and string theory models that predict many other particles very similar to the QCD axions and called axion-like particles. In general, the properties of the more general class of ALPs are very similar to that of the QCD axions and are determined by their coupling with two photons. The main difference between the two categories is that the mass of the QCD axions is related to the coupling parameter, while this is not necessarily the case for the APLs, making the corresponding parameter space larger. In addition to their properties that make them very viable candidates for the dark matter, their ability to solve other problems in physics from different theoretical origins makes the study of the phenomenology of axions and ALPs an extraordinary exciting subject for the current research. More technical details about the theoretical origin and the characteristics of such particles will be discussed in the following two chapters.

\subsection{Fuzzy dark matter} 

Fuzzy dark matter (FDM) \cite{hu2000fuzzy}, motivated by string theory, has recently become a hot candidate for dark matter that may resolve some of the current small scale problems of the cold dark matter model. It is also known as ultralight axions, ultralight scalar dark matter, or wave dark matter. The main idea of such a model is that the dark matter particles are made of ultra-light bosons in Bose-Einstein condensate state.  The rest mass of FDM is believed to be about $10^{-22} \ \text{eV}$, and the corresponding de Broglie wavelength is $\sim 1 \ \text{kpc}$. Therefore, the quantum effect of FDM plays an important role in structure formation.

\subsection{Sterile neutrinos}

Dark matter could also be a special form of neutrino, which is tricky because neutrinos are already special. As we know, the standard model neutrinos come in three different flavors, electron, muon, and tau neutrinos. They can shift between flavors as they travel through space \cite{gonzalez2008phenomenology}. Dark matter neutrinos, if they exist, they are predicted to interact with ordinary matter only when they flip between flavors. What supports this hypothesis is that neutrinos already fulfill a lot of the qualifications of a dark matter candidate. They have mass and are weakly interacting. However, given their relic density, they are not massive enough to contribute to the total amount of dark matter in the universe significantly. Furthermore, their low rate of interaction and low mass corresponds to a long free streaming length, meaning that they could not account for the structure formation on scales below $\sim 40 \ \text{Mpc}$. Therefore the standard model neutrinos are excluded. However, sterile neutrinos, meaning that they do not interact via the weak force, with a mass of at least $\sim 10 \ \text{keV}$, have been proposed as a candidate.

\subsection{Supersymmetric candidates}

Supersymmetry is one of the most popular proposals for physics beyond the standard model and is flexible enough to offer several dark matter candidates \cite{jungman1996supersymmetric, roszkowski1993supersymmetric}. Supersymmetric theories extend the symmetry properties of spacetime in ordinary quantum field theory to provide a link between bosons and fermions, \ie between particles with integer spins and particles with half-integer spins. If supersymmetry exists in nature, then each standard model particle should have a corresponding superpartner with the opposite spin-statistics. So far, there has been no unbroken supersymmetry observed in nature, and it is thus obvious that if nature is described by supersymmetry, it has to be broken at a certain high energy scale. Indeed, if supersymmetry is realized at low energy world like the one in which we live, it must be broken. It is common in supersymmetric theories with conserved R-parity that the supersymmetry breaking allows for a mass difference between the superpartners. In this framework, the lightest superpartner (LSP) is absolutely stable and only weakly interacting, and therefore it happens to be a very suitable dark matter candidate. The most studied supersymmetric dark matter candidates are listed downward.

\begin{itemize}
\item {\bf Gravitinos.} The first dark matter candidate has been proposed by supersymmetry was the gravitino \cite{pagels1982supersymmetry, feng2003superweakly1, feng2003superweakly2}. It is a spin $3/2$ particle and the superpartner of the graviton in local supersymmetry, \ie supergravity. If the gravitino is the LSP in some models, it is often quite light, \ie in order of $\text{keV}$, and would thus be considered to be a warm dark matter candidate. The overproduction of gravitinos is somewhat problematic in cosmology, though not insurmountably so. Noting that gravitinos should not be sufficiently heavy as in this scenario, they decay during or after the BBN giving rise to the famous ``gravitino problem’’. For these reasons, gravitinos can not account for the entire relic density of the dark matter composition.

\item {\bf Neutralinos.} The lightest neutralino is the most favored supersymmetric dark matter candidate. Several supersymmetric models contain four neutralinos characterized as electroweakly interacting particles with spin $1/2$ \cite{weinberg1982cosmological, goldberg1983constraint}. If the lightest neutralino is the LSP, it is considered as a perfect dark matter candidate. The successes of such models depend on the fact that the very light neutralinos are sufficiently suitable to account for the proper relic abundance of the dark matter compositions.

\item {\bf Sneutrinos.} Another interesting possibility for supersymmetric dark matter candidates is the partners of the neutrinos. However, this candidate is quite disfavored because one can not explain a relic density consistent with the dark matter as the sneutrinos annihilate very effectively, and this required their masses to be above $500 \ \text{GeV}$. Giving their quite large elastic scattering cross sections on nuclei, they would be easily detectable in current experiments, which is not the case for the dark matter yet \cite{hagelin1984perhaps, ibanez1984scalar, falk1994heavy}.

\item {\bf Axinos.} The axino is the supersymmetric partner of the axion and shares many of the same properties.  The theory does not predict the mass of the axino, and it could, in principle, be the lightest supersymmetric particle rather than the neutralino. Axinos may be produced in decays of heavier supersymmetric particles and thereby achieve a relic density appropriate to a dark matter candidate. They might be either warm or cold dark matter depending on the conditions in the early universe \cite{rajagopal1991cosmological}.

\item {\bf Q-balls.} Supersymmetric theories predict the existence of stable non-topological solitons known as Q-balls. They are characterized by their relatively strong self-interactions and can be absolutely stable if large enough. Therefore, they are suggested as promising candidates for the collisional dark matter \cite{kusenko1998supersymmetric}.

\item {\bf Split SUSY.} It is argued recently that the successes of supersymmetric models hinge on the fact that the gauginos have masses at the weak scale while having the scalar superpartners at the weak scale is somewhat problematic. The null results of all low energy supersymmetric searches lead to the development of Split SUSY models, which are based only on gauge unification and dark matter as guiding principles. If the Higgs fine-tuning problem is simply ignored, the scalar superpartners can be made very heavy, keeping light gauginos and Higgsinos. This seems to be an exciting scenario to new supersymmetric dark matter candidates \cite{wells2003implications, arkani2005supersymmetric}.
\end{itemize}

\subsection{Lightest Kaluza-Klein particle}

There are exotic dark matter candidates that arise in models with universal extra dimensions. According to such models, our four-dimensional spacetime may be embedded in a higher-dimensional space at high energy or short distance. This has some interesting theoretical ramifications such as the unification of forces, introduced as a concept by Kaluza in 1921 when he unified electromagnetism and gravity in a 5-dimensional theory \cite{appelquist2001bounds}. The additional spatial dimensions could be compactified, meaning that they are finite and probably very small. The simplest case would be that a compactified dimension is in the shape of a circle. In such a circle, the momentum traveling through it would be quantized. In contrast, the standard model physics would exist in the lowest state, with no momentum going in the direction of the compactified dimension, there would exist a possibility of excitations, holding a ladder of excited Kaluza-Klein (KK) states \cite{antoniadis1990possible}. This would correspond to additional particles outside the standard model, and potentially more possible dark matter candidates.

\subsection{Chaplygin gas}

One very exciting way was proposed in an attempt to unify dark energy and dark matter in a class of a simple cosmological model based on the use of a peculiar perfect fluid dubbed Chaplygin gas \cite{chaplygin1944gas, bilic2002unification, fabris2002density}. This perfect fluid is characterized by the following exotic equation of state $p = -A/\rho$, where $A$ is a positive real parameter. The pure Chaplygin gas has been extended to the so-called generalized Chaplygin gas with the following equation of state $p = -A/\rho^{\alpha}$,  where $\alpha$ is another positive real parameter \cite{kamenshchik2001alternative, bento2002generalized}. This type of fluid can arise in certain string-inspired models involving d-branes. The interesting feature of this model is that it naturally provides a universe that undergoes a transition from the decelerating phase driven by dust-like matter at a moderate redshift to the accelerating universe at later stages of its evolution.

\subsection{Mirror matter}

The concept of hidden sectors of matter is the modern version of the old idea of a mirror copy of the world$;$ see for review \cite{foot2004mirror}. In this scenario, the dark matter could just be ordinary matter in the mirror world and, thus, the universe consisting of our knowen world and the mirror world. The matter in the mirror wold is constrained to interact with our world only very weakly, and the only significant communication is gravitational. This scenario can be constructed in a braneworld context, where our world and a mirror world are two branes in a higher-dimensional space.

\subsection{Branons}

String theory naturally contains objects of many different dimensions, called branes \cite{cembranos2003brane}. These would naturally have fluctuations characterizable as particles, the so-called branons. These branons are massive and weakly interacting particles that have been proposed to explain the missing matter problem within the so-called brane-world models. In this framework, these branons can be made into suitable cold dark matter candidates, both thermal and non-thermal.

\subsection{WIMPzillas}

This scenario grounds on the presence of non-thermally produced superheavy particles called WIMPzillas. These particles may be produced from gravitational interactions at the end of inflation. For mass scales of $10^{13}$ GeV, if these particles are stable or at least longlived, they may account for the dark matter \cite{chung1998nonthermal, chung1998superheavy}. In addition, such particles might decay with a lifetime much longer than the age of the universe, providing a source of ultra high energy cosmic rays$;$ this is the so-called ``top-down’’ scenario for UHECR production.

\subsection{Primordial black holes}

A viewable option to explain the nature of dark matter is that it is composed of small-mass black holes called primordial black holes (PBHs), which might have been formed under the right conditions in the very early universe \cite{chapline1975cosmological}. The production of these PBHs is enhanced during periods where the equation of state softens ($p < \rho/3$), such as during a first-order phase transition. This is easy to understand as when the pressure support lessens, objects collapse quickly and form the PBHs. The last such phase transition in the universe is the quark-hadron phase transition, at a temperature $T \sim 100$ MeV. The mass contained in the horizon at this epoch is very roughly a solar mass.  Since the PBHs are produced before the BBN, they are considered non-baryonic, non-relativistic, and effectively collisionless and thus could be a very promising candidate for the cold dark matter in the universe.

\subsection{Self-interacting dark matter}

This is an increasingly popular option, assuming that dark matter is not just one particle but a collection of particles \cite{carlson1992self}. Just as the ordinary matter has a whole bunch of different particles, so too could dark matter. But because ordinary particles and dark particles do not interact much, we may never know. Perhaps the only way we could observe these particles would be indirectly, through their gravitational effect on the evolution of the cosmos.

\section{Dark matter searches} \label{Sec.25}

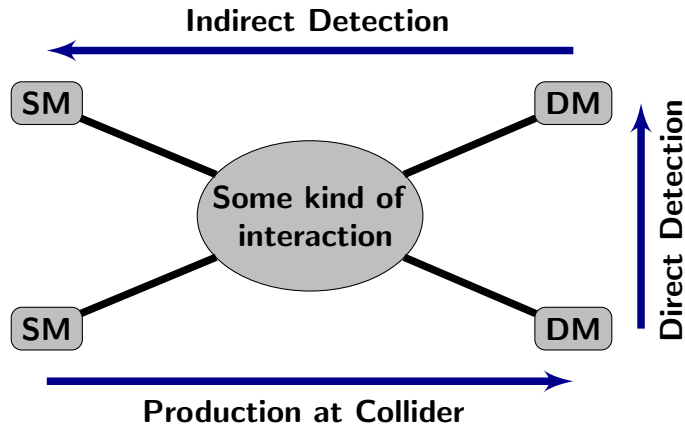
\begin{figure} [t!]
\centering
\begin{tikzpicture} 
[square/.style={draw, rounded corners, minimum width=width("#1"),  minimum height=width("#1")+2*\pgfshapeinnerysep, node contents={#1}}]

\draw [fill=lightgray]   (0,0) ellipse (1.5cm and 1cm);  

\node at (3.5,1.5) (v100) [square={\bf{DM}}, fill=lightgray]; 
\node at (3.5,-1.5) (v100) [square={\bf{DM}}, fill=lightgray];
\node at (-3.5,1.5) (v100) [square={\bf{SM}}, fill=lightgray]; 
\node at (-3.5,-1.5) (v100) [square={\bf{SM}}, fill=lightgray];              

{
\newcounter{tmp}  
\foreach \s in {latex'} { \stepcounter{tmp} \begin{scope}[yshift=-\thetmp cm] \draw[darkblue,line width=1.0mm,   arrows={-\s}] (-3.5,-1.2)-- (3.5,-1.2); \end{scope}}
\foreach \s in {latex'} { \stepcounter{tmp} \begin{scope}[yshift=-\thetmp cm] \draw[darkblue,line width=1.0mm,   arrows={-\s}] (4.4,0.5) --+ (0,3.0); \end{scope}}
\foreach \s in {latex'} { \stepcounter{tmp} \begin{scope}[yshift=-\thetmp cm] \draw[darkblue,line width=1.0mm,   arrows={-\s}] (3.5,5.2) -- (-3.5,5.2); \end{scope}}
}

\draw[black, line width=1.0mm] (-3.05,-1.3)--(-1.25,-0.55);
\draw[black, line width=1.0mm] (1.25,0.55)--(3,1.3);
\draw[black, line width=1.0mm] (3.0,-1.3)--(1.25,-0.55);
\draw[black, line width=1.0mm] (-1.25,0.55)--(-3.05,1.3);

\draw[black] (1.4,0.25) node[above,left] {\bf{Some kind of}};
\draw[black] (1.25,-0.25) node[above,left] {\bf{interaction}};
\draw[black] (2.2,-2.6) node[above,left] {\bf{Production at Collider}};
\draw[black] (2.0,2.6) node[above,left] {\bf{Indirect Detection}};
\draw[black] (4.8,1.7) node[above,left, rotate=90] {\bf{Direct Detection}};
  
\end{tikzpicture}
\caption{Schematic diagram of different strategies to detect a dark matter particle based on the distinction interaction processes.}
\label{Fig.2.61}
\end{figure}

So far, we have presented the most popular possibilities for the dark matter candidates. If dark matter exists, tremendous numbers of such dark matter particles are expected to fill the entire universe and be moving around everywhere and could even be passing through Earth each moment. Indeed, a given dark matter candidate must meet a bevy of conditions to be consistent with astrophysical and cosmological observations$;$ however, some of these conditions might be relaxed in the context of a suitable scenario. Although such observations constrain the measurement of impressive large scale quantities, they do not allow for the precise determination of the properties of the dark matter particles. Therefore, many experiments are currently working to detect the dark matter particles to probe their properties and the fundamental theory or at least the existence of one of them. However, because of their weak interactions with ordinary matter, they are very hard to detect. The diagram used to group all possible interactions between dark matter and the standard model particles is shown in figure \ref{Fig.2.61}. Based on the distinction interaction processes, there are three different complementary ways to detect dark matter$;$ direct detection, indirect detection, and collider detection. These different categories are discussed in this section based on materials elaborated in detail in references \cite{bertone2010particle, baer2005direct, klasen2015indirect}.

\subsection{Direct detection}

The direct detection is one of the most promising ways to search for dark matter particles. Currently,  there are numerous direct dark matter experiments are running around the world. Cosmic rays that are constantly coming towards the Earth collide with the upper atmosphere and create a shower of particles reaching the surface of the Earth. Thus it is essential to construct these dark matter detectors deep underground to avoid the background from these cosmic rays. 

The main idea for the direct detection experiments is to search for observable signals based on the effects of galactic dark matter particles passing through the detector. These effects may be in the form of elastic or inelastic scatters of the dark matter particles with atomic nuclei, or with electrons in the detector material. The result of this type of interaction could be the production of heat or light, which can be measured. Such interaction can be represented as
\begin{equation}
\text{DM} + \text{SM} \rightarrow \text{DM} + \text{SM} \:.
\end{equation}
Especially of interest, there are some experiments that use different materials in their detectors, and various detection techniques have claimed to found some hints for a dark matter particle. However, their results are not entirely consistent with each other and also are in disagreement with results from several different experiments that have not found any observable signal for dark matter. This makes these results still controversial and so must be clarified by new experimental data.

\subsection{Indirect detection}

In parallel to the direct detection experiments, a wide range of indirect detection projects have been developed to search for dark matter particles. The dark matter may annihilate or decay under certain conditions to produce standard model products, including charged particles, neutrinos, and photons. The detection of such products would constitute an indirect detection of dark matter. This type of interaction is described by
\begin{equation}
\text{DM} + \text{DM} \rightarrow \text{SM} + \text{SM} \:.
\end{equation}
For example, the annihilation of two dark matter particles can produce gamma rays that can be detected by space or ground-based gamma-ray telescopes. This example is of quite an interest because gamma-ray photons can travel across long astrophysical distances and would allow the identification of annihilation sources. Therefore, there is a wide array of cosmic-ray and gamma-ray observatories searching for indirect dark matter signals from astrophysical sources.

\subsection{Collider detection}

It may also be possible to produce dark matter particles and detect it in the laboratory. The idea here is allowing the collision between ordinary standard model particles coming from opposite directions at tremendous energies and very high speed in the hopes that heretofore undiscovered particles will emerge from the collisions. The search on this case is based on a possible process that is the inverse of the annihilation process
\begin{equation}
\text{SM} + \text{SM} \rightarrow \text{DM} + \text{DM} \:.
\end{equation}
The famous example based on this technique is the experiments in the Large Hadron Collider (LHC) at CERN that collides two beams of protons together at the highest energies and then looks for dark matter signals. Unfortunately,  none of these experiments have yet been able to detect any effect attributed to dark matter. In addition, there is no guarantee that any given particle observed at colliders is the astrophysical dark matter particle, even if it has the same properties listed above. In order to resolve this issue, the particle properties acquired from collider signals must be verified with those obtained from signals arising from astrophysical sources. 
\chapter{\textbf{The Strong CP Problem and Axions}} \label{ch3}

The discussion in this chapter provides an overview of the problem of CP violation in the theory of strong interactions and its solution$;$ for a thorough review of this topic, see, for example, references \cite{peccei1989strong, peccei2008strong, kim2010axions, dine2000tasi, dine2017axions}.  We start by introducing the $U(1)_A$ problem and its solution, which in turn begets the strong CP violation problem. After a brief look at other proposed solutions, we will describe the generally preferred Peccei-Quinn solution to the strong CP violation problem, and the resulting axion, together with some experimental considerations. The ability of axions to solve the strong CP problem as well as their predicted properties make them extraordinary exciting candidates for the CDM particle in the universe.

\section{The chiral QCD symmetry and the \boldmath$U(1)_A$ problem}

As we discussed in section \ref{Sec.1.2}, the quantum chromodynamics sector is the gauge theory that describes the strong interactions in the standard model of particle physics. The classical QCD Lagrangian for $N_f$ flavors reads \cite{peskin2018introduction}
\begin{equation}
\L_{\text{QCD}} =  - \frac{1}{4}  G^a_{\mu \nu} G_a^{\mu \nu}  + \sum_f \bar{q}_f (i \slashed{D}_s - m_f) q_f  \:,
\end{equation}
where $q_f$ are the quark fields with flavor $f$ , which runs over all presently known six flavors$;$ $u, d, s, c, b, \ \text{and}\ t$, with the corresponding masses $m_f$, and  $\slashed{D}_s$ is the covariant derivative contains the coupling between the quarks and gauge fields, which is defined as
\begin{equation}
 \slashed{D}_s = \gamma^{\mu} (\partial_{\mu} + ig_s \tau^a G_{\mu}^{a}) \:,
\end{equation}
where $g_s$ is the strong interaction coupling strength, $\tau^a$ are the generators of the $SU(3)_C$ group, and $G_{\mu}^{a}$ represents the $8$ gluon fields of the strong interaction.  The classical QCD Lagrangian possesses local $SU(N_c)$ color symmetry, with $N_c = 3$. In the limit that all quark masses vanish, the left-handed and right-handed fermions decouple, and the model shows a further exact global $G= U_L(N_f) \otimes U_R(N_f)$ chiral symmetry. More technically, all transformations which treat left-handed and right-handed fields separately are chiral transformations. We can then conclude that the global $U(2)\times U (2)$ flavor symmetry is an exact symmetry of only the massless theory since the mass term breaks chiral symmetry explicitly. Therefore, the left-handed and right-handed charges are decoupled and operate separately. Each of them generates an $SU(N_f)$ group of transformations. The whole chiral group is then decomposed into the direct product of two $SU(N_f)$ groups, which will be labeled with the subscripts L and R, respectively. For example, if we consider the two flavor theory with the $u$-quark and $d$-quark, the chiral symmetry implies
\begin{equation}
\left( \begin{matrix} u \\ d \end{matrix} \right) \rightarrow (U_L P_L + U_R P_R) \left( \begin{matrix} u \\ d \end{matrix} \right) \:,
\end{equation}
where $P_{L, R} \equiv \frac{1}{2} (1\pm \gamma^5)$ are the usual Dirac projection operators that produce left and right projections when operating on the Dirac spinors $u$ and $d$. In this case, $U_{L, R}$ are unitary two-by-two matrices, and $\gamma^5$ are the gamma matrices. 

Before going further in our argument, some important notes have to be mentioned. It is possible that symmetry can be exact or approximate. If symmetry is clearly realized in all states seen in nature, this is referred to as an exact symmetry. And if symmetry is valid only under certain conditions, it is known as an approximate symmetry. Another possibility is that symmetry may be spontaneously or explicitly broken.  According to the Goldstone theorem \cite{goldstone1961field}, the spontaneous breaking of an exact symmetry gives rise to massless scalar bosons, known as Goldstone bosons or sometimes also as Nambu-Goldstone bosons (NGBs). Moreover, the Goldstone theorem states that the number of massless particles is equal to the number of generators of the spontaneously broken symmetry. In contrast, the explicit breaking of an exact symmetry gives rise to a kind of pseudo-Nambu-Goldstone bosons (PNGBs), which are massive but light. Equivalently the spontaneous breaking of an approximate symmetry gives rise as well to PNGBs.

In fact, the assumption of massless quarks is very sensible if we only consider the two lightest quarks$;$ $u$-quark, and $d$-quark, because their masses are much smaller than the typical QCD scale $\mathrm{\Lambda}_{\text{QCD}}$. Even considering three flavors would be acceptable, nevertheless worse, approximation, since the mass of the $s$-quark introduced in theory is comparable to the $\mathrm{\Lambda}_{\text{QCD}}$.

We are now getting back to our main argument. The chiral QCD symmetry is exact in the quarks massless limit and spontaneously broken by the vacuum to an approximate large global symmetry $U(N_f)_v \otimes U(N_f)_A$ when we consider their real values. The first part of this symmetry $[U(N_f)_v = SU(N_f)_I \otimes U(1)_B]$, represents a vectoral symmetry term that consists of isospin symmetry times the global baryonic symmetry respectively. Both of these two symmetries have been realized in nature. The second part corresponds to the axial term $[U(N_f)_A = SU(N_f)_A \otimes U(1)_A$. Here, $SU(N_f)_A$ denotes the axial transformation symmetry, which is spontaneously broken in nature. PNGBs arise as a consequence of this breaking and are identified with the pions, the kaons, and the $\eta$ meson. However, the exact axial symmetry $U(1)_A$ is also not realized in nature, and it is expected to be broken. But no corresponding suitable PNGBs can be found yet.

To summarize, the classical QCD Lagrangian shows a $U(1)_A$ symmetry$;$ however, it has never been observed in nature, and this implies that it merely has to be broken, but at the same time, if such symmetry is broken, there must be a PNGB associated to the symmetry breaking which has yet to be identified. This issue is known as the $U(1)_A$ problem \cite{hooft1976computation, hooft1986instantons}.

\section{The resolution of the \boldmath$U(1)_A$ problem}

The resolution of the $U(1)_A$ problem begins from an incredibly brief remark taken by Weinberg, that, somehow, $U(1)_A$ is not a genuine symmetry of QCD, albeit, in the massless quark limit, it seems to be present. Then 't Hooft \cite{hooft1976symmetry} came up with a sufficiently reasonable explanation to the problem. He realized that the QCD vacuum has a more complicated structure, which in effect, makes $U(1)_A$, not a true symmetry of the strong interactions, even though it is an apparent symmetry of the QCD Lagrangian in the limit of vanishing quark masses. Thus, this might appear to have explained the $U(1)_A$  problem$;$ there is no mystery surrounding the missing PNGB.

Mathematically, to resolve the $U(1)_A$ problem, an extension to the classical QCD Lagrangian has to be supplied. This extension is known as the QCD $\theta$-term. Now let us explain how to provide this term. If $U(1)_A$ was indeed obeyed, the associated Noether current \cite{adler1969axial}
\begin{equation}
J^{\mu}_5 = \sum_q \bar{q} \gamma^{\mu} \gamma^5 q \:,
\end{equation}
where the sum is over light quarks and $\gamma^\mu$ are the Dirac matrices, would be conserved, \ie $\partial_\mu J^{\mu}_5 = 0$. However, it turns out that the divergence of the axial current $J^{\mu}_5$ gets quantum corrections from the triangle graph, which connects it to two gluon fields with quarks going around the loop. Therefore, as a consequence of this anomaly, the divergence of the axial current obtains nonzero quantum corrections and fails to be conserved
\begin{equation}
\partial_\mu J^{\mu}_5  = \frac{g_s^2 N_f}{32 \pi^2} G_a^{\mu \nu} \tilde{G}_{\mu \nu}^a \:,
\end{equation}
where $G_a^{\mu \nu}$ is the gluon field strength and  $\tilde{G}_{\mu \nu}^a = \frac{1}{2} \epsilon_{\mu \nu \alpha \beta} G_a^{\alpha \beta}$ is its dual. Here $\epsilon_{\mu \nu \alpha \beta}$ is the antisymmetric Levi-Civita symbol in four dimensions. Hence, in the limit of vanishing quark masses, although formally QCD is invariant under a $U(1)_A$ transformation\newpage \noindent
\begin{equation}
q_f \rightarrow e^{i \alpha \gamma^5 / 2} q_f \:,
\end{equation}
where $\alpha$ is an arbitral parameter, the chiral anomaly affected the action
\begin{equation}
\delta S= \alpha \int d^4 x \: \partial_{\mu} J^{\mu}_5 = \alpha \frac{g_s^2 N_f}{32 \pi^2} \int d^4 x \:  G_a^{\mu \nu} \tilde{G}_{\mu \nu}^a \:.
\end{equation}
However, as it turns out, there is a further complication here. The pseudo-scalar density entering in the anomaly can be shown to be a total derivative
\begin{equation}
G_a^{\mu \nu} \tilde{G}_{\mu \nu}^a  = \partial_{\mu} K^{\mu} \:,
\end{equation}
where $K^{\mu} $ is another current
 \begin{equation}
 K^{\mu} = \epsilon^{\mu \alpha \beta \gamma} A_{a \alpha} (G^a_{\beta \gamma} - \frac{g_s}{3} f_{abc} A_{b \beta}A_{c \gamma}) \:.
 \end{equation}
Here, $A_a^{\mu}$ are the gluon fields, and $f_{abc}$ are the QCD structure constant. These identities imply that $\delta S$ is a pure surface integral
\begin{equation} \label{eq.3.10}
\delta S = \alpha \frac{g_s^2 N_f}{32 \pi^2} \int d^4x \: \partial_{\mu} K^{\mu} =  \alpha \frac{g_s^2 N_f}{32 \pi^2} \int d\sigma_{\mu} K^{\mu} \:.
\end{equation} 
Of course, if the anomaly does not contribute to the action, the system will still be invariant under $U(1)_A$, and we will have our problem back. In fact, this is the case when using the naive boundary conditions that $A_a^{\mu}=0$ at spatial infinity, since it leads to a zero total contribution to the action due to the vanishing term  $\int \sigma_{\mu} K^{\mu}=0$. However, 't Hooft realized that there are actually topologically non-trivial field configurations, called instantons, that contribute to this operator, and thus it can not be neglected. Further, he showed that $A_a^{\mu}=0$ is not the correct boundary condition to be used, and instead, the appropriate choice is to take $A_a^{\mu}$ as a pure gauge field at spatial infinity, \ie this means $A_a^{\mu}$ should be either zero or a gauge transformation of zero at spatial infinity.  Hence, with these boundary conditions, it turns out that there are gauge configurations for which $\int \sigma_{\mu} K^{\mu} \neq 0$ and, therefore, the system is not anymore invariant under $U(1)_A$. Thus,  $U(1)_A$ is not a true symmetry of QCD.

Let us now discuss some important details of how instantons give rise to non-trivial contribution to the integral term in the action equation \eqref{eq.3.10}. Instantons (sometimes referred to as pseudo-particles) are classical solutions to the equation of motion in Euclidean spacetime rather than Minkowski spacetime, with finite non-zero action. Instantons are a non-perturbative effect either in quantum mechanics or quantum field theory, which can not be seen in perturbation theory. Instantons describe tunneling effects between different vacua of a theory, which changes the structure of the quantum vacuum qualitatively. These effects can lead to the dynamical breaking of $U(1)_A$ symmetry in QCD.

For more clarification, let us consider that we have a Yang-Mills theory, and for simplicity, we restrict ourselves with $SU(2)$ QCD. In this gauge, there are only spatial gauge fields $A_a^{\mu}$. Under a gauge transformation, these fields transform as\newpage \noindent
\begin{equation}
\frac{1}{2} \sigma_a A_a^{\mu} \equiv  A^{\mu} \rightarrow \mathrm{\Omega} A^{\mu} \mathrm{\Omega}^{-1} + \frac{i}{g_s} \nabla^{\mu} \mathrm{\Omega} \mathrm{\Omega}^{-1} \:,
\end{equation}
with $\sigma_a$ and $\mathrm{\Omega}$ represent here the Pauli matrices and an element of the gauge group $SU(2)$, respectively. This means that, the vacuum configuration either vanish or have the form $i g_s^{-1} \nabla^{\mu} \mathrm{\Omega} \mathrm{\Omega}^{-1}$. In the $A_a^{0}=0$ gauge, these vacuum configurations can be classified by how $\mathrm{\Omega}$ goes to unitary as $\vert x \vert \rightarrow \infty$
\begin{equation}
\mathrm{\Omega}_n \rightarrow e^{i2\pi n} \quad \text{as} \quad  \vert x \vert \rightarrow \infty \quad \text{with} \quad n= 0, \pm 1, \pm 2, \dots \:.
\end{equation} 
The quantum number $n$ is known as winding number and define as follows
\begin{equation}
n = \frac{i g_s^3}{24 \pi^2} \int d^3x \: \tr (\epsilon_{ijk} A_n^i A_n^j A_n^k) \:.
\end{equation}
The winding number $n$ characterizes the homotopy class for the $S_3 \rightarrow SU(2)$ mapping, \ie a mapping from the three-dimensional Euclidean space to the $SU(2)$ space.  It is an integer for a pure gauge field. In contrast, for a field that vanishes at spatial infinity, it can be expressed in terms of a non-zero surface integral. This implies a non-zero vacuum-vacuum transition amplitude. Keeping in mind that each vacuum is characterized by a distinct winding number, so we can refer to them as n-vacua, $\vert n \rangle$ that denotes the pure gauge configurations. Furthermore, since the vacua are degenerate and instantons allow transitions between them, the physical vacuum state must be written as a superposition of these $n$-vacua. Therefore the true vacuum is known as the theta vacuum and written as
\begin{equation}
\vert \theta \rangle = \sum_n e^{-in\theta} \vert n \rangle \:,
\end{equation}
where $\theta$ is an unknown, $2\pi$-periodic number referred to as the vacuum angle. We may note here that the $\theta$ vacuum is nothing but the Fourier transform of the $n$-vacua. Coupling this note with the following important point that, a gauge transformation that transforms the field configuration $\vert n \rangle \rightarrow \vert n+1 \rangle $  has a well-defined solution, and such a tunneling event is called an instanton. The effect of these mutually distinct theta vacua is that effective action gains an additional term. The path integral formulation of the vacuum to vacuum transition amplitude involves an effective action which is dependent on the vacuum angle $\theta$
\begin{equation}
S_{\text{eff}} [A] = S_0 [A] + \theta \frac{g_s^2}{32 \pi^2}  G_a^{\mu \nu} \tilde{G}^a_{\alpha \beta} \:,
\end{equation}
with $ S_0 [A] $ being the usual QCD action. This means the QCD Lagrangian now has an addition $\theta$-term \cite{peccei2008strong}
\begin{equation} 
\L_\theta = \theta \frac{g_s^2}{32 \pi^2} G_a^{\mu \nu} \tilde{G}^a_{\alpha \beta}  \:.
\end{equation}
Since quarks are not massless, and if we consider the weak as well as the strong interactions, then we have a general quarks mass term in the Lagrangian, which can be written as
\begin{equation}
\L_{\text{mass}} = \bar{q}_{iR} M_{ij} q_{jL} + h.c \:.
\end{equation}
where $M_{ij}$ represents the complex quark mass matrix. Consequently, the $U(1)_A$ transformations lead to an additional phase when we diagonalize the mass matrix. Thus the vacuum angle picks up contributions both from QCD and the electroweak sector. So in the full theory, the value of the $\theta$-term gains an additional contribution equal to the argument of the determinant of the matrix $M_{ij}$, and is denoted as $\arg \det M$. Hence, the total physical vacuum angle becomes \cite{peccei2008strong}
\begin{equation} \label{eq.3.18}
\bar{\theta} = \theta + \arg \det M
\end{equation}
This new vacuum term is called the effective vacuum angle, $\bar{\theta}$, and generally, it is non-zero. Hence, the additional $\theta$-term to the QCD Lagrangian  has to be rewritten in the following form \cite{peccei2008strong}
\begin{equation}
\L_\theta = \bar{\theta} \frac{g_s^2}{32 \pi^2} G_a^{\mu \nu} \tilde{G}^a_{\alpha \beta} \:.
\end{equation}
By introducing this extra term in the QCD Lagrangian, we have successfully allowed for non-vanishing contributions, by the anomalous current, to the action. Thus the $U(1)_A$ is in no way symmetry of QCD, and we no longer expect a thereto associated preserved Noether current or Goldstone mode. The $U(1)_A$ problem has indeed been solved.

\section{The strong CP problem}

As we discussed above, the $U(1)_A$ problem can be explained due to the complex structure of the QCD vacuum \cite{callan1976structure}. The mathematical resolution of the problem implies including the $\theta$-term into the total QCD Lagrangian. However, solving the $U(1)_A$ problem leads to a new fine-tuning problem referred to as the strong CP problem. The additional $\theta$-term violates the PT (\ie parity-time) symmetry, but conserve C (\ie charge conjugation) symmetry. This makes it a source of CP violation in the strong interaction. Although the total QCD Lagrangian includes such a CP-violating term, there is no experimental indication of CP violation in the strong interactions. For example, the electric dipole moments (EDM) for the neutron is one of the most sensitive CP-violating observables to the value of $\theta$. An estimate the neutron EMD obtained within chiral perturbation theory can be conveniently expressed in the form \cite{crewther1979chiral}
\begin{equation} \label{eq.3.20}
d_n = - \frac{m_u m_d}{m_u + m_d} \bar{\theta} (\bar{q} i \gamma^5 q) \:.
\end{equation}
Accordingly, the calculation obtained the following estimation 
\begin{equation}
\vert d_n \vert \sim ( 2.4 \pm 1.0)  \times 10^{-16} \bar{\theta} \ e \ \text{cm} \:,
\end{equation}
where $e$ is the electric charge. The current experimental bound is given by \cite{baker2006improved}
\begin{equation}
\vert d_n \vert < 2.9 \times 10^{-26} \ e \ \text{cm} \:.
\end{equation}
Such an experimental  result implies that the parameter $\theta$ is about of order $\vert \bar{\theta} \vert \lesssim 10^{-10}$. Thus we are tempted to think that $\bar{\theta}$ is zero. Nevertheless, there is no natural reason to expect $\bar{\theta}$ to be this small. In principle, the $\bar{\theta}$ parameter is a free parameter, and it may take a value anywhere in the range from $0$ to $2\pi$. Further, CP violation occurs in the standard model by allowing the quark masses to be complex, and thus the natural value of $\bar{\theta}$ is expected to be of order one. Thus it seems there is an unnatural cancellation between the two parameters of equation \eqref{eq.3.18} that are not related. In fact, we would like to understand the reason why the sum of two contributions with very different physical origins is extremely small, if not zero. This is the strong CP problem.

\section{The resolution of the strong CP problem}

As we saw in the previous section, the strong CP problem is clearly a very serious issue. There are three scenarios to solve this problem. In this section, we briefly discuss these possible solutions. Then in the next section, we discuss in detail the most preferable solution$;$ the axion solution.

\begin{itemize}
\item {\bf Massless quark solution.} The first suggestion is to set the up quark mass to zero, so the RHS of equation \eqref{eq.3.20} vanishes. Thus the neutron EDM vanishes with no constraints in this case. However, the massless quark solution is excluded, as seen in reference \cite{di2016qcd}. The lattice QCD simulations suggest a non-zero up quark mass. This is indeed experimentally supported by the other properties of the known mesons, including pions, require all of the quarks to have some sort of bare mass.

\item {\bf Solution with spontaneous CP breaking.} The second possibility might be constructed if CP is a symmetry of nature, which is spontaneously broken, and hence there would be no strong CP problem at all \cite{nelson1984naturally, barr1984solving}. Interestingly, even though we observe CP violation in nature, we can make use of this fact if CP is truly a fundamental symmetry of nature at high energies that have been spontaneously broken, such that at low energies, we do not observe it to be a symmetry. In this way, one starts with a theory that is CP invariant with $\bar{\theta}= 0$ at the Lagrangian level. Then, CP is spontaneously broken by the vacuum expectation value (VEV) of a CP-odd scalar. The interactions of this scalar may be engineered such that it can generate both the required CP-violating phase and $\bar{\theta}< 10^{-10}$. Although several models exist where this is successfully achieved, they require rather disturbing features, \eg complex Higgs VEVs, which cause further problems. Moreover, the biggest drawback for this possible solution is that experimental data is in excellent agreement with the CKM Model, which includes CP explicitly not spontaneously broken.

\item {\bf The axion solution.} The Peccei-Quinn or the axion solution is the most widely accepted solution to explain the smallness of the parameter $\bar{\theta}$ \cite{peccei1977constraints, peccei1977cp}. The model is based on the idea of introducing an additional chiral symmetry to the standard model Lagrangian, which effectively rotates the $\theta$-vacua away. In essence, this should effectively allow the theory to be insensitive to any additional source of CP violation generated at higher energies and accordingly solve the strong CP problem. The description of the mechanism of this solution is the subject of the next section.

\end{itemize}

\section{The Peccei-Quinn mechanism}

R. Peccei and H. Quinn proposed this theory in 1977 to solve the strong CP problem. They assumed that the standard model Lagrangian has an additional global chiral $U(1)$ symmetry, which was named after them as the Peccei-Quinn (PQ) symmetry $U(1)_{\text{PQ}}$ \cite{peccei1977constraints, peccei1977cp}. This symmetry is necessarily spontaneously broken at some large energy scale $f_a$, and this indeed gives rise to a massless NGB, which is called the axion. The consequences of introducing such symmetry into the theory are effectively promoting the vacuum angle from static CP-violating constant to a dynamical CP-conserving field, which is the axion field. As we will describe below, when the effective potential of the axion is minimized, the $\bar{\theta}$ dependence cancels, and the CP problem is no more. Hence, when we add the axion field to the standard model Lagrangian, and in order to make it invariant under the new $U(1)_{\text{PQ}}$ symmetry, it has to take the form
\begin{align} \label{eq.3.23}
\L_{\text{total}} &= \L_{SM} + \L_{\bar{\theta}} + \L_{a} \nonumber \\
&=  \L_{SM} + \bar{\theta} \frac{g_s^2}{32 \pi^2} G_b^{\mu \nu} \tilde{G}^b_{\mu \nu} - \frac{1}{2} \partial_{\mu} a \partial^{\mu} a + \L_{\text{int}} [\partial^\mu a/f_a , \psi] + \xi \frac{a}{f_a} \frac{g_s^2}{32 \pi^2} G_b^{\mu \nu} \tilde{G}^b_{\mu \nu} \nonumber \\
&= \L_{SM} - \frac{1}{2} \partial_\mu a \partial^\mu a +\L_{\text{int}} [\partial^\mu a/f_a , \psi] + \left( \bar{\theta} + \xi \frac{a}{f_a} \right) \frac{g_s^2}{32 \pi^2} G_b^{\mu \nu} \tilde{G}^b_{\mu \nu} \:,
\end{align}
where $\L_a$ refers to the Lagrangian of the new axion field. It is easy once we look at the previous equation to realize that the axion part of the Lagrangian contains the usual kinetic and interaction terms in addition to another term which is required to ensure that the Noether current associated to the new $U(1)_{\text{PQ}}$ symmetry has a chiral anomaly
\begin{equation}
\partial_\mu J^{\mu}_{\text{PQ}} =\xi \frac{g_s^2}{32 \pi^2} G_b^{\mu \nu} \tilde{G}^b_{\mu \nu} \:,
\end{equation}  
where $\xi$ is a coefficient. Since the axion field, $a(x)$ is the NGB of the broken symmetry$;$ it is invariant under a $U(1)_{\text{PQ}}$ transformation
\begin{equation}
a(x) \rightarrow a(x) + \alpha f_a \:,
\end{equation}
where $\alpha$ is the phase of the field, and $f_a$ represents here the order parameter associated with the breaking of the $U(1)_{\text{PQ}}$ symmetry. In such a way, we see that an axial transformation can shift $a$ to make it able to remove the $\bar{\theta}$ dependence of the theory and thus provide a dynamical solution to the strong CP problem. This is very important, as it means that the physical vacuum angle is actually $\bar{\theta}+\xi \langle a \rangle / f_a$, where $\langle a \rangle$ signifies the VEV of $a$, and $f_a$ is now the scale of the spontaneous breaking of the $U(1)_{\text{PQ}}$ symmetry.

Because of the complicated structure of the vacuum, the PQ symmetry must be explicitly instead of spontaneously broken. Therefore, the axion becomes a PNGB and picks up a small mass. This also means that the axion gains a nontrivial effective potential
\begin{equation}
\left \langle \frac{\partial V_{\text{eff}}}{\partial a} \right \rangle = - \frac{\xi}{f_a} \frac{g_s^2}{32 \pi^2}  \left \langle  G_b^{\mu \nu} \tilde{G}^b_{\mu \nu} \right \rangle \Big \vert_{\langle a \rangle = - \bar{\theta} f_a / \xi} =0 \:.
\end{equation}
If we were to neglect the nontrivial vacuum structure, \ie the instanton effects, the circle of minimal potential would be parallel to the plane, with degenerate ground states, and all the values $0 \leq \xi \frac{\langle a \rangle}{f_a} \leq 2 \pi $ are allowed. In this situation, the breaking of PQ symmetry remains spontaneous. But in the case of a nontrivial vacuum structure, the instanton effects must be taken into account, and then the circle of minimum potential becomes tilted. Hence in this scenario, the PQ symmetry becomes explicitly broken, and an axion mass $m_a \neq 0$ is generated. In this case, the mechanism through which the $\bar{\theta}$-term can be eliminated from the theory is to allow the effective potential for the axion field to get its minimum. This is known as  Peccei-Quinn solution and occurs at $\langle a \rangle = - \bar{\theta} f_a / \xi$. To understand the core of the solution, we may take a look at equation \eqref{eq.3.23}. Then, we easily can realize that generating potential for the axion field that is periodic in the effective vacuum angle $\bar{\theta} + \langle a \rangle \xi / f_a$ requires
\begin{equation}
V_{\text{eff}} \sim \cos \left( \bar{\theta}  + \xi \frac{\langle a \rangle}{f_a}\right) \:.
\end{equation}
In order to minimize the potential $V_{\text{eff}}$, Peccei and Quinn showed that taking the term inside the brackets to be zero (and not $\pi$) is the correct choice.  Hence, we get the following condition
\begin{equation}
\bar{\theta} +  \frac{\langle a \rangle}{f_a} = 0 \quad \Leftrightarrow \quad  \langle a \rangle= - \frac{f_a}{\xi} \bar{\theta} \:,
\end{equation}
and as the axion field evolves, and the potential minimum is reached, the CP-violating term from the Lagrangian \eqref{eq.3.23} is removed. The strong CP problem is solved. What has happened is that we have essentially switched the fixed-parameter $\bar{\theta}$for a dynamical variable with a CP-conserving minimum, the axion field. As the field evolves, it effectively relaxes the CP-violating term to zero.

Now, by defining the physical axion as $a_{\text{phys}} \equiv a-\langle a \rangle$, we can rewrite the Lagrangian \eqref{eq.3.23} in terms of  $a_{\text{phys}}$  which is now no longer has a CP-violating $\bar{\theta}$-term. It obviously takes the following form
 \begin{equation} \label{eq.3.29}
 \L_{\text{total}} =  \L_{\text{SM}} - \frac{1}{2} \partial_\mu a \partial^\mu a + \L_{\text{int}} [\partial^\mu a/f_a , \psi] + a \frac{g_s^2}{32 \pi^2} G_b^{\mu \nu} \tilde{G}^b_{\mu \nu} \:.
 \end{equation}
 
\section{The axion dynamics and models}

As we discussed in the previous section, the presence of the QCD anomaly is necessary to induce the axion potential whose minimum is located at $\bar{\theta}_{\text{eff}}=0$. Expanding the effective potential $V_{\text{eff}}$ for the axion field at the minimum gives the axion $a$ mass
\begin{equation}
m_a^2 = \left \langle \frac{\partial^2 V_{\text{eff}}}{\partial a^2} \right \rangle  = - \frac{\xi}{f_a} \frac{g_s^2}{32 \pi^2}  \left \langle  \dfrac{\partial}{\partial a} G_b^{\mu \nu} \tilde{G}^b_{\mu \nu} \right \rangle \Big \vert_{\langle a \rangle = - \bar{\theta} f_a / \xi} \propto \frac{1}{f_a} \:.
\end{equation}
The estimation of standard axion mass has been calculated using several methods such as current algebra techniques \cite{bardeen1978current} and effective Lagrangian \cite{bardeen1987constraints} approaches. We can easily realize that these calculations imply that the axion mass and interactions are characterized by the scale parameter $f_a$ of the spontaneous breaking of PQ symmetry. In the original model, the $U(1)_{\text{PQ}}$ symmetry breakdown coincided with that of electroweak breaking $f_a = v_F$ , with $v_F \simeq 250$ GeV. This identified the visible axion with a mass of order $100 \ \text{keV}$ to $1 \ \text{MeV}$, and the associated models then have become known as visible axion models. If axions had this mass, then they would have couplings large enough to be experimentally detected. As all the results were null, astrophysical, and experimental searches, have ruled out this type of axions.

Nevertheless, the initial assumption that the value of the scale parameter $f_a$ is close to the electroweak scale was not necessary. Indeed, the original realization of the axion has been excluded experimentally, and now $f_a$ is thought to lie much higher. When $f_a \gg v_F$, then the axion is very light, very weakly coupled, and very long-lived. Models, where this occurs, have become known as invisible axion models. These invisible, light, and weakly interacting axions are, therefore, very promising dark matter candidates. In this section, we will shortly review the properties of the standard weak-scale axions and then generalize the discussion to the invisible axion models.

\subsection{The visible axion models}

The original model of the axion was proposed by Weinberg and Wilczek \cite{weinberg1978new, wilczek1978problem}, based on the idea of Peccei and Quinn \cite{peccei1977constraints, peccei1977cp}. This is called the Peccei-Quinn-Weinberg-Wilczek (PQWW) model, or the visible axion model. Since the basic ingredient of the axion model is a global chiral $U(1)$-symmetry of the standard model Lagrangian, it is clear that an extension of the standard model Lagrangian is required. In principle, the axion is embedded in the phase of the Higgs field in the usual standard model. But, one Higgs doublet is not enough to give rise to the axion because the other three Goldstone modes are absorbed by the longitudinal degrees of the standard model gauge bosons, and the remaining Higgs boson has a potential. Therefore, the simplest extension that can make the standard model Lagrangian invariant under the $U(1)_{\text{PQ}}$ symmetry is to consider a model with the Higgs sector contains at least two scalar field doublets. This minimal model introduces exactly two Higgs fields, $\phi_1$ and $\phi_2$, to absorb independent chiral transformations of the quarks and leptons. Hence, the presence of the Higgs doublets fields can allow the Lagrangian density to be invariant under the desired chiral $U(1)_{\text{PQ}}$ transformations
\begin{align}
\qquad \qquad \qquad  \qquad
\begin{split}
a & \rightarrow a + \alpha \, v_F \:, \\
\phi_{1} & \rightarrow e^{2  \alpha / x} \phi_{1} \:, \\
\phi_{2}  & \rightarrow  e^{2  \alpha x} \phi_{2} \:, 
\end{split}
\begin{split}
u_{Rj} & \rightarrow  e^{-i \alpha x} u_{Rj}   \:,  \\
 d_{Rj} & \rightarrow e^{-i \alpha / x} d_{Rj}  \:, \\
\ell_{Rj} &  \rightarrow  e^{-i \alpha / x} \ell_{Rj}  \:, 
 \end{split}
 \qquad \qquad \qquad  \qquad
\end{align}
Then, the relevant parts of the Yukawa-sector of the standard model Lagrangian involving the Higgs doublet fields can be defined as follows
\begin{equation} \label{eq.3.32}
\L_{\text{Yukawa}} = y_{ij}^u \bar{q}_{Li} u_{Rj} \phi_1 + y_{ij}^d \bar{q}_{Li} d_{Rj} \phi_2 + y_{ij}^u \bar{l}_{Li} \ell_{Rj} \phi_2 + h.c \:.
\end{equation}
The usual Yukawa coupling between these Higgs scalars and the Fermions spontaneously breaks the flavor symmetry and gives masses to the quarks and leptons. Because of this spontaneous symmetry breaking, the two the Higgs doublets $\phi_1$ and $\phi_2$ develop the following nonzero VEVs
\begin{equation}
\langle \phi_1^0 \rangle = v_1 \quad \text{and} \quad \langle \phi_2^0 \rangle = v_2 \:,
\end{equation}
where $\phi_1^0$ and $\phi_2^0$ are the neutral components of $\phi_1$ and $\phi_2$, respectively. The VEVs of the Higgs doublets break both the electroweak symmetry and the $U(1)_{\text{PQ}}$ symmetry at a scale $f_a =\sqrt{v_1^2 +v_2^2} $. In the simplest model, the scale of $U(1)_{\text{PQ}}$ symmetry breaking scale is taken to be the same as the electroweak symmetry breaking scale. Corresponding to the two Higgs doublets, there are now four physical Higgs scalars and four Nambu-Goldstone (NG) modes. Three of the NG modes will give masses to the $W^{\pm}, Z$ bosons, and the remaining one becomes the axion field. The symmetries of the Lagrangian shows that we have a $U (1)$ phase for each Higgs doublet. The axion in this model is the common phase field of the Higgs doublets $\phi_1$ and $\phi_2$ orthogonal to weak hypercharge. If the ratio of the VEVs of the Higgs doublets is defined as $x=v_2/v_1$, then it is easy to isolate the axion content in $\phi_1$ and $\phi_2$ as
\begin{equation}
\phi_1 = \frac{v_1}{\sqrt{2}} \left( \begin{matrix} 1 \\ 0 \end{matrix} \right) e^{i x a / f} \quad \text{and} \quad \phi_1 = \frac{v_1}{\sqrt{2}} \left( \begin{matrix} 0 \\ 1 \end{matrix} \right) e^{i a /x f} \:.
\end{equation}
One of the two linear combinations of these phases is the electroweak hypercharge degree of freedom, which is absorbed by the $Z$ boson, and the other degree of freedom  is then the axion field
\begin{equation}
a = \frac{1}{f_a} (v_1 \Im \phi_1^0 - v_2 \Im \phi_2^0 ) \:.
\end{equation}
Since the $U(1)_{\text{PQ}}$ symmetry is spontaneously broken, the corresponding NGB, the axion, will be massless. But the QCD gluon anomaly will break this symmetry explicitly, and hence, the axion will become a PNGB with a small mass. The mass for the standard axion was estimated to be given by
\begin{equation}
m_a^{st} = \frac{m_\pi f_\pi}{v} N_g (x+\frac{1}{x}) \frac{\sqrt{m_u m_d}}{m_u + m_d} \simeq 25 N_g (x+\frac{1}{x}) \ \text{Kev} \:, 
\end{equation}
where $m_\pi$ is the mass of the pion meson, $f_\pi \simeq 92 \ \text{Mev}$ is pion decay constant, and $N_g$ is the number of quark generators. The axion thus directly acquires its mass by mixing with $\pi^0$, which occurs with the gluon coupling. The $\pi^0$-mixing then induces another coupling of the axion to two photons. Writing the interaction Lagrangian describing this coupling as
\begin{equation} \label{eq.3.37}
\L_{a \gamma \gamma} = \frac{\alpha}{4 \pi} K_{a \gamma \gamma} \frac{a_{\text{phys}}}{f_a} G^{\mu \nu} \tilde{G}_{\mu \nu} \:,
\end{equation}
where $K_{a \gamma \gamma}$ is the axion-two-photon coupling constant of $\bm{\mathcal{O}}(1)$ and is defined as
\begin{equation}
K_{a \gamma \gamma} = N_g (x + \frac{1}{x}) \frac{m_u}{m_u + m_d} \:.
\end{equation}

Then, for all axion models, the mass of the axion and its $\pi$ and $\eta$ couplings, respectively, can be characterized by
\begin{equation}
m_a = \lambda_m m_a^{\text{st}} \left( \frac{v}{f_a} \right) \:, \quad
\xi_{a \pi} = \lambda_3 \frac{f_\pi}{f_a} \:, \quad \xi_{a \pi} = \lambda_0 \frac{f_\eta}{f_a} \:, 
\end{equation}
where $\lambda_m$, $\lambda_3$, and $\lambda_0$ are model parameters of $\bm{\mathcal{O}}(1)$. It seems to be important now to mention that all axion models can be characterized by the model-dependent parameter $K_{a \gamma \gamma}$ together with parameters $\lambda_m$, $\lambda_3$, and $\lambda_0$, which are calculated for the PQ model and written as
\begin{align}
\begin{split}
\lambda_m &= N_g \left( x + \frac{1}{x} \right) \:, \\
\lambda_3 &= \frac{1}{2} \left[ \left( x + \frac{1}{x} \right)  - N_g \left( x + \frac{1}{x} \right) \frac{m_d-m_u}{m_d + m_u} \right] \:, \\
\lambda_0 &= \frac{1}{2} (1-N_g) \left( x + \frac{1}{x} \right) \:.
\end{split}
\end{align}

\subsection{The invisible axion models}

In the visible axion models, when the axion mass is less than the electron mass, the axions can decay into two photons obeying the interaction giving by equation \eqref{eq.3.37} and hence, the axion becomes long-lived. When the axion mass is more than the electron mass, it can decay rapidly into two electrons and become short-lived
\begin{equation}
\tau (a \rightarrow e^{+} e^{-}) = \frac{8 \pi v^2 v_2^2}{m_e^2 v_1^2 \sqrt{m_a^2 - 4 m_e^2}} \:.
\end{equation}
Unfortunately, both these possibilities have been ruled out experimentally. In addition, the strongest constraints to rule out the existence of the visible axion came from the nonobservation  of the K-decay to axion, which has been estimated to be
\begin{equation}
Br (K^{+} \rightarrow \pi^{+} + a) \simeq 3 \times 10^{-5} (x+\frac{1}{x})^2 \:.
\end{equation}
This is well above the KEK bounds$;$ however, the experimental  constraints  to the K-decay implied
\begin{equation}
Br (K^{+} \rightarrow \pi^{+} + \text{nothing}) \leqslant 3.8 \times 10^{-8}  \:,
\end{equation}
where ``nothing’’ in the decay products includes the long-live axions, which would escape detection. We conclude from this argument that the original PQ model with $f_a = v_F$ and its associated visible axion are ruled out by experimental results.

However, on the other hand, invisible axion models when $f_a \gg v_F$  are still viable. These models avoided the problem of the original PQWW model by allowing the $U(1)_{\text{PQ}}$ symmetry breaking to occur at scale $f_a$ much higher than the electroweak scale since the coupling of axions with other particles are suppressed by $1/f_a$. Therefore, the axion in such models is characterized as very light, very weakly coupled, and very long-lived.

More technically, the essence of the invisible axion models is introducing a new complex scalar field $\phi$, but $SU(2)_L \times U(1)_Y$ is a singlet. This field is called the Peccei-Quinn field, carries only a PQ  charge, and does not participate in the electroweak interactions. The $U(1)_{\text{PQ}}$ symmetry then must be broken by the VEV of this field, which has a scale much larger than the one set by the electroweak interactions. Let us now define the change in the PQ field $\phi$ under the $U(1)_{\text{PQ}}$ transfiguration as
\begin{equation}
\phi \rightarrow e^{i \alpha} \phi \:.
\end{equation}
If we impose the potential for the PQ field $\mathrm{\Phi}$
\begin{equation}
V(\mathrm{\Phi}) = \frac{\lambda}{4} (\vert \mathrm{\Phi} \vert^2 - f_a^2)^2 \:,
\end{equation}
it acquires the VEV $\vert \langle \mathrm{\Phi} \rangle \vert = f_a \gg v_F$. Then, instead of defining the axion to be the phase direction of the standard model Higgs doublets in the visible axion models, the invisible axion field is the, essentially, the phase of the PQ field $\phi$
\begin{equation}
\phi = \frac{f_a}{\sqrt{2}} e^{i a / f_a} \:.
\end{equation}

There are two main classes of such invisible axion models, depending on whether or not they have a direct coupling to leptons. These two classifications are the Kim-Shifman-Vainshtein-Zakharov (KSVZ)-type and the Dine-Fischler-Srednicki-Zhitnitsky (DFSZ)-type models. We shall briefly review here both two types.

\subsubsection*{The KSVZ model}

The KSVZ model \cite{kim1979weak, shifman1980can} suggests the existence of some new heavy quarks $Q$ that carry the PQ charge, while the ordinary quarks and leptons do not. Then, the QCD anomaly is obtained via the Yukawa coupling between such heavy quarks $Q$ and the PQ fields, which can be written as
\begin{equation}
\L_{\text{KSVZ}} = - y_{Q} \bar{Q}_L \phi Q_R + h.c \:.
\end{equation}
Under the $U(1)_{\text{PQ}}$ symmetry, the heavy quark $Q$ transforms as
\begin{align}
\qquad \qquad \qquad  \qquad
\begin{split}
Q_L & \rightarrow e^{i \alpha/2} Q_L \:, 
\end{split}
\begin{split}
Q_R & \rightarrow e^{-i \alpha/2} Q_R \:.
 \end{split}
 \qquad \qquad \qquad  \qquad
\end{align}

Calculations analogous to the one of the PQWW model obtained the following axion characteristic parameters for the KSVZ
\begin{equation}
\lambda_m =1 \:, \quad \lambda_3 = - \frac{1}{2} \frac{m_d - m_u}{m_d + m_u} \:, \quad \lambda_0 = - \frac{1}{2} \:,
\end{equation}
and
\begin{equation}
K_{a \gamma \gamma} = 3 q_{Q}^2 -  \frac{4 m_d - m_u}{3(m_d + m_u)} \:,
\end{equation}
where $q_{Q}$ is the electrical charge of the heavy quarks $Q$.

\subsubsection*{The DFSZ model}

The DFSZ model \cite{dine1981simple, zhitnitskii1980possible} is an extension of the PQWW model, and similarly, it realizes the QCD anomaly without the need to introduce a heavy quark. As in the PQWW model, both the quarks and leptons carry PQ charge, and hence two standard model Higgs doublets $\phi_1$ and $\phi_2$ are still required. In addition, a complex scalar field $\phi$ is added. The trick now is that the quarks and leptons directly couple to the Higgs doublets $\phi_1$ and $\phi_2$ through the usual Yukawa terms \eqref{eq.3.32}, but do not couple to the PQ field $\phi$. However, the quarks and leptons feel the effects of the axion through the interactions of the PQ filed $\phi$ with the two Higgs doublets $\phi_1$ and $\phi_2$. The PQ field couple the two Higgs doublets through the scalar potential
\begin{align}
V(\phi_1, \phi_2, \phi) =  \frac{\lambda_1}{4}  (\phi_1^{\dagger} \phi_1 - v_1^2)^2  &+ 
\frac{\lambda_2}{4} (\phi_2^{\dagger} \phi_2 - v_2^2)^2 + \frac{\lambda}{4} (\vert \phi \vert^2 - f_a^2)^2  + (a \phi_1^{\dagger} \phi_1 + b \phi_2^{\dagger} \phi_2 ) \vert \phi \vert^2  \nonumber \\
&+ c(\phi_1 \cdot \phi_2 \phi^2 + h.c.) + d \vert \phi_1 \cdot \phi_2 \vert^2 + e \vert \phi_1^{\dagger} \phi_2 \vert^2 \:.
\end{align}
The Lagrangian density is invariant under the $U(1)_{\text{PQ}}$ symmetry transformation
\begin{align}
\qquad \qquad \qquad  \qquad
\begin{split}
\phi_{1} & \rightarrow e^{-i  \alpha} \phi_{1} \:, \\
\phi_{2}  & \rightarrow  e^{-i  \alpha} \phi_{2} \:, \\
\phi & \rightarrow e^{i \alpha} \phi  \:,
\end{split}
\begin{split}
u_{Rj} & \rightarrow  e^{i \alpha} u_{Rj}   \:,  \\
 d_{Rj} & \rightarrow e^{i \alpha} d_{Rj}  \:, \\
\ell_{Rj} &  \rightarrow  e^{i \alpha} \ell_{Rj}  \:, 
 \end{split}
 \qquad \qquad \qquad  \qquad
\end{align}
The axion is then a linear combination of the phase of the three scalar fields $\phi_1^0$, $\phi_2^0$, and $\phi$. Defining $X_1= 2 v_2^2/v^2$ and $X_2= 2 v_1^2/v^2$, the contribution of the axion field $a$ in $\phi_1$ and $\phi_2$ can be isolated as
\begin{equation}
\phi_1 = \frac{v_1}{\sqrt{2}} \left( \begin{matrix} 1 \\ 0 \end{matrix} \right)e^{i X_1 a /f} \:, \quad \text{and} \quad  \phi_2 = \frac{v_2}{\sqrt{2}} \left( \begin{matrix} 0 \\1 \end{matrix} \right)e^{i X_2 a /f}  \:.
\end{equation}
The axion characteristic parameters for the DFSZ axion model have been calculated and obtained as follows
\begin{equation}
\lambda_m =1 \:, \quad \lambda_3 = \frac{1}{2} \frac{X_1 - X_2}{2 N_g} - \frac{m_d - m_u}{m_d + m_u} \:, \quad \lambda_0 =  \frac{1-N_g}{2N_g} \:,
\end{equation}
and
\begin{equation}
K_{a \gamma \gamma} = \frac{3}{4} -  \frac{4 m_d + m_u}{3(m_d + m_u)} \:.
\end{equation}

\section{Properties of the invisible axion}

As we discussed above in the previous section, the existence of the standard visible axion has been ruled out experimentally$;$ therefore, we will concentrate here in this section on describing the properties of the invisible axions. In general, the axion is a neutral pseudo-scalar particle, with a very light mass and very weak interactions with matter, since it has not been detected yet. The most significant feature of the axion properties is that they depend on the energy scale of the spontaneous breaking of the PQ symmetry $f_a$. These properties are the axion mass $m_a$, and the coupling constant of the axion to other particles denoted by $i$, both of them are inversely proportional to $f_a$
\begin{equation}
m_a \propto \frac{1}{f_a} \:, \quad \text{and} \quad g_{ai} \propto \frac{1}{f_a} \:.
\end{equation} 
In principle, the axion interacts mainly with gluons and photons, and it could also interact with fermions. The different axion models describe these interactions and to express their coupling strengths as a function of $f_a$. Some features of these interactions will be discussed below, see references \cite{kaplan1985opening, srednicki1985axion} for a detailed discussion on these topics.

\subsection{Coupling to gluons}

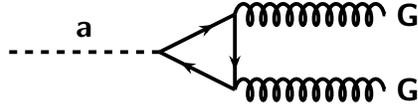
\begin{figure}[ht!]
\begin{center}
\begin{tikzpicture}
\draw[scalarr, black, line width=.5mm] (0,0)-- (2,0);
\draw[fermion, black, line width=.5mm] (2,0)--(3,0.5);
\draw[fermion, black, line width=.5mm] (3,-0.5)--(2,0);
\draw[fermion, black, line width=.5mm] (3,0.5)--(3,-0.5);
\draw[gluon, black, line width=.5mm] (3,0.5)--(5,0.5);
\draw[gluon, black, line width=.5mm] (3,-0.5)--(5,-0.5);

\node at (1,0.3) {\bf{a}};	
\node at (5.3,0.5) {\bf{G}};
\node at (5.3,-0.5) {\bf{G}};
\end{tikzpicture}
\caption[Feynman diagram of the axion coupling with gluon.]{Feynman diagram of the axion coupling with gluon.}
\label{Fig.3.1}
\end{center}
\end{figure}

The coupling of axions with gluons is described by the interaction term in the total Lagrangian density \eqref{eq.3.29} and given by the expression
\begin{equation}
\L_{aG} = \frac{\alpha_s}{8 \pi f_a} \ a \ G^a_{\mu \nu} \tilde{G}_a^{\mu \nu} \:,
\end{equation}
expressed as a function of the strong fine-structure constant $\alpha_s$, and $a$ refers to the axion field. Figure \ref{Fig.3.1} shows the Feynman diagrams for the axion-gluons interaction.

\subsection{Mass of axions}

The coupling of axion with gluons makes possible the mixing with pions as well, and the obtained axion mass can be then expressed as
\begin{equation} \label{eq.3.58}
m_a = \frac{m_\pi f_\pi}{f_a} \frac{\sqrt{m_u m_d}}{m_u+m_d} \simeq 6 \times 10^{-6} \ \text{eV} \ \left( \frac{10^{12} \ \text{GeV}}{f_a} \right) \:,
\end{equation}
with astrophysical and cosmological observations imply that $f_a \sim 10^{9} \text{--} 10^{12} \ \text{GeV}$, hence, the axion has a very tiny mass $m_a \sim 10^{-6} \text{--} 10^{-3} \  \text{eV}$.

\subsection{Coupling to photons}

\begin{figure}[ht!]
\begin{center}
\begin{tikzpicture}
\pgfmathsetmacro{\CosValueee}{cos(30)}
\pgfmathsetmacro{\SinValueee}{sin(30)}
\draw[scalarr, black, line width=.5mm] (0,0)-- (2,0);
\draw[fermion, black, line width=.5mm] (2,0)--(3,0.5);
\draw[fermion, black, line width=.5mm] (3,-0.5)--(2,0);
\draw[fermion, black, line width=.5mm] (3,0.5)--(3,-0.5);
\draw[vector, black, line width=.5mm] (3,0.5)--(5,0.5);
\draw[vector, black, line width=.5mm] (3,-0.5)--(5,-0.5);

\node at (1,0.3) {\bf{a}};	
\node at (5.3,0.5) {\bf{G}};
\node at (5.3,-0.5) {\bf{G}};
\node at (8,0) {\bf{a} $\hspace{10mm}  \pi^{0} $};
\draw[scalarr, black, line width=.5mm] (9,0)-- (10.5,0);
\draw[vector, black, line width=.5mm] (10.5,0)--(10.5+2*\CosValueee,2*\SinValueee);
\draw[vector, black, line width=.5mm] (10.5,0)--(10.5+2*\CosValueee,-2*\SinValueee);
\draw[fermionn, black, line width=.5mm] (7.6,-0.05)--(8.2,-0.05);
\draw[fermionn, black, line width=.5mm] (8.3,-0.05)--(7.7,-0.05);
\node at (10.4,0)  {\Large  $\bullet$};
\node at (10.8+2*\CosValueee,2*\SinValueee) {$\gamma$};
\node at (10.8+2*\CosValueee,-2*\SinValueee) {$\gamma$};
\end{tikzpicture}
\caption[Feynman diagrams of the axion-photon coupling$:$ coupling of the axion to two photons via a triangle loop through fermions carrying PQ and electric charges (left panel), and axion-pion mixing producing the generic coupling of axions to photons (right panel).]{Feynman diagrams of the axion-photon coupling$:$ coupling of the axion to two photons via a triangle loop through fermions carrying PQ and electric charges (left panel), and axion-pion mixing producing the generic coupling of axions to photons (right panel).}
\label{Fig.3.2}
\end{center}
\end{figure}
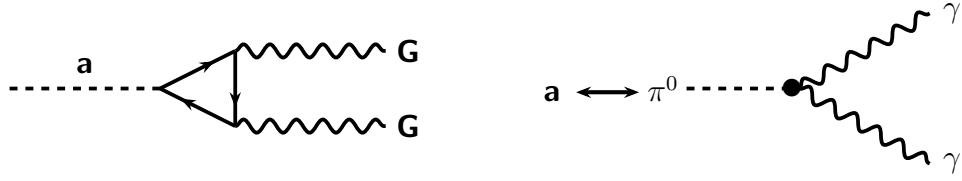

The mixing of axions with pions, due to the coupling to gluons, permits the description of the interactions of axions with photons by the Primakoff effect. Figure \ref{Fig.3.2} shows the Feynman diagrams for the axion-photon interaction for the two contributions from a triangle loop through fermions carrying PQ and electric charges and axion-pion mixing producing the generic coupling of axions to photons. The coupling of an axion to two photons and the contribution of such interaction to the Lagrangian density is given by the expression
\begin{equation}
\L_{a \gamma \gamma} = - \frac{1}{4} g_{a \gamma \gamma} \ a \ F_a^{\mu \nu} \tilde{F}^a_{\mu \nu} = g_{a \gamma \gamma} \ \mathbf{E} \cdot \mathbf{B} \ a \:,
\end{equation} 
where $g_{a \gamma \gamma}$ is the axion-two-photon coupling constant, $F^{\mu \nu}$ is the photon field strength tensor, $\tilde{F}_{\mu \nu}$ denotes its dual, $\mathbf{E}$ is the electric field, and $\mathbf{B}$ is the magnetic field. The magnitude of the coupling constant $g_{a \gamma \gamma}$ is parameterized by
\begin{equation} \label{eq.3.60}
g_{a \gamma \gamma}  = \frac{\alpha}{2 \pi f_a} C_{a \gamma \gamma} \:,
\end{equation}
where $\alpha = e^2/2\pi$ is the fine-structure constant, and $C_{a \gamma \gamma} $ is a numerical coefficient given by
\begin{equation}
C_{a \gamma \gamma} = \frac{E}{N} - \frac{2}{3} \frac{4+Z}{1+Z} \:,
\end{equation} 
where $Z \equiv m_u/m_d$ gives the ratio between the up-quark and down-quark masses, and  $E/N$ gives the ratio between the electromagnetic and color anomalies. The value of $E/N$ is zero in the KSVZ model, while it depends on the charge assignment of leptons in the DFSZ model.
 
\subsection{Coupling to fermions}

\begin{figure}[ht!]
\begin{center}
\begin{tikzpicture}
\pgfmathsetmacro{\CosValue}{cos(30)}
\pgfmathsetmacro{\SinValue}{sin(30)}
\pgfmathsetmacro{\CosValuee}{cos(60)}
\pgfmathsetmacro{\SinValuee}{sin(60)}
\pgfmathsetmacro{\CosValueee}{cos(40)}
\pgfmathsetmacro{\SinValueee}{sin(40)}
\pgfmathsetmacro{\CosValueeee}{cos(70)}
\pgfmathsetmacro{\SinValueeee}{sin(70)}
\pgfmathsetmacro{\CosValueeeee}{cos(190)}
\pgfmathsetmacro{\SinValueeeee}{sin(190)}
\draw[fermion, black, line width=.5mm] (0,0)-- (2*\CosValue,2*\SinValue);
\draw[fermion, black, line width=.5mm] (2*\CosValue,2*\SinValue)-- (4*\CosValue,4*\SinValue);
\draw[scalarrr, black, line width=.5mm] (2*\CosValue,2*\SinValue)-- (2*\CosValue+2*\CosValuee,2*\SinValue+2*\SinValuee);
\node at (2*\CosValue,2*\SinValue)  {\Large  $\bullet$};
\node at (-0.2,-0.1)  {$\bm{e}^{-}$};
\node at (4*\CosValue+0.3,4*\SinValue+0.1)  {$\bm{e}^{-}$};
\node at (2*\CosValue+2*\CosValuee+0.2,2*\SinValue+2*\SinValuee+0.2) {\bf{a}};
\node at (2*\CosValue+0.2,2*\SinValue-0.3)  {$\bm{g_{ae}}$};
\draw[fermion, black, line width=.5mm] (5,0)-- (5+2*\CosValue,2*\SinValue);
\draw[fermion, black, line width=.5mm] (5+2*\CosValue,2*\SinValue)-- (5+4*\CosValue,4*\SinValue);
\draw[fermion, black, line width=.5mm] (5,0)-- (5+4*\CosValue,4*\SinValue);
\draw[scalarrr, black, line width=.5mm] (6.4,1.8)-- (6.4+2*\CosValueee,1.8+2*\SinValueee);
\draw[vector, black, line width=.5mm] (5+0.72*2*\CosValue,0.72*2*\SinValue)-- (5+0.72*2*\CosValue+0.666*\CosValueeee,0.72*2*\SinValue+0.666*\SinValueeee);
\draw[antivector, black, line width=.5mm] (5+1.4*2*\CosValue,1.4*2*\SinValue)-- (5+1.4*2*\CosValue+0.666*\CosValueeeee,1.4*2*\SinValue-0.666*\SinValueeeee);
\draw[fermionnn, black, line width=.5mm] (5+0.72*2*\CosValue+0.666*\CosValueeee,0.72*2*\SinValue+0.666*\SinValueeee)  -- (6.4,1.8);
\draw[fermionnn, black, line width=.5mm] (6.85,1.55) -- (6.4,1.8);
\draw[fermionnn, black, line width=.5mm] (5+0.72*2*\CosValue+0.666*\CosValueeee,0.72*2*\SinValue+0.666*\SinValueeee) --(5+1.4*2*\CosValue+0.666*\CosValueeeee,1.4*2*\SinValue-0.666*\SinValueeeee);
\node at (5+0.72*2*\CosValue,0.72*2*\SinValue)  {\Large  $\bullet$};
\node at (5+1.4*2*\CosValue,1.4*2*\SinValue)  {\Large  $\bullet$};
\node at (6.4,1.8)  {\Large  $\bullet$};
\node at (5+0.72*2*\CosValue+0.666*\CosValueeee,0.72*2*\SinValue+0.666*\SinValueeee)  {\Large  $\bullet$};
\node at (5+1.4*2*\CosValue+0.666*\CosValueeeee,1.4*2*\SinValue-0.666*\SinValueeeee)  {\Large  $\bullet$};
\node at (5-0.2,-0.1)  {$\bm{e}^{-}$};
\node at (5+4*\CosValue+0.3,4*\SinValue+0.1)  {$\bm{e}^{-}$};
\node at (6.6+2*\CosValueee,2.0+2*\SinValueee) {\bf{a}};
\node at (5+2*\CosValue+0.2,2*\SinValue-0.3)  {$\bm{g_{ae}}$};
\draw[fermion, black, line width=.5mm] (10,0)-- (10+2*\CosValue,2*\SinValue);
\draw[fermion, black, line width=.5mm] (10+2*\CosValue,2*\SinValue)-- (10+4*\CosValue,4*\SinValue);
\draw[scalarrr, black, line width=.5mm] (10+2*\CosValue,2*\SinValue)-- (10+2*\CosValue+2*\CosValuee,2*\SinValue+2*\SinValuee);
\node at (10+2*\CosValue,2*\SinValue)  {\Large  $\bullet$};
\node at (10-0.2,-0.1)  {\bf{N}};
\node at (10+4*\CosValue+0.3,4*\SinValue+0.1)  {\bf{N}};
\node at (10+2*\CosValue+2*\CosValuee+0.2,2*\SinValue+2*\SinValuee+0.2) {\bf{a}};
\node at (10+2*\CosValue+0.2,2*\SinValue-0.3)  {$\bm{g_{aN}}$};
\end{tikzpicture}
\caption[Feynman diagrams of axion direct coupling with electrons, higher-order coupling of axion and electron, and axion-to-nucleon coupling, respectively.]{Feynman diagrams of axion direct coupling with electrons, higher-order coupling of axion and electron, and axion-to-nucleon coupling, respectively.}
\label{Fig.3.3}
\end{center}
\end{figure}
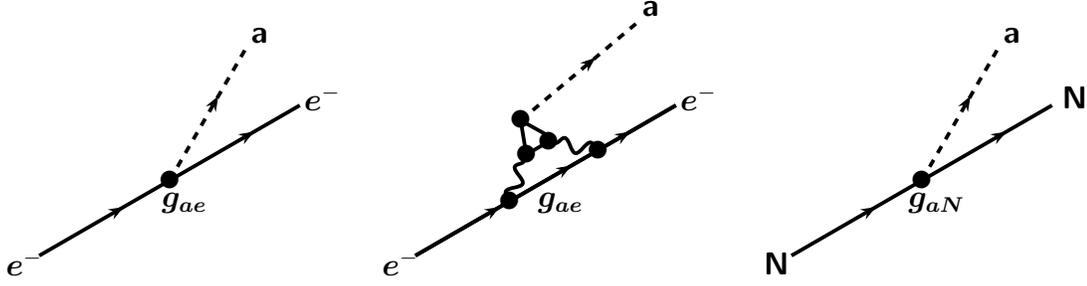

Fermions like electrons and quarks would show Yukawa coupling to axions if they carry the PQ charge. The Feynman diagrams of axion direct coupling with electrons, higher-order coupling of axion and electron, and axion-to-nucleon coupling, respectively, are shown in figure \ref{Fig.3.3}. The coupling of axion to fermions contributes to the Lagrangian density with a term equal to
\begin{equation} 
\L_{a f} = \frac{g_{a f}}{2 m_f} (\bar{\psi_f \gamma^\mu \gamma_5 \psi_f}) \partial_\mu \ a \:,
\end{equation} 
where $\psi_f$ is the fermion field, $f_m$ is its mass, and  $g_{a f}$ represents the axion-fermion coupling constant, that can be written as
\begin{equation} \label{eq.3.63}
g_{a f} = \frac{C_f m_f}{f_a} \:.
\end{equation}   
The dimensionless coefficient plays the role of a Yukawa coupling with an effective PQ charge $C_f$ and a fine-structure constant of the axion $\alpha_{\alpha f} = g_{a f}^2 / 4 \pi$ can be defined.

Some axion models calculated the interaction coefficient $C_e$ that describes the axion effective coupling with electrons at the tree level. In addition, and despite there are no free quarks exists below the QCD scale $\mathrm{\Lambda}_{\text{QCD}} \approx 200 \ \text{ MeV}$, the coupling of axions to light quarks at tree level and the mixing of axions with pions lead to calculate the interaction coefficient of axions with proton $C_p$ and neutron $C_n$ and therefore derive the effective axion-to-nucleon coupling.

\subsection{Further processes}

There are some other processes involving axioms quite relevant in the frame of astrophysics. The predominant emission process of axion in stars would be the Primakoff effect $\gamma + Ze \rightarrow Ze + a$, in which a photon can be converted into an axion in the presence of strong electromagnetic fields. This process is quite relevant in axion searches because axions could be reconverted into photons (and detected) inside a strong magnetic field via the inverse Primakoff effect.

Further processes in axion models with tree level coupling to electrons might dominate the axion emission in white dwarfs and red giants. These processes are the Compton $\gamma + e^{-} \rightarrow e^{-} + a$, and the Bremsstrahlung emission $ e^{-} + Ze \rightarrow Ze +  e^{-} + a$. Also, the axion-nucleon Bremsstrahlung $N+N \rightarrow N+N+a$, could be relevant in supernova explosion \cite{garcia2015solar}.

\subsection{Lifetime of axions}

Due to the coupling with photons, the axion would decay into two photons with a decay rate that can be expressed as follows
\begin{align}
\mathrm{\Gamma}_{a \rightarrow \gamma \gamma} &= \frac{g_{a \gamma \gamma}^2 m_a^3}{64 \pi} = \frac{\alpha^2}{256 \pi^3} C_{a \gamma \gamma} \frac{m_a^3}{f_a^2} \nonumber \\
& \simeq 2.2 \times 10^{-51} {s}^{-1} \left( \frac{10^{12} \ \text{GeV}}{f_a} \right)^5 \:,
\end{align}
where we used here the axion-to-photon coupling from \eqref{eq.3.60}, the axion mass from \eqref{eq.3.58}, and $C_{a \gamma \gamma}=1$ for simplicity. This gives the axion a lifetime of $ \tau_a \equiv \mathrm{\Gamma}^{-1}_{a \rightarrow \gamma \gamma} \sim 4.5 \times 10^{50} \ \text{s}$. This lifetime of the axion exceeds the age of the universe $\sim 4.3 \times 10^{17} \ \text{s}$ for $f_a \gtrsim 10^5 \ \text{GeV}$. Hence, the invisible axion is almost stable, which motivate us to consider it as the dark matter of the universe.

\section{Status of axion search} 

In the visible axion models, the PQ symmetry was broken along with the electroweak symmetry and the axion couples to matter directly.  This model had been ruled out by laboratory limits shortly after their conception. The invisible axion models, on the other hand, considered the scale of PQ symmetry breaking to be much higher than the electroweak scale. Hence, these models predict a very light axion with very weak coupling to ordinary matter. Such weak coupling of the axion would allow it to evade existing experimental searches, and hence, this invisible axion model is allowed by present experiments.

Although the invisible axions are very light, very weakly coupled, and have not yet been discovered, they are not necessarily to be totally invisible. The allowed range of parameters is highly constrained from astrophysical and cosmological considerations and also from laboratory measurements. We will not discuss here how the invisible axions might affect astrophysics and cosmology, and instead, we will illustrate the astrophysical and cosmological bounds on the invisible axions in the next chapter.
\chapter{\textbf{Axion Dark Matter and Cosmology}} \label{ch4}

We discussed in the previous chapter that the axion is a PNGB, which appears after the spontaneous breaking of the PQ symmetry, which was proposed to solve the strong CP violation problem. In this chapter, we will study the possible role that axion and axion-like particles can play to explain the mystery of dark matter. Then, to discuss whether they correctly explain the present abundance of the dark matter, we will investigate their production mechanism in the early universe. Afterward, we will discuss the recent astrophysical, cosmological, and laboratory bounds on the axion coupling with the ordinary matter.

\section{Axion as a dark matter candidate}

As mentioned in chapter \ref{ch2}, only a small fraction of the total matter content of the universe is made of baryonic matter, while the vast majority is constituted by dark matter \cite{aghanim2018planck}. It is usually assumed to consist of one or several new species of weakly interacting particles \cite{carlson1992self}. It is characterized to be electrically neutral and interacting with ordinary matter only through gravity \cite{bertone2005particle}. In principle, interactions under the weak and strong forces can be allowed, depending on the model. Dark matter can not be constituted by any large fraction of baryonic matter, as the baryon density in the universe is tightly constrained by, among other things, CMB measurements. If the QCD axions exist, they would be considered a very promising and well-motivated candidate for dark matter, as they have the main characteristics to be dark matter. They are nearly collisionless, electrically neutral, very stable, non-baryonic, and weakly interacting with the standard model particles \cite{marsh2016axion}. Depending on the production process, thermal or non-thermal, the axions could constitute either hot dark matter or cold dark matter. We will explain this point as we progress during this chapter.

The non-baryonic forms of dark matter are usually subdivided into three classes$;$ hot, warm, and cold dark matter \cite{kauffmann1993formation, bond1984dark, primack1984dark}. This pertains to objects or particles that have large, intermediate, and small free-streaming lengths, respectively. The free-streaming length is a characteristic length scale for dark matter particle and defines as the possible distance of travel due to random velocities in the early universe, adjusted for the expansion of the universe. Since the particles can propagate freely over this distance scale, any smaller fine-scale structure is smoothed out.

\begin{itemize}

\item {\bf Hot dark matter (HDM).} HDM requires nearly massless particles that they move at relativistic velocities. This is achieved if the dark matter particle's mass is satisfying $m \lesssim T_d$, which means that they are still relativistic by the time of their decoupling. In addition, HDM is defined as particles with $m \lesssim 1 \ \text{eV}$, which is the temperature at which the universe energy density moves from being radiation dominated to matter-dominated. HDM particles have long free-streaming lengths, $\gtrsim 1 \ \text{Mpc}$. A couple of examples of such candidates are neutrinos and HDM axions.

\item {\bf Warm dark matter (WDM).} WDM consists of particles with masses satisfying $m \lesssim T_d$ as for HDM but in this case $m \gtrsim 1 \ \text{eV}$. Therefore, these particles were relativistic at their decoupling temperature, but nonrelativistic at the time of matter-radiation equality. The free-streaming size is of dwarf-galaxy scale, or $\sim 1$ Mpc. There are no clear candidates for WDM, but non-thermally produced gravitinos, sterile neutrinos, and light neutralinos could be viable ones.

\item {\bf Cold dark matter (CDM).} CDM refers to objects that are sufficiently massive and moving at sub-relativistic velocities. This is achieved if the dark matter particle's mass is much greater than the temperature at which it decouples from the cosmological plasma, $m \gtrsim T_d$. Hence, early in time after the big bang at the point of decoupling, these particles become nonrelativistic, massive, and have a short free-streaming length, $\lesssim 1$ Mpc. The major CDM candidate is WIMPs, gravitinos, MACHOs, CDM axions particles.

\end{itemize}

\subsection{Axion-like particles}

The visible and invisible axion models are not the only extensions to the standard model introducing new symmetries. In fact, there are plenty of theories have been proposed to extend the standard model by introducing new symmetries and new particles, for example, supersymmetric theories and string theory models \cite{ringwald2012exploring}. If any of these new symmetries is global and gets spontaneously broken, then a Goldstone boson or PNGB is obtained. If this boson is a light scalar or pseudo-scalar, it can share qualitative properties with the QCD axion. This means that, in addition to invisible QCD axions, there can be many other axion-like particles (ALPs). These ALPs are usually predicted by string theory-driven models \cite{arvanitaki2010string, cicoli2012type, anselm1982second}. In particular, ALPs naturally arise from string theory due to the compactification of extra dimensions \cite{svrcek2006axions, witten1984some}.

In general, it is expected that ALPs couple to ordinary matter in the same way as the QCD axions do. Thus, the coupling constants for the ALPs interactions with photons and fermions would be similar to the ones in equations \eqref{eq.3.60} and \eqref{eq.3.63}, respectively. To adapt those equations to ALPs, one would remove the terms containing z, as they come from the mixing between axions and mesons, and replace $f_a$ with the ALP decay constant. Another difference is that, while axions had originally been proposed with the aim of solving the strong CP problem, ALPs can be considered as pure predictions of theories beyond the standard model. They are not needed for any specific purpose but can be considered potential candidates for the particles of dark matter.

\subsection{HDM axions}

Axion, and more generally, ALPs may be copiously produced in the early universe, including via thermal processes. Therefore, relic axions and ALPs constitute an HDM component. Hadronic axions that do not couple directly to charged leptons would be produced by Primakoff reactions with the quarks in the primordial quark-gluon plasma (QGP). After the temperature of the universe drops below $\mathrm{\Lambda}_{\text{QCD}}$ and confinement occurs, the dominant thermalization process is $\pi + \pi \leftrightarrow a + \pi$ \cite{chang1993hadronic}.

\subsection{CDM axions}

Since HDM, no matter what the constituents are, seemingly can not make up more than a small part of the total dark matter density, CDM axions \cite{dine1983not, preskill1983cosmology, abbott1983cosmological, stecker1982evolution} are usually more seriously considered. They are produced non-thermally in the early universe by the misalignment mechanism and under certain circumstances, also via the decay of topological defects such as axion strings and domain walls. We will discuss the contribution of each process to the axion abundance in section \ref{sec.4.3}.

\section{Thermal production of axions}

If the temperature of the primordial plasma is sufficiently high, axions are created and annihilated during interactions among the standard model particles in the thermal bath of the QCD plasma. The produced axions from such processes are known as thermal axions and described by the standard freeze-out scenario \cite{kolb1990early, turner1986thermal, masso2002axion}. The number density $n_a^{th}(t)$ of thermal axions obeys the Boltzmann equation
\begin{equation} \label{eq.4.3}
\dfrac{d n_a^{\text{th}}}{dt} + 3 H n_a^{\text{th}} = \mathrm{\Gamma} (n_a^{\text{eq}}-n_a^{\text{th}}) \:,
\end{equation}
where $H$ is the Hubble parameter, and $\mathrm{\Gamma}$ is the  interaction  rate at which axions are created and annihilated in the plasma
\begin{equation}
\mathrm{\Gamma} = \sum_i n_i \langle \sigma_i v \rangle \:.
\end{equation}
Here, the sum is over all processes involving axions $a + i\leftrightarrow 1 + 2$, with $1$, and $2$ refer to other particles species, $n_i$ is the number density of particles species $i$, and $\langle \sigma_i v\rangle$ indicates averaging over the momentum distribution of the particles involved, with $\sigma_i$ in the corresponding cross-section, and $v$ is the relative velocity between the particle $i$ and the axion. The term $n_a^{\text{eq}}$ represents the number density of axions at thermal equilibrium, which is obtained by using the Bose-Einstein distribution
\begin{equation}
n_a^{\text{eq}} = \frac{\zeta(3)}{\pi^2} T^3 \:,
\end{equation}
where $\zeta (3) = 1.202 \cdots$ is the Riemann zeta function of argument $3$, and $T$ is the temperature of the plasma. Using the conservation of the number density at equilibrium
\begin{equation}
\dfrac{dn_a^{\text{eq}}}{dt} + 3 H n_a^{\text{eq}} =0 \:,
\end{equation}
we can rewrite equation \eqref{eq.4.3} as follows,
\begin{equation} 
\dfrac{d}{dt} \left[ R^3 (n_a^{\text{eq}}-n_a^{\text{th}})  \right] = - \mathrm{\Gamma}  R^3 (n_a^{\text{eq}}-n_a^{\text{th}}) \:,
\end{equation}
where $R(t)$ is the scale factor of the universe. The solution of this equation implies that the axion would be in thermal equilibrium as long as its interaction rate is faster than the Hubble expansion, $\mathrm{\Gamma}(T) > H(T) $. Therefore, the thermal population of axions was produced when this condition was satisfied for a few times at some point in the early universe until the axions decouple from the plasma at the decoupling temperature $T_{\text{th}}$ which define with $\mathrm{\Gamma}(T_{\text{th}})= H(T_{\text{th}})$. The thermal population of axions thus established did not subsequently get diluted away by inflation or some other cause of huge entropy release \cite{kuster2007axions}.

In the thermal bath, axions couple differently to fermions depending on the axion model, whereas their coupling with gluons is model-independent. The thermal average of the interaction rate $\mathrm{\Gamma}$ is calculated, including the following three elementary processes for thermalizing axions in the early universe
\begin{center}
\begin{inparadesc}
\item[(a)] $a+q \leftrightarrow g + q \:, \quad$  
\item[(b)] $a+g \leftrightarrow q + \bar{q} \:, \quad$ and $\quad$
\item[(c)] $a+g \leftrightarrow g+g\:.$
\end{inparadesc}
\end{center}
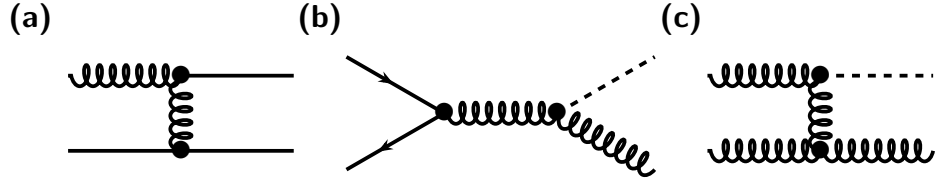
\begin{figure}[ht!]
\begin{center}
\begin{tikzpicture}
\pgfmathsetmacro{\CosValueee}{cos(30)}
\pgfmathsetmacro{\SinValueee}{sin(30)}
\draw[fermionnn, black, line width=.5mm] (0.5,0)-- (2,0);
\draw[fermionnn, black, line width=.5mm] (2,0)-- (3.5,0);
\draw[gluon, black, line width=.5mm, rotate=180] (-2,-1.0)--(-0.5,-1.0);
\draw[fermionnn, black, line width=.5mm] (2,1.0)-- (3.5,1.0);
\draw[gluon, black, line width=.5mm, rotate=360] (2,0)--(2,1.0);
\node at (2,0)  {\Large  $\bullet$};
\node at (2,1)  {\Large  $\bullet$};
\draw[gluon, black, line width=.5mm, rotate=180] (-7,-0.5)--(-5.5,-0.5);
\draw[fermionin, black, line width=.5mm] (5.5-1.5*\CosValueee,0.5+1.5*\SinValueee)--(5.5,0.5);
\draw[fermion, black, line width=.5mm] (5.5,0.5)-- (5.5-1.5*\CosValueee,0.5-1.5*\SinValueee);
\draw[scalarr, black, line width=.5mm] (7,0.5)-- (7+1.5*\CosValueee,0.5+1.5*\SinValueee);
\draw[gluon, black, line width=.5mm, rotate=360] (7+1.5*\CosValueee,0.5-1.5*\SinValueee)--(7,0.5);
\node at (5.5,0.5)  {\Large  $\bullet$};
\node at (7,0.5)  {\Large  $\bullet$};
\draw[gluon, black, line width=.5mm, rotate=180] (-10.5,0)--(-9,0);
\draw[gluon, black, line width=.5mm, rotate=180] (-12,0)--(-10.5,0);
\draw[gluon, black, line width=.5mm, rotate=180] (-10.5,-1.0)--(-9,-1.0);
\draw[scalarr, black, line width=.5mm] (10.5,1.0)-- (12,1.0);
\draw[gluon, black, line width=.5mm, rotate=360] (10.5,0)--(10.5,1);
\node at (10.5,0)  {\Large  $\bullet$};
\node at (10.5,1)  {\Large  $\bullet$};
\node at (0,1.75) {\bf{(a)}};	
\node at (3.85,1.75) {\bf{(b)}};
\node at (8.65,1.75) {\bf{(c)}};
\end{tikzpicture}
\caption[Feynman diagrams of the processes which produce thermal axions in the early universe.]{Feynman diagrams of the processes which produce thermal axions in the early universe.}
\label{Fig.4.1}
\end{center}
\end{figure}

The Feynman diagrams of the processes which produce thermal axions in the early universe are shown in figure \ref{Fig.4.1}. These processes have a cross-section of the order of $\sigma_{\text{agg}}=\alpha_s^3/f_a^2$, and accordingly, the interaction rate becomes\newpage \noindent
\begin{equation}
\mathrm{\Gamma}(T) \sim \frac{\alpha_s^3}{f_a^2} T^3 \sim \frac{10^5}{f_a^2} \:,
\end{equation}
where $\alpha_s = g_s^2/4\pi^2$. Assuming that axions thermalize at temperature $T_{th}$, their number density today is obtained by assuming the conservation of the comoving number density,
\begin{equation}
n_a^{\text{th}} (T_0) = n_a^{\text{th}} (T_{\text{th}}) \left( \frac{a (T_{\text{th}})}{a (T_0)} \right)^3 = 7.5 \ \text{cm}^{-3} \frac{107.75}{g_{\ast} (T_{\text{th}})} \:,
\end{equation}
where $g_{\ast}(T)$ denotes the relativistic number of degrees of freedom at temperature $T$. The current abundance from thermal production can be then estimated to be
\begin{equation}
\mathrm{\Omega}_a = 10^{-8} \left( \frac{100}{g_{\ast}} \right) \frac{10^{12} \ \text{GeV}}{f_a} \:.
\end{equation}
For $f_a \sim 10^{12} \ \text{GeV}$, the axion production from the thermal processes is not efficient. If $f_a$ is smaller, the coupling of the axion increases, and the axion decouples later, helping to get more relic abundance. It is possible to get the correct relic abundance for $f_a \sim 10^{6} \ \text{GeV}$, but this value is normally excluded by astrophysics bounds. However, the astrophysical bounds do not apply if the axion has an anomalously small coupling to photons. In this scenario, the axion forms a hot dark matter candidate, but bounds from Planck already exclude the axion as hot dark matter. Therefore, taking into account that hot dark matter can not be the dominant dark matter component can lead to putting an upper bound on the axion mass. A recent study involving CMB anisotropy measurements, halo power spectrum data, and Hubble constant measurements provides an approximate lower limit of the axion mass $m_a \sim 0.7 \ \text{eV}$ \cite{hannestad2010neutrino}.

\section{Non-thermal production of axions} \label{sec.4.3}

The most efficient mechanism for axion production is non-thermal, and there are three types such processes, the misalignment mechanism \cite{preskill1983cosmology, dine1983not, abbott1983sikivie}, axion string decay, and axion domain walls decay \cite{davis1986cosmic, lyth1992estimates}. The abundance of the non-thermally produced axions depends on two important scales. The first is the temperature at which the axion mass, arising from non-perturbative QCD effects, becomes significant. The second scale is the temperature $T_{\text{PQ}}$, at which the PQ phase transition occurs when the  $U(1)_{\text{PQ}}$ symmetry spontaneously breaks. As we will see below, depending on whether this temperature $T_{\text{PQ}}$ is greater or less than the inflationary reheating temperature $T_R$, the contribution of each of the three production mechanisms to the cold axion population is determined.
 
At high temperatures, the QCD effects are not significant, and the axion mass is negligible. Gradually the axion acquires its mass due to the non-perturbative QCD effect. At a critical time, $t_1$, when $m_a  t_1 \sim 1$, the axion mass becomes important and the corresponding temperature of the universe at $t_1$ is $T_1 \simeq 1 \ \text{GeV}$. The $U(1)_{\text{PQ}}$ symmetry is then unbroken at early times and temperatures greater than $T_{\text{PQ}}$. At $T_{\text{PQ}}$, it breaks spontaneously, and the axion field, proportional to the phase of the complex scalar field acquiring a vacuum expectation value, may have any value. The phase varies continuously, changing by order one from one horizon to the next. At that time, axion strings may appear as topological defects.

Now we have to distinguish between two cases. The first one occurs if the reheating temperature after inflation is below the PQ transition temperature $T_{\text{PQ}} > T_R$. In this case, the axion field is homogenized over vast distances, and the string density is diluted because of inflation, to a point where it is extremely unlikely that our visible universe contains any axion strings. When the axion mass turns on at $t_1$, the axion field starts to oscillate. The amplitude of this oscillation is determined by how far from zero the axion field is when the axion mass turns on. The axion field oscillations do not dissipate into other forms of energy and hence contribute to the cosmological energy density today. Such a contribution is called the misalignment mechanism.  Since the axion string density is diluted by inflation, thus the misalignment mechanism is the only process that contributes significantly to the density of cold axions. In this scenario, the non-thermal production of axions is then estimated by investigating the evolution of the background filed. The equation of motion for a homogeneous axion field in an FRW universe is described by
\begin{equation}
\ddot{a} + 3H\dot{a}+m_a^2 a=0 \:.
\end{equation}
When the axion mass is constant, at some time after the QCD phase transition, the approximate solution to this equation is
\begin{equation}
a \sim a_{\text{init}} \frac{1}{R^{3/2}} \cos(m_a t) \:,
\end{equation}
where $a_{\text{init}}$ is the initial axion value chosen arbitrarily between $-\pi$ and $\pi$ at the moment of the PQ phase transition. The energy density is then given by
\begin{equation}
\rho =\frac{1}{2} (\dot{a}^2 + m_a^2 a^2) \:,
\end{equation}
scales with $1/R^3$ like non-relativistic matter. This leads us to expect the CDM axion energy density to be \cite{sikivie2008axion}
\begin{equation} \label{eq.4.12}
\mathrm{\Omega}_a \sim 0.15 \left( \frac{f_a}{10^{12} \ \text{GeV}} \right)^{7/6} \left(\frac{ 0.7}{h} \right)^2 \theta_i^2 \:,
\end{equation}
where $\theta_i$ is the initial random value of the vacuum angle, the misalignment angle,  and $h$ is the reduced Hubble parameter.

The other case occurs if $T_{\text{PQ}} < T_R$, meaning the reheating temperature after inflation is above the PQ transition temperature. In such a case, the axion field is not homogenized, and strings radiate cold, massless axions until non-perturbative QCD effects become significant at temperature $T_1$. When the universe cools to $T_1$, the axion strings become the boundaries of $N$ domain walls. The network of the domain walls bounded by the axion strings is unstable, and therefore it rapidly radiates cold axions and decay. Here, in addition to the contribution of the vacuum realignment mechanism to the density of cold axions, there are significant contributes from both the string decay and the wall decay. The three contributions give a total CDM axion energy density \cite{sikivie2008axion}
\begin{equation} \label{eq.4.13}
\mathrm{\Omega}_a \sim 0.7 \left( \frac{f_a}{10^{12} \ \text{GeV}} \right)^{7/6} \left(\frac{ 0.7}{h} \right)^2 \:.
\end{equation}
Comparing this to the measured CDM density provides an approximate lower limit of the axion mass $m_a \sim 10 \ \mu \text{eV}$. However, equations \eqref{eq.4.12} and \eqref{eq.4.13}  are subject to many sources of uncertainty, aside from the uncertainty about the contribution from string decay \cite{sikivie2008axion}. Consequently, it is not yet clear with sufficient precision how much of these axions are produced.

\section{Bounds on axion coupling} \label{sec.4.4}

Axions and ALPs can be searched for in astrophysics, cosmology, and laboratory experiments, based on the theoretical predictions discussed in the previous chapter. The interaction between axions and ordinary matter can be exploited, and bounds on the axion couplings can be obtained as a result of both observations and simulations. In this section, we present bounds on axion properties coming from observations and experiments, see references \cite{raffelt2008astrophysical, asztalos2006searches, abbott1983cosmological, marsh2016axion} for a thorough review. Note that recent research based on both observations and experiments still aiming to improve the current limits on axions and ALPs couplings. In this context,  we show in chapter \ref{ch5} some techniques in which different astrophysical environments are used to probe new limits on the coupling of ALPs with photons.

\subsection{Astrophysical axion bounds}

The existence of the axions and ALPs would slightly affect physics related to the star's evolution and well established physical processes. The axion properties can then be bounded by the discrepancies that some axion parameters would entail in deeply studied astrophysical processes. The main argument resides in the fact that axions can be produced in hot and dense environments like in stars and other galactic objects. The axion production, which depends on $f_a$, provides an additional energy loss channel for the source. In the following, we summarize some of these astrophysical bounds.

Consequently, stars have been used to derive some constraints on the axions and ALPs coupling parameters. Photons in the stellar interior would convert into axions or ALPs by the Primakoff process. This would be a very efficient energy loss mechanism for the star. The most obvious star that can be exploited to constrain axion parameters is our sun. The energy loss by solar axion emission requires enhanced nuclear burning and increases the solar ${}^8B$ neutrino flux \cite{raffelt2008astrophysical}. The observation of ${}^8B$ neutrino flux gives a bound $g_{a \gamma \gamma} \lesssim 7 \times 10^{-10} \ \text{GeV}^{-1}$. The solar neutrino flux constraint can also be applied to the axion-electron coupling, giving $g_{a ee} < 2.8 \times 10^{-11} \ \text{GeV}^{-1}$, where $g_{a ee} \equiv C_e m_e /f_a$, and $m_e$ is the electron mass.

A more restrictive bound comes from the observations of globular clusters. The helium-burning lifetimes of horizontal branch (HB) stars give a bound for axion-photon coupling $g_{a \gamma \gamma} \lesssim 0.6 \times 10^{-10} \ \text{GeV}^{-1}$. Moreover, the delay of helium ignition in red-giant branch (RGB) stars due to the axion cooling gives $g_{a ee} < 2.5 \times 10^{-13} \ \text{GeV}^{-1}$.

In addition, the energy loss rate of the supernova 1987A leads to another very strong constraint on the axion-nucleon coupling. For a small value of the coupling, the mean free path of axions becomes larger than the size of the supernova core (so-called the ``free streaming’’ regime). In this regime, the energy loss rate is proportional to the axion-nucleon coupling squared, and one can obtain the limit $f_a \gtrsim 4 \times 10^{8} \ \text{GeV}$. On the other hand, for a large value of the coupling, axions are ``trapped’’ inside the supernova core. In this regime, by requiring that the axion emission should not have a significant effect on the neutrino burst, one can obtain another bound $f_a \lesssim \bm{\mathcal{O}}(1) \times 10^6 \ \text{GeV}$. However, in this ``trapped’’ regime, it was argued that the strongly-coupled axions with $f_a \lesssim \bm{\mathcal{O}}(1) \times 10^5 \ \text{GeV}$ would have produced an unacceptably large signal at the Kamiokande detector, and hence they were ruled out \cite{raffelt2008astrophysical}.

Furthermore, the axion-electron coupling is constrained by the observation of white-dwarfs. The cooling time of white-dwarfs due to the axion emission gives a bound $g_{a ee} < 4 \times 10^{-13} \ \text{GeV}^{-1}$. Recently, it is reported that the fitting of the luminosity function of white-dwarfs is improved due to the axion cooling, which implies $g_{a ee} \simeq (0.6\text{--}1.7) \times 10^{-13} \ \text{GeV}^{-1}$. Also, observed pulsation period of a ZZ Ceti star can be explained by means of the cooling due to the axion emission, if $g_{a ee} \simeq (0.8-2.8) \times 10^{-13} \ \text{GeV}^{-1}$. These observations might imply the existence of the $\text{meV}$ mass axion, but require further discussions.

Additionally, the axion with mass $m_a \sim \bm{\mathcal{O}}(1) \ \text{eV}$ in galaxy clusters makes a line emission due to the decay into two photons, whose wavelength is $\lambda_a \simeq 24800 \ \mathrm{\AA}/[m_a/ 1 \ \text{eV}]$. This line emission gives observable signature in telescopes. Such a line has not been observed in any telescope searches, which excludes the mass range $3 \ \text{eV} \lesssim m_a \lesssim 8 \ \text{eV}$.

\subsection{Cosmological axion bounds}

According to the astrophysical axion bounds, the spontaneous breaking of the PQ symmetry is expected to happen at a very high energy scale. As a consequence of this, axions start playing quite a relevant role in cosmology. Axion cosmological effects depend on the value of the PQ symmetry breaking scale $f_a $, which is the value related to the potential dark matter nature of axions.

In the previous section, we discussed the conditions for axions to be cold dark matter candidates. This happens if axions are non-thermally produced via the misalignment mechanism and the decay of topological strings and domain walls. The resulting cosmological energy density is expressed as a function of $f_a$. In order to obtain the energy densities observed in the universe, if a cold dark matter axion population exists, then the energy scale has to be $f_a \sim 10^{12} \ \text{GeV}$.

The other possibility is to have hot dark matter axions, thermally produced in the early universe via axion-pion conversion ($\pi + \pi \leftrightarrow \pi + a$). Massive thermal axions would have a similar impact on cosmological observables as massive neutrinos. The cosmological bound on the breaking energy scale, in this case, is $f_a \lesssim 10^{12} \ \text{GeV}$, while the most stringent limit on the thermal axion mass has been placed using the Planck 2015 data, and is $m_a < 0.529 \ \text{eV}$ \cite{di2016cosmological}.

The astrophysical and cosmological bounds on axion couplings narrow the parameter space where axions could still exist. Moreover, these bounds drive the experimental searches aiming at axion detection.

\subsection{Laboratory axion bounds}

We have mentioned that bounds on axion mass came from laboratory experiments at the very early stage of axion history. When Weinberg and Wilczek proposed a $100 \ \text{keV} \ \text{--} \ 1 \ \text{MeV}$ mass axion as a Peccei-Quinn boson, such a candidate was soon ruled out, as an axion with a mass of this order would have had large enough couplings to be observed in several laboratory experiments. This is the start of the invisible axions era. Since then, there have been other axion searches in laboratory-based experiments. They are mainly magnetometry searches for spin-dependent forces mediated by axion exchange. The results were generally weaker than the bounds set by both astrophysical and cosmological studies.

Another approach to axion-like particle detection is a microwave light shining through the wall experiment, which exploits the Primakoff effect twice \cite{betz2012status}. A photon source is fired against a wall, and some of the photons are converted to axion-like particles by the Primakoff effect. The wall blocks the unconverted photons but not the ALPs that reach a receiving cavity where the reciprocal conversion takes place via the Primakoff effect. The photons thus obtained can be detected.

Additionally, a very interesting recent publication searches for ultra low-mass axion dark matter, based on the proposed interaction between an oscillating axion dark matter field and gluons or fermions. Assuming that such an interaction would induce oscillating electric dipole moments of nucleons and atoms, they set the first laboratory bound on the axion-gluon coupling and also improved on previous laboratory constraints on the axion-nucleon interaction.

In the next chapters, we will provide more information about the possible role of axions and axion-like particles in cosmology and astrophysics. The interest of the scientific community in this kind of search is growing fast, and their discovery would have a strong impact on particle physics and beyond.

\chapter{\textbf{Phenomenology of ALPs Coupling with Photons in the Jets of AGNs}} \label{ch5}

Interestingly there are many string theory models of the early universe that motivate the existence of a homogeneous cosmic ALP background analogous to the cosmic microwave background, arising via the decay of string theory moduli in the very early universe. In this chapter, we study the phenomenology of the coupling between the CAB ALPs and photons in the presence of an external electromagnetic field that allows the conversion between ALPs and photons. Based on this scenario, we examine the detectability of signals produced by ALP-photon coupling in the highly magnetized environment of the relativistic jets produced by active galactic nuclei. Then, we test the cosmic ALP background model that is put forward to explain the soft X-ray excess in the Coma cluster due to CAB ALPs conversion into photons using the M87 jet environment. Moreover, we demonstrate the potential of the active galactic nuclei jet environment to probe low-mass ALP models and to potentially constrain the model proposed to better explain the Coma cluster soft X-ray excess.

\section{Introduction} \label{sec.5.1}

If ALPs really exist in nature, they are expected to couple with photons in the presence of an external electric or magnetic field through the Primakoff effect with a two-photon vertex \cite{sikivie1983experimental}. This coupling gives rise to the mixing of ALPs with photons \cite{raffelt1988mixing}, which leads to the conversion between ALPs and photons. This mechanism serves as the basis to search for ALPs$;$ in particular, it has been put forward to explain a number of astrophysical phenomena, or to constrain ALP properties using observations. For example, over the last few years, it has been realized that this phenomenon would allow searches for the ALPs in the observations of distant active galactic nuclei in radio galaxies \cite{bassan2010axion, horns2012probing}. Since photons emitted by these sources can mix with ALPs during their propagation in the presence of an external magnetic field and this might reduce photon absorption caused by extragalactic background light \cite{harris2014photon}. This scenario might lead to a suitable explanation for the unexpected behavior for the spectra of several AGNs \cite{mena2013hints}. Furthermore, because only photons with polarization in the direction of the magnetic field can couple to ALPs, this coupling can lead to change in the polarization state of photons. This effect can also be useful to search for ALPs in the environment of AGNs by looking for changes in the linear degree of polarization from the values predicted by the synchrotron self-Compton model of gamma ray emission.

Over the last few years, it has been realized that this phenomenon would allow searches for the ALPs in the observations of distant AGNs in radio galaxies \cite{bassan2010axion, horns2012probing}. Since photons emitted by these sources can mix with ALPs during their propagation in the presence of an external magnetic field, and this might reduce photon absorption caused by extragalactic background light \cite{harris2014photon}. Recent observations of blazars by the Fermi Gamma-Ray Space Telescope \cite{abdo2011fermi} in the $0.1\text{--}300 \ \text{GeV}$ energy range show a break in their spectra in the 1-10 GeV range. In their paper \cite{mena2013hints, mena2011signatures}, Mena, Razzaque, and Villaescusa-Navarro have modeled this spectral feature for the flat-spectrum radio quasar 3C454.3 during its November 2010 outburst, assuming that a significant fraction of the gamma rays convert to ALPs in the magnetic fields in and around the large scale jet of this blazar.
 
Furthermore, many galaxy clusters show a soft X-ray excess in their spectra below $1 \textup{--} 2 \ \text{keV}$ on top of the extrapolated high-energy power \cite{bonamente2002soft, durret2008soft}. This soft excess was first discovered in 1996 from the Coma and Virgo clusters, before being subsequently observed in many other clusters \cite{lieu1996discovery, lieu1996diffuse, bowyer1996extreme}. However, the nature of this component is still unclear. There are two astrophysical explanations for this soft X-ray excess phenomenon, for review, see \cite{durret2008soft, angus2014soft}. The first is due to emission from a warm $T \approx 0.1 \ \text{keV}$ gas. The second is based on inverse-Compton scattering of $\gamma \sim 300 \textup{--} 600$ non-thermal electrons on the cosmic microwave background. However, both of these two explanations face difficulties in clarifying the origin of the soft X-ray excess \cite{angus2014soft}. Recent work \cite{conlon2013excess} proposed that this soft excess is produced by the conversion of a primordial cosmic ALP background with $0.1 \textup{--} 1 \ \text{keV}$ energies into photons in the magnetic field of galaxy clusters. The existence of such a background of highly relativistic ALPs is theoretically well-motivated in models of the early universe arising from compactifications of string theory to four dimensions. Also, the existence of this CAB can be indirectly probed through its contribution to dark radiation, but this is beyond the scope of this work and we confine ourselves here by highlighting how this can change the ALP bounds.

The main aim of the work presented in this chapter is to follow the approach of \cite{mena2013hints} to test the CAB model that is put forward in \cite{angus2014soft} to explain the soft X-ray excess in the Coma cluster based on the conversion between CAB ALPs and photons in the presence of an external magnetic field using the astrophysical environment of the M87 jet. We aim as well to demonstrate the potential of the AGNs jet environment to probe low-mass ALP models and to potentially constrain the model proposed to explain the soft X-ray excess in the Coma cluster better.
 
The structure of this chapter is as follows. In section \ref{sec.5.2}, we review the theoretical model that describes the ALP-photon mixing phenomenon. In section \ref{sec.5.3'},  we describe some aspects of the astrophysical environment of the active galactic nuclei and their jets where these interactions may take place. In section \ref{sec.5.3}, we study the effects of ALPs conversion into photons to reproduce spectral curvature in the radio quasar 3C454.3 to constrain fundamental parameters of the ALP-photon conversion mechanism. In section \ref{sec.5.4}, we briefly discuss the motivation for the existence of the CAB. Then in section \ref{sec.5.5}, we check whether the soft X-ray excess in the environment of the M87 AGN jet can be explained due to CAB ALPs conversion into photons in the jet magnetic field. In section \ref{sec.5.6}, the results of a numerical simulation of the ALP-photon coupling model are discussed and compared with observed soft excess luminosities in observations. Finally, our conclusion is provided in section \ref{sec.5.7}.

\section{ALP-photon coupling model} \label{sec.5.2}

We first outline the theory of the conversion between ALPs and photons in an external magnetic field within the environment of the jets of AGNs following \cite{mena2013hints, mena2011signatures}. In the presence of a background magnetic field, the coupling of an ALP with a photon is described by the effective Lagrangian \cite{sikivie1983experimental, raffelt1988mixing, anselm1988experimental}

\begin{equation} \label{eq.5.1}
\mathrm{\ell}_{a\gamma} = - \frac{1}{4} g_{a\gamma} \mathrm{F}_{\mu \nu} \tilde{\mathrm{F}}^{\mu \nu} a = g_{a\gamma} \ \mathbf{E} \cdot \mathbf{B} \ a \:,
\end{equation}
where $g_{a\gamma}$ is the ALP-photon coupling parameter with dimension of inverse energy, $\mathrm{F}_{\mu \nu}$ and $\tilde{\mathrm{F}}^{\mu \nu}$ represent the electromagnetic field tensor and its dual respectively, and $a$ donates the ALP field. While $\mathbf{E}$ and $\mathbf{B}$ are the electric and magnetic fields, respectively. Then, we consider a monochromatic and linearly polarized ALP-photon beam of energy $\omega$ propagating along the $z$-direction in the presence of an external and homogeneous magnetic field. The equation of motion for the ALP-photon system can be described by the coupled Klein-Gordon and Maxwell equations arising from the Lagrangian equation \eqref{eq.5.1}. For very relativistic ALPs when $\omega \gg m_a$, the short-wavelength approximation can be applied successfully and accordingly, the beam propagation can be described by the following Schr$\ddot{\text{o}}$dinger-like form \cite{raffelt1988mixing, bassan2010axion}\newpage \noindent
\begin{equation} \label{eq.5.2}
\left( i \dfrac{d}{dz} + \omega + \bm{\mathcal{M}} \right)
  \left( \begin{matrix} A_{\perp}(z) \\ A{\parallel}(z) \\ a(z) \end{matrix} \right) =0 \:,
\end{equation}
where $A_{\perp}$ and $A_{\parallel}$ are the photon linear polarization amplitudes along the $x$ and $y$ axis, respectively, and $a(z)$ donates the ALP amplitude. Here,  $\bm{\mathcal{M}}$ represents the mixing matrix of the ALP field with the photon polarization components. Since only photons with polarization parallel to the magnetic field couple to ALP, so for simplicity, we restrict our attention to the case of magnetic field transverse $\mathbf{B}_T$ to the beam direction (\ie in the x-y plane). Therefore, if we choose the y-axis along $\mathbf{B}_T$, so that $\mathbf{B}_x$ vanishes and the mixing matrix can be written as
\begin{equation} \label{eq.5.3}
\bm{\mathcal{M}} = \left( \begin{matrix} 
\mathrm{\Delta}_{\perp} & 0 & 0 \\ 0 &  \mathrm{\Delta}_{\parallel} & \mathrm{\Delta}_{a\gamma} \\ 0 & \mathrm{\Delta}_{a\gamma} & \mathrm{\Delta}_{a} \end{matrix} \right) \:.
\end{equation}
As expressed in references \cite{bassan2010axion, mena2013hints, mena2011signatures}, the elements of $\bm{\mathcal{M}}$ and their references values are given as below$:$
\begin{align}  \label{eq.5.4}
\mathrm{\Delta}_{\perp}  &\equiv 2 \ \mathrm{\Delta}_{\text{QED}} + \mathrm{\Delta}_{\text{pl}} \:,  \nonumber \\[10pt]
\mathrm{\Delta}_{\parallel}  &\equiv \frac{7}{2} \ \mathrm{\Delta}_{\text{QED}} + \mathrm{\Delta}_{\text{pl}} \:,  \nonumber \\
\mathrm{\Delta}_{a\gamma} &\equiv \frac{1}{2} \ g_{a\gamma} B_T \simeq 1.50 \times 10^{-11} \ \left( \frac{g_{a\gamma}}{10^{-10} \ \text{GeV}^{-1}} \right) \left( \frac{B_T}{10^{6} \ \text{G}} \right) \ \text{cm}^{-1} \:,  \nonumber \\
\mathrm{\Delta}_a &\equiv  -\frac{m_a^2}{2\omega} \simeq -2.53 \times10^{-13} \ \left( \frac{\omega}{\text{keV}} \right)^{-1} \left( \frac{m_a}{10^{-7} \ \text{eV}} \right)^{2} \ \text{cm}^{-1} \:.
\end{align}
The two terms $\mathrm{\Delta}_{\text{QED}}$ and $\mathrm{\Delta}_{\text{pl}}$, account for the QED vacuum polarization and the plasma effects, and they are determined as
\begin{align} \label{eq.5.5}
\mathrm{\Delta}_{\text{QED}}  &\equiv \frac{\alpha \omega}{45 \pi} \left( \frac{B_T}{B_{cr}} \right)^{2} \simeq 1.34 \times10^{-12} \ \left( \frac{\omega}{\text{keV}} \right) \left( \frac{B_T}{10^{6} \ \text{G}} \right)^{2} \ \text{cm}^{-1} \:,  \nonumber \\
\mathrm{\Delta}_{\text{pl}} &\equiv - \frac{\omega^2_{pl}}{2\omega}  \simeq -3.49 \times 10^{-12} \ \left( \frac{\omega}{\text{keV}} \right)^{-1} \left( \frac{n_e}{10^{8} \ \text{cm}^{-3}} \right) \ \text{cm}^{-1} \:.
\end{align}
Here, $\alpha$ is the fine structure constant and $B_{\text{cr}}=4.414 \ \text{G}$ is the critical magnetic field. In a plasma, the photons acquire an effective mass given in term of the plasma frequency  $\omega^2_{\text{pl}}=4\pi \alpha n_e / m_e$, where $n_e$ is the plasma electron density.

For a general case, when the transverse magnetic field $\mathbf{B}_T$ makes an angle $\xi$, (where $0 \leq \xi \leq 2\pi$), with the $y$-axis in a fixed coordinate system. A rotation of the mixing matrix (equation \eqref{eq.5.3}) in the x-y plane and the evolution equation of the ALP-photon system can then  be written as
\begin{equation} \label{eq.5.6}
\resizebox{.9\hsize}{!}{$
i \dfrac{d}{dz} \left( \begin{matrix} A_{\perp}(z) \\ A{\parallel}(z) \\ a(z) \end{matrix} \right) = -  \left( \begin{matrix} 
\mathrm{\Delta}_{\perp} \cos^2 \xi + \mathrm{\Delta}_{\parallel} \sin^2 \xi & \cos \xi \sin \xi (\mathrm{\Delta}_{\parallel}+\mathrm{\Delta}_{\perp}) & \mathrm{\Delta}_{a\gamma} \sin \xi  \\ 
\cos \xi \sin \xi (\mathrm{\Delta}_{\parallel}+\mathrm{\Delta}_{\perp})  & \mathrm{\Delta}_{\perp} \sin^2 \xi + \mathrm{\Delta}_{\parallel} \cos^2 \xi  & \mathrm{\Delta}_{a\gamma} \cos \xi  \\
 \mathrm{\Delta}_{a\gamma} \sin \xi & \mathrm{\Delta}_{a\gamma} \cos \xi & \mathrm{\Delta}_{a} \end{matrix} \right)
  \left( \begin{matrix} A_{\perp}(z) \\ A{\parallel}(z) \\ a(z) \end{matrix} \right) \:.$}
\end{equation}
As we can see from the model equations, the conversion proportionality between ALPs and photons is very sensitive to the transverse magnetic field $\mathbf{B}_T$ and the plasma electron density $n_e$ profiles. However, their configurations in the AGN jets are not fully clear yet. In our work here, we adopted the following $\mathbf{B}_T$ and $n_e$ profiles
\begin{equation} \label{eq.5.7}
B_T(r,R) =J_s(r) \cdot  B_{\ast} \left( \frac{R}{R_{\ast}} \right)^{-1} \text{G} \:, \quad \text{and} \quad
n_e(r,R) = J_s(r) \cdot n_{e,\ast} \left( \frac{R}{R_{\ast}}  \right)^{-s} \ \text{cm}^{-3} \:.
\end{equation}
Where $r$ is the distance from the jet axis, $R$ is the distance along the jet axis from the central supermassive black hole (SMBH), believed to be at the center of the AGN, and $R_{\ast}$ represents a normalization radius. The function $J_s(r)$ is the exponentially scaled modified Bessel function of the first kind of order zero \cite{abramowitz1965handbook, arfken1985mathematical}, used to scale the magnetic field and the electron density profiles with our choice here for the scale length to be three times the Schwarzschild radius of the central supermassive black hole. The normalization parameters $B_{\ast}$ and $n_{e,\ast}$ can be found in by fitting observational data for a given environment with the suggested magnetic field and electron density profiles. The parameter $s$ is model-dependent and takes different values of $s=1, 2, \text{and}, 3$. Note that in this work, we study two scenarios related to distinguished environments. The first one is an attempt to explain the Fermi spectrum of the radio quasar 3C454.3 based on ALPs conversions to photons. In contrast, the other one uses independent measurements to constrain the contribution of the ALP-photon concertions to the M87 emissions.

Using the set of parameters discussed above, the evolution equations \eqref{eq.5.6} can be numerically solved to find the two components of the photon linear politicization$;$ $A_{\perp}$ and $A_{\parallel}$. If we consider the initial state is ALPs only, the initial condition is $(A_{\perp}, A_{\parallel}, a)^t = (0,0,1)$ at the distance $R = R_{\text{min}}$ where the ALPs enter the jet. Then the probability for an ALP to convert into a photon after traveling a certain distance in the magnetic field inside the AGN jet can be defined as \cite{raffelt1988mixing, mirizzi2008photon}
\begin{equation} \label{eq.5.9}
P_{a\rightarrow \gamma} (E) =  \vert A_{\parallel}(E) \vert^2 + \vert A_{\perp}(E) \vert^2  \:.
\end{equation}
Notice here that we follow the authors of reference \cite{mena2013hints} in assuming the jet field to be coherent along the studied scale, and ALP-photon conversion probability is independent of the coherence length of the magnetic field. In addition, in \cite{marsh2017new}, the authors find coherence lengths in M87 in excess of longest scales studied here. We will discuss later how the ALP-photon conversion probability is affected by the jet geometry and the direction of the beam propagation inside the jet. Moreover, the observations can be compared with the ALP-photon mixing model results by plotting the energy spectrum ($\nu F_{\nu} \equiv E^2 dN/dE$) as a function of the energy of the photons with $\omega \equiv E(1+z)$ where $z$ is the AGN redshift. The final photon spectrum is obtained by multiplying the photon production probability $P_{a\rightarrow \gamma} (E)$ with the CAB Band spectrum $S(E)$ \cite{band1993batse}
\begin{equation} \label{eq.5.10}
E^2 \dfrac{dN}{dE} = P_{a\rightarrow \gamma} (E) \cdot S(E) \:.
\end{equation}

Hence, the ALP-photon mixing for the AGN jet model includes six free parameters$:$ the normalization for the magnetic field $B_{\ast}$, the normalization for the electron density $n_{e,\ast}$, the ALP mass $m_a$, the ALP-photon coupling parameter $g_{a\gamma}$, and two additional geometric parameters $\theta$ and $\phi ;$ we will discusses their role later in section \ref{sec.5.4}.

\section{The astrophysical environment of AGNs and their jets} \label{sec.5.3'}

Before we go on with our discussion of the effects of ALP-photon conversion, we describe in this section some aspects of the astrophysical environment of the active galactic nuclei and their jets where these interactions may take place. AGNs are compact cores of a special category of active galaxies that are characterized by being among the most massive and luminous compact objects in the universe, see for review \cite{begelman1984theory, peterson1997introduction}. They are distinguished also by emitting intense amounts of radiation at any range of the electromagnetic spectrum due to the accretion onto the central supermassive black holes that are considered to be the most likely engine of the activity in these sources. According to the accepted evolutionary models, galaxies gather their masses through a sequence of accretion and merger phases. As a result of the process, million-to-billion solar mass SMBHs formed at the center of the massive galaxies. An accretion disk may form around the SMBH due to the non-zero angular momentum of the infalling matter. As substantial energy is released in the accretion disk due to its efficient mass-energy conversion, the compact center of the galaxy becomes an AGN. The existence of the SMBH in the center of AGN explains not only the large energy output based on the release of gravitational energy through accretion phenomena but also the small size of the emitting regions and connected to it the short variability time scales of AGN. General relativistic modeling of magnetized jets from accreting black hole showed that provided the central spin is high enough, a pair of relativistic jets are launched from the immediate vicinity of the black hole, composed mostly by electrons and positrons (in light jets), or electrons and protons (in heavy jets). These jets could be powered either by the rotational energy of the central SMBH or by the magnetized accretion disk wind accelerated by magneto-centrifugal forces.

The presence of jets affects the spectrum of AGNs through the emission of synchrotron radiation and Inverse Compton scattering of low energy photons, thus leading to a prominent non-thermal spectrum, sometimes extending from radio frequencies all the way up to $\gamma$-ray energies. Particles are accelerating on helical paths along the magnetic field lines, emitting synchrotron radiation. In the Fourier transform of the continuum spectrum of the synchrotron radiation, the most powerful frequency is the critical frequency. The electron energies and magnetic field strengths typical to AGN render the critical frequency into the radio regime, so the collimated jets are observed as radio-loud AGN. Apart from the synchrotron radiation in radio to optical (and in some cases X-ray) wavelengths, AGNs and their jets emit energetic particles in the X-ray and $\gamma$-ray bands due to the inverse Compton scattering.

Observation of synchrotron radiation from the jets of AGNs implies that the material in the jet is a magnetized plasma. The content of the plasma may be electrons and protons or electrons and positrons or a mixture of them.  In general, the emission spectra of the emitting particles (mostly electrons) can be characterized by the power-law energy distribution. Since the jet broadens with distance $R$ from the core, the magnetic field $B$ and the density number of the emitting particles $n$ decrease as $B(R) \propto R^{-b}$ and $n(R) \propto R^{-a}$, where the exponent $a$ and $b$ are positive numbers \cite{chatterjee2010multi}. It is well-established that galaxy clusters feature strong magnetic fields with typical field strengths ranging from a few $0.1 \ \mu\text{G}$ up to several $10 \ \mu\text{G}$ for the most massive one\cite{dubois2009influence}. Such fields are extended over megaparsec distances and have kiloparsec coherence scales. However, the values of the jet main physical parameters$;$ such as the matter density and composition inside the jets, are still very poorly understood or even constrained and required more investigations in order to gain insight into the main physical properties of the AGNs.  In the current literature, the typical electron density within the environment of the AGNs jets is estimated to be ranging from tens to a few thousands $\text{cm}^{-3}$ \cite{kakkad2018spatially}. 

\section{Signatures of ALP-photon coupling in the jets of AGNs} \label{sec.5.3}

The Fermi Large Area Telescope in the period from 2010 September 1st to December 13th, reported observations of the radio quasar 3C454.3 at a redshift of z=0.895, constitutes of four epochs \cite{abdo2011fermi}$:$ (i) A pre-flare period, (ii) A 13 day long plateau period, (iii) A 5-day flare, and (iv) A post-flare period. The ALP-photon mixing model then has been used in \cite{mena2013hints} to fit these observation data by plotting the $\gamma$-ray energy spectra ($\nu F_{\nu} \equiv E^2 dN/dE$) as a function of the energy of the photon $E$ to get some constraints on the ALP parameters. To validate the results in \cite{mena2013hints}, we replicated the model analyses considering the photons to be initially unpolarized, and the following initial condition has been applied$:$ $(A_{\parallel}(E), A_{\perp}(E), a(E))=(\frac{1}{\sqrt{2}}, \frac{1}{\sqrt{2}}, 0)$ at $z \equiv R = 10^{18} \ \text{cm}$. Besides, the angel $\xi$ has been fixed to be $\pi/4$ during the whole calculations. The evolution equations \eqref{eq.5.2} have been solved numerically to produce the spectral feature of the blazar 3C454.3 in its four epochs for three different cases of the electron profile density.

\begin{figure}[t!]
\centering
\includegraphics[width=18.2pc]{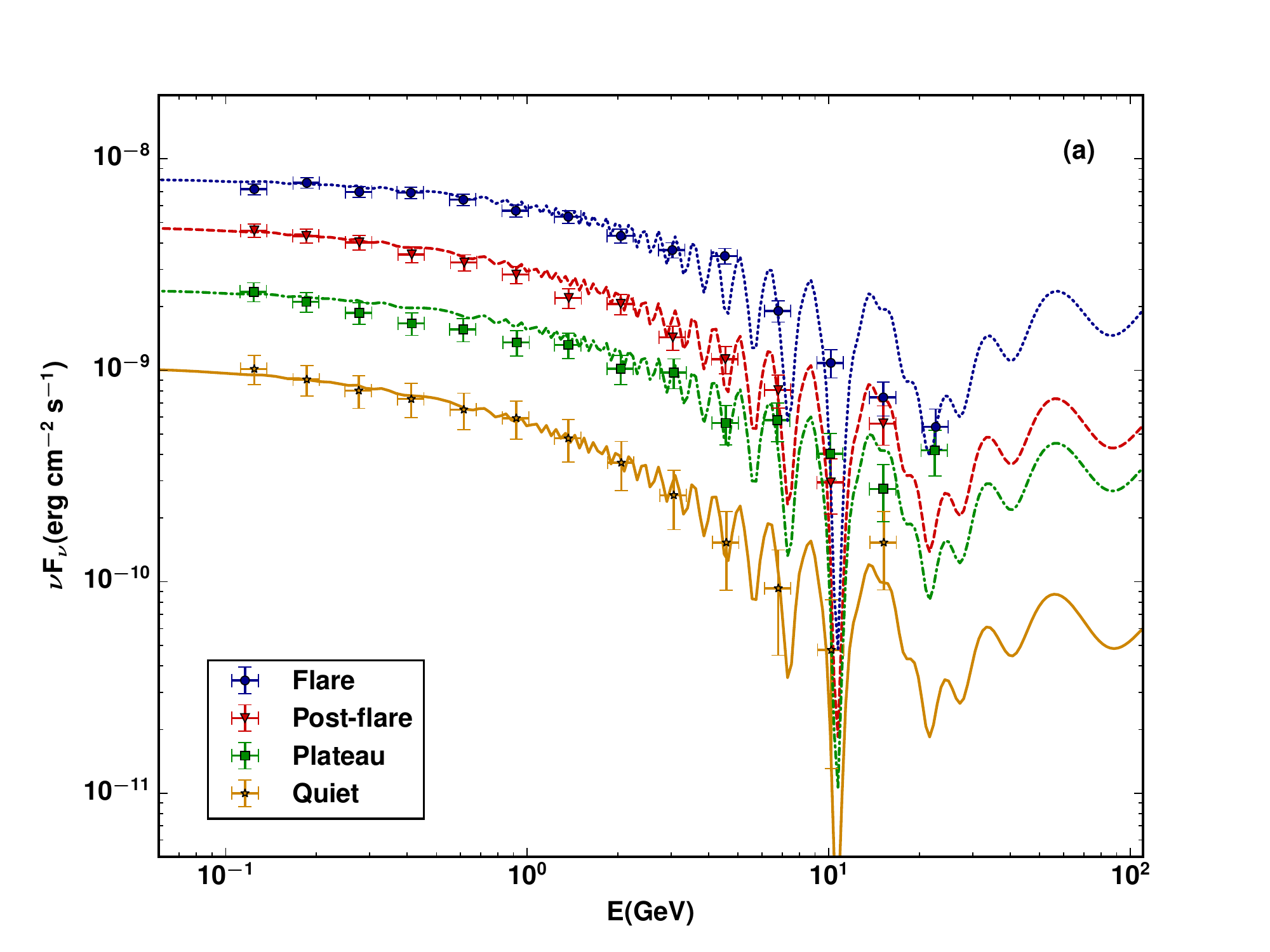}
\includegraphics[width=18.2pc]{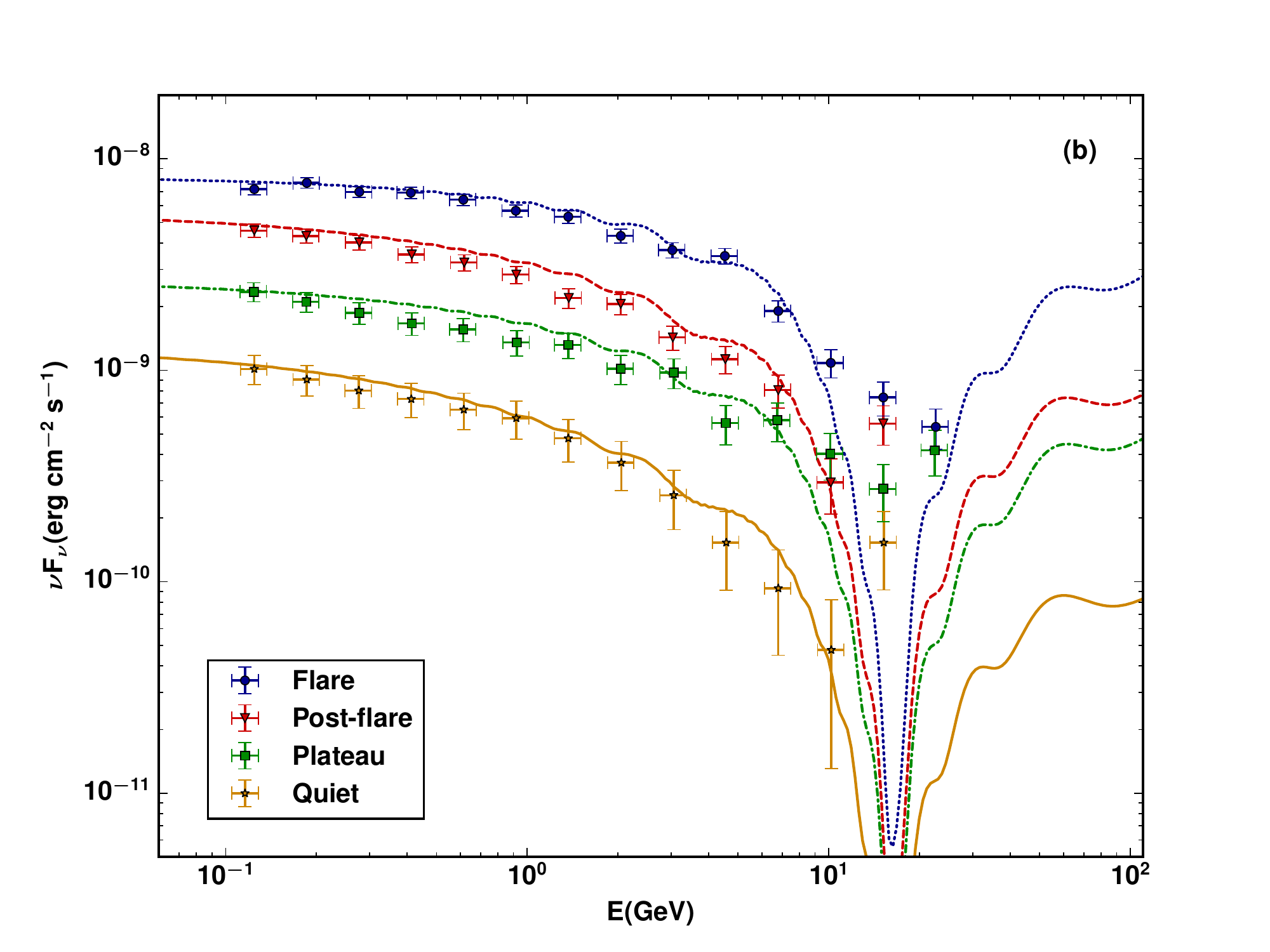} 
\includegraphics[width=18.2pc]{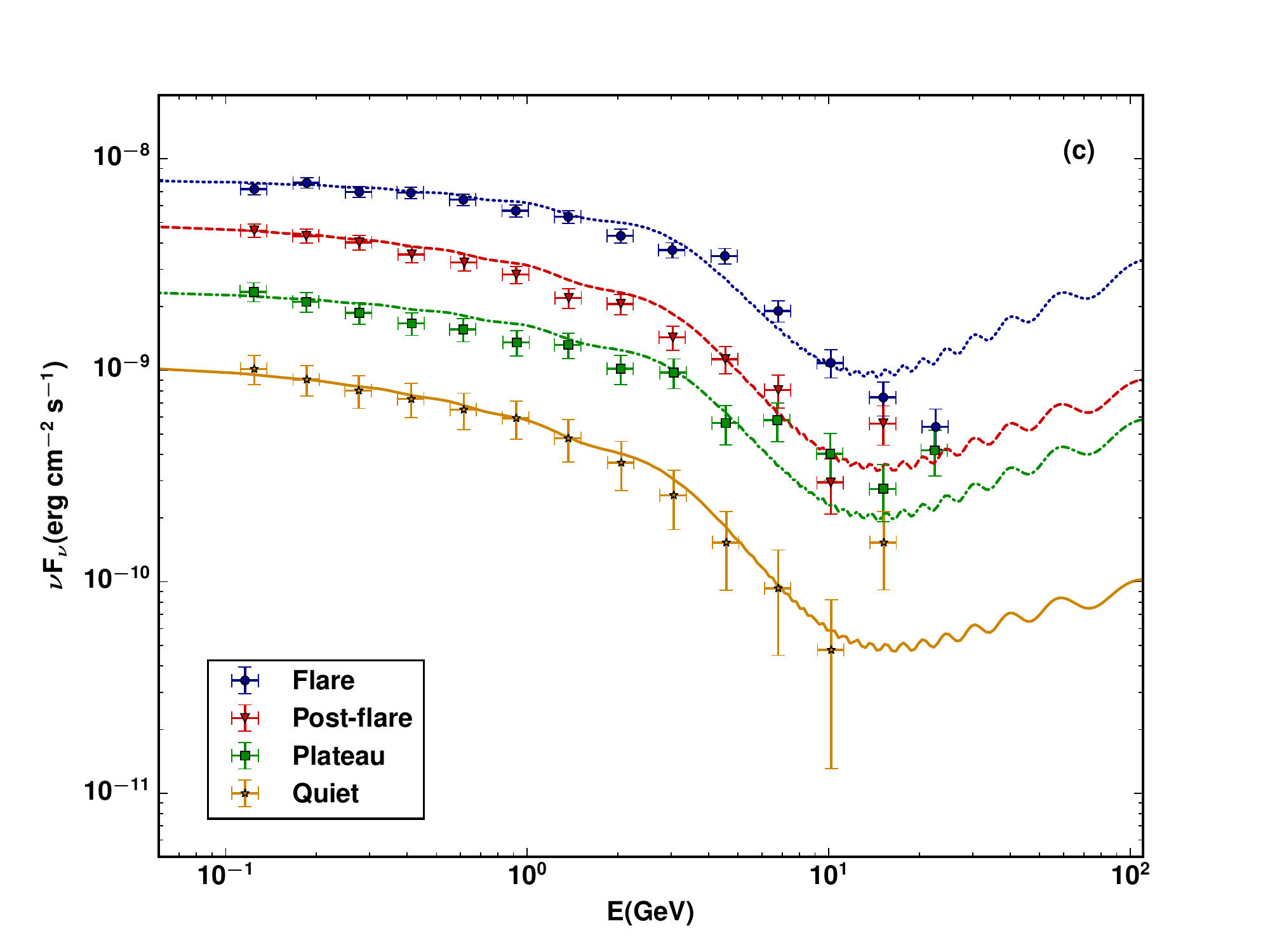}
\caption{The numerical simulation of the spectral energy fitting observations of blazar 3C454.3 for four different epochs produced using ALP-photon mixing model with electron profile density$:$ (a) $n_e \propto R^{-1}$, (b) $n_e \propto R^{-2}$  and (c) $n_e \propto R^{-3}$.}
\label{fig.5.1}
\end{figure}

Figures \ref{fig.5.1} show the numerical simulation of the spectral energy fitting observations to blazar 3C454.3 for four different epochs (flare, post-flare, plateau, and quiet) using the ALP-photon mixing model with the three different electron density profiles with the following values of the model-dependent parameter $s=1$ (a), $2$ (b), and $3$ (c). The two spectral parameters $C$ and $\mathrm{\Gamma}$ have been varied from epoch to another as they affected by the $\gamma$-ray emission region, while the other four parameters$;$ $\phi$, $\eta$, $m_a$, and $g_{a\gamma}$ have been kept fixed. Comparing our results with the one published in \cite{mena2013hints}, allows us to deduce that the spectral energy obtained using the numerical solutions to the evolution equations \eqref{eq.5.2} show a good agreement with these published results. The best fitting of the ALP-photon mixing model for the observation data of blazar jet has been achieved  when the transition between photons to ALPs take place over different radii, $R \sim 10^{18}-10^{21} \ \text{cm}$ for $\phi \sim 10^{-2} $, $\eta \sim 10^9$, $m_a \sim 10^{-7} \ \text{eV}$, and $g_{a\gamma}\sim 10^{-10} \ \text{GeV}^{-1}$.

So far, we have examined the ALP-photon model developed by Mena and Razzaque to fit the spectral features of the flat-spectrum radio quasar 3C454.3 during its November 2010 outburst. This allowed the aforementioned authors to derive constraints on the ALP parameters by assuming that a significant fraction of the gamma rays convert to ALPs in the presence of an external magnetic field in the large scale jet of this blazar. We reproduced their results that have a very good agreement with the observation data for the 3C454.3 blazar. Indeed, this makes us very confident that our simulation is robust. As a step forward in the following sections, we will discuss how to use the environment of the M87 AGN jet to test whether the CAB ALP conversion to photons, which is proposed to explain the Coma cluster X-ray excess, survives comparison with X-ray data in M87. Before moving on to the discussion of the cosmic axion background in the next section, it is worth mentioning the following two notes. Discussing the example of 3C454.3 blazar was to verify the parameters used for the ALP-photon mixing model in reference \cite{mena2013hints} and to ensure that the simulation is reproduced in the correct way. Then, we used the suitable parameters related to the M87 environment and the soft X-ray excess CAB model for the Coma cluster, because our next aim is to use the M87 environment to test the CAB Coma model. The motivation of considering the M87 AGN is because it is the best characterized AGN in the literature and the availability of information and data about it \cite{macchetto1997supermassive, gebhardt2009black, event2019first}.

\section{Cosmic axion background} \label{sec.5.4}

In this section, we follow reference \cite{conlon2013excess}$;$ the authors motivate the existence of a homogeneous cosmic axion background with $0.1 \textup{--} 1$ keV energies arising via the decay of string theory moduli in the very early universe. The origin of this idea came from the four-dimensional effective theories arising from compactifications of string theory. A generic feature of these effective theories is the presence of massive scalar particles called moduli, coming from the compactification of massless fields. Theoretical arguments indicate that the existence of such moduli should be responsible for the reheating of the standard model degrees of freedom. Regardless of the details of the inflation model, moduli are usually displaced from their final metastable minimum during inflation and begin to oscillate at the end of inflation. As they are characterized by their very weak, typically suppressed by Planck-mass interactions, the moduli are long-lived.  The oscillating moduli fields subsequently come to dominate the energy density of the universe, which enters a modulus-dominated stage lasts until the moduli decay into visible and hidden sector matter and radiation, thus leading to reheating. The visible sector decays of the modulus rapidly thermalize and initiate the hot big bang. The gravitational origin of the moduli implies that moduli can also decay to any hidden sector massless particles with extremely weak interactions, such as ALPs. Two-body decays of a modulus field $\mathrm{\Phi}$ into ALPs are induced by the Lagrangian coupling $\frac{\mathrm{\Phi}}{M_P} \partial_\mu a \partial^\mu a$, resulting in ALPs with an initial energy $E_a=m_{\mathrm{\Phi}}/2$. 

Considering these ALPs arising from moduli decays at the time of reheating are weakly interacting, they do not thermalize, and the vast majority of ALPs propagate freely to the present day. This implies that they would linger today, forming a homogeneous and isotropic CAB with a non-thermal spectrum determined by the expansion of the universe during the time of moduli decay. Furthermore, because they are relativistic, they contribute to the dark radiation energy density of the universe, but this is beyond the scope of this work. For moduli masses $m_{\mathrm{\Phi}} \approx$ GeV the present energy of these axions is $E_a \sim 0.1 \textup{--} 1$ keV. The suggestion being that the natural energy for such a background lies between $0.1$ and $1$ keV. Furthermore, in \cite{conlon2013cosmophenomenology} it was shown that such CAB would have a quasi-thermal energy spectrum with a peak dictated by the mass of the ALP. It is also worth noting here that the ALP number density in the CAB and consequently their flux with energies between $E$ and $E+dE$ is directly related to the shape of the CAB energy spectrum which in general depend on the mean ALP energy $\langle E_a \rangle$ with some model-dependent constants that may be measured by the CAB contribution to the number of neutrino species $\mathrm{\Delta} N_{\text{eff}}$ which is directly related to the CAB energy density $\rho_a$ \cite{day2016cosmic}. This CAB is also invoked in \cite{angus2014soft} to explain the soft X-ray excess on the periphery of the Coma cluster with an ALP mass of $1.1\times 10^{-13}$ eV and a coupling to the photon of $g_{a\gamma} = 2 \times 10^{-13} \ \text{GeV}^{-1}$. 

In this work, we will assume, for convenience, that the CAB has a thermal spectrum with an average energy of $\langle E_a \rangle = 0.15 \ \text{keV}$. We then normalize the distribution to the typical example quoted in \cite{conlon2013cosmophenomenology}. We use the thermal distribution as an approximation, as the exact shape of the distribution will not substantially affect the conclusions we draw. We can then determine the fraction of CAB ALPs converted into photons within the environment of the M87 AGN jet and use this to determine a resulting photon flux. Under these assumptions, the predicted photon flux depends on the value of the CAB mean energy $\langle E_a \rangle$. This flux can be compared to X-ray measurements to see if such environments can constrain low-mass ALPs and put limits on the ALP explanation of the X-ray excess.

\section{The soft X-ray excess CAB in the environment of M87 AGN jet} \label{sec.5.5}

It is established that most of the baryonic mass of the galaxy clusters are composed of a hot ionized intracluster medium (ICM) with temperatures about of $T \approx 10^8 \ \text{K}$ (corresponding to $\omega \approx 7 \ \text{keV}$) and number densities in the range of  $n \sim 10^{-1} \textup{--} 10^{-3} \ \text{cm}^3$. The ICM generates diffuse X-ray emission through thermal bremsstrahlung. A thermal bremsstrahlung spectrum gives a good approximation of a constant emissivity per unit energy at low energies. However, observations of many galaxy clusters show a soft X-ray excess in their spectra at low energies around $1 \textup{--} 2 \ \text{keV}$, which is above that from the hot ICM, and the origin of this component is still unclear. In this work, we have adopted the scenario that this soft X-ray excess is produced by the conversion of a primordial cosmic ALP background into photons in the cluster magnetic field. The central M87 AGN of the Virgo cluster is the best characterized AGN in the literature due to its proximity \cite{macchetto1997supermassive, gebhardt2009black, event2019first}. In this respect, we numerically solve the ALP-photon mixing model that is described in section \ref{sec.5.2} and use it to study the photon production probability from CAB ALPs in the environment of the M87 AGN jet. Then, we test the model to reproduce the X-ray emission for the M87 AGN from the ALP-photon conversion with very low ALP mass and very small ALP to photon coupling.

The M87 AGN is a radio galaxy at a luminosity distance of $16.7 \pm 0.2$ Mpc \cite{mei2007acs} and a redshift of z = 0.00436. Based on its radio images and the modeling of its interaction with the surrounding environment, it is suggested that the M87 jet is misaligned with respect to the line of sight \cite{biretta1999hubble, biretta1995detection}.  Therefore, we consider the situation when there is a misalignment between the ALP-photon beam propagation direction and the AGN jet direction. Accordingly, we have to take into account the geometry of the jet of the AGN and the direction of the ALP-photon beam propagation. In this case, two more parameters may play an important role in the study of the ALP-photon conversion probability are the misalignment angle $\theta$ between the jet direction and the line of sight and the AGN jet opening angle $\phi$, which define the jet geometry. When $\theta=0$, the ALP-photon beam propagate parallel to the R-direction, and we expect that the ALP-photon conversion probability is not affected by the jet geometry represented by the jet opening angle $\phi$. However, when the ALP-photon beam crosses the jet diagonally, making a misalignment with the R-direction, the ALP-photon conversion probability may be affected by the misalignment angle $\theta$ as well as the jet opening angle $\phi$. The opening angle $\phi$ for the jet of the M87 AGN near the base is less than about 5 degrees \cite{biretta1995detection}, and the misalignment angle $\theta$ is less than about 19 degrees \cite{walker2008vlba, hada2017structure}.

In this study, we make our choices such that the magnetic field and the electron density profiles used for M87 AGN are consistent with the obtained values in \cite{park2019faraday}. In this perspective, we use an electron profile density profile $n_e \propto R^{-1}$ with the model-dependent parameter $s=1$ in equation \eqref{eq.5.7}. For our case of varying $\mathbf{B}_T$ and $n_e$ with $R$ as in equation \eqref{eq.5.7}, transition of ALPs to photons take place over different distances,  $R \sim 10^{16} \textup{--} 10^{17} \ \text{cm}$, with normalization radius $R_{\ast} = 6 \times 10^{20} \ \text{cm}$. The environmental parameters are taken as $B_{\ast}= 1.4 \times 10^{-3}$ G and $n_{e,\ast}=0.3$ cm$^{-3}$ at the distance $R_{\ast}$. Note that as $n_e$ values are only available at larger distances from the base \cite{park2019faraday}, we assume that we can extrapolate its values down to small radii. In addition, we set the ALP mass to be $1.1\times 10^{-13}$ eV and we start with ALP-photon coupling around $g_{a\gamma}  \sim 2 \times 10^{-13}$ GeV$^{-1}$ in agreement with the models derived in \cite{angus2014soft} to explain the soft X-ray excess on the periphery of the Coma cluster.

\section{Probing low-mass ALP models within the jets of AGNs} \label{sec.5.6}

In this section, we discuss the results of the numerical simulation of the conversion probabilities and present a description of the predictions of the scenario of CAB ALPs conversion to photons for the M87 AGN soft X-ray excess. To obtain our results, we apply the ALP-photon mixing model to study the probability of CAB ALPs to convert to photons in the intergalactic magnetic field on the jet of M87 AGN with the initial state of ALPs only at $R_{\text{min}}= 10^{16} \ \text{cm}$. Figure \ref{fig.5.1} shows the ALP-photon conversion probability $P_{a \rightarrow \gamma} (E)$ as a function of energy for different values of the misalignment angle $\theta$ and the jet opening angle $\phi$. The different curves on the left panel correspond to $\theta = 5^\circ, 10^\circ, 15^\circ,$ and $20^\circ$ at fixed $\phi=4^\circ$, while the different curves on the right panel correspond to $\phi = 4^\circ, 8^\circ,$ and $12^\circ$ at fixed $\theta=20^\circ$. It seems to be clear from the two graphs that the maximum conversion probability occurs when the misalignment angle $\theta$ is more close to the opening angle of $\phi$. This might be explained due to the relation between the ALP-photon beam direction and jet geometry. For the beam to cross the jet diagonally from one side to another, making an arbitrary angle with the magnetic field between zero and $\pi$, the condition for the beam to make the longest path is that the misalignment angle $\theta$ to be very close (but not equal) to the opening angle $\phi$. We have to remark here that the longest path is less than the maximum distance $R_{\text{max}} = 10^{17} \ \text{cm}$ when there is no misalignment. The misalignment at a given opening angle controls the point at which the ALP-photon beam would leave the jet and, therefore, the total distance that the beam travels inside the jet. Since we selected an electron density profile $n_e \propto R^{-1}$, this defines the relationship between the misalignment and the electron density profile that sets the critical energy at which stronger mixing starts as shown by the plots where less misaligned cases have higher critical energies and thus less overlap with the CAB spectrum.
\begin{figure}[th!]
\centering
\includegraphics[width=.495\textwidth]{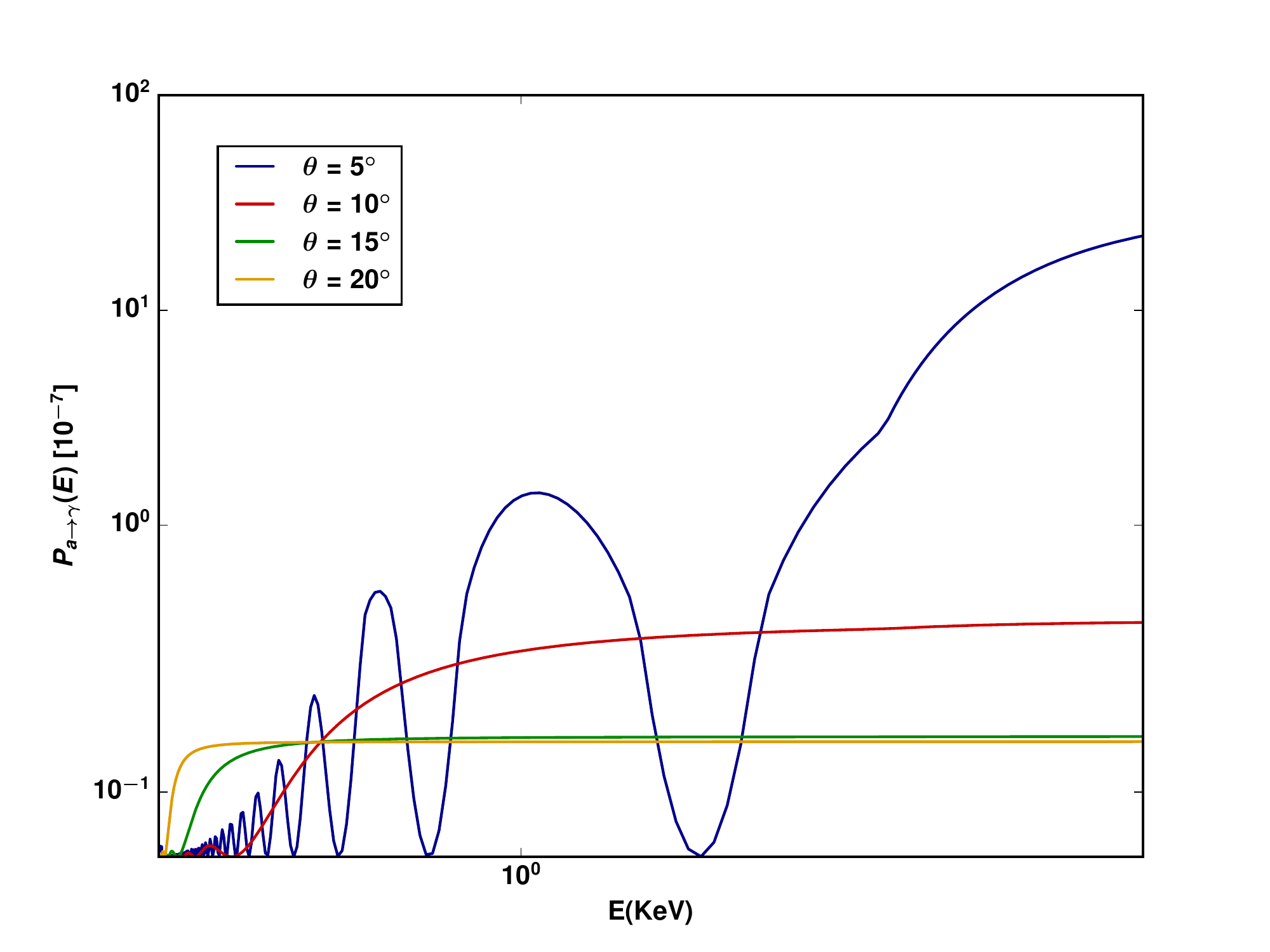}
\hfill
\includegraphics[width=.495\textwidth]{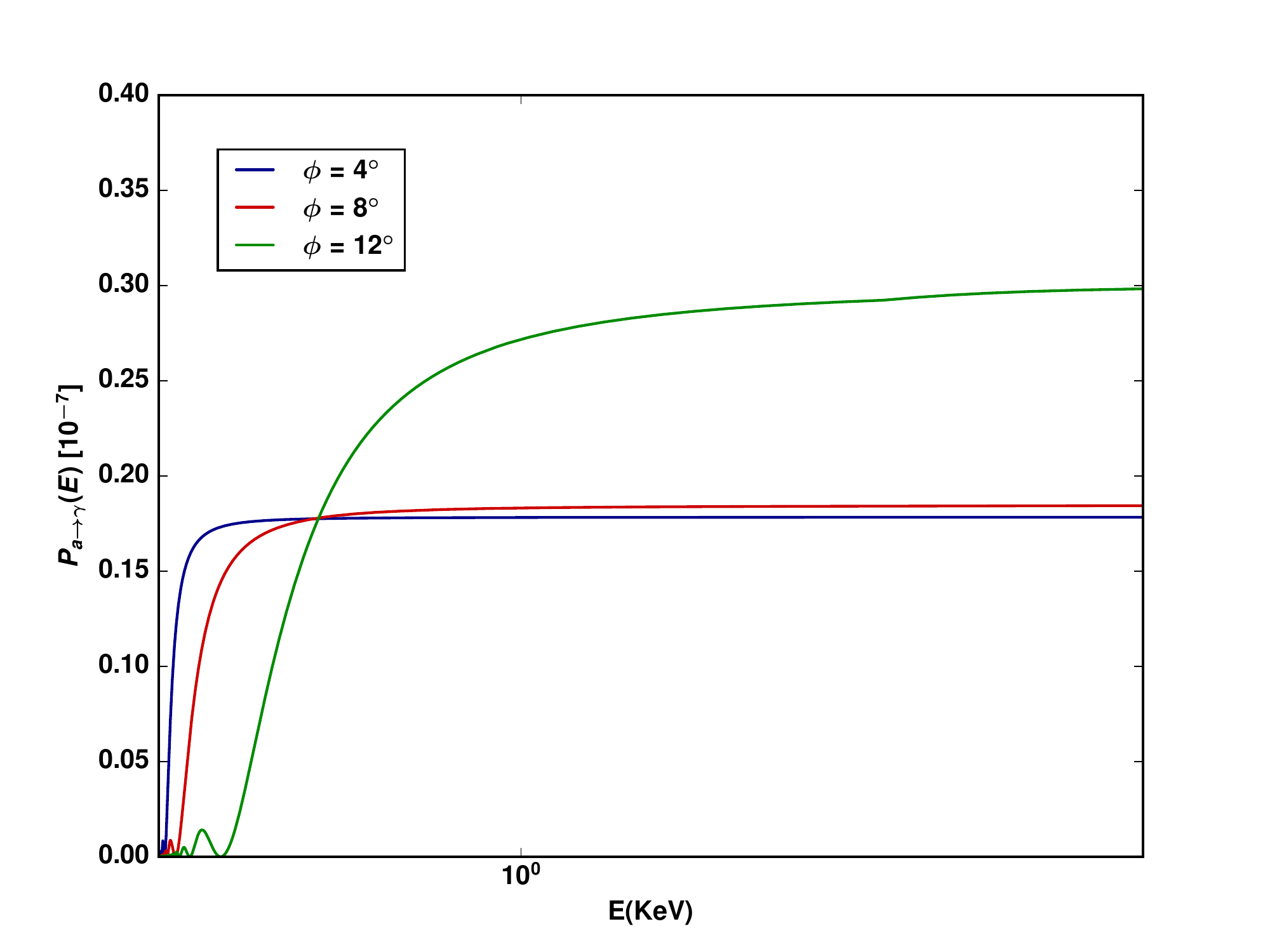}
\caption{Plot of the ALP-photon conversion probability $P_{a \rightarrow \gamma} (E)$. Left panel$:$ The different curves correspond to $\theta = 5^\circ, 10^\circ, 15^\circ,$ and $20^\circ$ at fixed $\phi=4^\circ$. Right panel$:$ The different curves correspond to $\phi = 4^\circ, 8^\circ,$ and $12^\circ$ at fixed $\theta=20^\circ$.}
\label{fig.5.1}
\end{figure}

\begin{figure}[th!]
\centering
\includegraphics[width=.495\textwidth]{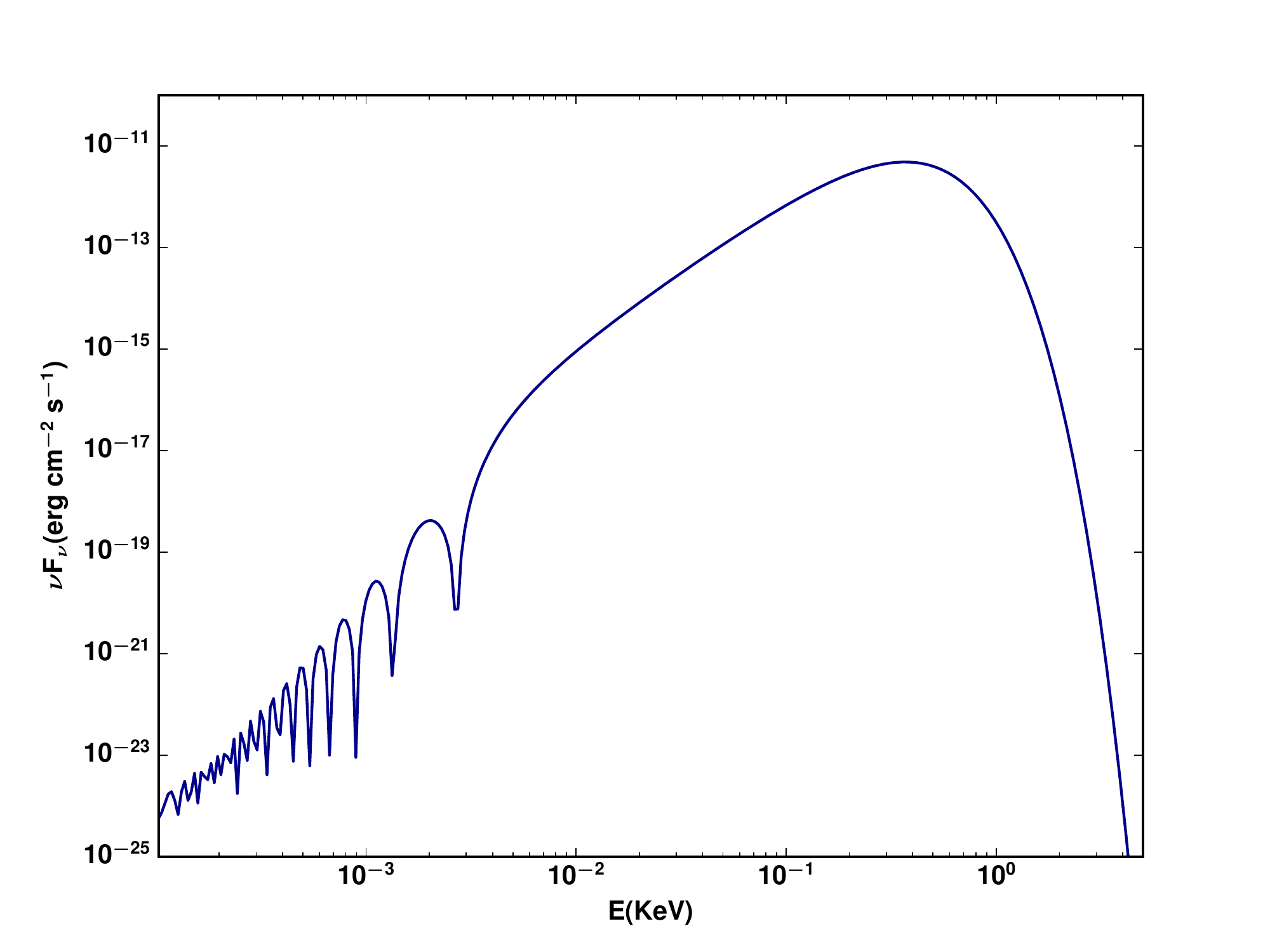}
\hfill
\includegraphics[width=.495\textwidth]{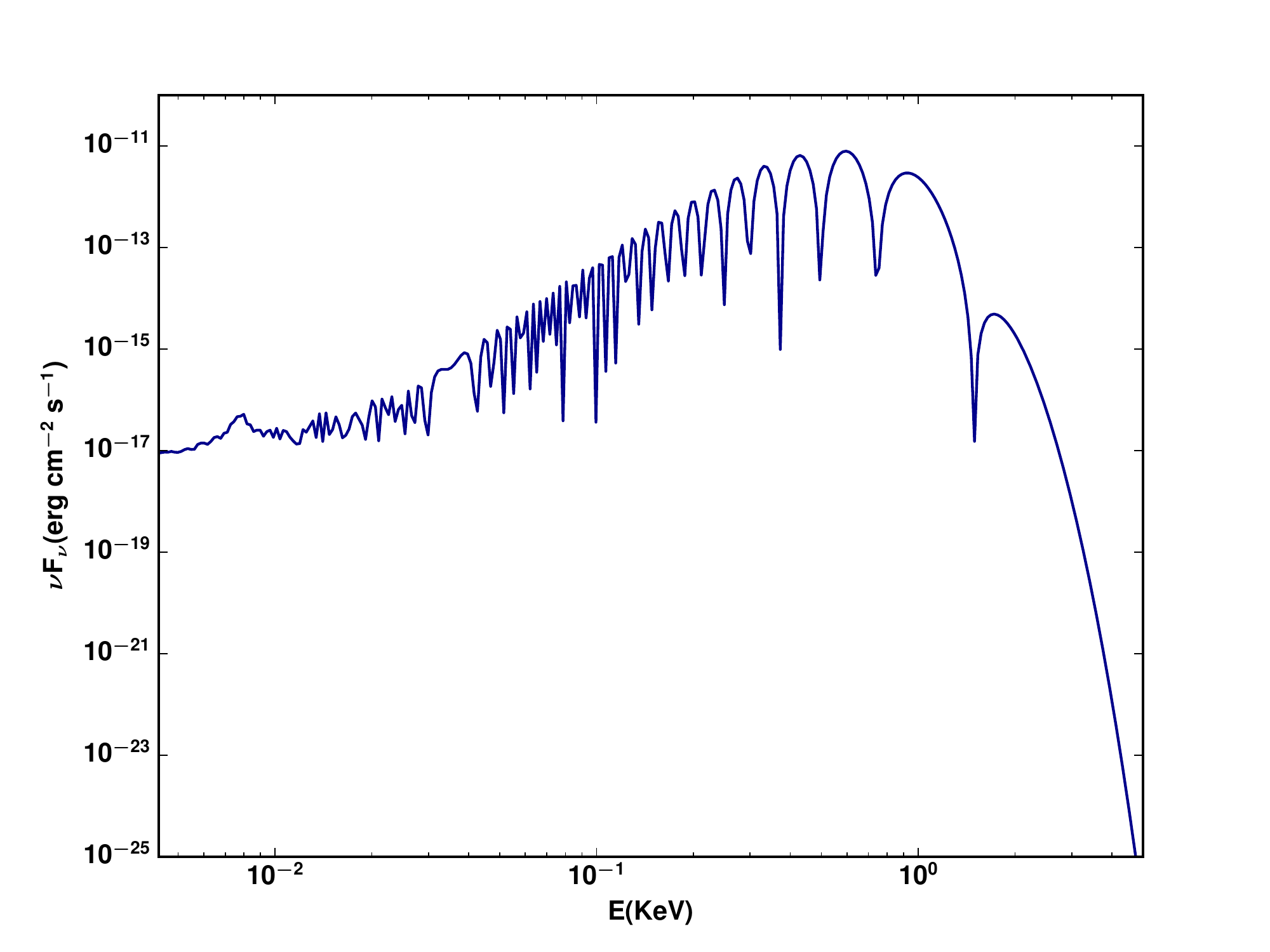}
\vfill
\includegraphics[width=.495\textwidth]{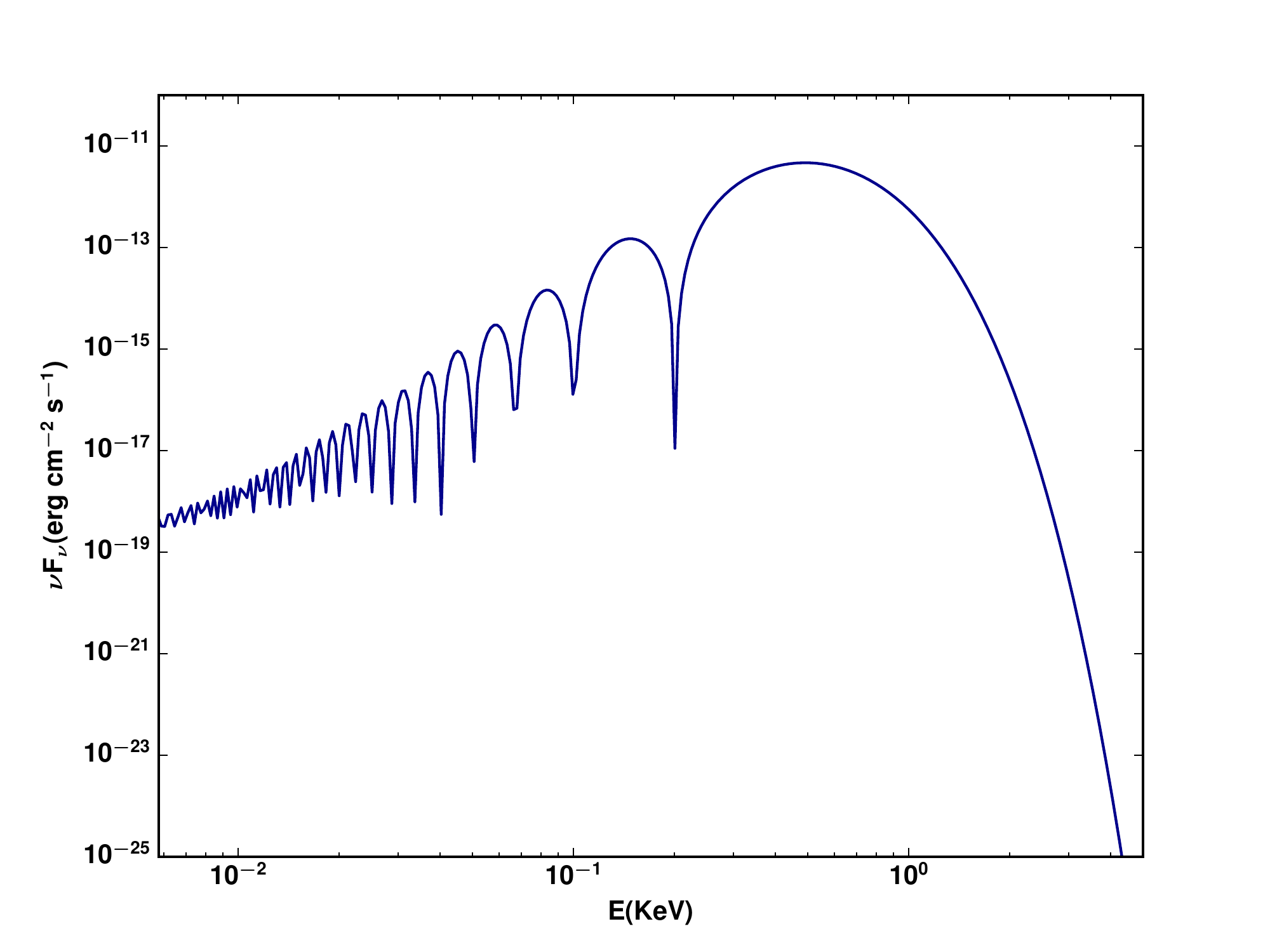}
\hfill
\includegraphics[width=.495\textwidth]{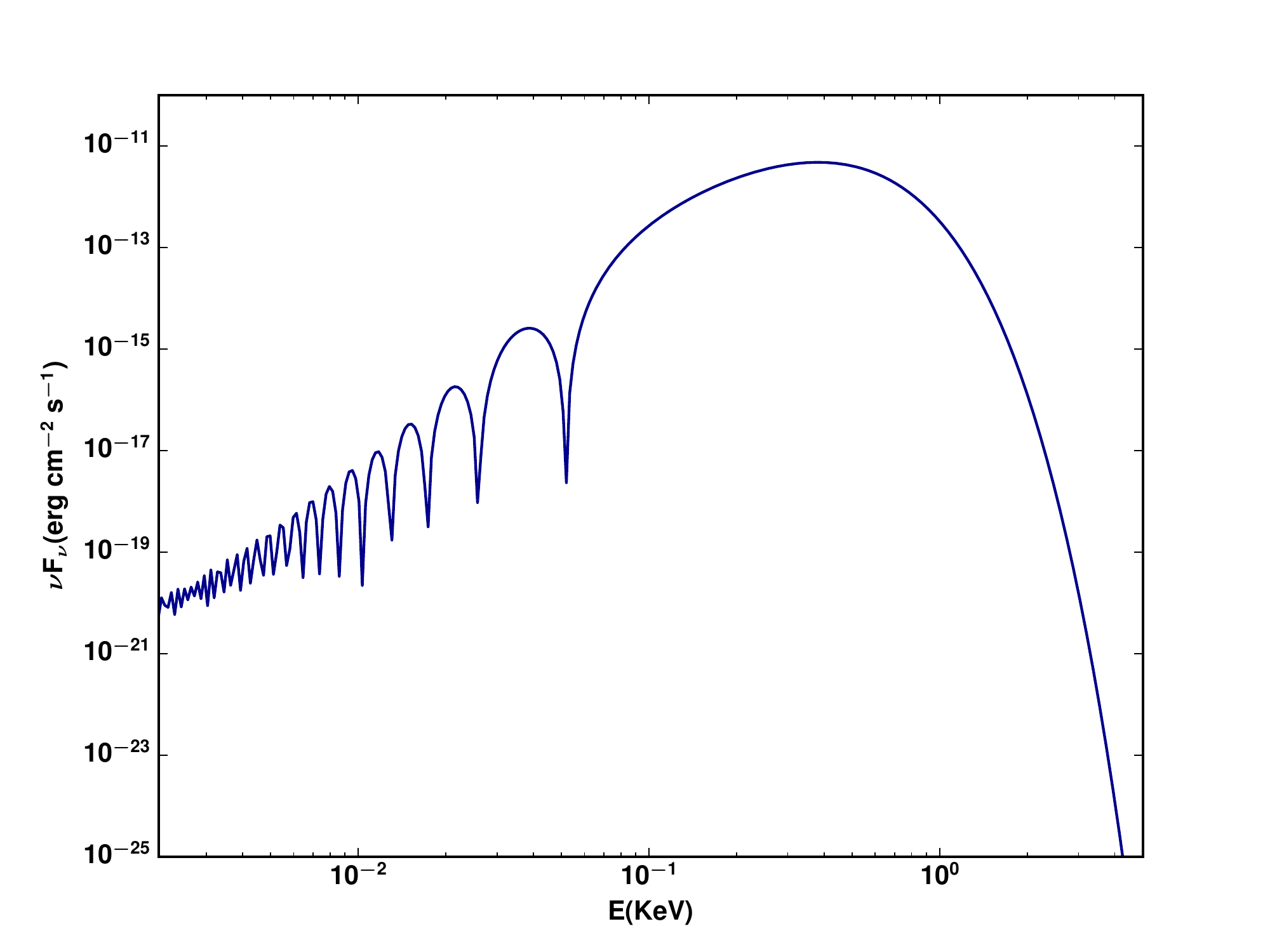}
\vfill
\includegraphics[width=.495\textwidth]{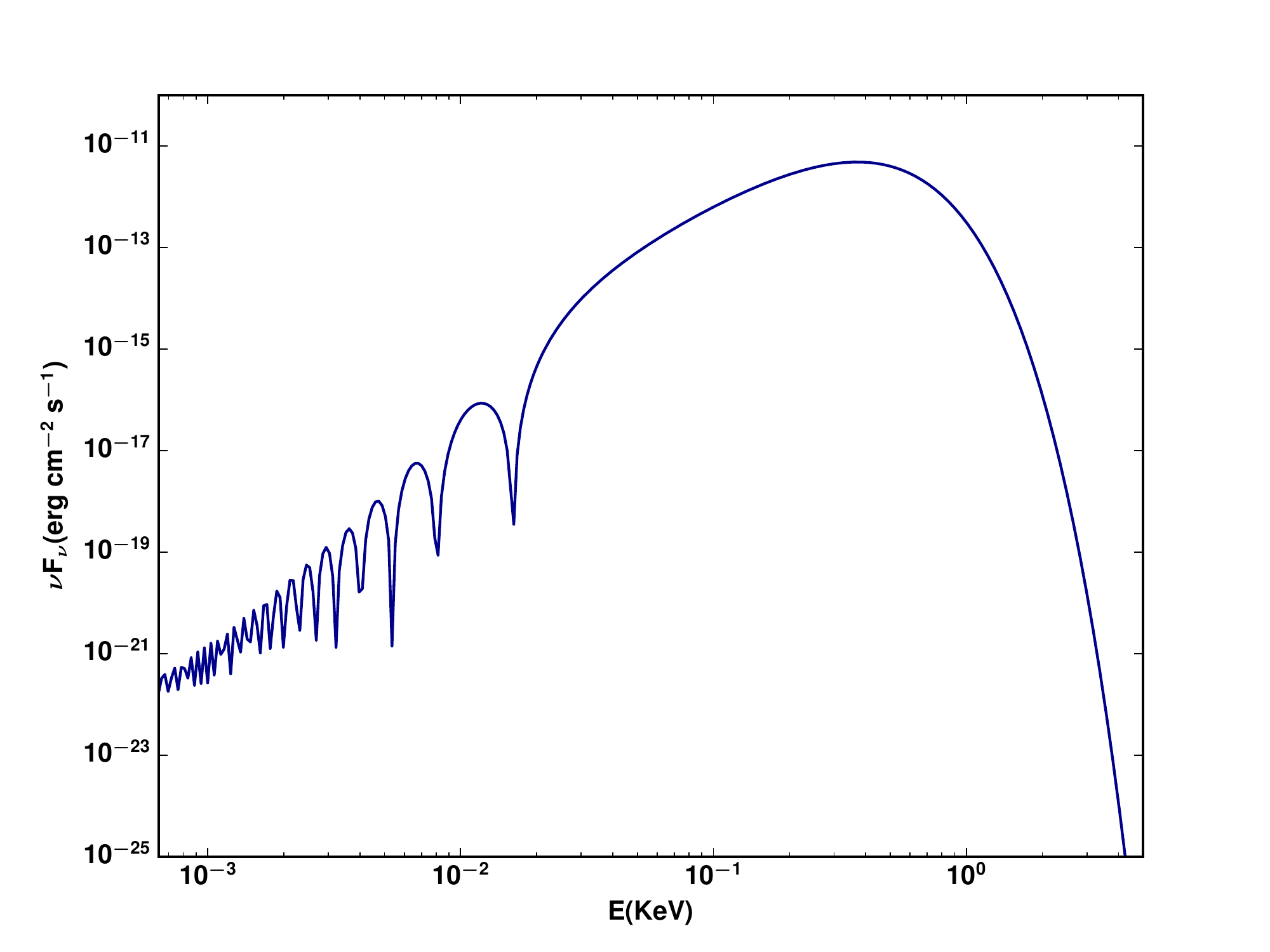}
\hfill
\includegraphics[width=.495\textwidth]{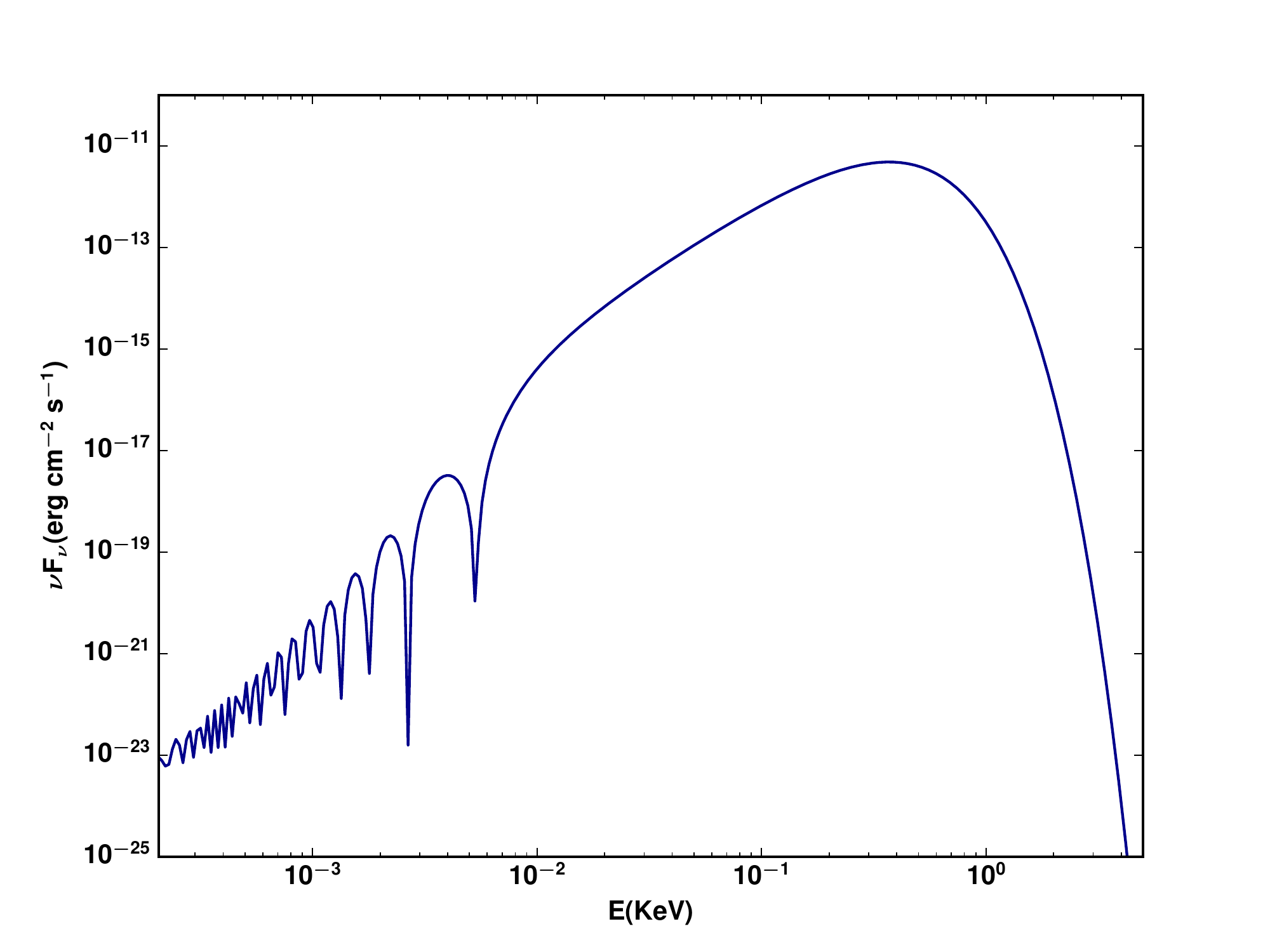}
\caption{The numerical simulation of the energy spectrum distributions from ALPs conversion to photons in the intergalactic magnetic field on the jet of M87 AGN at fixed opening angle $\phi =4^\circ$ and different values of the misalignment angle $\theta$. Top left panel, $\theta=0^\circ$. Top right panel, $\theta=5^\circ$. Middle left panel, $\theta=10^\circ$. Middle right panel, $\theta=15^\circ$. Bottom left panel, $\theta=20^\circ$. Bottom right panel, $\theta=25^\circ$.}
\label{fig.5.2}
\end{figure}

At this stage, we are ready to present the output of the model and check to put some new constants on the acceptable limit of the value of the ALP-photon coupling parameter $g_{a\gamma}$. Figure \ref{fig.5.2} shows our results for the energy spectrum distributions obtained from the numerical simulation for the ALPs conversion to photons in the intergalactic magnetic field of the jet of M87 AGN. For these plots, we kept using plasma electron density profile $n_e \propto R^{-1}$ as taking the parameter $s=1$ in equation \eqref{eq.5.7} as well as using opening angle $\phi=4^\circ$ for the jet of M87 AGN. The different plots then represent the energy spectrum distributions at different values of the misalignment angle that the ALP-photon beam makes with the jet direction $\theta = 0^\circ, 5^\circ, 10^\circ, 15^\circ, 20^\circ,$ and $25^\circ$. For each case, we find the maximum value for the ALP to Photon coupling $g_{a\gamma}$ such that we do not exceed the observed flux $\sim 3.76 \times 10^{-12} \ \text{erg} \ \text{cm}^{-2} \ \text{s}^{-1}$ from the M87 AGN between $0.3$ and $8$ $\text{keV}$ \cite{m87chandra}. Table \ref{tab.5.1} shows six different cases of  $\theta = 0^\circ, 5^\circ, 10^\circ, 15^\circ, 20^\circ, 25^\circ$ at fixed $\phi=4^\circ$ and three different cases of $\phi = 4^\circ, 8^\circ, 12^\circ$ at fixed $\theta=20^\circ$ with the corresponding constraints the model put on the ALP to Photon coupling to produce the correct flux that is compatible with observations. The summary of the results presented in the table shows that the model put constraints on the value of ALP to photon coupling to be about $\sim 7.50 \times 10^{-15}  \textup{--} 6.56 \times 10^{-14} \ \text{GeV}^{-1}$ for ALP masses $m_a \lesssim  10^{-13} \ \text{eV}$ if there is a misalignment between the AGN jet direction and the line of sight less than 20 degrees.

It is important to note that the bounds on $g_{a\gamma}$ we derive in this work are stronger than those found in \cite{marsh2017new}, which also simulates the effects of ALP-photon conversion in the environment of M87. The advantage in terms of constraints comes because we specifically study a CAB, whereas the authors in \cite{marsh2017new} consider the loss of photons to ALP interactions in general, rather than the addition of photons from a cosmic ALP flux. The very large magnitude of the background ALP flux is what allows us to achieve such strong constraints. Additionally, we have to note that \cite{marsh2017new} derive more universal limits, as they do not require a CAB to exist for their limits to be valid.

\begin{table}[ht!]
\centering
\begin{tabular}{|c|c|c|c|}
\hline
$\:$ $\theta$ (${}^\circ$), $\phi= 4^\circ$ $\:$& $\qquad$ $g_{a\gamma}$  $(\text{GeV}^{-1})$ $\qquad$ &$\:$ $\phi$ (${}^\circ$), $\theta = 20^\circ$ $\:$ & $\qquad$ $g_{a\gamma}$  $(\text{GeV}^{-1})$ $\quad$ \\
\hline
$0$ &  $\lesssim 3.91 \times 10^{-13}$ &$4$ &  $\lesssim 6.56 \times 10^{-14}$\\
$5$ & $\lesssim 9.17 \times 10^{-15}$&$8$ &  $\lesssim 2.32 \times 10^{-14}$\\
$10$ & $\lesssim 7.50 \times 10^{-15}$&$12$ &  $\lesssim 7.99 \times 10^{-15}$\\
$15$ & $\lesssim 2.08 \times 10^{-14}$&&\\
$20$ & $\lesssim 6.56 \times 10^{-14}$&&\\
$25$ & $\lesssim 1.98 \times 10^{-13}$&&\\
\hline
\end{tabular}
\caption{The ALP to photon coupling $g_{a\gamma}$ corresponds to different values of the misalignment angle $\theta$ and the jet opening angle $\phi$ for the M87 AGN at which we produce the correct flux that is compatible with observations.}
\label{tab.5.1} 
\end{table}

\section{Conclusion} \label{sec.5.7}

\begin{figure}[t!]
\centering
\includegraphics[scale=1.035]{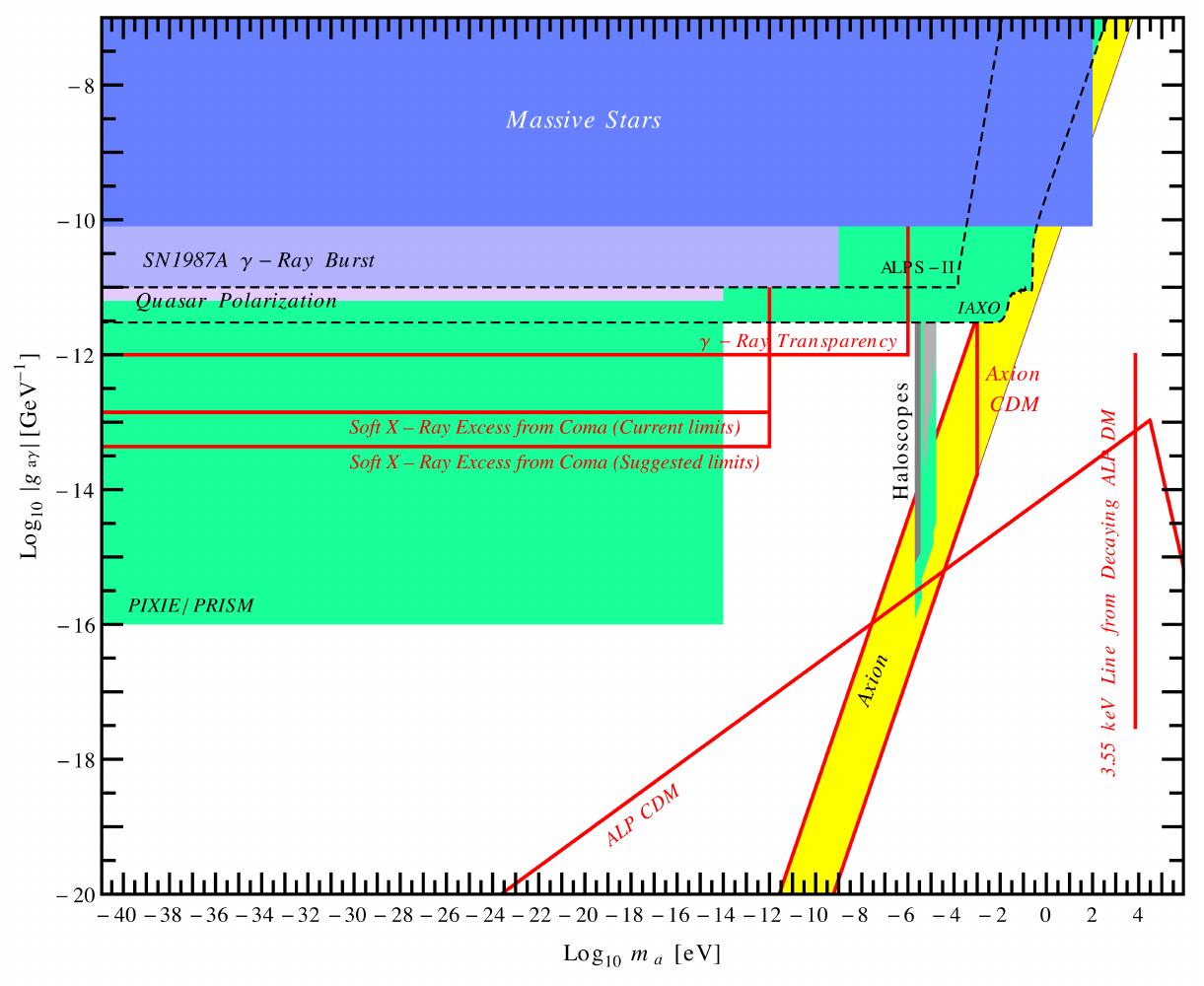} 
\caption[The allowed regions of the ALP mass-coupling plane. The bands for the ALP-photon coupling $g_{a\gamma}$ where the conversion of CAB ALPs with $m_a \lesssim 10^{-13} \ \text{eV}$ to photons can explain the soft X-ray excess of the Coma cluster are shown with the current limits and the suggested new limits.]{The allowed regions of the ALP mass-coupling plane. The bands for the ALP-photon coupling $g_{a\gamma}$ where the conversion of CAB ALPs with $m_a \lesssim 10^{-13} \ \text{eV}$ to photons can explain the soft X-ray excess of the Coma cluster are shown with the current limits and the suggested new limits. This figure is extended from \cite{dias2014quest}.}
\label{Fig.5.4}
\end{figure}  

In this chapter, we studied the conversion between ALPs and photons in the presence of external magnetic fields using the ALP-photon mixing model developed in \cite{mena2013hints, mena2011signatures}. Then examined the detectability of signals produced by ALP-photon coupling in the highly magnetized environment of the relativistic jets produced by active galactic nuclei. Further, we tested the CAB model argued in \cite{angus2014soft} to explain the soft X-ray excess on the Coma cluster periphery. We did this by calculating the X-ray emission due to CAB propagation in the jet of the M87 AGN. It is evident in these results that the overall X-ray emission for the M87 AGN, between $0.3$ and $8$ keV, can be reproduced via the photons production from CAB ALPs with coupling $g_{a\gamma}$ in the range of $\sim 7.50 \times 10^{-15}  \textup{--} 6.56 \times 10^{-14} \ \text{GeV}^{-1}$ for ALP masses $m_a \sim 1.1 \times  10^{-13} \ \text{eV}$ if there is a misalignment between the AGN jet direction and the line of sight less than 20 degrees, since the M87 jet has been found to be misaligned by less than 19 degrees in \cite{walker2008vlba, hada2017structure}. These values are up to an order of magnitude smaller than the current best fit value on the ALP-photon coupling $g_{a\gamma} \sim 2 \times 10^{-13} \ \text{GeV}^{-1}$ obtained in soft X-ray excess CAB model for the Coma cluster \cite{angus2014soft}. This casts doubt on the current limits of the largest allowed value of the ALP-photon coupling as a universal background would need to be consistent with observations of any given environment. Thus, our results exclude the current best fit value on the ALP-photon coupling for the Coma soft X-ray excess CAB model from \cite{angus2014soft} and instead place a new constraint that $g_{a\gamma} \lesssim 6.56 \times 10^{-14} \ \text{GeV}^{-1}$ when a CAB is considered.  Figure \ref{Fig.5.4} summarizes the range of the allowed ALP-photon couplings together with other constraints on the ALP parameter space. The green regions are the projected sensitivities of the light-shining-through-wall experiment ALPS-II, of the helioscope IAXO, of the haloscopes ADMX and ADMX-HF, and of the PIXIE or PRISM cosmic microwave background observatories. Axions and ALPs with parameters in the regions surrounded by the red lines may constitute all or a part of cold dark matter, explain the cosmic $\gamma$-ray transparency, and the soft X-ray excess from Coma. Our results using the environment of the M87 AGN jet show that we can exclude the area between the red line labeled ``Soft X-Ray Excess from Coma (Current limits)'' and the other red line labeled ``Soft X-Ray Excess from Coma (Suggested limits)'' from the allowed parameter space. This is a strong motivation to improve the sensitivity of the current and future versions of the ALP searches in both observations and laboratory experiments down to such a range of the ALP-photon coupling.
 
\chapter{\textbf{Potential of SKA to Detect CDM ALPs with Radio Astronomy}} \label{ch6}

Recently, it has been pointed out that ALPs may form a Bose-Einstein condensate and, through their gravitational attraction and self-interactions, they can thermalize to spatially localized clumps. The coupling between ALPs and photons allows the spontaneous decay of ALPs into pairs of photons. For ALP condensates with very high occupation numbers, the stimulated decay of ALPs into photons is also possible, and thus the photon occupation number can receive Bose enhancement and grows exponentially. In this chapter, we study the evolution of the ALPs field due to their stimulated decays in the presence of an electromagnetic background, which exhibits an exponential increase in the photon occupation number by taking into account the role of the cosmic plasma in modifying the photon growth profile. In particular, we focus on quantifying the effect of the cosmic plasma on the stimulated decay of ALPs as this may have consequences on the detectability of the radio emissions produced from this process by the forthcoming radio telescopes such as the Square Kilometer Array telescopes with the intention of detecting the CDM ALPs.

\section{Introduction} \label{sec.6.1}

As we discussed before, the coupling between ALPs and photons is possible due to the ALP-two-photon interaction vertex \cite{sikivie1983experimental}. This ALP-photon coupling allows for the Primakoff conversion between ALPs and photons in the presence of an external electric or magnetic field, as well as for the radiative decay of the ALPs into pairs of photons. These two processes provide the theoretical basis for the majority of the recent mechanisms to search for ALPs in both laboratory experiments and astrophysical environments. Interactions with photons, electrons, and hadrons are also possible in the literature for the QCD axion. However, the main focus lies only on the effect of ALP-photon coupling because of two reasons. The first is owing to the coupling with photons is the most common characteristic between the QCD axion and the generic ALPs. The second is due to the fact that photon is the is only bosonic standard model particle we know precisely that it is lighter than the generic dark matter ALPs \cite{alonso2020wondrous}. However, increasing the effects of other couplings such as the self-couplings is doable through modifications to the standard ALPs. Although the effects of the ALPs self-interactions remain very weak, they can become important in the thermal equilibrium of ALPs condensate.

An essential consequence in the context of ALPs decay is the fact that ALPs are identical bosons$;$ their very low mass indicates that their density and occupation numbers can be very high. Therefore it is suggested that ALPs may form a Bose-Einstein condensate with only short-range order \cite{sikivie2009bose, davidson2013bose, davidson2015axions}. This ALPs BEC can then thermalize through their gravitational attraction and self-interactions to spatially localized clumps \cite{chang1998studies, sikivie2009bose, erken2012cosmic}. By clump, we mean gravitationally bound structures composed of axions or ALPs, which may present in the dark matter halos around many galaxies$;$ for more detail about the properties of these BEC clumps, see references \cite{schiappacasse2018analysis, hertzberg2018scalar}. Of particular importance here is that the system in such high occupancy case is well described by a classical field \cite{guth2015dark}. Indeed, the spontaneous decay rate of ALPs is very small as a result of both their very low mass and their very weak coupling. However, the rapid expansion of the early universe would lead to an extremely homogeneous and coherent initial state of the ALP field. Hence, all ALPs are in the same state with very high occupation number. Therefore, the stimulated decay of ALPs $a \rightarrow \gamma \gamma$ with a very high rate is very likely in the presence of an electromagnetic background with a certain frequency. During this process, the electromagnetic wave will be greatly enhanced and Bose enhancement effects seem plausible. In contrast, in a scenario of an empty and non-expanding universe, the resulting stimulated emission would induce extremely rapid decay of ALPs into photons, invalidating most of the interesting parameter space \cite{alonso2020wondrous}. Nevertheless, the rapid expansion of the early universe and the plasma effects can disrupt this extremely fast process. These effects are crucial as they modify the evolution of the ALPs field that is resulting from their stimulated decays in the presence of an electromagnetic background. Indeed, this is an exciting approach to look for dark matter ALPs using the current and near-future experiments along with astrophysical observations.

In the past years, there have been many attempts to detect signals from dark matter ALPs using radio telescopes, most of which were based on the Primakoff conversion of ALPs into photons \cite{caputo2018looking, caputo2019detecting}. Recently it was shown that the stimulated ALP decays in the astrophysical environments may also generate radio signals comparable with that of the Primakoff conversion \cite{caputo2018looking, sigl2017astrophysical}. In the work presented in this chapter, we study the evolution of the ALPs field due to their stimulated decays in the presence of an electromagnetic background, which exhibits an exponential increase in the photon occupation number by taking into account the role of the cosmic plasma in modifying the photon growth profile. In particular, we attempt to focus on investigating the plasma effects in modifying the detectability that results from the stimulated emissions from ALP clumps. Based on this scenario, we explore the potential of the near future radio telescopes such as the Square Kilometer Array \cite{dewdney2009square} telescopes to detect an observational signature of ALPs decay into photons in suitable astrophysical fields.

The outline of this chapter is as follows. The discussion in the previous paragraphs sets the plan for the rest of this chapter. In the following section \ref{sec.6.2}, we review the theoretical basics for ALPs and their spontaneous and stimulated decay. In section \ref{sec.6.3}, we used the classical equation of motion to describe the evolution of the electromagnetic field in ALPs background. In section \ref{sec.6.4}, the corrections to the equation of motion due to plasma effects are considered. In section \ref{sec.6.5}, an approximate estimation of the plasma density distribution in the universe is performed. In section \ref{sec.6.6}, we discuss the effect of plasma density on preventing the stimulated decay of ALPs based on the numerical solutions for the ALP-photon decay. Then the possibility of detecting radio signature from the stimulated decay of ALPs using the SKA radio telescopes is discussed in section \ref{sec.6.7}. Finally, our conclusion is provided in section \ref{sec.6.8}.

\section{{Interactions between ALPs and photons}} \label{sec.6.2}

As we discussed before, most of the phenomenological implications of ALPs are due to their feeble interactions with the standard model particles that take place through the Lagrangian \cite{sikivie1983experimental, raffelt1988mixing, anselm1988experimental}
\begin{equation} \label{eq.6.2.1}
\mathrm{\ell}_{a} = \frac{1}{2} \partial_{\mu} a \partial^{\mu} a - V(a) - \frac{1}{4} \mathrm{F}_{\mu \nu} \tilde{\mathrm{F}}^{\mu \nu}  - \frac{1}{4} g_{a\gamma} a \mathrm{F}_{\mu \nu} \tilde{\mathrm{F}}^{\mu \nu} \:.
\end{equation}
The specific form of the potential $V(a)$ is model dependent. For the QCD axion, it results from non-perturbative QCD effects associated with instantons, which are non-trivial to compute with accuracy. For a more simple and general case, the ALPs potential can be written in terms of the ALP mass $m_a$ and the energy scale of PQ symmetry breaking $f_a$ in the simple form
\begin{equation} \label{eq.6.2.2}
V(a) = m_a^2 f_a^2 \left[ 1- \cos \left( \frac{a}{f_a} \right) \right] \:.
\end{equation}
Since we shall only be interested in the non-relativistic regime for ALPs, we focus on very small field configurations $a \ll f_a$. For a high ALPs density when we can not neglect the ALPs self-interactions. In this case, the potential can be expanded as a Taylor series with the dominant terms\newpage \noindent
\begin{equation} \label{eq.6.2.3}
V(a) \approx \frac{1}{2} m_a^2 a^2 - \frac{1}{4 !} \frac{m_a^2}{f_a^4} a^4+ \dots \:.
\end{equation}
Note that the specific values of the ALPs mass $m_a$ and the energy scale $f_a$ are model dependent. The coupling strength of ALP to two-photons can be given by the relation
\begin{equation} \label{eq.6.2.4}
g_{a \gamma} = \frac{\alpha}{2 \pi f_a} \ C\:,
\end{equation}
where and $C$ is a dimensionless model-dependent coupling parameter, usually thought to be of order unity.

\subsection{Spontaneous decay rate of ALPs} \label{sec.6.2.1}

The ALP-two-photon interaction vertex allows for the Primakoff conversion between ALPs and photons in the presence of an external electric or magnetic field, as well as for the radiative decay of the ALPs into pairs of photons. The majority of the ALPs searches in both the laboratory experiments and in astrophysical environments are based on these two processes. The spontaneous decay rate of an ALP with mass $m_a$ in vacuum into pair of photons, each with a frequency $\omega=m_a / 4\pi$, can be expressed from the usual perturbation theory calculations in term of its mass and ALP-photon coupling $g_{a \gamma}$ \cite{kelley2017radio}. The lifetime of ALPs can be given by the inverse of their decay rate as 
\begin{equation} \label{eq.6.2.5}
\tau_a \equiv \mathrm{\Gamma}_{\text{pert}}^{-1} = \frac{64 \pi}{m_a^3 g_{a\gamma}^2} \:.
\end{equation}
However, this is the specific form for the QCD axions, not for the generic ALPs$;$ we still can use it as a reasonable estimation. For the typical QCD axions with mass $m_a \sim 10^{-6} \ \text{eV}$ and coupling with photons $g_{a\gamma} \sim 10^{-12} \ \text{GeV}^{-1}$, one can insert these limits to equation \eqref{eq.6.2.4} and evaluate the perturbative decay time for general ALPs which found to be about $\sim 2 \times 10^{44} \ \text{s}$. This lifetime for ALPs is much larger than the present age of the universe $\sim 4.3 \times 10^{17} \ \text{s}$. Therefore ALPs seem to be super stable in the cosmic scale, and perhaps this is the main reason for neglecting the ALPs decay in the literature. According to this scenario, the spontaneous decay of ALPs can not be responsible for producing any observable signal that can be detectable by the current or near-future radio telescopes. 

\subsection{Stimulated decay rate of ALPs} \label{sec.6.2.2}

Up to this point, we ignored the fact that ALPs are bosons$;$ their very low mass indicates that their density and occupation numbers can be very high. Therefore it is suggested that ALPs may form a Bose-Einstein condensate with only short-range order. Note that the authors of \cite{guth2015dark} claimed that an ALP BEC of long-range correlations is unjustified, as it is driven by attractive interactions. This ALPs BEC can then thermalize through their gravitational attraction and self-interactions to spatially localized clumps. These clumps are gravitationally bound structures composed of axions or ALPs in the form of either solitons or Bose stars and, they may present in the dark matter halos around many galaxies \cite{guth2015dark}. Indeed, if ALPs present during inflation, their field is expected to be totally homogenized within the horizon. Therefore, all ALPs are expected to be in the same state that means their occupation number is very huge
\begin{equation} \label{eq.6.2.6}
f_a \sim \frac{\rho_a}{m_a H^3} \sim 10^{40} \left( \frac{\text{eV}}{m_a} \right)^2 \left( \frac{A}{10^{11} \text{GeV}} \right)^2 \left( \frac{m_a}{H} \right)^3 \:,
\end{equation}
where $m_a$ represents the mass of the ALP field, and $A$ donates the typical field amplitude. Such a very huge occupation number makes the effects of Bose enhancement seem to be possible, as we now describe. However, this is not sufficient to lead to an enhanced decay rate. Bose enhancement is ensured by enormous occupation numbers in the final state as well. In this case, we are studying the population of photons, and the ALPs decay into two-photons is a two-body decay that produces identical photons with the same energy. Therefore, it is expected for these photons to quickly accumulate with an enormous occupation number resulting in Bose enhancement to ensue. In this case, the system with such a high occupancy case is well described by a classical field. The stimulated decay of ALPs $a \rightarrow \gamma \gamma$ with a very high rate is possible in the presence of an electromagnetic wave with a certain frequency. During this process, the electromagnetic wave will be greatly enhanced. Given the very low mass of the ALPs and their incredibly weak coupling, the stimulated emission in an empty and non-expanding universe would induce ALPs to decay very rapidly into photons, nullifying most of the interesting parameter space. The decay time obtained in \cite{alonso2020wondrous} from the equation of motion for ALPs condensates is just about $\sim 10^{-7} \ \text{s}$, which is dramatically small comparing to the perturbative decay time $\sim 2 \times 10^{44} \ \text{s}$. Indeed this scenario is much problematic as it can not lead to ALPs stable enough to be accounted for the dark matter content in the universe.

In a recent paper \cite{alonso2020wondrous}, the authors have looked for an explanation for the significant divergence in the decay time of ALPs using the classical and the standard perturbative calculations. At first glance, with taking Bose enhancement into account, the calculations would lead to a dramatically small decay time of ALPs. Indeed this would be the scenario if the stimulated decay occurs in an empty and non-expanding universe. However, it is argued in \cite{abbott1983cosmological, preskill1983cosmology, dine1983not} that there are two effects that can reduce the rate of the stimulated decay of ALPs into photons. The rapid expansion of the early universe redshifts the target photon population and therefore it can discourage enhanced decay rate of ALPs. In addition to the expansion of the universe, the plasma effects also are crucial as it modifies the photon's propagation of and prevents the early decay of ALPs. Inside the plasma, photons have an effective mass that kinematically forbids the decay of lighter particles, including ALPs or at least a portion of it. Consequently, the redshifting of the decay products due to the expansion of the universe as well as the effective plasma mass of the photon can prevent the extremely fast decay rate of ALPs into photons. Indeed, this is an exciting approach to look for dark matter ALPs using the current and near-future experiments along with astrophysical observations. Therefore in the following section, we used the classical equation of motion to account for the Bose enhancement of ALPs in the cosmic plasma.

Let us now consider the enhancement of the ALP decay rate by a stimulation effect in the presence of a photon background. The number density of ALPs $n_a$ obeys the Boltzmann equation that could be written in the following form
\begin{equation} \label{eq.6.2.7}
\dot{n}_a \simeq - n_a \mathrm{\Gamma}_{\text{pert}} (1+2 f_{\gamma}) \:,
\end{equation}
where $f_{\gamma}$ indicates the photon occupation number. Therefore, we have an effective ALP decay rate
\begin{equation} \label{eq.6.2.8}
\mathrm{\Gamma}_{\text{eff}}  = \mathrm{\Gamma}_{\text{pert}} (1+2 f_{\gamma}) \:.
\end{equation}
The term between the brackets is the stimulation factor due to the background photon radiation. It seems to be clear that only in the limits of very low photon occupation $f_{\gamma} \ll 1$, the effective decay rate of ALP indicated from the previous formula should be identical to the spontaneous decay rate $\mathrm{\Gamma}_{\text{pert}}$ given by equation \eqref{eq.6.2.5}. Otherwise, if the final state photon modes have been significantly populated by previous decays, then $\mathrm{\Gamma}_{\text{eff}}$ can become much larger by stimulated emission. Further, the number of photons generated of modes with momentum around $k = m_a/2$ are given by
\begin{equation} \label{eq.6.2.9}
f_\gamma (k=m_a/2) = \frac{4 \pi^2 a_0}{m_a^2 g_{a \gamma}} \frac{n_\gamma}{n_a} \:.
\end{equation}
Therefore, the required condition for the occupation number of photons $f_\gamma$ to be greater than unity and thus leading to a stimulated emission becomes
\begin{equation} \label{eq.6.2.10}
n_\gamma > \frac{g_{a \gamma} m_a^2}{4 \pi^2 a_0} n_a \:.
\end{equation}
By using the spontaneous decay rate formula (2.5), the effective ALP decay rate reads
\begin{equation} \label{eq.6.2.11}
\mathrm{\Gamma}_{\text{eff}} = \frac{g^2_{a \gamma} m_a^3}{64 \pi} \left( 1+ \frac{8 \pi^2 a_0}{m_a^2 g_{a \gamma}} \frac{n_\gamma}{n_a} \right) n_a \:.
\end{equation}
Then, the Boltzmann equation for the photons number density can be written as
\begin{equation} \label{eq.6.2.12}
\dot{n}_\gamma = 2 \mathrm{\Gamma}_{\text{eff}} \ n_a = \frac{g^2_{a \gamma} m_a^3}{32 \pi} \left( 1+ \frac{8 \pi^2 a_0}{m_a^2 g_{a \gamma}} \frac{n_\gamma}{n_a} \right) n_a \:.
\end{equation}
If the condition in equation \eqref{eq.6.4.4} is fulfilled, the previous Boltzmann equation has an exponentially growing solution of the general form
\begin{equation} \label{eq.6.2.13}
n_\gamma = \exp [\tilde{\mu}  t] n_\gamma (0) \:,
\end{equation}
with growth rate
\begin{equation} \label{eq.6.2.14}
\tilde{\mu} = \pi \frac{g_{a \gamma} m_a a}{4} \:.
\end{equation}
We notice here that the stimulated decay leads to a growth rate increases exponentially with a factor of $g_{a \gamma}$ instead of a growth rate proportional to a factor of $g^2_{a \gamma}$ in the spontaneous decay.

\section{Evolution of electromagnetic field in ALPs background} \label{sec.6.3}

To investigate the stimulated decay of the ALP condensates, we study the growth of the electromagnetic field using a simple model that the ALP background field is uniform, and the density is time-independent. The stable ALP condensate solutions are non-relativistic, with huge occupancy numbers \cite{schiappacasse2018analysis, hertzberg2018scalar}. In the non-relativistic regime it is useful to express the real ALP field $a(x,t)$ in terms of a complex Schr$\ddot{\text{o}}$dinger field $\phi(x,t)$ according to
\begin{equation} \label{eq.6.3.1}
a(x,t) = \frac{1}{\sqrt{2 m_a}} \left[ e^{-im_a t} \phi(x,t) + e^{im_a t} \phi^\ast(x,t) \right] \:.
\end{equation}
In general, such a homogeneous ALP background field is unstable due to gravitational and self-interactions, which lead this configuration to collapse and form ALP condensate clumps. For simplicity, we ignore this more realistic situation and restrict by treating the ALPs as homogeneous and oscillating periodically in the classical field limit. This approximately represents a harmonic oscillation for small field amplitudes
\begin{equation} \label{eq.6.3.2}
a(t) = a_0 \cos(\omega_0 t) \:,
\end{equation}
where the amplitude of oscillation is $a_0$. For the non-relativistic limit for ALPs, the frequency has an excellent approximation that $\omega_0 \approx m_a$. We now assume that the amplitude $a_0$ is independent of time and position. Then the dynamics of the ALP field is given by the standard non-relativistic Hamiltonian
\begin{equation} \label{eq.6.3.3}
H = \int d^3x \left[ \frac{1}{2} \left( \dfrac{\partial a}{\partial t} \right)^2 + \frac{1}{2} (\nabla a)^2 + \frac{1}{2} m_a^2 a^2 \right] \:. 
\end{equation}
From the Hamiltonian, one can derive the ALPs density that contributes to the total energy density of the universe as
\begin{equation} \label{eq.6.3.4}
\rho_a = \frac{1}{2} m_a^2 a_0^2 \:. 
\end{equation}
Thus, the equation of motion for the ALP field in the Friedmann-Lema{\^\i}tre-Robertson-Walker (FLRW) cosmological background takes the familiar form
\begin{equation} \label{eq.6.3.5}
\ddot{a} + 3 H \dot{a} - \frac{\nabla^2}{R^2(t)} a+ V'(a) = 0 \:. 
\end{equation}
Here,  $V' \equiv dV/da$. Since the equation of motion \eqref{eq.6.2.4} is given by the dimensionless form, the dynamics of the ALP does not depend on the values of $m_a$ and $f_a$. For the homogeneous case of the ALP field $a$, the equation of motion reduces to
\begin{equation} \label{eq.6.3.6}
\ddot{a} + 3 H \dot{a} + m_a^2 a = 0 \:.
\end{equation}

As we discussed above, the classical field equations are enough to describe the regime of ALP condensates with a very high occupancy number. Therefore we consider a quantized four-vector potential $\hat{\mathrm{A}}^{\mu}=(\hat{\mathrm{A}}_0, \hat{\accbm{\mathrm{A}}})$ in a classical background given by the ALP field $a$. We vary the above Lagrangian \eqref{eq.6.2.1} with respect to $\hat{\mathrm{A}}^{\mu}$ to obtain the Heisenberg equation of motion. Assuming a non-relativistic and homogeneous ALP field is given as in equation \eqref{eq.6.3.1}, the space derivatives of the ALP field $a$ can be neglected when compared with $\partial a / \partial t$ at least at the scale of the photon momentum and translational invariance. For the electromagnetic field, we deal with the Lorenz gauge, $\partial_\mu \hat{\accbm{\mathrm{A}}}^\mu = 0$, and use the remaining gauge freedom to set $\hat{\accbm{\mathrm{A}}}_0 = 0$. Then the resulting equations of motion for the ALP field $a$ propagating in the electromagnetic field $\hat{\mathrm{A}}^{\mu}$ are
\begin{align}
\ddot{a} + m_a^2 a - g_{a \gamma} \dot{\hat{\accbm{\mathrm{A}}}} \cdot ( \nabla \times \hat{\accbm{\mathrm{A}}} ) &= 0 \:, \label{eq.6.3.7} \\ \label{eq.6.3.8}
\ddot{\hat{\accbm{\mathrm{A}}}} - \nabla^2 \hat{\accbm{\mathrm{A}}}  + g_{a \gamma} \dot{a} \, (\nabla \times \hat{\accbm{\mathrm{A}}} )&=  0 \:.
\end{align}
Now we consider an incoming monochromatic electromagnetic wave with frequency $\omega > 0$. We are assuming that the electromagnetic field is weak. Then we can choose that $a_0$ does not change, and the ALP field can be regarded as a background field.  Now because we consider  the ALP field to be time-dependent, it is useful to make a Fourier transform with respect to the spatial coordinates only
\begin{equation} \label{eq.6.3.10}
\hat{\accbm{\mathrm{A}}}(t,\hat{z}) = \int \dfrac{d^3k}{(2\pi)^3} e^{-i\hat{k} \cdot \hat{z}} \hat{\accbm{\mathrm{A}}}_{\bm{k}}(t)   \:.
\end{equation}
Then we can neglect the term $(\nabla \times \hat{\accbm{\mathrm{A}}})$ in equations \eqref{eq.6.3.7} and \eqref{eq.6.3.8}, by Fourier transforming the vector potential as
\begin{equation} \label{eq.6.3.9}
\hat{\accbm{\mathrm{A}}}_{\bm{k}} (t)= \sum_{\lambda = \pm} \left[ \hat{a}_{\bm{k}, \lambda} \bm{\epsilon}_{\bm{k}, \lambda} s_{\bm{k}, \lambda} (t)+\hat{a}^{\dagger}_{\bm{k}, \lambda} \bm{\epsilon}^{\ast}_{\bm{k}, \lambda} s^{\ast}_{\bm{k}, \lambda} (t) \right] \:,
\end{equation}
where $\bm{\epsilon}_{\bm{k}, \lambda=\pm}$ are the photon circular-polarization vectors and $ \hat{a}_{\mathbf{k}, \lambda}$ and $\hat{a}^{\dagger}_{\bm{k}, \lambda}$ are annihilation and creation operators, respectively. Here $s_{\mathbf{k}, \lambda}$ correspond to the mode functions that have to be solved. Whereas $i \bm{k} \times \bm{\epsilon}_{\bm{k}, \lambda} = k \bm{\epsilon}_{\mathbf{k}, \lambda}$, the polarizations of the mode functions  $s_{\bm{k}, \lambda}$ decouple and the equations of motions \eqref{eq.6.3.7} and \eqref{eq.6.3.8} can rewritten as follows
\begin{align}
\ddot{a}+ \left[ k^2 +m_a^2 \right] a - g_{a \gamma} k \dot{s}_{\bm{k},+} s_{\bm{k},+} &= 0 \:, \label{eq.6.3.11} \\ \label{eq.6.3.12}
\ddot{s}_{\bm{k},\pm} +\left[ k^2 \mp g_{a \gamma} k \omega_0 a_0 \sin(\omega_0 t)\right] s_{\bm{k},\pm}&= 0  \:.
\end{align}
The last equation takes the form of the well-known Mathieu equation that is one of the equations that describe the parametric oscillators\footnote{Parametric system refers to a system at which the motion depends on a time-dependent parameter.}. The Mathieu equation can be solved analytically in particular cases or exactly using numerical methods. In general, it has a periodic solution, and its parametric space characterizes by a band structure of stable regions correspond to oscillatory solutions and unstable (resonant) regions correspond to exponentially growing solutions. The term inside the square brackets represents the frequency $\omega_k^2(t)=\omega_k^2(t+T)$, where $T=2\pi / \omega_0$ is the period of oscillations of the condensate. The  coupling between the mode functions $s_{\bm{k},\pm}$ and the ALPs field depends on $k$. In regimes of ALPs background with low density and weak coupling $(k/\omega_0) \gg (g_{a \gamma} a_0 /2)$, the periodicity of $\omega_k(t)$ leads to a spectrum of narrow resonant bands equally spaced at $k^2 \approx (n/2)^2 \omega_0^2$ for positive integer numbers of $n$. The resonance solutions corresponds to the stimulated decay process $a \rightarrow \gamma \gamma$ and lead to mode functions $s_{\bm{k},\pm}$ grow exponentially with time. In the limit of the small amplitude of the photon field, it is useful to expand the solution for the mode functions $s_{\bm{k},\pm}$ as a harmonic expansion
\begin{equation} \label{eq.6.3.13}
s_{\bm{k},\omega,\pm} = \sum_{\omega= - \infty}^{\infty} f_{\omega,\pm} (t) e^{-i \omega t} \:,
\end{equation}
where $f_{\omega,\pm} (t)$ are slowly varying functions, and sum runs only over integer multiplies of $\omega_0/4$. In vacuum, we take the dispersion relation $\omega = k$ for electromagnetic waves. In the first instability band, only the lowest frequencies $\omega =\pm \omega_0 /2$ are dominating. Inserting this expansion \eqref{eq.6.3.13} into the Mathieu equation \eqref{eq.6.3.12} and dropping all fast varying terms we obtain
\begin{equation}
\dot{f}_{\omega, \pm} + \frac{1}{2 i \omega} (\omega^2 - k^2) f_{\omega, \pm} \mp \frac{g_{a \gamma} a_0 \omega_0 k}{4 \omega} \left( f_{\omega - \omega_0, \pm} - f_{\omega + \omega_0, \pm}\right) =0 \:.
\end{equation} 
For the lowest frequency modes corresponds to $\omega =\pm \omega_0 /2$, we get
\begin{align}
\dot{f}_{\frac{\omega_0}{2}, \pm} + \frac{1}{i \omega_0} (\frac{\omega_0^2}{4} - k^2) f_{\frac{\omega_0}{2}, \pm} \mp \frac{g_{a \gamma} a_0 k}{2}  f_{\frac{-\omega_0}{2}, \pm} &= 0 \:, \label{eq.6.3.15} \\ \label{eq.6.3.16}
\dot{f}_{\frac{- \omega_0}{2}, \pm} - \frac{1}{i \omega_0} (\frac{ \omega_0^2}{4} - k^2) f_{\frac{- \omega_0}{2}, \pm} \mp \frac{g_{a \gamma} a_0 k}{2}  f_{\frac{\omega_0}{2}, \pm}  &= 0  \:.
\end{align}
The last two equations represent the classical equations of motion for the electromagnetic field in vacuum. At this stage, we have ignored in our analysis the fact that the universe is permeated by an ionized plasma that has a severe impact on the propagation of photons. Therefore, before we go further in discussing the possible solutions to the classical equations of motion of the photon field, we want to consider the more realistic case that the universe is full of plasma. In this scenario, the density of the cosmic plasma might have a significant role in modifying the photon growth profile and their effect has to be determined to investigate whether it is worth to be taken into account. For this reason, we will analyze in the next section the corrections that can be made to the equations of motion of photons due to the effect of the plasma.

\section{Corrections due to plasma effects} \label{sec.6.4}

This far, we discussed the equations of motion for the electromagnetic field in vacuum. In this case, the photons have no mass. To be more realistic, we have to take into account the effect of the plasma that fills the universe. Photons propagating in a plasma acquires an effective mass equal to the plasma frequency according to \cite{carlson1994photon}
\begin{equation} \label{eq.6.4.1}
\omega_p^2 = \frac{4 \pi \alpha n_e}{m_e} \:,
\end{equation}
where $m_e$ and $n_e$ are the mass and number density of the free electrons, respectively. The interactions between the ALPs and photons can be quite large if they have the same effective mass. This explains the reason that the decay processes $a \rightarrow \gamma \gamma $ are kinematically forbidden in the very early universe. As the number density of the free electrons is so high that the plasma frequency and, consequently, the effective mass of the photons are very high. However, the number density of the free electrons decreases in the late universe, and the process becomes easily kinematically allowed. Dark matter usually resides in galactic halos, where the typical electron number density $n_e$ is about $0.03 \ \text{cm}^{-3}$ \cite{hertzberg2018dark}. If one use equation \eqref{eq.6.4.1} and the giving value for the $n_e$, it is found that the photon has an effective mass of $\bm{\mathcal{O}}(10^{-12}) \ \text{eV}$. Therefore, the plasma mass correction is small enough to consider its effect as a small modification. The dispersion relation for the photon propagating in a plasma modifies into $\omega^2 = k^2 + \omega^2_{\text{pl}}$. Consequently, the equation of motion for the electromagnetic modes can be rewritten in the form
\begin{equation} \label{eq.6.4.2} 
\ddot{s}_{\bm{k},\pm} +\left[ k^2 + \omega_p^2 \mp g_{a \gamma} k \omega_0 a_0 \sin(\omega_0 t)\right] s_{\bm{k},\pm} = 0  \:.
\end{equation}
Dropping all higher-order terms and fast varying terms for the lowest frequency modes $\omega = \pm \omega_0/2$, the modified equation of motions for the electromagnetic field under the presence of plasma become
\begin{align}
\dot{f}_{\frac{\omega_0}{2}, \pm} + \frac{1}{i \omega_0} (\frac{\omega_0^2}{4} - k^2 - \omega_p^2) f_{\frac{\omega_0}{2}, \pm} \mp \frac{g_{a \gamma a_0 k}}{2}  f_{- \frac{\omega_0}{2}, \pm} &= 0 \:, \label{eq.6.4.3} \\ \label{eq.6.4.4}
\dot{f}_{\frac{- \omega_0}{2}, \pm} - \frac{1}{i \omega_0} (\frac{ \omega_0^2}{4} - k^2 - \omega_p^2) f_{\frac{- \omega_0}{2}, \pm} \mp \frac{g_{a \gamma a_0 k}}{2}  f_{\frac{\omega_0}{2}, \pm}  &= 0  \:.
\end{align}
Note that for simplicity, we consider the plasma has a uniform density with stable frequency $\omega_p$.  In this case, the resonance will not be stopped because, in astrophysical background, the incoming wave has a continuous spectrum. However, the density of plasma is, in fact, not uniform, and plasma frequency is a function of time $\omega_p(t)$. In this case, the above equations are a good enough approximation for the equations of motion in a single area with almost the same plasma frequency, and we have different resonant frequencies for each area.

For the QCD axion with mass $m_a \sim 10^{-5} \ \text{eV}$ and coupling with photon $g_{a \gamma} \sim 6 \times 10^{-11} \ \text{GeV}^{-1}$, the authors of \cite{hertzberg2018dark} found that the ratio $\omega_p^2/(g_{a \gamma} a_0 k)$ is of $\bm{\mathcal{O}}(10^{-4})$. Accordingly, they claimed to neglect the plasma effects and indicated that the massless photon approximation is reasonable to consider. But, for ALPs with lower masses $m_a \sim 10^{-13} \ \text{eV}$ and coupling $g_{a \gamma} \sim 10^{-12} \ \text{GeV}^{-1}$, the ratio $\omega_p^2/(g_{a \gamma} a_0 k)$ would be of $\bm{\mathcal{O}}(10^{14})$. Besides, the density of the cosmic plasma is a function of the cosmic time and we might have to take its evolution into account as well. For these reasons, we can not ignore the plasma effects in the such case for ALPs.

Solving equations \eqref{eq.6.4.3} and \eqref{eq.6.4.4} would lead to an instability associated with an exponentially growing solutions $f \sim \exp(\mu t)$. The parameter $\mu$ is called Floquet exponent and gives the growth rate of the exponential function. Note that the resonance phenomenon occurs only for real Floquet exponents. One explicitly finds
\begin{equation} \label{eq.6.4.5} 
\mu_k = \sqrt{\frac{g_{a \gamma}^2 a_0^2 (k^2+ \omega_p^2)}{4} - \frac{1}{\omega_0^2} \left( k^2 + \omega_p^2 - \frac{\omega_0^2}{4} \right)^2} \:.
\end{equation}
During the resonance, the electromagnetic wave grows exponentially. However, this growth can not persist forever because the density of the axion field decreases as axions decay to photons and dissatisfy the resonant condition. The maximum growth obtained for $k = \sqrt{\omega_0^2 - 4 \omega_p^2} /2$ is
\begin{equation} \label{eq.6.4.6} 
\mu_{\text{max}} = \frac{g_{a \gamma}  a_0 \omega_0}{4}  \:.
\end{equation}
The edges of the instability band are given by values of $k$ at which $\mu = 0$, as the resonance occurs only for the real part of the Floquet exponent. The left and the right edges read as
\begin{equation} \label{eq.6.4.7} 
k_{L/R} = \sqrt{\frac{g_{a \gamma}^2 a_0^2 \omega_0^2}{16} + \frac{\omega_0^2}{4}} \pm \frac{g_{a \gamma} a_0 \omega_0}{4} \:,
\end{equation}
and it is straightforward to find that the bandwidth is
\begin{equation} \label{eq.6.4.8} 
\mathrm{\Delta} k = \frac{g_{a \gamma} \omega_0 a_0}{2} \:.
\end{equation}

\begin{figure}[t!]
\centering
\includegraphics[width=0.49\textwidth]{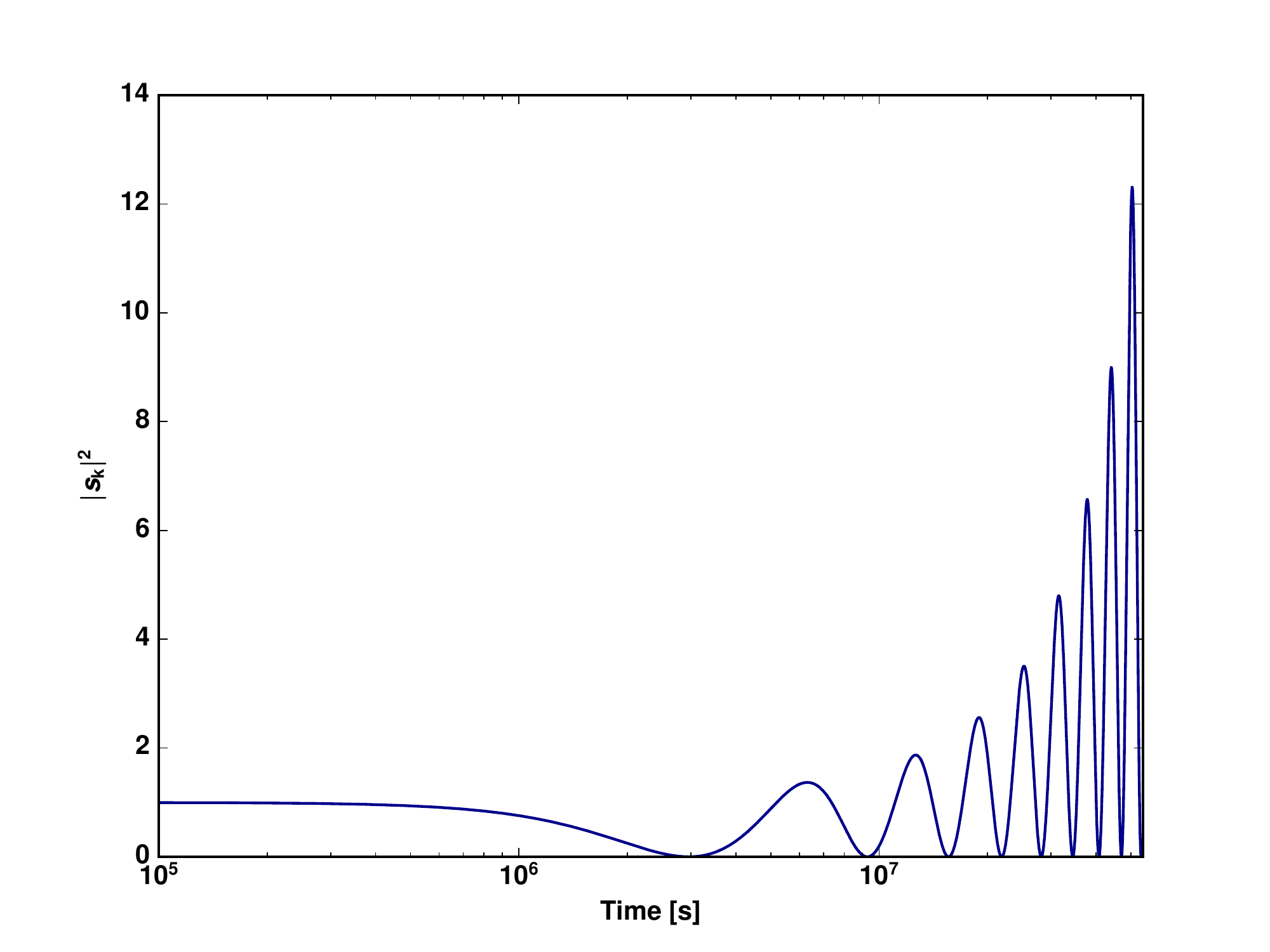}
\includegraphics[width=0.49\textwidth]{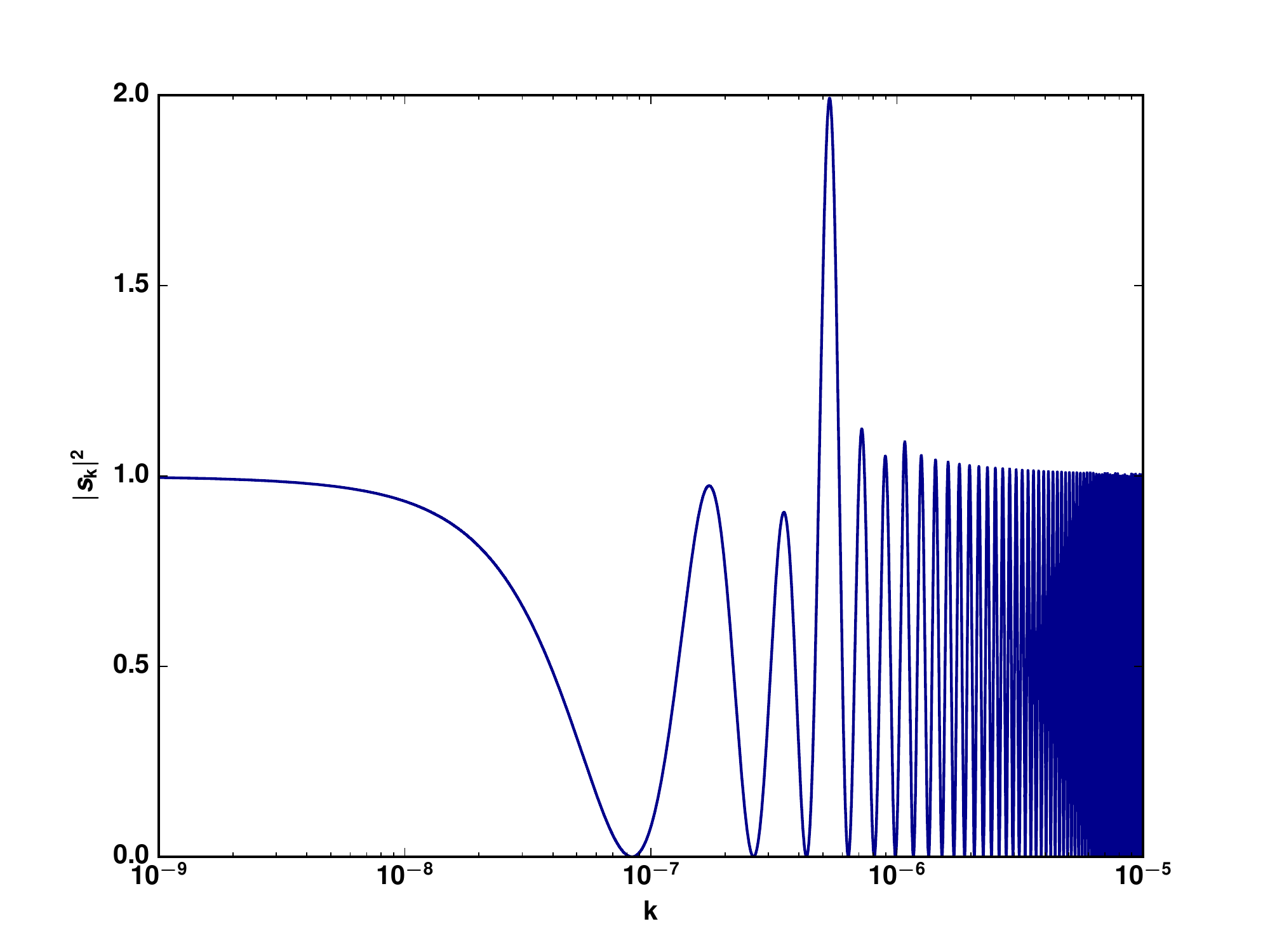}
\caption{Left panel$:$ Numerical solution of the Mathieu equation \eqref{eq.6.4.2} showing its instability bands as function of cosmic time. Right panel$:$ The first instability band of the Mathieu equation as function of the wavenumber $\boldmath{k}$ at early time.}
\label{fig.6.1}
\end{figure}

The Mathieu equation \eqref{eq.6.4.2} can be solved numerically to obtain the evolution of the electromagnetic field due to the decay of the ALP field and the instability bands of resonant enhancement. Among our calculations in order to obey the cosmological stability condition as claimed in \cite{alonso2020wondrous}, we normalize the factor $g_{a \gamma} a_0$ to be less than unity. Figure \ref{fig.6.1}, left panel, describes the electromagnetic field enhancement released from the ALPs decay as a function of the cosmic time when the wavenumber $k= \omega_0/2$. In contrast figure \ref{fig.6.1}, right panel, shows the instability bands by plotting the numerical solution $\vert s_{\bm{k}}  \vert^2$ of the Mathieu equation at a late time as a function of $k$ at early time. This solution of the Mathieu equation can be interpreted as a parametric resonance, occurring when $k = \omega_0/2$ in the absence of the plasma effect. The parametric resonance occurs at $k = \sqrt{\omega_0^2 - 4 \omega_p^2} /2$ when the effect of plasma is taken into account. This growth rate can be compared with the one found in equation \eqref{eq.6.2.14}, where we recall that $n_k \sim \vert s_{\bm{k}} \vert^2$. One realizes that the two rates agree apart from a numerical factor. This discrepancy can be explained due to the overestimation of the stimulated decay discussed in subsection \ref{sec.6.2.2} that leads to producing a larger photon growth rate. Thus, it is a better approximation to interpret the stimulated decay as a narrow parametric resonance of the Mathieu equation, and vice-versa.

\section{Plasma density evolution} \label{sec.6.5}

\begin{figure}[ht!]
\centering
\includegraphics[width=0.75\textwidth]{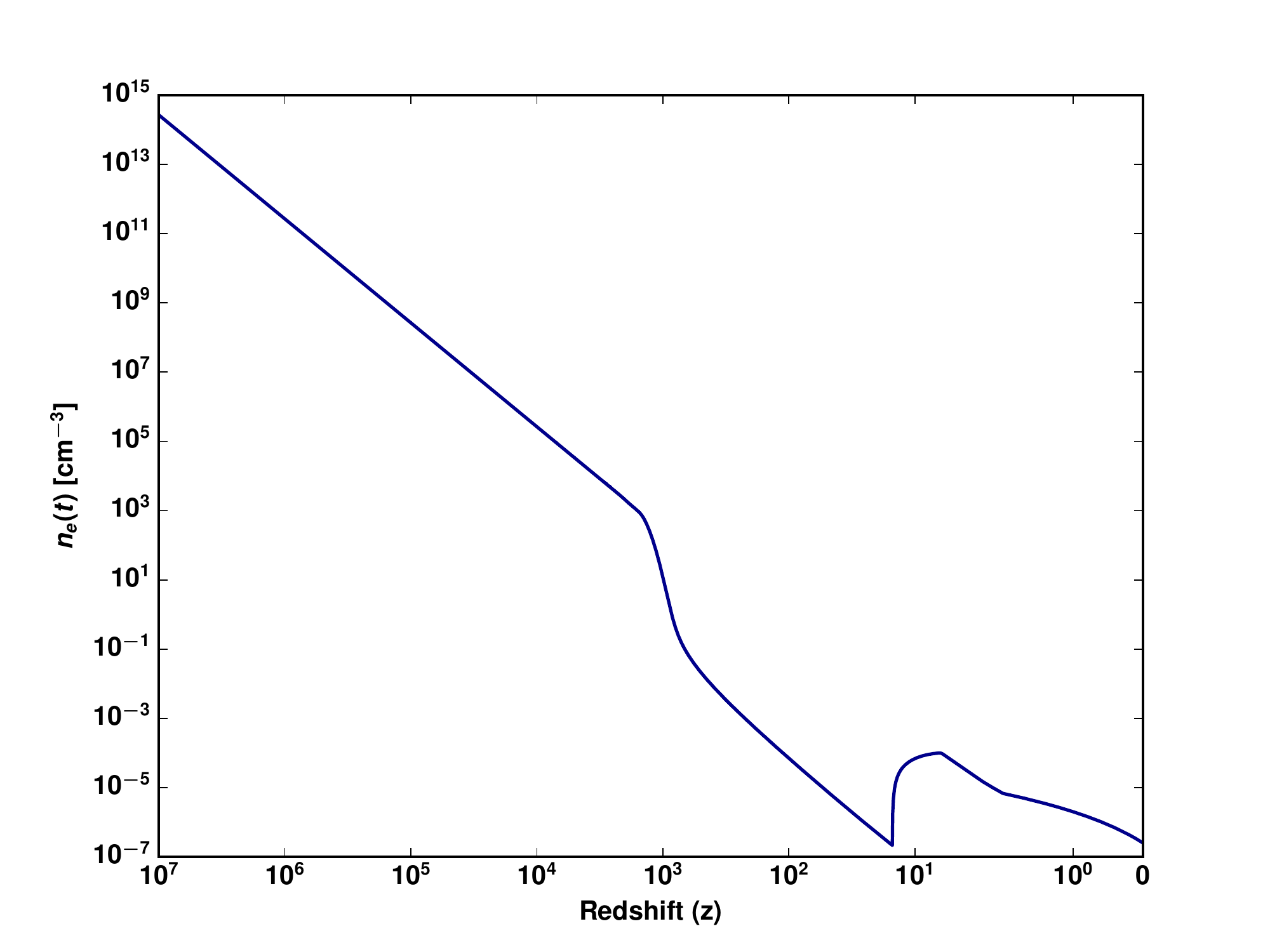}
\caption{Electron number density as a function of redshift as in equation \eqref{eq.6.5.1}.}
\label{fig.6.2}
\end{figure}

Plasma density seems to be one of the important parameters that may play a crucial role in characterizing the ALPs stimulated decay. Therefore, before discussing the effect of plasma density on preventing the stimulated decay of ALPs, at least an approximate estimation of the plasma density distribution in the universe must be performed. A useful reference density is the typical density of plasma in the galactic halos, where dark matter usually resides. The typical value of the electron number density $n_e$ in the galactic halos is about $0.03 \ \text{cm}^{-3}$ \cite{hertzberg2018dark}. Generally, the electron number density in the universe can be described by the formula \cite{ryden2017introduction}
\begin{equation} \label{eq.6.5.1}
n_e =\eta n_{\gamma} X_e \:,
\end{equation}
where $\eta = (6.1 \pm 0.06) \times 10^{-10}$ is the baryon-to-photon ratio, $n_{\gamma} = 0.244 \times (kT/\hbar c)^{3}$ is the number density of photons, and $X_e$ ionized fraction of atoms \cite{puchwein2019consistent}. The temperature of the CMB at any redshift $z$ is given as $T=T_0 (1+z)$ with $T_0= 2.725 \ \text{K}$ corresponds to the CMB temperature in the present day. Hence, using equation \eqref{eq.6.5.1} we can estimate the plasma density evolution a function of redshift as shown in figure \ref{fig.6.2}. It seems to be clear from the graph that the electron number density gets to a minimum value of about $n_e \sim 2 \times 10^{-7} \ \text{cm}^{-3}$, which expected to happen before the reionization era at redshift $z \sim 15$. The today electron number density expected to be about $n_e \sim 2.37 \times 10^{-7} \ \text{cm}^{-3}$.

\section{Plasma density prevents the stimulated decay of ALPs} \label{sec.6.6}

\begin{figure}[ht!]
\centering
\includegraphics[width=0.75\textwidth]{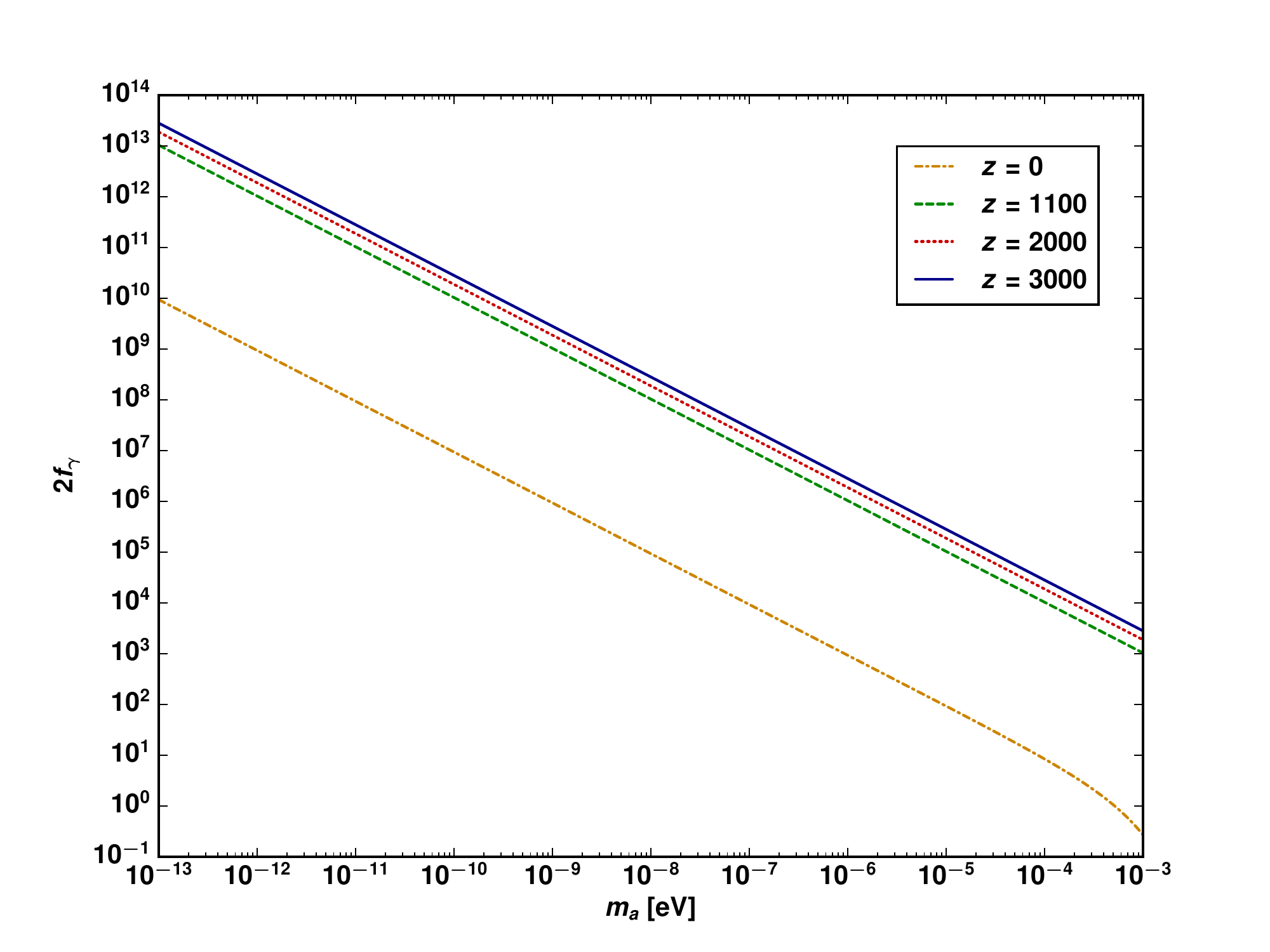}
\caption{The stimulated emission factor arising from the CMB and at different redshifts as in equation \eqref{eq.6.6.1}.}
\label{fig.6.3}
\end{figure}

\begin{figure}[t!]
\centering
\includegraphics[width=0.49\textwidth]{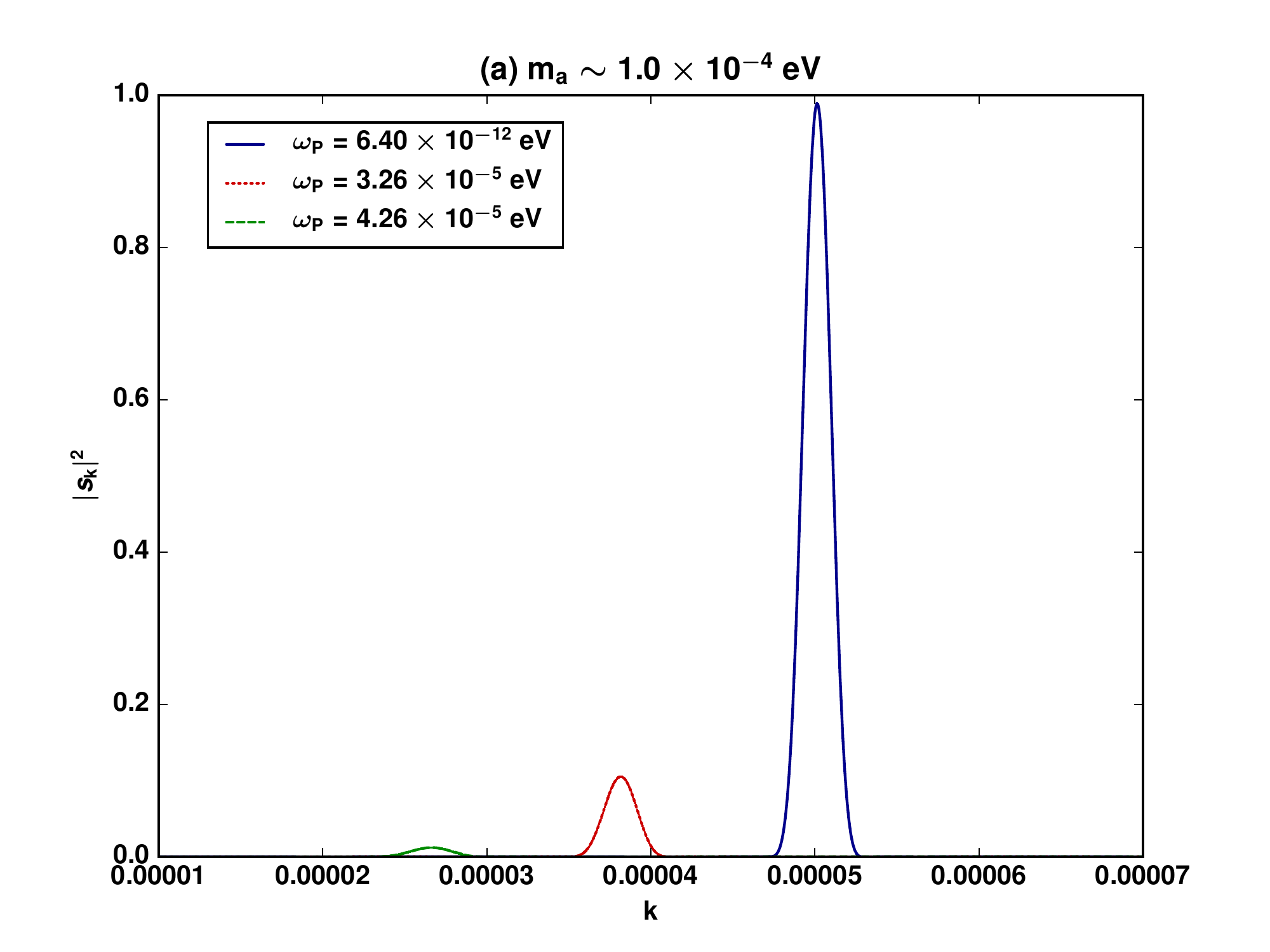}
\includegraphics[width=0.49\textwidth]{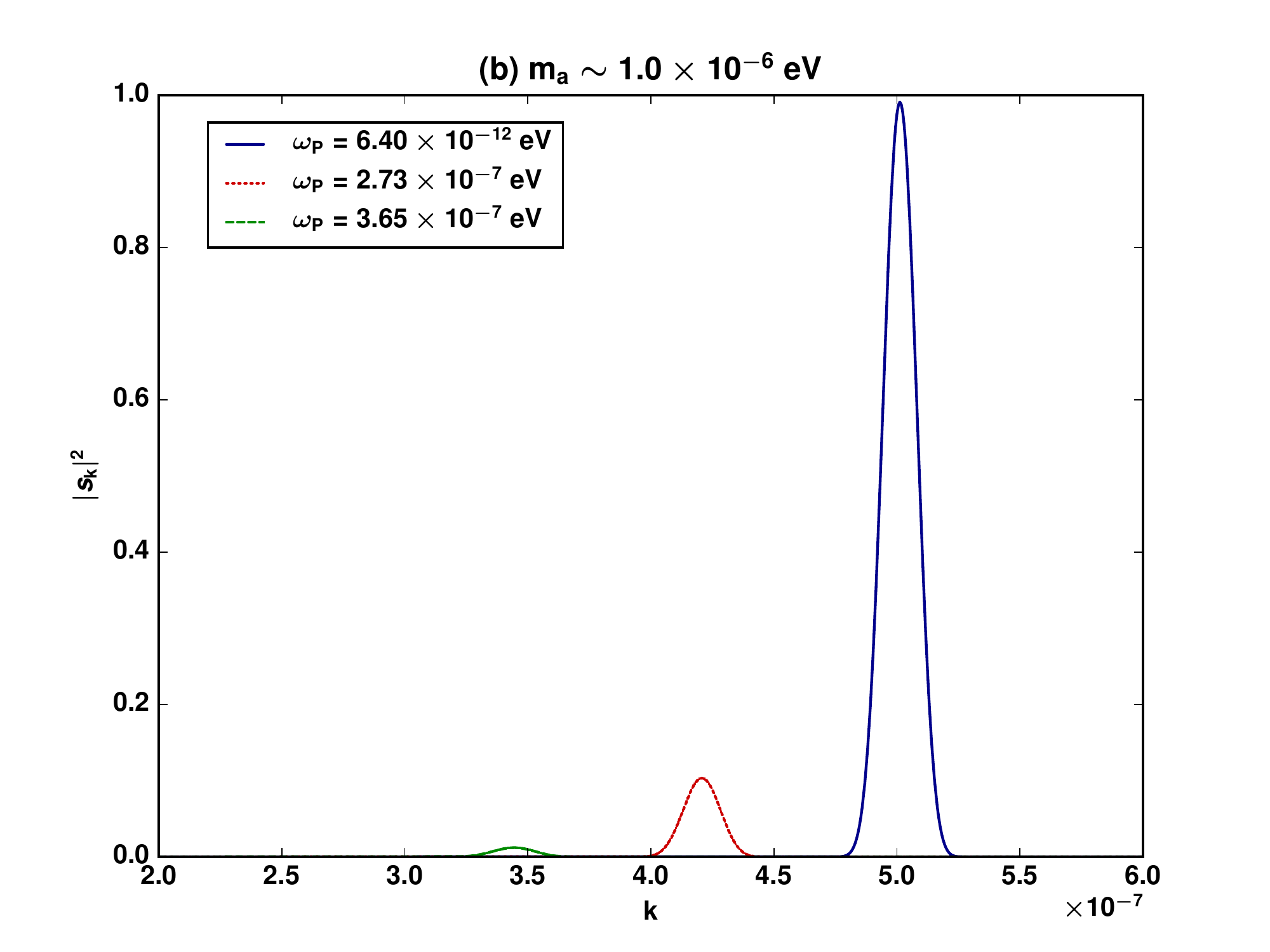}
\includegraphics[width=0.49\textwidth]{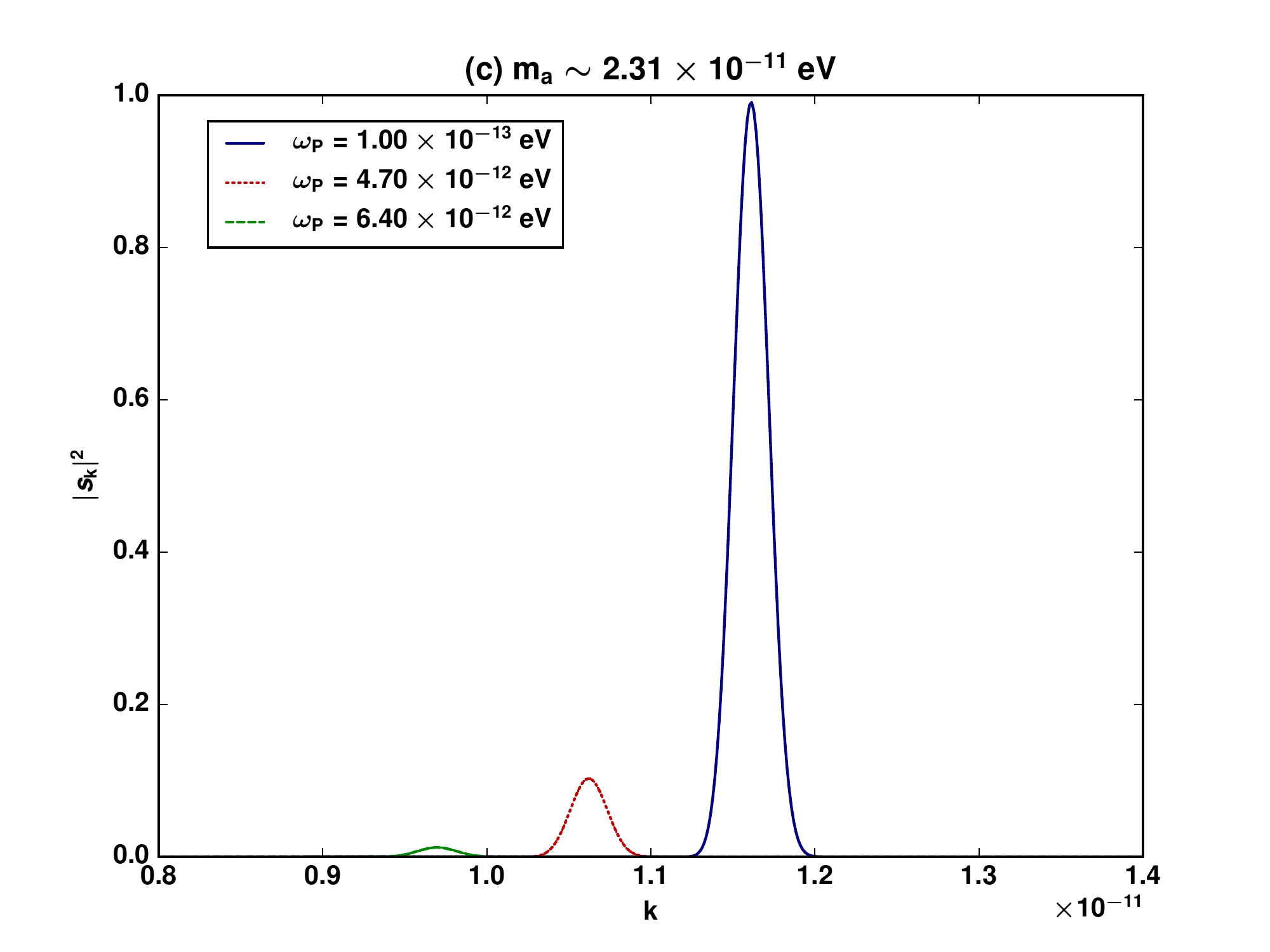}
\includegraphics[width=0.49\textwidth]{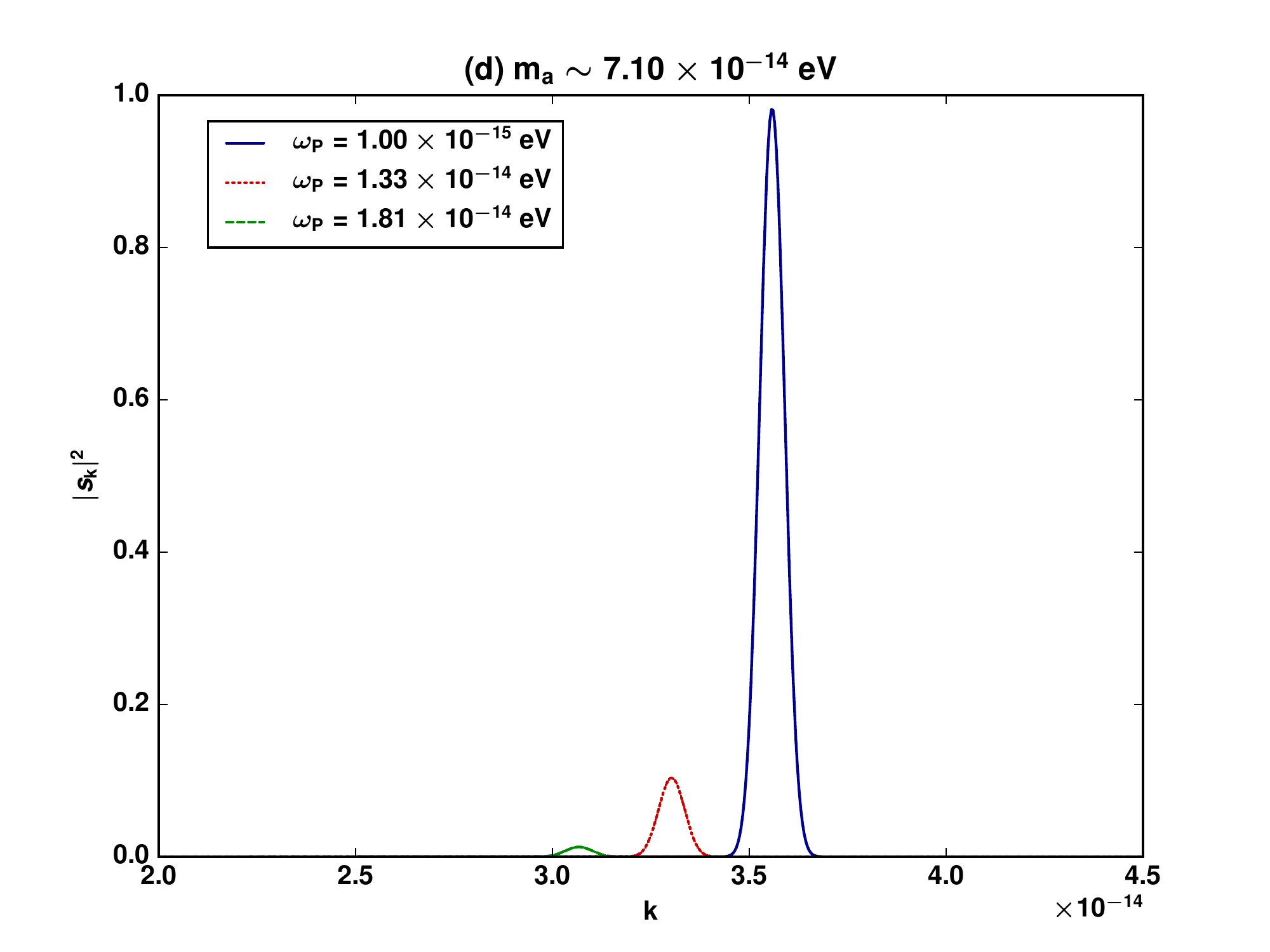}
\caption{Effect of plasma on destroying the stimulated ALPs decay for different ALPs masses, obtained from the numerical solutions of the Mathieu equation \eqref{eq.6.4.2}.}
\label{fig.6.4}
\end{figure}

Below in this section, we present the projected effect of the plasma density in preventing the stimulated decay of ALPs. The effect of stimulated emission manifests in terms of the stimulated emission factor $2 f_{\gamma}$ in equation \eqref{eq.6.2.8}. Regardless of the astrophysical environment, we consider that the stimulated decay produces an enhancement of the ALPs decay rate by factors arising only from the CMB as in reference \cite{caputo2019detecting}
\begin{equation} \label{eq.6.6.1}
f_{\gamma, \text{CMB}} (m_a)= \frac{1}{e^{(E_{\gamma}/k_B T)} -1} \:,
\end{equation}
where $E_\gamma = m_a/2$, $k_B$ is the Boltzmann constant, and $T$ is the temperature of the CMB on redshift $z$. Note that, the factor $f_{\gamma}$ suppose to be a linear combination over all the sources which contribute to the photon bath with the same energy as that produced in the ALP decay. However, we only consider here the contribution from the CMB to extrapolate $f_{\gamma}$ corresponds to ALPs with low masses down to $m_a \sim 10^{-13} \ \text{eV}$. Figure \ref{fig.6.3} shows the stimulated emission factor $2 f_{\gamma}$ arising from the CMB at different redshifts $z=0, 1100, 2000,$ and $3000$. In  table \ref{tab.6.1} we present the numerical values of the stimulated emission factor $2 f_{\gamma}$ on high redshift $z \sim 1100$ for a set of ALP masses, the perturbative decay time $\tau_{a}$, and the stimulated correction to the decay time $\tilde{\tau}_{a}$. Then we discuss the required plasma density that is able to significantly prevent this enhancement. Figure \ref{fig.6.4} illustrates the effect of plasma on the stimulated decay of ALPs, assuming four different categories of their masses, obtained from the numerical solutions of the Mathieu equation \eqref{eq.6.4.2}.

\begin{table}[h]
\centering
\scalebox{0.8}{
\begin{tabular}{|c|c|c|c|}
\hline
ALP mass $m_a \ [\text{eV}]$ & Enhancement factor $2 f_{\gamma}$  & Perturbative decay time $\tau_{a} \ [\text{s}]$ & Stimulated correction $\tilde{\tau}_{a} \ [\text{s}]$   \\
\hline
$\sim 1.00  \times 10^{-4}$ & $\sim 1.03 \times 10^{4}$ & $\sim 2.01 \times 10^{38}$ & $\sim 1.95 \times 10^{34}$ \\
$\sim 1.00  \times 10^{-6}$ & $\sim 1.03 \times 10^{6}$ & $\sim 2.01 \times 10^{44}$ & $\sim 1.95 \times 10^{38}$ \\
$\sim 2.31  \times 10^{-11}$ & $\sim 4.46 \times 10^{10}$ & $\sim 1.61 \times 10^{58}$ & $\sim 3.59 \times 10^{47}$ \\
$\sim 7.10  \times 10^{-14}$ & $\sim  1.46 \times 10^{13}$ & $\sim 5.62 \times 10^{65}$ & $\sim 3.85 \times 10^{52}$\\
\hline
\end{tabular}}
\caption{A set of ALP masses and the corresponding stimulated emission factor arising from the CMB, perturbative decay time, and stimulated correction to the decay time.}
\label{tab.6.1} 
\end{table}

The first case (a) is when we assume ALPs with mass $m_a \sim 10^{-4} \ \text{eV}$. For the plasma in the galactic halos, the today value of its electron number density is about $0.03 \ \text{cm}^{-3}$ and using equation \eqref{eq.6.4.1} this is corresponding to plasma frequency about $6.4 \times 10^{-12} \ \text{eV}$. Plasma with such frequency has no effect on the stimulated decay of ALPs in this case. A significant effect of plasma in this scenario requires at least plasma with a frequency of the order of $4.26 \times 10^{-5} \ \text{eV}$ and this is corresponding to electron number density $n_e \sim 1.32 \times 10^{12} \ \text{cm}^{-3}$ to effectively reduce the amplitude of the resonance. In our results, we consider reducing the enhancement factor for the stimulated decay of ALPs to about $1\%$ of its maximum value. Note that reducing the amplitude in this way seems sufficient as the plasma density required to vanishing the amplitude with more accuracy is not much different from the values presented here. It seems to be clear from figure \ref{fig.6.2} that this very high plasma density only exists at very high redshifts $z \gtrsim 1.77 \times 10^6$. The second case (b) is corresponding to ALPs with mass $m_a \sim 10^{-6} \ \text{eV}$. The typical plasma in the galactic halos has no significant effect on the ALPs stimulated decay in this case as well. However, plasma with a frequency of about $\sim 3.65 \times 10^{-7} \ \text{eV}$, which corresponds to a plasma density of about $n_e \sim 9.66 \times 10^{7} \ \text{cm}^{-3}$ at redshifts $z \gtrsim 7.4 \times 10^4$, is able to significantly prevent the ALPs stimulated decay. Because these two cases are expected to be valid only in the very early universe, this makes us rule out the possibility of any current significant plasma effect on the stimulated decay of ALPs with these masses. However, one may worry that such plasma density can be found at low redshifts in astrophysical environments with a very high density. Indeed, this is highly unlikely, but this does not preclude that it is important to look carefully at the astrophysical environment when considering the stimulated decay of ALPs with this range of masses in the late universe eras.

In the third case (c) we consider ALPs with mass about $m_a \sim 2.31 \times 10^{-11} \ \text{eV}$. This scenario is quite interesting since we look at the effect of plasma with number densities comparable to the current typical value $n_e \sim 0.03 \ \text{cm}^{-3}$ in the galactic halos where dark matter expected to be with a high amount. From figure \ref{fig.6.2} we might also note that the cosmic plasma density at redshift $z \sim 5.54 \times 10^2$ is equivalent to the current plasma density in the galactic halos that is at redshift $z \sim 0$. Such type of plasma is only able to prevent the stimulated decay of ALPs with masses $m_a \lesssim 2.31 \times 10^{-11} \ \text{eV}$. The fourth case (d) is for low-mass ALPs at $m_a \lesssim 7.10 \times 10^{-14} \ \text{eV}$. We notice in this scenario that the current cosmic plasma number density about $n_e \sim 2.37 \times 10^{-7} \ \text{cm}^{-3}$, which correspond to plasma frequency about $\sim 1.81 \times 10^{-14} \ \text{eV}$, is able to prevent the stimulated decay of ALPs with such low-mass. Since this plasma with such density is prevalent in the universe at present with redshift $z \sim 0$, we claim that at the current time the plasma should always be able to prevent or at least significantly reduce the stimulated decay of ALPs with masses $m_a \lesssim 7.10 \times 10^{-14} \ \text{eV}$. From figure \ref{fig.6.2}, one can see that the cosmic plasma density around the end of the recombination epoch at redshifts $z \lesssim 16.25$ is also comparable to the current plasma density. This makes the stimulated decay of ALPs with this range of masses is allowed at this epoch as well.

The summary of these results is presented in table \ref{tab.6.2} and shows a list of plasma frequencies and their corresponding plasma densities that are required to protect the stimulated decay of ALP with different ranges of masses. Hence, plasma with higher densities beyond these values for each case does not automatically protect the ALPs from stimulated emission cascades. This sets bounds on the mass ranges for the ALPs that can be a subject for the search for observational signals based on their stimulated decay as we will discuss in the following section. In table \ref{tab.6.2} we summarize as well the redshifts at which the plasma density reached the required threshold values to suppressing the stimulated decay of ALPs with each mass category. In addition, we provide the photon frequencies expected from the decay of ALPs with the given masses.

\begin{table}[h]
\centering
\scalebox{0.75}{
\begin{tabular}{|c|c|c|c|c|}
\hline
ALP mass $m_a \ [\text{eV}]$ & Plasma frequency $\omega_p \ [\text{eV}]$  & Plasma density $n_e \ [\text{cm}^{-3}]$ &Redshift [z]& Photon frequency $ [\text{Hz}]$   \\
\hline
$\sim 1.00 \times 10^{-4}$ & $\lesssim 4.26 \times 10^{-5}$ & $\lesssim 1.32 \times 10^{12}$& $\lesssim  1.77 \times 10^{6}$& $\sim 1.21 \times 10^{10}$ \\
$\sim 1.00 \times 10^{-6}$ & $\lesssim 3.65 \times 10^{-7}$ & $\lesssim 9.66 \times 10^{7}$ &$\lesssim 7.40 \times 10^{4}$& $\sim 1.21 \times 10^{8}$ \\
$\sim 2.31 \times 10^{-11}$ & $\lesssim  6.40 \times 10^{-12}$& $\lesssim 2.97 \times 10^{-2}$&$\lesssim  5.54 \times 10^{2}$ & $\sim 2.80 \times 10^{3}$ \\
$\sim 7.10 \times 10^{-14}$ & $\lesssim 1.81 \times  10^{-14}$& $\lesssim 2.37 \times 10^{-7}$ &$\lesssim  16.25$& $\sim 8.52$\\
\hline
\end{tabular}}
\caption{List of plasma frequencies and their corresponding plasma densities that are required to protect the stimulated decay of ALP with different ranges of masses.}
\label{tab.6.2} 
\end{table}

\section{Detecting radio signature from the stimulated decay of ALPs} \label{sec.6.7} 

In recent years, there is increasing interest in using radio telescopes to search for a radio signal produced by cold dark matter. The SKA is considered to be the most sensitive radio telescope ever \cite{colafrancesco2015probing}. This makes it the most aspirant radio telescope to unveil any expected signature of dark matter. Therefore, we investigate in this section whether interesting to take the stimulated decay of ALPs into account for searching for radio signals produced using the SKA telescopes$;$ other related work includes references \cite{tkachev2015fast, braaten2017emission, caputo2018looking, caputo2019detecting}. In this work, in particular, we clarify the role that the cosmic plasma plays in modifying the detectability of radio emissions results from the stimulated decay of ALPs.

Based on the scenario of ALPs stimulated decay that we discussed in the previous sections, we explore here the potential of the near future radio telescopes, in particular the SKA telescopes, in an attempt to detect an observational signature of CDM ALPs decay into photons in astrophysical fields. As shown in the previous section, regardless of the astrophysical environment, by considering the enhancement of the stimulated decay produces only by the CMB, our results imply the following. From figure \ref{fig.6.3}, one can deduce that the decay rate of ALPs with large enough masses $m_a \gtrsim 10^{-4} \ \text{eV}$ does not receive a remarkable enhancement. Therefore, in this case, it is not expected that the stimulated decay has an essential role, and we can not count on the spontaneous decay of ALPs for producing a significant observational signal because of the low rate of this process. Consequently, this can be considered as an upper limit for the ALP mass that can be accountable for any significant radio signature produced from the stimulated decay of ALPs. For ALP masses $m_a \lesssim 7.10 \times 10^{-14} \ \text{eV}$, the current cosmic plasma, which is prevalent in the universe, can prevent or at least significantly reduce the stimulated decay of ALPs. In contrast, the typical plasma at present time in the galactic halos where dark matter is expected to be with a high amount can prevent the stimulated decay of ALP with masses $m_a \lesssim 2.31 \times 10^{-11} \ \text{eV}$. This puts a lower bound on the ALP mass at which the stimulated decay of ALPs is allowed by the plasma effects. Hence, with the today plasma density, the stimulated decay of ALP in the $10^{-11} \text{--} 10^{-4} \ \text{eV}$ mass range is allowed by the plasma effect. In principle, this makes it possible to consider a radio emissions due to the stimulated decay of ALPs with this mass range.

Within the allowed range by the plasma effects, the search for ALPs in the $10^{-6}  \text{--} 10^{-4} \ \text{eV}$ mass range seems to be the most exciting scenario. In this case we would expect some form of low-frequency radio background due to the stimulated decay of ALPs peaked in the frequency range about $120 \ \text{MHz} \ \text{--} \ 12 \ \text{GHz}$. Fortunately, considering the upcoming reach radio frequency range of the SKA telescopes from $50 \ \text{MHz}$ to $20 \ \text{GHz}$, it would be able to detect photons produced from the decay of ALPs with such range of masses. For comparison, the results presented here are consistent with the results presented in \cite{alonso2020wondrous} that the current plasma frequency in the range of $\sim 10^{-14}\text{--}10^{-13} \ \text{eV}$ sets a lower bound on the ALP mass that can be subject to stimulated decay. Our results also agree with the results presented in \cite{caputo2019detecting} that suggested the search for a radio signal based of the stimulated decay of axion dark matter in the $10^{-7}  \text{--} 10^{-3} \ \text{eV}$ mass range in the near-future radio observations by SKA or using forthcoming axion search experiments, such as ALPS-II and IAXO. Indeed, this process was not possible in the early universe eras because the effective plasma mass of the photons was significantly higher than this range of masses for the ALPs and consequently it was able to prevent their stimulated decays as claimed in \cite{alonso2020wondrous}.

Finally, the possible signature produced from the stimulated decay of ALPs is expected to be a large flux of photons with a frequency close to half the ALP mass as shown in table \ref{tab.6.2}. This signal supposed to appear as a narrow spectral line, broadened by the ALPs velocity dispersion as discussed in \cite{caputo2019detecting}. Since dark matter expected to exist with high amounts in the halos around many galaxies, these structures offer an interesting astrophysical environment for testing the CDM ALPs decay scenario. Under this hypothesis, the decay of an ALP will produce two photons each with energy $m_a/2$. Hence, the energy flux density, \ie the power per unit area per unit frequency, is given by
\begin{equation} \label{eq.6.7.1}
S_a = \frac{m_a \mathrm{\Gamma}_\text{eff}}{4 \pi d_L^2} \:,
\end{equation}
which accounts for luminosity distance $d_L$ from observer to the halo at redshift $z$, and $\mathrm{\Gamma}_\text{eff}$ is the effective ALP decay rate as given in equation \eqref{eq.6.2.7} at redshift $z$. Then, the total energy flux density of the radio signal from a smooth ALPs background at an energy $E=m_a/2$ can be expressed as
\begin{equation} \label{eq.6.7.2}
S_{\text{decay}} = \frac{m_a \mathrm{\Gamma}_\text{eff}}{4 \pi d_L^2} \,  n_\text{ac} dV_c \:,
\end{equation}
where $n_\text{ac}$ is the comoving number density of the ALPs in the dark matter halos and the term $dV_c$ is the comoving volume element per unit redshift. In figure \ref{fig.6.4} we estimate using the last expression the total energy flux density $S_{\text{decay}}$ of the radio signal arising from the stimulated decay of ALPs as a function of the redshift for the same set of ALPs masses that we examine in this study.  These fluxes are significantly small comparing to the energy flux density of the CMB Which can be well estimated at any redshift $z$ using the expression  $S_{\text{CMB}}= 0.065 (1+z)^3 \ \text{eV} \, \text{cm}^{-2} \, \text{s}^{-1}$ \cite{dermer2009high, ambaum2010thermal}. However, the standard paradigm of hierarchical structure formation predicts that small structures of CDM form first and then merging into larger ones. This leads to a clumpy distribution of CDM inside the galactic halos. Therefore, radio signals arising from the stimulated decay of ALPs in the galactic halos are expected to be enhanced due to the substructures of the galactic halos. This enhancement is known as the boost factor and should be about a few orders of magnitude \cite{strigari2007precise}. Including this boost factor, the radio signals arising from the stimulated decay of ALPs in the galactic halos could be comparable to the CMB signal. Indeed, this contribution should be taken into account in modifying the background radiation field and hence offering an exciting scenario to the search for a possible signature from the CDM as well as a viable explanation for the EDGES 21 cm anomaly as claimed in \cite{feng2018enhanced, mirocha2019does}.
\begin{figure}[ht!]
\centering
\includegraphics[width=0.75\textwidth]{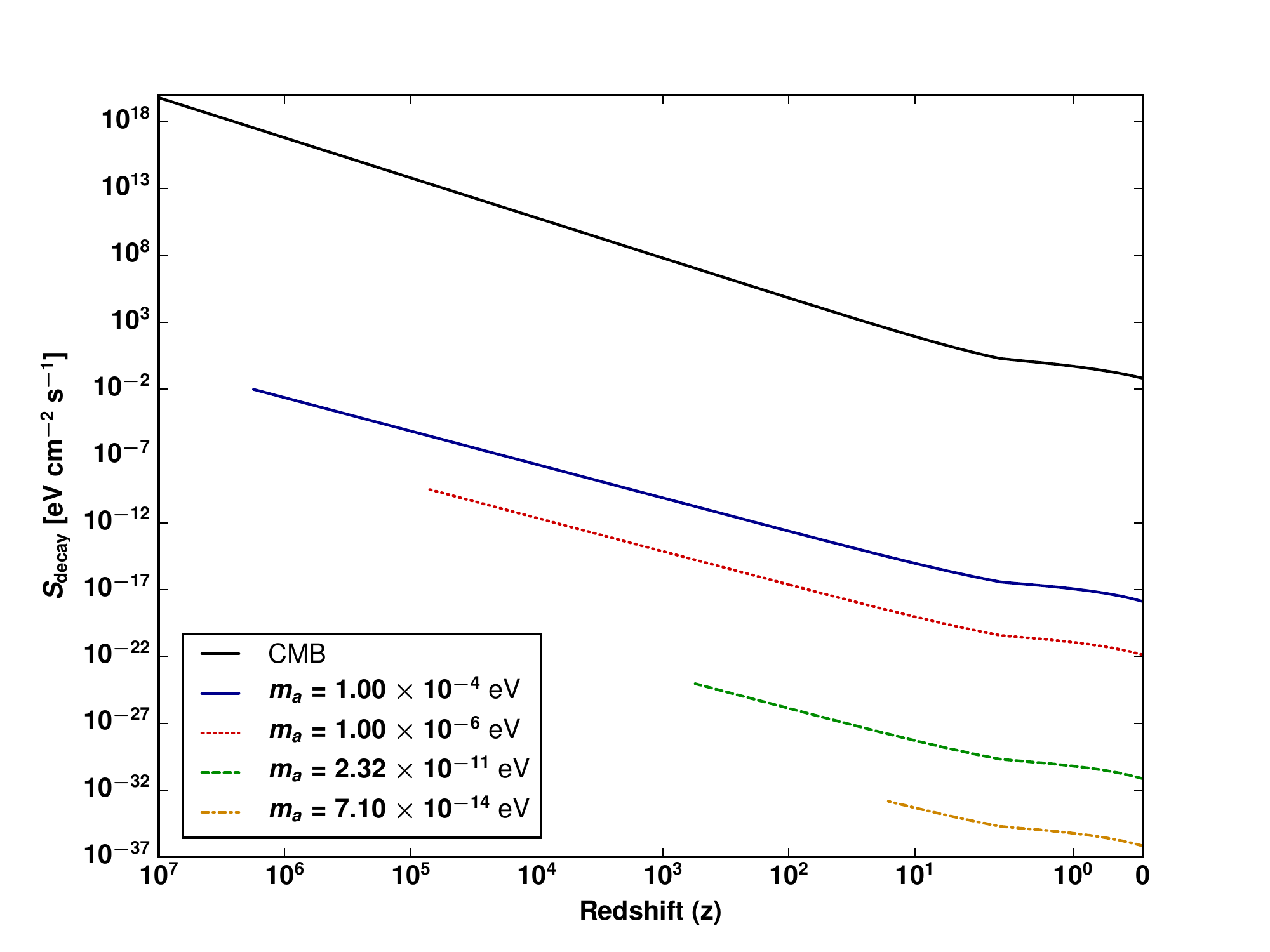}
\caption{Estimating the radio flux arising from the stimulated decay of ALPs as in equation \eqref{eq.6.7.2}.}
\label{fig.6.5}
\end{figure}

An interesting discussion on the sensitivity of the SKA radio telescopes to detect such signals from some astrophysical targets can be found in \cite{caputo2019detecting, hertzberg2020merger}. It has been illustrated in these works that with near-future radio observations by SKA, it will be possible to increase sensitivity to the ALP-photon coupling by a few orders of magnitude. In our results here we have shown that neither the current cosmic plasma nor the plasma in the galactic halos can prevent the stimulated decay of ALP for the given range of masses. However, the near-future SKA radio telescopes would be only able to reach sensitivities as low as the $10^{-17} \ \text{eV} \ \text{cm}^{-2} \ \text{s}^{-1}$ orders of magnitude \cite{dewdney2013ska1}. This is a few orders of magnitude higher than the most promising radio signal predicted in figure \ref{fig.6.5} from the stimulated decay of ALPs, which puts more challenges on the way to detect any observable signal. Leaving this aside, these results support the idea that the stimulated decay process can increase the sensitivity of the SKA radio telescopes to detect the radio emissions produced by the decay of ALPs, and indeed taking it into account is a big step forward towards achieving this goal in future.

\section{Conclusion} \label{sec.6.8}

In this chapter, we studied the possibility of detecting an observable signature produced due to the decay of CDM ALPs into photons at radio frequencies. For ALPs with masses and coupling with photons allowed by astrophysical and laboratory constraints, the ALPs are very stable on the cosmological scale, and their spontaneous decay can not be responsible for producing any detectable radio signal. Since ALPs are identical bosons, they may form a Bose-Einstein condensate with very high occupation numbers. The presence of an ambient radiation field leads to a stimulated enhancement of the decay rate. Depending on the astrophysical environment and the ALP mass, the stimulated decay of ALPs BEC in an expanding universe and under the plasma effects can be counted for producing observable radio signals enhanced by several orders of magnitude. 

We have examined that the stimulated decay of ALPs arising from the presence of the ambient background of the CMB photons results in a large enhancement of the decay rate. However, the decay rate of ALPs with large enough masses does not receive a remarkable enhancement produces by the CMB. This puts an upper limit for the ALP mass that can be accountable for any significant radio signature produced from the stimulated decay of ALPs. Further, the cosmological stability of ALPs dark matter is ensured by a combination of the expansion and the plasma effects and the latter is the most restrictive of the two. In our results, we found the plasma densities that are required to suppress the stimulated decay of ALP with different ranges of masses. This puts a lower bound on the ALP mass at which the stimulated decay of ALPs is allowed by the plasma effects. The results showed as well at which redshift the plasma density reached the required threshold values to suppressing the stimulated decay of ALPs with each mass category. This decides whether the ALP abundance in the modern epoch would be sufficient for a detectable signal and/or constituting all of dark matter. For ALPs in the $10^{-11}  \text{--} 10^{-4} \ \text{eV}$ mass range in astrophysical environments with large radio emission, this emission might potentially increase by a few orders of magnitude, and neither the current cosmic plasma nor the plasma in the galactic halos can prevent the stimulated decay of the ALPs with this range of masses. Interestingly, the stimulated decay of ALPs in the $10^{-6}  \text{--} 10^{-4} \ \text{eV}$ mass range can, in principle, lead to a signal that can be detectable by the near-future radio telescopes such as the SKA. We should be noted that this stimulated decay is only allowed at redshift window $z \lesssim 10^{4}$, and thus the ALPs can not decay efficiently at higher redshift due to the effects of the plasma. Finding such a signal consistent with the predictions for the stimulated decay of the CDM ALPs in such a very low mass range will make this technique the essential method to understanding the properties of dark matter, such as its spatial distribution and clustering in cosmological structures. In addition, this might offer an exciting scenario to explain several unexpected astrophysical observations, \eg the EDGES 21 cm anomaly. Indeed, this should depend on other parameters like astrophysical environments and radio telescopes sensitivity. That is worth being a subject of intense future work.
\chapter{\textbf{Summary and Conclusion}} \label{ch7}

The standard model of particle physics, together with the standard model of cosmology, provides the best understanding of the origin of matter and the most acceptable explanation to the behavior of the universe. However, the shortcomings of the two standard models to solve some problems within their framework are promoting the search for new physics beyond the standard models. This is why, recently, there are several studies underway at the interface between cosmology, particle physics, and field theory. In this context, the dark matter searches remain elusive and pose one of the highest motivating scenarios to go beyond the standard models.

The existence of stable cold dark matter is established based on many astrophysical and cosmological observations such as the cosmic microwave background, the large scale structure, the galactic rotation curves, and beyond. In particular, the Planck 2018 data predicts that the visible universe contains non-baryonic dark matter, which is estimated at more than five times greater than the ordinary baryonic matter. However, the content and properties of the dark matter remain one of the most pressing challenges in cosmology and particle physics. Although there are many suggested candidates for the dark matter content, there is currently no evidence of any of them. In this thesis, we focused on understanding the nature of dark matter by studying the phenomenology of axions and axion-like particles, highly viable dark matter candidates. The typical axions are identified in the Peccei-Quinn mechanism as pseudo-Nambu-Goldstone bosons appear after the spontaneous breaking of the PQ symmetry that is introduced to explain the absence of CP violation for the strong interaction, which is theoretically obvious from the Lagrangian of quantum chromodynamics. Furthermore, many extensions of the standard model of particle physics, including string theory models, generalized this concept to predict more such axion-like particles, which may arise as pseudo-scalar particles from the breaking of various global symmetries. The phenomenology of ALPs is the same as of QCD axions characterized by their coupling with two photons. The main difference between them is that the QCD axion coupling parameter is determined by the axion mass$;$ however, this is not necessarily the case for ALPs. Although the masses of the generic ALPs are expected theoretically to be very tiny, nowadays they are accounted for as the most leading candidates to compose a major part, if not all, of the dark matter content of the universe. This is motivated by the theoretical predictions for their properties that are determined by their low mass and very weak interactions with the standard model particles. It is also worthwhile to mention that it is believed that the main mechanisms to generate the dark matter ALPs in the early universe are the misalignment mechanism and the decay of strings and domain walls.

In the first part of this thesis, we started by reviewing the current status of the searches for the dark matter. In particular, we briefly explained the first hints that dark matter exists, elaborated on the strong evidence physicists and astronomers have accumulated in the past years, discussed possible dark matter candidates, and described the various detection methods used to probe the dark matter's mysterious properties. Then, the theoretical backgrounds about the QCD axion, including the strong CP problem, the Peccei-Quinn solution, and the phenomenological models of the axion, were described. This was followed by a brief discussion of the main properties of the invisible axions. After that, the topic of the possible role that axions and ALPs can play in explaining the mystery of dark matter was illustrated. Further, to look at the question of whether they correctly explain the present abundance of the dark matter, we investigated their production mechanism in the early universe. Later, we discussed the recent astrophysical, cosmological, and laboratory bounds on the axion coupling with the ordinary matter.

In the second part of this thesis,  we considered a homogeneous cosmic ALP background analogous to the cosmic microwave background and motivated by many string theory models of the early universe. The coupling between the CAB ALPs traveling in cosmic magnetic fields and photons allows ALPs to oscillate into photons and vice versa. Using the M87 jet environment, we tested the CAB model that is put forward to explain the soft X-ray excess in the Coma cluster due to CAB ALPs conversion into photons. Then we demonstrated the potential of the active galactic nuclei jet environment to probe low-mass ALP models and to potentially exclude the model proposed to explain the Coma cluster soft X-ray excess. We found that the overall X-ray emission for the M87 AGN requires an ALP-photon coupling $g_{a\gamma}$ in the range of $\sim 7.50 \times 10^{-15} \textup{--} 6.56 \times 10^{-14} \ \text{GeV}^{-1}$ for ALP masses $m_a \sim 1.1 \times 10^{-13} \ \text{eV}$ as long as the M87 jet is misaligned by less than about 20 degrees from the line of sight. These values are up to an order of magnitude smaller than the current best fit value on $g_{a\gamma} \sim 2 \times 10^{-13} \ \text{GeV}^{-1}$ obtained in soft X-ray excess CAB model for the Coma cluster. The results presented in this part cast doubt on the current limits of the largest allowed value of $g_{a\gamma}$ and suggest a new constraint that $g_{a\gamma} \lesssim 6.56 \times 10^{-14} \ \text{GeV}^{-1}$ when a CAB is assumed. This might bring into question whether the CAB explanation of the Coma X-ray excess is even viable.
 
In the third part of this thesis,  we turned our attention to consider a scenario in which ALPs may form Bose-Einstein condensate, and through their gravitational attraction and self-interactions, they can thermalize to spatially localized clumps. The coupling between ALPs and photons allows the spontaneous decay of ALPs into pairs of photons. For ALP condensates with very high occupation numbers, the stimulated decay of ALPs into photons is possible, and thus the photon occupation number can receive Bose enhancement and grows exponentially. We studied the evolution of the ALPs field due to their stimulated decays in the presence of an electromagnetic background, which exhibits an exponential increase in the photon occupation number with taking into account the role of the cosmic plasma in modifying the photon growth profile. In particular, we focus on quantifying the effect of the cosmic plasma on the stimulated decay of ALPs as this may have consequences on the detectability of the radio emissions produced from this process by the forthcoming radio telescopes such as the Square Kilometer Array telescopes with the intention of detecting the CDM ALPs. The results presented in this part argue that neither the current cosmic plasma nor the plasma in the galactic halos can prevent the stimulated decay of ALP with the $10^{-11}  \text{--} 10^{-4} \ \text{eV}$ mass range. Interestingly, the radio signal produced via the stimulated decay of ALPs in the range $10^{-6}  \text{--} 10^{-4} \ \text{eV}$ mass range is expected to be within the reach of the next-generation of the SKA radio telescopes. It is worth noting that this stimulated decay is only allowed at redshift window $z \lesssim 10^{4}$ and it is not efficient at higher redshift due to the effects of the plasma. The detection of an indication related to this technique might provide essential priors and predictions that are of paramount importance to understanding the properties of dark matter and offer an exciting scenario to explain several unexpected astrophysical observations.

Finally, in future work, it is worth extending the study of this PhD project to examine whether ultra-light ALPs can solve the core-cusp problem and the capability of the CAB to explain the EDGES 21 cm anomaly. Since there is still a lot of exciting theoretical aspects about the nature of dark matter from different perspectives, the expected impact of studying these topics would give rise to solve the mystery of dark matter. This, in addition, may have a future role in the discovery of new physics beyond the standard model of particle physics, and it possibly changes our main understanding of fundamental physics.

We can sum up that, with the work presented here, we point out that the research on axions and ALPs will be one of the leading frontiers in the near future since the discovery of these particles can solve some of the common unresolved problems between particle physics and cosmology and take us a step forward towards understanding nature. For completeness, some useful notations and conversion relations are broadly outlined in the appendix.
\markboth{Summary and Conclusion}{Summary and Conclusion}
\chapter*{\textbf{Appendix A}}
\addcontentsline{toc}{chapter}{\textbf{Appendix A}}
\begin{appendices}
\renewcommand{\thechapter}{\arabic{chapter}}
\renewcommand{\thesection}{A.\arabic{section}}
\renewcommand{\theequation}{A.\arabic{equation}}
\setcounter{equation}{0}

For convenience, we summarize in this appendix some useful units and conversion relations that are often used in this thesis. 
\section{Fundamental constants} \label{AppendixA.1}
The explicit numerical values of some of the universal constants in nature are given downward here$:$
\begin{itemize}
\item Electron rest mass $m_e= 9.109 \times 10^{-31} \ \text{kg} \:,$
\item Electron electric charge $e= 1.6022 \times 10^{-19} \ \text{C} \:,$
\item Speed of light in free space $c=2.9979 \times 10^8  \ \text{m} \ \text{s}^{-1} \:,$
\item Planck's constant $h= 6=6.626 \times 10^{-34} \ \text{kg} \ \text{m}^2 \text{s}^{-1} \:,$
\item Reduced Planck's constant $\hbar=h/2 \pi=1.0546 \times10^{-34} \ \text{kg} \ \text{m}^2 \text{s}^{-1}=6.6 \times 10^{25} \ \text{GeV} \ \text{s} \:,$
\item Boltzmann's constant $k_B =  8.6 \times 10^{14} \ \text{GeV} \ \text{K}^{-1} \:,$
\item Permittivity of free space $\epsilon_0= 8.854 \times 10^{-12} \ \text{F} \ \text{m}^{-1} \:,$
\item Fine structure constant $\alpha = e^2 /(4\pi \epsilon_0 \hbar c)= (137.04)^{-1} \:,$
\item Gravitational constant $G=6.673 \times 10^{-11} \ \text{N} \ \text{m}^2 \ \text{kg}^{-2} \:.$
\end{itemize}

\section{Units and conventions} \label{AppendixA.2}
If not stated otherwise, through out this thesis we use the natural units system in which $c=\hbar=k_B=1$. In this system there is only one fundamental dimension the energy in units of electron-volt $\text{eV}$. Note that the unit of mass, in which the relation $E=mc^2$ is used implicitly, therefore one has the important conversion $1 \ \text{eV} = 1.602 \times 10^{-19} \ \text{kg} \ \text{m}^2 \ \text{s}^{-2}$. This implies the following conversion rules from the international system of units (SI) to the natural system of units$:$
\begin{align}
1 \ \text{s} & = 1.52 \times 10^{24} \ \text{GeV}^{-1} \:,\\
1 \ \text{m} & = 5.10 \times 10^{15} \ \text{GeV}^{-1} \:,\\
1 \ \text{kg} & = 5.62\times 10^{26} \ \text{GeV} \:,\\
1 \ \text{K} & = 8.62 \times 10^{-14} \ \text{GeV} \:.
\end{align}
Spatial distances in astrophysics and cosmology are often measured in parsec, abbreviated with pc. In SI units, it is
\begin{equation}
1 \ \text{pc} = 3.1 \times 10^{16} \ \text{m} \:.
\end{equation}
In natural units, we can express a parsec via an inverse energy, and the following conversion rule holds
\begin{equation}
1 \ \text{Mpc} = 1.55 \times 10^{38} \ \text{GeV}^{-1} \:.
\end{equation}
A handy measure for masses of astrophysical objects is the mass of our sun $\mathrm{M}_{\odot}$. The solar mass is about
\begin{equation}
\mathrm{M}_{\odot} \ \text{Mpc} = 1.99 \times 10^{30} \ \text{kg} \:,
\end{equation}
or equivalently
\begin{equation}
\mathrm{M}_{\odot} \ \text{Mpc} = 1.11 \times 10^{57} \ \text{GeV} \:.
\end{equation}  
\end{appendices}
\chapter*{\textbf{Appendix B}}
\addcontentsline{toc}{chapter}{\textbf{Appendix B}}
\begin{appendices}
\renewcommand{\thechapter}{\arabic{chapter}}
\renewcommand{\thesection}{B.\arabic{section}}
\renewcommand{\theequation}{B.\arabic{equation}}
\setcounter{equation}{0}

In this appendix, we collect the basic notations and formulas that are often used in this thesis.
\section{Notations} \label{AppendixA.1}

A dot represents the time derivative, while a prime represents the spatial derivative, \ie $ \dot{f}= d F/d t$, and $F'= d F/d x$.

A double dots represents the time derivative of second-order, while a double primes represents the spatial derivative of second-order, \ie $ \ddot{F}= d^2 F/d t^2$, and $F''= d^2 F/d x^2$.

Partial derivative of $F$ with respect to $\mu$ is given as $\partial_{\mu} F \equiv \partial F^{\mu}$.

In an inertial system, we set
\begin{equation}
x^1 := x, \quad x^2 := y, \quad x^3 := z, \quad x^0 := ct \:,
\end{equation}
where $x, y, z$ are right-handed Cartesian coordinates, t is time, and c is the speed of light in a vacuum. Generally,
\begin{itemize}
\item Latin indices run from 1 to 3 (\eg, $i, j = 1, 2, 3$), and
\item Greek indices run from 0 to 3 (\eg, $\mu, \nu = 0, 1, 2, 3$).
\end{itemize}
In particular, we use the Kronecker symbols
\begin{equation}
\delta_{ij}= \delta^{ij} = \delta^i_j := \begin{cases} 1 &\mbox{if} i = j \:, \\ 0 & \mbox{if} i\neq j \:, \end{cases}
\end{equation}
and the Minkowski symbols
\begin{equation}
\eta_{\mu \nu}= \eta^{\mu \nu}  := \begin{cases} 1 &\mbox{if} \mu = \nu =0 \:, \\ -1 &\mbox{if} \mu = \nu =1,2,3 \:, \\ 0 & \mbox{if}\mu\neq \nu \:. \end{cases}
\end{equation}

In the context where the cosmic expansion is taken into account, we work in spatially flat Friedmann-Robertson-Walker (FRW) universe with a metric
\begin{equation}
ds^2 = g_{\mu \nu} dx^{\mu} dx^{\nu}= - dt^2 + R^2(t) [dx^2 + dy^2 +dz^2] \:,
\end{equation}
where R(t) is the scale factor of the universe. We denote the cosmic time as $t$ and the conformal time as $\tau$, where $d\tau= dt/R(t)$.

\section{Einstein's summation convention} \label{AppendixB.2}

In the Minkowski space-time, we always sum over equal upper and lower Greek (resp. Latin) indices from $0$ to $3$ (resp. from 1 to 3). For example, for the position vector, we have
\begin{equation}
\mathbold{X} = x^j \mathbold{e}_j \sum^3_{j=1} x^j \mathbold{e}_j \:,
\end{equation}
where $\mathbold{e}_1, \mathbold{e}_2, \mathbold{e}_3$ are orthonormal basis vectors of a right-handed orthonormal system. Moreover,
\begin{equation}
\eta_{\mu \nu} x^{\nu}= x^j \mathbold{e}_j \sum^3_{\nu=0} \eta_{\mu \nu} x^{\nu} \:.
\end{equation}
Greek indices are lowered and lifted with the help of the Minkowski symbols. That
is,
\begin{equation}
x_{\mu} := \eta_{\mu \nu} x^{\nu} \:, \quad x_j = - x^j \:, \quad j=1,2,3 \:.
\end{equation}

For the indices $\alpha, \beta, \gamma, \delta = 0, 1, 2, 3$, we introduce the antisymmetric symbol $\epsilon^{\alpha \beta \gamma \delta}$ which is normalized by
\begin{equation}
\epsilon^{0123} := 1 \:,
\end{equation}
and which changes sign if two indices are transposed. In particular,  $\epsilon^{\alpha \beta \gamma \delta}=0$  if two indices coincide. For example, $\epsilon^{0213}=-1$ and $\epsilon^{0113}=0$. Lowering of indices yields $\epsilon_{\alpha \beta \gamma \delta} := - \epsilon^{\alpha \beta \gamma \delta}$. For example, $\epsilon_{0123}=-1$.

\end{appendices}

\markboth{References}{References}
{\nocite{*}}
\bibliographystyle{unsrt}
\bibliography{Introductory/References}
\addcontentsline{toc}{chapter}{\textbf{References}}
\end{document}